\documentclass[a4paper,11pt]{report}

\usepackage{amsmath,amsfonts,amssymb}
\usepackage{amsthm}
\usepackage{graphicx}
\usepackage{a4wide}
\usepackage{hyperref}
\usepackage[bf,small,center]{titlesec}

\usepackage{Insert} % My package for boxed "Inserts"

\titleformat{\chapter}[block]
	{\bfseries\Huge\filcenter}
	{\thechapter\\}
	{8pt}
	{\bfseries\huge\filcenter}
\titleformat*{\subsection}{\itshape\filcenter}

\DeclareMathOperator{\tr}{Tr} % Trace
\DeclareMathOperator{\vol}{Vol} % Volume
\newcommand{\Ne}[1]{\ensuremath{\mathcal{N}={#1}}} % Supersymmetry: N = ?
\newcommand{\ls}{\ensuremath{{l_s}}} % String length
\newcommand{\gs}{\ensuremath{{g_s}}} % String coupling
\newcommand{\ket}[1]{\lvert #1 \rangle} % ket
\newcommand{\hyp}[2]{\ensuremath{(\mathbf{#1},\mathbf{#2})}} % Quiver representations
\newcommand{\ads}{{\ensuremath{AdS_5\times S^5}}} % AdS_5 x S^5
\newcommand{\gym}{{\ensuremath{g_{\text{YM}}^2}}} % g^2_{YM}
\newcommand{\tym}{{\ensuremath{\theta_{\text{YM}}}}} % \theta_{YM}
\newcommand{\un}{\ensuremath{\mathbf{1}}} % Unit matrix
\newcommand{\abs}[1]{{\ensuremath{\left\lvert #1 \right\rvert}}} % Absolute value
\newcommand{\hd}[1]{\ensuremath{\phantom{ }^{\star_{#1}}}} % Hodge dual in d dimensions
\newcommand{\hde}{\ensuremath{\phantom{ }^{\hat{\star}}}} % Hodge dual in 11 dimensions

\newcommand{\talpha}{{\tilde{\alpha}}}
\newcommand{\tpsi}{{\tilde{\psi}}}
\newcommand{\ttheta}{{\tilde{\theta}}}
\newcommand{\tphi}{{\tilde{\phi}}}

\newcommand{\cZ}[1]{{\ensuremath{\mathbb{C}^2/\mathbb{Z}_{#1}}}} % C^2 / Z_?
\newcommand{\cZZ}[1]{{\ensuremath{\mathbb{C}^3/\mathbb{Z}_{#1}\times\mathbb{Z}_{#1}}}} % C^3 / Z_? x Z_?
\newcommand{\cz}{\cZ{2}} % C^2 / Z_2
\newcommand{\czz}{\cZZ{2}} % C^3 / Z_2 x Z_2

\newcommand{\hodge}[6]{\begin{tabular}{ccccccc}
&&&#1&&&\\
&&#2&&#2&&\\
&#3&&#4&&#3&\\
#5&&#6&&#6&&#5\\
&#3&&#4&&#3&\\
&&#2&&#2&&\\
&&&#1&&&\\
\end{tabular}} % Hodge diamond: h_0,0 h_1,0 h_2,0 h_1,1 h_3,0 h_2,1
\newcommand{\hodgeCY}[2]{\hodge{1}{0}{0}{#1}{1}{#2}} % Hodge diamond for a CY: h_1,1 h_2,1
\newcommand{\ghodge}{\begin{tabular}{ccccccc}
&&&$h^{0,0}\hspace{-0.3cm}$&&&\\
&&$h^{1,0}\hspace{-0.3cm}$&&$h^{0,1}\hspace{-0.3cm}$&&\\
&$h^{2,0}\hspace{-0.3cm}$&&$h^{1,1}\hspace{-0.3cm}$&&$h^{0,2}\hspace{-0.3cm}$&\\
$h^{3,0}\hspace{-0.3cm}$&&$h^{2,1}\hspace{-0.3cm}$&
&$h^{1,2}\hspace{-0.3cm}$&&$h^{0,3}\hspace{-0.3cm}$\\
&$h^{3,1}\hspace{-0.3cm}$&&$h^{2,2}\hspace{-0.3cm}$&&$h^{1,3}\hspace{-0.3cm}$&\\
&&$h^{3,2}\hspace{-0.3cm}$&&$h^{2,3}\hspace{-0.3cm}$&&\\
&&&$h^{3,3}\hspace{-0.3cm}$&&&\\
\end{tabular}} % Most general Hodge diamond
\newcommand{\hodgesm}[4]
{\begin{tabular}{ccccc}
&&#1&&\\
&#2&&#2&\\
#3&&#4&&#3\\
&#2&&#2&\\
&&#1&&\\
\end{tabular}} % Hodge diamond in d=2: h_0,0 h_1,0 h_2,0 h_1,1
\newcommand{\hodgeK}{\hodgesm{1}{0}{1}{20}} % Hodge diamond for K3
\newcommand{\ghodgesm}
{\begin{tabular}{ccccc}
&&$h^{0,0}\hspace{-0.3cm}$&&\\
&$h^{1,0}\hspace{-0.3cm}$&&$h^{0,1}\hspace{-0.3cm}$&\\
$h^{2,0}\hspace{-0.3cm}$&&$h^{1,1}\hspace{-0.3cm}$&&$h^{0,2}\hspace{-0.3cm}$\\
&$h^{2,1}\hspace{-0.3cm}$&&$h^{1,2}\hspace{-0.3cm}$&\\
&&$h^{2,2}\hspace{-0.3cm}$&&\\
\end{tabular}} % Most general Hodge diamond in d=2

\begin{document}

% Title page and contents

\thispagestyle{empty}

\begin{flushright}
{\small
ITP-UU-03/67\\
SPIN-03/43\\
hep-th/0312070\\
}
\end{flushright}

\begin{center}

\vspace{1cm}

{\Huge The Gauge/String Correspondence}\\

\vspace{0.5cm}

{\Huge Towards Realistic Gauge Theories}
{\huge \protect\footnote{Based on the author's Ph.D. thesis ``Studies of the Gauge/String Theory Correspondence'', University of Torino, Italy, October 2003.}}

\vspace{1cm}

% Authors
{\Large Emiliano Imeroni}

\vspace{0.5cm}

% Institutions
{\large \emph{Institute for Theoretical Physics \& Spinoza Institute}}

\vspace{0.2cm}

{\large \emph{Utrecht University, Postbus 80.195, 3508 TD Utrecht, The Netherlands}}

\vspace{0.4cm}

{\large email: \verb|E.Imeroni@phys.uu.nl|}\\
\end{center}

\vspace{2cm}

\centerline{{\bf Abstract}}

\vspace{0.3cm}

This report presents some studies of the gauge/string theory correspondence, a deep relation that is possible to establish between quantum field theories with local gauge symmetry and superstring theories including gravity. In its original version, known as AdS/CFT duality, the correspondence involves \Ne{4} Super Yang--Mills theory in four space-time dimensions, which is a superconformal theory with a high degree of supersymmetry, thus very far from describing the physical world.

We explore extensions of the correspondence towards less supersymmetric and non-conformal gauge theories. Specifically, we study gauge theories in three and four dimensions, with eight or four preserved supersymmetries and exhibiting a scale anomaly, by means of supergravity solutions describing D-brane configurations of type II string theory. We show how relevant information on these gauge theories can be extracted from the dual classical solutions, both at the perturbative (e.g. running coupling constant, chiral anomaly) and non-perturbative level (e.g. effective superpotential).

\newpage

\tableofcontents

% "Introduction"

\chapter{Introduction}\label{c:introduction}

The physics of elementary particles, in the range of energies accessible to experiments, is described with very high precision by the Standard Model of particle physics. The Standard Model is a \emph{quantum gauge theory}, that is, a quantum field theory with local gauge symmetry.

This report deals with the study of quantum gauge theories in a somewhat  non-standard way, making use of different concepts and tools drawn from the fascinating arena of string theory, supersymmetry and gravity. This is what goes under the name of \emph{gauge/string theory} or \emph{gauge/gravity} correspondence.

It is a well known fact that string theory emerged as an attempt to explain strong interactions, and later evolved into a proposal for a theory able to incorporate all forces of nature into a coherent quantum framework. String theory has a lot of appealing properties for this role: just to give a brief list, it automatically includes gravity, does not present problems related to ultraviolet divergences, has no free adjustable parameters except for the scale, is a highly constrained and essentially unique theory, and includes many ideas which have been put forward in many attempts towards a unified description of nature, such as extra dimensions and supersymmetry. 

However, the interpretation of string theory as a putative ``theory of everything'' only makes the requirement of addressing gauge theories even stronger. In particular, if any sense for the description of the physical world should be given to string theory, it cannot avert from its ability to correctly describe the gauge theory interactions which are at the core of the Standard Model.

In fact, the possibility of using string theory as an efficient tool for computing gauge theory amplitudes has been known since the beginning of its history. A single diagram of perturbation theory of the open string reduces, in the field theory limit where the tension of the string is taken to infinity, to a sum of Feynman diagrams of a gauge theory, and this opens the way to many possible simplifications or alternative derivations of many gauge theory results.

During the last decade, the interplay between gauge theory and string theory has become even richer and fruitful. The discovery of \emph{Dirichlet branes}~\cite{Polchinski:1995mt}, or D-branes, that are new non-perturbative dynamical extended objects of diverse dimensions naturally present in type I and II string theories, has led to a new perspective on the way gauge theory is embedded in the superstring (besides of course making available a great tool to explore the landscape of connections between string theories via the exploitation of various perturbative and non-perturbative dualities).
%:Figure: Strings and branes
\begin{figure}
\begin{center}
\includegraphics[scale=.7]{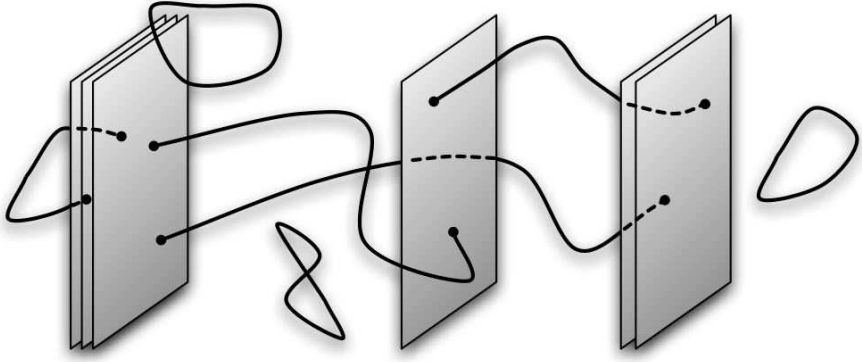}
\caption{{\small Strings and D-branes.}}
\label{f:dbranes}
\end{center}
\end{figure}

This new perspective on gauge theories from the string theory point of view has its central point in the \emph{double nature} of D-branes.
\begin{itemize}
\item On the one hand, D-branes admit a description in terms of perturbative \emph{open string theory}. They are defined as $p$-dimensional hyperplanes on which open string endpoints lie. The excitations of the open strings attached to a D-brane thus describe its dynamics. In particular, the massless states of the open string spectrum both represent the collective coordinates of a D-brane and give rise to a $(p+1)$-dimensional \emph{gauge theory} living on its world-volume.
\item On the other hand, D-branes can be regarded as sources of \emph{closed string} fields. They are non-perturbative \emph{classical solitons} of string theory, solutions of the low-energy supergravity equations of motion.
\end{itemize}
The deep consequences of the above complementarity of descriptions, which in some sense is a modern refinement of the older concept of \emph{open/closed string duality}, are the central theme of this report.

There is a particular example in which the twofold role of D-branes has proven extremely powerful for establishing a deep connection between gauge and string theory: it is the case of D3-branes in flat space, whose world-volume supports at low energies \Ne{4} superconformal Super Yang--Mills theory in four space-time dimensions.

A particular ``near-horizon'' limit of the geometry generated by the D3-branes has the surprising consequence of decoupling the bulk degrees of freedom from the ones living on the branes, while transforming the geometry into the product of five-dimensional Anti-de Sitter space times a five-sphere, where branes are replaced by Ramond-Ramond flux.

This is the basis of the Maldacena conjecture, or \emph{AdS/CFT duality}, which states that \emph{Type IIB string theory on \ads{} is dual to \Ne{4} Super Yang--Mills theory}~\cite{Maldacena:1998re,Gubser:1998bc,Witten:1998qj}. This is an exact duality, in the sense that the partition functions of the two theories are defined to be equivalent. This breakthrough achievement then relates two drastically different theories, a gauge theory and a gravitational theory, opening new perspectives on both of them, and is the deepest known example of gauge/gravity correspondence. The importance of this duality also resides in the fact that the strong coupling regime of the gauge theory has its dual closed string description in low-energy classical supergravity, thus allowing the opening of a computable window into non-perturbative gauge physics.

The AdS/CFT duality has been thoroughly tested in the recent years with great success. However, it is also worth pointing out one important limitation. The duality deals with \Ne{4} Super Yang--Mills theory, that is very far from providing a realistic description of the physical world. The main features which make \Ne{4} Super Yang--Mills theory ``unrealistic'' are:
\begin{itemize}
	\item it preserves a \emph{high amount of supersymmetry} (sixteen supercharges);
	\item it is a \emph{superconformal theory}.
\end{itemize}

Our interest here will be reporting on some attempts that have been pursued in order to generalize the correspondence beyond the above limitation. We will then study extensions of the gauge/string theory correspondence towards \emph{less supersymmetric and non-conformal gauge theories}. From what we said, it should be clear that this is a necessary step if we have in mind that ultimately the gauge/string correspondence should be able to give us relevant information on gauge theories which have something to do with the Standard Model which describes our physical world.

It is fair to say that, at the moment of writing, it has not yet proven possible to establish a full, exact duality between a gravitational and a gauge theory in the case in which the latter is non-conformal and preserves less than sixteen superysmmetries.

However, this does not mean that one cannot fruitfully use gravity to study these more ``realistic'' gauge theories. In fact, here we try giving a coherent, and reasonably self-contained, review of a particular approach to these extensions of the gauge/gravity correspondence. This approach is based on finding D-brane configurations in type II string theory that host on their world-volume some gauge theories of interest, and to study these theories by means of a dual gravitational description. There are some nice reviews of this and related approaches~\cite{Aharony:2002up,Bertolini:2003iv,Bigazzi:2003ui,DiVecchia:2003ne}, and of course more detailed references will be given in the following chapters.

How does one engineer configurations of D-branes supporting gauge theories with the requested properties? From what we said above, it should be clear that in order to pursue this program we have to act in two directions:
\begin{enumerate}
	\item \emph{Reduce supersymmetry}: the first step is to break part of the sixteen supercharges preserved by D-branes in flat space. This is achieved by appropriately choosing a \emph{closed string background} which breaks some amount of supersymmetry. For instance, one can consider type II strings on orbifolds or Calabi--Yau manifolds.
	\item \emph{Break conformal invariance}: in general, the theory living on a standard D-brane in such a background does not present any scale anomaly yet. In order to make it scale-anomalous, one has to act on the \emph{open string} side, by engineering configurations with particular properties, as for example D-brane having part of their world-volume wrapped on a non-trivial cycle of the ambient manifold.
\end{enumerate}
We will study in detail several D-brane setups in type II string theory, such as ``fractional'' D-branes on orbifolds and on the conifold and D-branes wrapped on supersymmetric cycles inside Calabi--Yau manifolds. The corresponding dual gauge theories are three- and four-dimensional scale-anomalous theories preserving eight or four supercharges. Among these, there are some gauge theories of more direct ``phenomenological'' interest, such as pure \Ne{1} and \Ne{2} Super Yang--Mills theory in four space-time dimensions.

Once a D-brane configuration yielding a gauge theory of interest has been constructed, the next step is finding the low-energy closed string description, in terms of a classical supergravity solution. We will see that the careful analysis of the solution yields interesting pieces of information on the dual gauge theory.  For example, depending on the context and the amount of unbroken supersymmetry, one can reproduce from classical supergravity relevant quantum gauge theory information such as:
\begin{itemize}
	\item \emph{at the perturbative level}: running gauge coupling constant, chiral anomaly, metric on the moduli space;
	\item \emph{at the non-perturbative level}: instanton action, gaugino condensation and chiral symmetry breaking, effective superpotential, tension of confining strings.
\end{itemize}
It is clear from the above list that the study of the generalizations of the gauge/string theory correspondence towards realistic gauge theories is an extremely rich subject, whose further study might very well lead to a deeper and better understanding of quantum gauge theories.

\subsection*{Organization of the paper}

This report is organized as follows. In chapter~\ref{c:dbranes}, after a brief summary of concepts and notations of perturbative string theory, we set the stage by introducing D-branes from different points of view. We introduce the perturbative notion of D-brane in open string theory, as a hypersurface where open strings end, and its interpretation as a solitonic solution of the low-energy closed string theory. This twofold nature of D-branes will be the central concept in all subsequent parts, and will be used to see how information on a gauge theory can be extracted from the classical geometry describing D-branes. This will be our first meeting with the gauge/gravity correspondence. Finally, concentrating on the case of D3-branes, we will consider the main example of the correspondence, by briefly introducing the duality between \Ne{4} Super Yang--Mills theory in four space-time dimensions and type IIB string theory compactified on  \ads.

Chapter~\ref{c:engineering} is devoted to the construction of brane configurations which allow obtaining on the world-volume of D-branes at low energies gauge theories with less than 16 supersymmetries and a running gauge coupling, namely, in the case of four-dimensional theories, broken conformal symmetry. The goal of these constructions is to get information on more ``realistic'' gauge theories. We will introduce some concepts about D-branes on non-trivial spaces, such as Calabi--Yau manifolds and orbifolds. In particular, we will extensively study, by means of simple examples, fractional D-branes on orbifolds and on the conifold, as well as D-branes wrapped on supersymmetric cycles. We will also establish some connections among these approaches, and between them and other approaches, such as stretched branes and ``geometric engineering''.

In chapter~\ref{c:8susy} we will present our first explicit constructions of gauge/gravity correspondence away from AdS/CFT. Two simple examples will be used to explain how one can obtain classical solutions corresponding to fractional D-branes on orbifolds and D-branes wrapped on a supersymmetric cycle inside a Calabi--Yau manifold, and extract from them information about the gauge theory living on the world-volume of the branes. The examples comprise a system of D4-branes wrapped on a two-cycle inside a Calabi--Yau twofold and a system of fractional D2/D6-branes on a \cz{} orbifold. In both cases, the dual gauge theory is scale-anomalous Super Yang--Mills theory in three space-time dimensions with 8 supercharges. After summarizing and generalizing the techniques derived with the aid of the examples, we apply them to two systems dual to \Ne{2} Super Yang--Mills theory in four space-time dimensions, namely a system of wrapped D5-branes and one of fractional D3/D7 branes.

Finally, chapter~\ref{c:4susy} reports on some attempts of constructing supergravity duals of \Ne{1} Super Yang--Mills theory in four dimensions, a theory with many qualitative similarities to non-supersymmetric QCD. In particular, we will see how relevant information can be extracted for the pure \Ne{1} theory from two classical solutions, one describing D5-branes wrapped on a two-cycle inside a Calabi--Yau threefold and one corresponding to fractional D-branes on the conifold. Finally, we will use another solution, corresponding to fractional D3-branes on a \czz{} orbifold, in order to study \Ne{1} SQCD with matter in the fundamental representation of the gauge group.

A final observation is that we will often use boxed ``inserts'', either in order to present material which is useful for the understanding, but not directly related to the logical development of the text, or to briefly sketch some topics when space does not allow more detailed explanations. As a first example, in insert~\ref{i:notations} on page~\pageref{i:notations} we summarize some of the notations and conventions we will be using throughout the paper.

%:Insert: Notations
\begin{Insert}{Notations}\label{i:notations}
We use the $(-\ +\ +\ \cdots\ +)$ signature of the metric, and set $\hbar=c=1$. We denote the string length as $\ls$ (the relation with the Regge slope is given by $\ls = \sqrt{\alpha'}$), and the closed string coupling as $\gs$.

When we write down space-time or world-volume actions, it is generally understood that the fields appearing in the action are \emph{fluctuations} around the vacuum expectation values of the fields. For instance, dilaton terms such as $e^{\Phi}$ do not contain the expectation value $\gs$.

We try to stick with the notation $X^\mu$ for coordinate fields in a world-sheet theory, and $x^\mu$ when referring generically to the coordinates of a space-time. However, especially when dealing with world-volume actions, some (usually non-confusing) mismatch of notation is unavoidable. In world-volume actions, we denote pull-backs of space-time fields with a hat.

We define the totally antisymmetric $\varepsilon$ symbols in $D$ dimensions as $\varepsilon^{012\dotsm (D-1)}=-\varepsilon_{012\dotsm (D-1)}=+1$. A differential $p$-form is defined as $\omega_p=\frac{1}{p!}\omega_{\mu_1 \dotsm\mu_p} dx^{\mu_1}\wedge\dotsm\wedge dx^{\mu_p}$, and its Hodge dual in $D$ dimensions is:
\begin{equation*}
	\hd{D}\omega_p=\frac{\sqrt{-\det G_D}}{p!(D-p)!} \varepsilon_{\nu_1\dotsm \nu_{D-p}\mu_1 		\dotsm\mu_p} \omega^{\mu_1\dotsm\mu_p}
		dx^{\nu_1}\wedge\dotsm\wedge dx^{\nu_{D-p}}\,.
\end{equation*}		
Moreover, $\hd{}$ denotes $\hd{10}$ and $\hde$ denotes $\hd{11}$. We will usually indicate the rank of a form with a subscript, as above, but we will often use just $B$ to denote the Neveu--Schwarz-Neveu--Schwarz two-form of string theory.

Finally, in order to describe configurations of D-branes, we often use diagrams as in the following example:
\begin{center}
\begin{tabular}{|c|c|c|c|c|c|c|c|c|c|c|}
\hline
&0&1&2&3&4&5&6&7&8&9\\
\hline
D$4$ &$-$&$-$&$-$&$\bigcirc$&$\bigcirc$&$\cdot$&$\cdot$&$\cdot$&$\cdot$&$\cdot$\\
\hline
\end{tabular}
\begin{tabular}{cl}
$-$&denotes a direction \emph{longitudinal} to the world-volume\\
$\bigcirc$&denotes a \emph{compact} direction longitudinal to the world-volume\\
$\cdot$&denotes a direction \emph{transverse} to the world-volume\\
\end{tabular}
\end{center}
\end{Insert}
% "D-branes and the Gauge/Gravity Correspondence"

\chapter{D-branes and the Gauge/Gravity Correspondence}\label{c:dbranes}

This chapter presents a survey of some ingredients about strings and Dirichlet branes that we will need. A complete treatment would require a lot of space, and ours would anyway be inferior to others presented in many standard references. Therefore, we will be a bit sketchy and limit ourselves to the presentation of some basic notions, emphasizing just the tools and concepts that will be essential for the analysis of later chapters.

\section{Perturbative strings}\label{s:strings}

We start by summarizing some facts about perturbative superstring theory. Our goal is just to set up some notation - for a much more complete treatment we refer for instance to the standard references~\cite{Polchinski:1998,Green:1987}. Our presentation will follow quite closely the one given in~\cite{DiVecchia:1999rh,DiVecchia:2003ne}.

The two-dimensional world-sheet action describing the space-time propagation of a supersymmetric string in the superconformal gauge is given by:
\begin{equation}\label{polyact}
	S = -\frac{T}{2}\int_{\mathcal{M}}  d\tau d\sigma \left( \eta^{\alpha\beta}
		\partial_\alpha X^\mu \partial_\beta X_\mu
		- i \bar{\psi}^\mu \rho^\alpha \partial_\alpha \psi_\mu \right)\,.
\end{equation}
Some observations on this action are given below:
\begin{itemize}
\item $T$ is the tension of the string, which is related to the string length via $T = \tfrac{1}{2\pi\ls^2}$;
\item $\mathcal{M}$ is the world-sheet of the string, parameterized by the coordinates $\xi^\alpha=(\tau,\sigma)$, where $\sigma$ has range $\sigma\in[0,s]$, and the flat metric $\eta^{\alpha\beta}$ has signature $(-,+)$;
\item $X^\mu$, $\mu=0,\ldots,9$ are ten world-sheet scalar fields corresponding to the coordinates of the target space-time;
\item $\psi^\mu$, $\mu=0,\ldots,9$ are world-sheet Majorana spinors, and the matrices $\rho^\alpha$ provide a representation of the Clifford algebra $\{\rho^\alpha,\rho^\beta\}=-2\eta^{\alpha\beta}$.
\item The action is invariant under the world-sheet supersymmetry transformations:
\begin{equation}\label{wssusy}
	\delta X^\mu = \bar{\epsilon}\psi^\mu\,,\qquad
	\delta \psi^\mu = -i \rho^\alpha \partial_\alpha X^\mu \epsilon\,,\qquad
	\delta \bar{\psi}^\mu = i \bar{\epsilon} \rho^\alpha \partial_\alpha X^\mu\,,
\end{equation}
where the supersymmetry parameter $\epsilon$ is a Majorana spinor satisfying $\rho^\beta\rho^\alpha \partial_\beta \epsilon = 0$. 
\end{itemize}

Let us start from the world-sheet bosons. The equations of motion and boundary conditions derived from~\eqref{polyact} read:
\begin{align}
	\partial^\alpha \partial_\alpha X^\mu &= 0\,,\label{steom}\\
	\partial_\sigma X \cdot \delta X \big\rvert_{\sigma=s}
		- \partial_\sigma X \cdot \delta X \big\rvert_{\sigma=0} &= 0\,.\label{stbc}
\end{align}
There are three different possibilities for satisfying~\eqref{stbc}:
\begin{itemize}
\item \emph{Neumann open string} (N) boundary conditions. We choose $s=\pi$ and impose:
\begin{equation}
	\partial_\sigma X^\mu \big\rvert_{\sigma=0,\pi} = 0\,.
\end{equation}
The most general solution of the equations of motion~\eqref{steom} when Neumann boundary conditions are imposed at both endpoints of the string is:
\begin{equation}
	X^\mu_{\text{NN}} (\tau,\sigma) = x^\mu + 2\ \ls^2 p^\mu \tau
		+ i \sqrt{2}\ \ls \sum_{n\neq 0} \left( \frac{\alpha_n^\mu}{n}
		e^{-in\tau} \cos n\sigma \right)\,.
\end{equation}
\item \emph{Dirichlet open string} (D) boundary conditions. We choose again $s=\pi$ and impose:
\begin{equation}
	\delta X^\mu \big\rvert_{\sigma=0,\pi} = 0\,.
\end{equation}
The most general solution of the equations of motion with Dirichlet boundary conditions at both endpoints is:
\begin{equation}
	X^\mu_{\text{DD}} (\tau,\sigma) = \frac{c^\mu (\pi-\sigma) + d^\mu \sigma}{\pi}
		- \sqrt{2}\ \ls \sum_{n\neq 0} \left( \frac{\alpha_n^\mu}{n}
		e^{-in\tau} \sin n\sigma \right)\,.
\end{equation}
\item \emph{Closed string} boundary conditions. We choose $s=2\pi$ and impose:
\begin{equation}
	X^\mu (\tau,0) = X^\mu(\tau,2\pi)\,.
\end{equation}
The most general solution of the equations of motion~\eqref{steom} in this case is:
\begin{equation}\label{Xcl}
	X^\mu_{\text{closed}} (\tau,\sigma) = x^\mu + \ls^2 p^\mu \tau
		+ i \frac{\ls}{\sqrt{2}} \sum_{n\neq 0} \frac{1}{n}
		\left( \alpha_n^\mu e^{-in(\tau-\sigma)} 
		+ \talpha_n^\mu e^{-in(\tau+\sigma)}\right)\,.
\end{equation}
\end{itemize}
Of course, in the case of the open string, one can satisfy~\eqref{stbc} in a variety of ways. First, one can choose different boundary conditions for different $\mu$. Moreover, one can also impose different conditions at the two endpoints of the open string, namely Neumann at one endpoint and Dirichlet at the other, and in this case one talks about mixed or ND boundary conditions.

Notice that the Dirichlet boundary conditions are the only ones which do not preserve space-time Poincar\'e invariance. In fact, as we will extensively discuss in the following, they describe an extended physical object in space-time.

Passing to the fermionic degrees of freedom, it is useful to introduce light-cone coordinates $\xi_{\pm} = \tau \pm \sigma$, in terms of which the equations of motion and boundary conditions read:
\begin{align}
	\partial_+ \psi_-^\mu = \partial_- \psi_+^\mu &= 0\,,\label{psieom}\\
	\left( \psi_+ \delta \psi_+ - \psi_- \delta \psi_-\right)\big\rvert_{\sigma=0,s} &= 0\,,
\end{align}
where $\psi_\pm = \tfrac{1 \mp \rho^0 \rho^1}{2} \psi^\mu$. In the case of the open string, we can satisfy the boundary conditions by imposing:
\begin{equation}
	\psi_- (\tau,0) = \eta_1 \psi_+ (\tau,0)\,,\qquad
	\psi_- (\tau,\pi) = \eta_2 \psi_+ (\tau,\pi)\,,
\end{equation}
where $\eta_{1,2}=\pm 1$. We therefore obtain two sectors of the open string spectrum:
\begin{itemize}
\item $\eta_1 = \eta_2$: \emph{Ramond} (R) sector;
\item $\eta_1 = -\eta_2$: \emph{Neveu--Schwarz} (NS) sector.
\end{itemize}
The general solution of~\eqref{psieom} satisfying these boundary conditions is:
\begin{equation}
	\psi_{\pm}^\mu \sim \sum_r \psi_r^\mu e^{-ir(\tau\pm\sigma)}\,,\qquad
	\text{where}\qquad \begin{cases}
		r\in\mathbb{Z}\quad \text{R sector}\\
		r\in\mathbb{Z}+\frac{1}{2}\quad \text{NS sector}
	\end{cases}
\end{equation}

In the case of the closed string, the fields $\psi_\pm$ are independent and can be either periodic or anti-periodic:
\begin{equation}
	\psi_- (\tau,0) = \eta_3 \psi_- (\tau,2\pi)\,,\qquad
	\psi_+ (\tau,0) = \eta_4 \psi_+ (\tau,2\pi)\,,
\end{equation}
so that we have in total four different sectors:
\begin{itemize}
\item $\eta_3 = \eta_4 = 1$: \emph{Ramond-Ramond} (R-R) sector;
\item $\eta_3 = \eta_4 = -1$: \emph{Neveu--Schwarz-Neveu--Schwarz} (NS-NS) sector;
\item $\eta_3 = -\eta_4 = 1$: \emph{Ramond-Neveu--Schwarz} (R-NS) sector;
\item $\eta_3 = -\eta_4 = -1$: \emph{Neveu--Schwarz-Ramond} (NS-R) sector.
\end{itemize}
The general solution now reads:
\begin{subequations}
\begin{align}
	\psi_{-}^\mu &\sim \sum_r \psi_r^\mu e^{-ir(\tau-\sigma)}\,,\quad
	\text{where}\quad \begin{cases}
		r\in\mathbb{Z}\quad \text{R sector (left-moving)}\\
		r\in\mathbb{Z}+\frac{1}{2}\quad \text{NS sector (left-moving)}
	\end{cases}\\
	\psi_{+}^\mu &\sim \sum_r \tpsi_r^\mu e^{-ir(\tau+\sigma)}\,,\quad
	\text{where}\quad \begin{cases}
		r\in\mathbb{Z}\quad \text{R sector (right-moving)}\\
		r\in\mathbb{Z}+\frac{1}{2}\quad \text{NS sector (right-moving)}
	\end{cases}
\end{align}
\end{subequations}

Let us now pass to the quantum theory. The oscillators $\alpha_n^\mu$, $\talpha_n^\mu$, $\psi_r^\mu$ and $\tpsi_r^\mu$ act as annihilation and creation operators on the Fock space. In fact, the canonical (anti)commutation relations between $X^\mu$ and $\psi^\mu$ and their conjugate momenta imply the following relations for the oscillators:
\begin{equation}
	[x^\mu, p^\nu] = i \eta^{\mu\nu}\,,\quad
	[\alpha_m^\mu,\alpha_n^\nu]
		= m \delta_{m,-n} \eta^{\mu\nu}\,,\quad
	\{\psi_r^\mu,\psi_s^\nu\}
		= \delta_{r,-s} \eta^{\mu\nu}\,,
\end{equation}
with analogous relations for the tilded oscillators, and all remaining (anti)commutators vanishing.

Let us start from the open string, whose vacuum $\ket{0,k}$ with momentum $k$ is defined as:
\begin{equation}
	p^\mu \ket{0,k} = k^\mu \ket{0,k}\,,\qquad\qquad
	\alpha_n^\mu \ket{0,k} = \psi_r^\mu \ket{0,k} = 0\,,\qquad \forall n,r > 0\,.
\end{equation}
Notice that while the NS sector has a unique ground state, in the R sector there are zero-modes satisfying the Dirac algebra $\{ \psi_0^\mu,\psi_0^\nu\}=\eta^{\mu\nu}$ that can thus be represented as $32\times32$ ten-dimensional gamma matrices. This implies that the R ground state transforms as a 32-dimensional Dirac spinor. To describe it in more detail, it is convenient to choose the new oscillator basis:
\begin{equation}
	d_i^\pm = \begin{cases}
		\frac{1}{\sqrt{2}} (\psi_0^1 \mp \psi_0^0)\qquad &i=0\\
		\frac{1}{\sqrt{2}} (\psi_0^{2i} \pm \psi_0^{2i+1})\qquad &i=1,2,3,4
		\end{cases}
\end{equation}
in terms of which the algebra becomes $\{ d_i^+ , d_j^- \}= \delta_{ij}$. The $d_i^\pm$ act as raising and lowering operators on the 32 Ramond ground states, denoted as:
\begin{equation}\label{Rgs}
	\ket{\mathbf{s}} = \ket{s_0,s_1,s_2,s_3,s_4}\,,
\end{equation}
where $s_i=\pm\tfrac{1}{2}$, a single $d_i^+$ raises $s_i$ from $-\tfrac{1}{2}$ to $\tfrac{1}{2}$ and:
\begin{equation}
	d_i^- \ket{-\tfrac{1}{2},-\tfrac{1}{2},-\tfrac{1}{2},-\tfrac{1}{2},-\tfrac{1}{2}} = 0\,.
\end{equation}

Because of the Lorentz signature of the metric, the Fock space we have just defined contains states with negative norm, which are physically unacceptable. To select the physical states, one introduces the energy-momentum tensor and the supercurrent and looks at their action on the states. In the light-cone coordinates, the energy-momentum tensor has the following non-vanishing components:
\begin{equation}
	T_{++} = \partial_+ X \cdot \partial_+ X + \frac{i}{2} \psi_+ \cdot \partial_+ \psi_+\,,\qquad
	T_{--} = \partial_- X \cdot \partial_- X + \frac{i}{2} \psi_- \cdot \partial_- \psi_-\,,
\end{equation}
while the supercurrent. namely the N\"other current associated to the world-sheet supersymmetry transformation~\eqref{wssusy}, is:
\begin{equation}
	J_+ = \psi_+ \cdot \partial_+ X\,,\qquad
	J_- = \psi_- \cdot \partial_- X\,.
\end{equation}
Starting with the mode expansions of $X^\mu$ and $\psi^\mu$ that we wrote above, one derives the expansions of $T$ and $G$. The Fourier components of the energy-momentum tensor are the Virasoro generators:
\begin{equation}
	L_n = \frac{1}{2} \sum_m \alpha_{-m} \cdot \alpha_{n+m}
		+ \frac{1}{2} \sum_r \left(\frac{n}{2}+r\right) \psi_{-r} \cdot \psi_{r+n}\,,
\end{equation}
where the normal ordering of the operators is understood, $m,n\in\mathbb{Z}$ and $r\in\mathbb{Z}$ in the R sector and $r\in\mathbb{Z}+\tfrac{1}{2}$ in the NS sector. In the case of $L_0$ after dealing with ordering problems one obtains:
\begin{equation}
	L_0 = \ls^2 p^2 + \sum_{n=1}^\infty \alpha_{-n} \cdot \alpha_n
		+ \sum_{r>0} r\ \psi_{-r} \cdot \psi_r\,.
\end{equation}
The Fourier components of the supercurrent are instead given by:
\begin{equation}
	G_r = \sum_{n\in\mathbb{Z}} \alpha_{-n} \cdot \psi_{r+n}\,.
\end{equation}
The above modes satisfy the following super-Virasoro algebra:%
\footnote{The algebra~\eqref{virasoro} holds in the NS sector, and also in the R sector with the redefinition $L_0\to L_0 + \tfrac{5}{8}$.}
\begin{equation}\label{virasoro}
\begin{split}
	[ L_m, L_n ] &= (m-n) L_{m+n} + \tfrac{5}{4}(m^3-m)\delta_{m,-n}\,,\\
	\{ G_r, G_s \} &= 2L_{r+s} + \tfrac{5}{4} (4r^2 - 1)\delta_{r,-s}\,,\\
	[ L_m, G_r ] &= \tfrac{1}{2} (m-2r) G_{m+r}\,.
\end{split}
\end{equation}
The physical states of the theory are the ones satisfying the following conditions:
\begin{equation}
	(L_0 - a) \ket{\psi} = 0\,,\qquad\qquad
	L_n \ket{\psi} = G_r \ket{\psi} = 0\,,\qquad \forall n>0, r\ge 0\,,
\end{equation}
where $a=\tfrac{1}{2}$ in the NS sector and $a=0$ in the R sector.

The mass spectrum of the theory that one obtains by deriving the Hamiltonian and expanding it in terms of oscillators is:
\begin{equation}
	M^2 = \frac{1}{\ls^2} \Big[ \sum_{n,r>0} \left( \alpha_{-n} \cdot \alpha_n
		+ r\ \psi_{-r} \cdot \psi_r \right) - a \Big]\,.
\end{equation}
Notice that the first state in the spectrum, namely the NS ground state, is tachyonic with $M^2 = - \tfrac{1}{2\ls^2}$.

Let us now consider the closed string. Everything is parallel to the case of the open string we just considered, except the fact that there are two sets of oscillators. One therefore needs two copies of the Virasoro algebra, two sets of physical state conditions, and so on. In particular, notice that the expression of $L_0$ differs from the open string case and one has:
\begin{equation}
	L_0 = \frac{\ls^2}{4} p^2 + \sum_{n=1}^\infty \alpha_{-n} \cdot \alpha_n
		+ \sum_{r>0} r\ \psi_{-r} \cdot \psi_r\,,\quad
	\tilde{L}_0 = \frac{\ls^2}{4} p^2 + \sum_{n=1}^\infty \talpha_{-n} \cdot \talpha_n
		+ \sum_{r>0} r\ \tpsi_{-r} \cdot \tpsi_r\,.
\end{equation}
The mass spectrum is given by:
\begin{equation}
	M^2 = \frac{2}{\ls^2} \Big[ \sum_{n,r>0} \left( \alpha_{-n} \cdot \alpha_n + \talpha_{-n} \cdot \talpha_n
		+ r\ \psi_{-r} \cdot \psi_r + r\ \tpsi_{-r} \cdot \tpsi_r \right) - a - \tilde{a} \Big]\,,
\end{equation}
and again we see that the lowest lying state is a tachyon with mass $M^2=-\tfrac{2}{\ls^2}$. In addition one has to impose the level-matching condition on the physical states:
\begin{equation}
	\left( L_0 - \tilde{L}_0 - a + \tilde{a} \right) \ket{\psi} = 0\,.
\end{equation}

The spectrum described above, both for the open and the closed string, contains a tachyonic state and is not space-time supersymmetric. In order to achieve a supersymmetric spectrum and remove the tachyon one has to implement a particular truncation of the theory, by performing the so-called GSO projection.

The GSO projection consists of retaining in the theory only the states which have even world-sheet fermion number $F$. For example, in the NS sector of an open string theory we can define the following projector:
\begin{equation}\label{GSONS}
	P_{\text{NS}} = \frac{1+ (-1)^{F_{\text{NS}}}}{2}\,,\qquad
		F_{\text{NS}} = \sum_{r=1/2}^{\infty} \psi_{-r} \cdot \psi_r - 1\,.
\end{equation}
Using the projector~\eqref{GSONS}, we see that the tachyonic ground state is removed from the spectrum of the string. In the R sector, the projection acts similarly on the non-zero modes, and acts on the zero-modes as a chirality projection, retaining the states~\eqref{Rgs} with:
\begin{equation}
	\sum_{i=0}^4 s_i = \text{even (odd).}
\end{equation}
Let us then examine the first level of the open string spectrum which is retained by the projection. From the NS sector, we get the following massless state:
\begin{equation}
	\epsilon_\mu(k) \psi^\mu_{-1/2}\ket{0,k}\,,\qquad
	k^2 = 0\,,\qquad k \cdot \epsilon = 0\,,
\end{equation}
which is a gauge vector field with transverse polarization, that has 8 on-shell degrees of freedom. The massless state coming from the R sector is instead:
\begin{equation}
	u_{\mathbf{s}} (k) \ket{\mathbf{s}}\,,\qquad
	k^2 = 0\,,\qquad k\cdot \Gamma_{\mathbf{s'}\mathbf{s}} u_{\mathbf{s}} = 0\,.
\end{equation}
Thanks to the GSO projection, this state becomes a Majorana--Weyl spinor, which has 8 on-shell degrees of freedom in ten dimensions. We have therefore shown that a necessary condition for supersymmetry, namely that on-shell bosonic and fermionic degrees of freedom match at each mass level, is realized at the massless level. In fact, one can show that the full spectrum of the string is supersymmetric, but we will not discuss this issue here.

\subsection*{Type II theories}

Let us instead pass to the closed superstring theories. Here, we have four different sectors, and we need to perform the GSO projection in each. In particular, we have the choice of performing the same or the opposite projection in the left-moving and right-moving R sectors. Depending on this choice, we obtain two inequivalent theories, containing the following sectors, where the signs denote the parity under $(-1)^F$:
\begin{center}
\begin{tabular}{lcccc}
Type IIB (chiral): & (NS+,NS+) & (R+, NS+) & (NS+,R+) & (R+,R+) \\
Type IIA (non-chiral): & (NS+,NS+) & (R+, NS+) & (NS+,R$-$) & (R+,R$-$) \\
\end{tabular}
\end{center}

We then analyze the massless part of the spectrum of the type II theories. The NS-NS sector has identical content in the two theories, and a generic massless state is:
\begin{equation}\label{NSmassless}
	\psi_{-1/2}^\mu \tpsi_{-1/2}^\nu \ket{0,k}\,,\qquad k^2 = 0\,.
\end{equation}
This state can be saturated with a traceless symmetric tensor:
\begin{equation}
	\epsilon_{\mu\nu}^{(h)} = \epsilon_{\nu\mu}^{(h)}\,,\qquad
	\epsilon_{\mu\nu}^{(h)}\eta^{\mu\nu}=0\,,\qquad
	k^\mu\epsilon_{\mu\nu}^{(h)} = 0\,,
\end{equation}
thus giving a physical state that can be identified with a graviton $h_{\mu\nu}$ in space-time. Alternatively, we can saturate~\eqref{NSmassless} with an antisymmetric tensor:
\begin{equation}
	\epsilon_{\mu\nu}^{(B)} = -\epsilon_{\nu\mu}^{(B)}\,,\qquad
	k^\mu\epsilon_{\mu\nu}^{(B)} = 0\,,
\end{equation}
that gives rise to a two-form gauge potential $B_2$. Finally, we can obtain a scalar field, the dilaton $\Phi$, by saturating~\eqref{NSmassless} with:
\begin{equation}
	\epsilon_{\mu\nu}^{(\Phi)} = \frac{1}{\sqrt{8}} (\eta_{\mu\nu} - k_\mu \ell_\nu - k_\nu\ell_\mu)\,,\qquad
	\ell^2 = k^2 = 0\,,\quad \ell\cdot k = 1\,.
\end{equation}
In the NS-R sector, we have the massless vector-spinor state:
\begin{equation}
	u_{\mu \mathbf{s}} \psi_{-1/2}^\mu \ket{0,\mathbf{\tilde{s}},k}\,,\qquad
	k^2 = k^\mu \tilde{u}_{\mu\mathbf{\tilde{s}}} 
		= k \cdot \Gamma_{\mathbf{\tilde{s}'}\mathbf{\tilde{s}}}
		\tilde{u}_{\mu \mathbf{\tilde{s}}}=0\,,
\end{equation}
that is reducible under the action of the Lorentz group. The decomposition gives a spin $\tfrac{3}{2}$ gravitino and a spin $\tfrac{1}{2}$ dilatino with opposite chiralities. Exactly the same analysis apply to the R-NS sector, and the theory has therefore two gravitini, hence the name ``type II''. Finally we pass to the R-R sector, which contains again bosonic states. The generic massless state reads:
\begin{equation}\label{Rmassless}
	u_{\mathbf{s}}\tilde{u}_{\mathbf{\tilde{s}}} \ket{\mathbf{s},\mathbf{\tilde{s}}}\,,\qquad
	k \cdot \Gamma_{\mathbf{s}'\mathbf{s}}
		u_{\mathbf{s}}
		= k \cdot \Gamma_{\mathbf{\tilde{s}'}\mathbf{\tilde{s}}}
		\tilde{u}_{\mathbf{\tilde{s}}}=0\,.
\end{equation}
In order to analyze the properties of the above state, it is easier to turn for a while to a conformal field theory formulation. In fact, string theory in the superconformal gauge is a two-dimensional superconformal field theory. We will not treat in detail this approach here (a full treatment can be found for example in~\cite{Polchinski:1998}), but just mention that in conformal field theory each state $\ket{\psi}$ of the theory is dual to a specific operator, called a \emph{vertex operator}, which has the property that $\ket{\psi} = \lim_{z,\bar{z}\to 0} \mathcal{V}^{(\psi)} (z,\bar{z})\ket{0}$, where $z = e^{i(\tau-\sigma)}$ and $\bar{z}=e^{i(\tau+\sigma)}$ are complex coordinates on the world-sheet. In our case, the R-R vacuum is obtained from the NS-NS vacuum via the action of the \emph{spin fields}:
\begin{equation}
	\ket{\mathbf{s},\mathbf{\tilde{s}}} = \lim_{z,\bar{z}\to 0} S^{\mathbf{s}} (z) \tilde{S}^{\mathbf{\tilde{s}}} (\bar{z}) \ket{0}\,.
\end{equation}
Using this together with~\eqref{Rmassless}, we can write the following decomposition:
\begin{multline}\label{Rdec}
	\lim_{z,\bar{z}\to 0} u_{\mathbf{s}}\tilde{u}_{\mathbf{\tilde{s}}} S^{\mathbf{s}} (z)
		\tilde{S}^{\mathbf{\tilde{s}}} (\bar{z}) \ket{0}\\
		= \frac{1}{32} \sum_{n=0}^{10} \frac{(-1)^{n+1}}{n!}
		u_{\mathbf{s}} (\Gamma_{\mu_1\dotsm \mu_n} C^{-1}) \tilde{u}_{\mathbf{\tilde{s}}}
		\lim_{z,\bar{z}\to 0} S^{\mathbf{s}} (z) 
		(C \Gamma^{\mu_1\dotsm \mu_n}) \tilde{S}^{\mathbf{\tilde{s}}} (\bar{z}) \ket{0}\,,
\end{multline}
where $C$ denotes the charge conjugation matrix. Now the analysis differs between the type IIB and type IIA theory, in which the two spinors in~\eqref{Rdec} have respectively the same or opposite chirality. In fact, it can be shown that in the case of type IIB theory, only terms with $n$ odd contribute to the sum in~\eqref{Rdec}, while the situation is reversed in the case of the type IIA theory, where only terms with even $n$ contribute. This means that the following fields:
\begin{equation}
	F_{\mu_1\dotsm\mu_n} = u \Gamma_{\mu_1\dotsm\mu_n} C^{-1} \tilde{u}\,,
\end{equation}
exist only if $n$ is odd in type IIB or even in type IIA theory. From the above expression and~\eqref{Rmassless}, one also finds that the $F$ fields satisfy the equations of motion and Bianchi identity appropriate for an $n$-form field-strength (instead of gauge potential):
\begin{equation}
	dF_n = d (\hd{} F)_{10-n} = 0
\end{equation}
(see insert~\ref{i:notations} on page~\pageref{i:notations} for some of our conventions). Notice that we can interpret this fact by saying that perturbative type II strings are not the elementary source of R-R fields. Therefore, type IIB theory contains the $p$-form potentials $C_0$, $C_2$ and $C_4$ (with self-dual field strength), while in type IIA we have $C_1$ and $C_3$. Higher form potentials are related via Hodge duality to the ones we just listed. In what follows, the bosonic part of the massless spectrum of the type II theories will have particular relevance, and we summarize it in table~\ref{t:spectrum}.
%:Table: Type II spectrum
\begin{table}
\begin{center}
\begin{tabular}{|c|c|c|}
\hline
& NS-NS sector & R-R sector \\
\hline
Type IIB & $G_{\mu\nu}$, $B_2$, $\Phi$ & $C_0$, $C_2$, $C_4$ \\
Type IIA & $G_{\mu\nu}$, $B_2$, $\Phi$ & $C_1$, $C_3$ \\
\hline
\end{tabular}
\end{center}
\caption{{\small Bosonic massless spectrum of type II superstring theories.}}
\label{t:spectrum}
\end{table}

Notice that we have presented in this section only two of the five known consistent superstring theories. This choice is dictated by the fact that we will only work with type II theories in the rest of this report. Apart from the type II theories, one has the type I $SO(32)$ open+closed unoriented string theory, and the $SO(32)$ and $E_8\times E_8$ heterotic string theories. 

\subsection*{Low-energy effective actions}

Another important ingredient that we need is given by the low-energy effective actions of the type II superstring theories that we just reviewed. By low-energy effective action we mean a classical field theory action which describes the dynamics of the massless states of the string in the field theory limit $\ls\to 0$, where the string reduces to a point particle. The fields appearing in the effective actions are the ones listed in table~\ref{t:spectrum}, and one can obtain the explicit expressions by computing string amplitudes and/or analyzing the renormalization group functions of the string sigma-model.

In either way, one obtains the \emph{type II supergravity} actions. The form in which they are directly related to string theory amplitudes is given in the so-called \emph{Einstein frame}, while the analysis of the sigma-model leads naturally to the \emph{string frame}. The bosonic part of the type IIB supergravity action in the string frame reads:
\begin{multline}\label{IIBst}
	S_{\text{IIB}}^{\text{(st)}} = \frac{1}{2\kappa^2} \Bigg\{ \int d^{10}x \sqrt{-\det G}\ e^{-2\Phi}R
		- \frac{1}{2} \int \Big[ - 8 e^{-2\Phi} d\Phi \wedge \hd{} d\Phi
		+ e^{-2\Phi}  H_3 \wedge \hd{}H_3\\
		+ F_1 \wedge \hd{}F_1 + \tilde{F}_3 \wedge \hd{}\tilde{F}_3
		+ \frac{1}{2} \tilde{F}_5 \wedge \hd{}\tilde{F}_5
		- C_4\wedge H_3 \wedge F_3
		\Big] \Bigg\}\,.
\end{multline}
where $\kappa$ is related to string quantities via $\kappa = 8\pi^{7/2}\gs\ls^4$ and:
\begin{equation}
	H_3 = dB_2\,,\qquad F_n = dC_{n-1}\,,\qquad
	\tilde{F}_3 = F_3 + C_0 \wedge H_3\,,\qquad
	\tilde{F}_5 = F_5 + C_2 \wedge H_3\,.
\end{equation}
Notice that the action~\eqref{IIBst} does not take into account the self-duality of $\tilde{F}_5$, that has to be imposed on-shell.%
\footnote{Writing a complete action for type IIB supergravity is a complicated task, and has been proven possible only in a complicated formalism with auxiliary fields.}
The bosonic part of the type IIA supergravity action in the string frame is:
\begin{multline}\label{IIAst}
	S_{\text{IIA}}^{\text{(st)}} = \frac{1}{2\kappa^2} \Bigg\{ \int d^{10}x \sqrt{-\det G}\ e^{-2\Phi}R
		- \frac{1}{2} \int \Big[ - 8 e^{-2\Phi} d\Phi \wedge \hd{} d\Phi
		+ e^{-2\Phi}  H_3 \wedge \hd{}H_3\\
		- F_2 \wedge \hd{}F_2 - \tilde{F}_4 \wedge \hd{}\tilde{F}_4
		+ B_2\wedge F_4 \wedge F_4
		\Big] \Bigg\}\,,
\end{multline}
where:
\begin{equation}
	H_3 = dB_2\,,\qquad F_n = dC_{n-1}\,,\qquad
	\tilde{F}_4 = F_4 - C_1 \wedge H_3\,.
\end{equation}
Notice that the above expressions for the low-energy actions have a non-standard Einstein-Hilbert term, due to the dilaton dependence. We can write them in a more standard way by passing to the Einstein frame, where the physical properties are clearer, with the redefinition:
\begin{equation}
	ds^2_{\text{(E)}} = e^{-\Phi/2} ds^2_{\text{(st)}}\,.
\end{equation}
The type II supergravity actions in the Einstein frame are then given by:
\begin{multline}\label{IIBE}
	S_{\text{IIB}}^{\text{(E)}} = \frac{1}{2\kappa^2} \Bigg\{ \int d^{10}x \sqrt{-\det G}\ R
		- \frac{1}{2} \int \Big[ d\Phi \wedge \hd{} d\Phi + e^{-\Phi} H_3 \wedge \hd{}H_3\\
		+ e^{2\Phi} F_1 \wedge \hd{}F_1 + e^{\Phi} \tilde{F}_3 \wedge \hd{}\tilde{F}_3
		+ \frac{1}{2} \tilde{F}_5 \wedge \hd{}\tilde{F}_5
		- C_4\wedge H_3 \wedge F_3
		\Big] \Bigg\}\,,
\end{multline}
and:
\begin{multline}\label{IIAE}
	S_{\text{IIA}}^{\text{(E)}} = \frac{1}{2\kappa^2} \Bigg\{ \int d^{10}x \sqrt{-\det G}\ R
		- \frac{1}{2} \int \Big[ d\Phi \wedge \hd{} d\Phi + e^{-\Phi} H_3 \wedge \hd{}H_3\\
		- e^{3\Phi/2} F_2 \wedge \hd{}F_2 - e^{\Phi/2} \tilde{F}_4 \wedge \hd{}\tilde{F}_4
		+ B_2\wedge F_4 \wedge F_4
		\Big] \Bigg\}\,.
\end{multline}

Another action that we will use in the following is the one of eleven-dimensional supergravity, which is supposed to describe the low-energy dynamics of the eleven-dimensional ``M-theory'', namely (in a restricted sense) the strong coupling limit of type IIA string theory. The bosonic field content of the theory comprises the metric and a three-form gauge potential $A_3$, and the action reads:
\begin{equation}\label{11dsugra}
	S_{11} = \frac{1}{2\kappa_{11}^2} \Bigg\{ \int d^{11}x \sqrt{-\det G}\ R
		+ \frac{1}{2} \int \Big[ F_4 \wedge \hde F_4
		- \frac{1}{3} A_3\wedge F_4 \wedge F_4
		\Big] \Bigg\}\,,
\end{equation}
where $F_4 = dA_3$ and $\kappa_{11}$ is related to the eleven-dimensional Planck length $l_{\text{P}}=\gs^{1/3}\ls$ via $\kappa_{11}^2 = 2^7 \pi^8 {l_{\text{P}}}^9$. Standard dimensional reduction on a space-like circle yields the type IIA action~\eqref{IIAE}.

\section{D-branes and open strings}\label{s:dbranes}

This section is devoted to the introduction of essential physical objects of string theory, the \emph{Dirichlet branes} or D-branes. In open string theory, they are simply defined as hyperplanes where open strings end, with Dirichlet boundary conditions in the appropriate directions. However, we will see that there is much more about D-branes than that, and they will play a fundamental role in all our subsequent analysis of the gauge/gravity correspondence. Many of the things we will cover are for instance reviewed in the recent book~\cite{Johnson:2003gi}.

\subsection*{Strings on a circle}

We start by considering the compactification of a closed string theory on a circle. The most general solution of the equations of motion~\eqref{steom} for a closed string can be written as:
\begin{equation}
	X^\mu (\tau,\sigma) = x^\mu + \frac{\ls}{\sqrt{2}}(\alpha_0^\mu+\talpha_0^\mu) \tau
		- \frac{\ls}{\sqrt{2}} (\alpha_0^\mu-\talpha_0^\mu) \sigma
		+ i \frac{\ls}{\sqrt{2}} \sum_{n\neq 0} \frac{1}{n}
		\left( \alpha_n^\mu e^{-in(\tau-\sigma)} 
		+ \talpha_n^\mu e^{-in(\tau+\sigma)}\right)\,.
\end{equation}
and the momentum is given by:
\begin{equation}
	p^\mu = \frac{1}{\sqrt{2}\ \ls} (\alpha_0^\mu + \talpha_0^\mu)\,.
\end{equation}
We saw that in the uncompactified case the two zero-modes must be identifed, $\alpha_0^\mu = \talpha_0^\mu = \tfrac{\ls}{\sqrt{2}} p^\mu$, due to the periodicity condition under $\sigma\to\sigma+2\pi$, and one gets the mode expansion~\eqref{Xcl} in terms of the momentum $p^\mu$.

Now, let us compactify a single direction of the target space-time, call it $X$ without indices, on a circle of radius $R$:
\begin{equation}
	X \simeq X + 2\pi R\,.
\end{equation}
In this case, the momentum along the compact direction must be quantized as:
\begin{equation}\label{quantmom}
	p = \frac{n}{R}\,,\qquad n\in\mathbb{Z}\,.
\end{equation}
Moreover, a closed string may wind around the compact direction $X$, which means that under $\sigma\to\sigma+2\pi$, $X$ needs not be single-valued, but we rather have:
\begin{equation}\label{wind}
	X(\tau,\sigma+2\pi) \simeq X(\tau,\sigma)+ 2\pi R w\,,
\end{equation}
$w$ being the winding number around the direction $X$. Taking the conditions~\eqref{quantmom} and~\eqref{wind} together, we get the following equations:
\begin{equation}
	\alpha_0 + \talpha_0 = \sqrt{2}\ \ls \frac{n}{R}\,,\qquad
	\alpha_0 - \talpha_0 = \frac{\sqrt{2}}{\ls} wR\,,
\end{equation}
implying:
\begin{equation}
	\alpha_0 = \frac{\ls}{\sqrt{2}} \left(\frac{n}{R}+\frac{wR}{\ls^2}\right)\,,\qquad
	\talpha_0 = \frac{\ls}{\sqrt{2}} \left(\frac{n}{R}-\frac{wR}{\ls^2}\right)\,.
\end{equation}
The bosonic part of the zero-mode Virasoro generators gets then modified as follows:
\begin{equation}
\begin{aligned}
	L_0 &= \frac{\ls^2}{4}p^2 + \frac{\ls^2}{4}\left( \frac{n}{R} + \frac{wR}{\ls^2} \right)^2
		+ \sum_{n=1}^{\infty} \alpha_{-n}\cdot \alpha_n\,,\\
	\tilde{L}_0 &= \frac{\ls^2}{4}p^2 + \frac{\ls^2}{4}\left( \frac{n}{R} - \frac{wR}{\ls^2} \right)^2
		+ \sum_{n=1}^{\infty} \talpha_{-n}\cdot \talpha_n\,,
\end{aligned}
\end{equation}
where $p$ now denotes the momentum in the uncompactified directions. By using these expressions, one sees that the bosonic part of the mass operator reads:
\begin{equation}\label{masscomp}
	M^2 = \left(\frac{n}{R}\right)^2 + \left(\frac{wR}{\ls^2}\right)^2
		+ \text{ oscillator part}\,.
\end{equation}
We therefore see that two new types of states have appeared to enrich the spectrum of the compactified theory. There are Kaluza--Klein modes contributing to the energy by $\tfrac{n}{R}$, and these would be present also in a compactified field theory. In addition, there are excitations generated by the winding modes of the string around the compact direction.

\subsection*{T-duality}

The formula~\eqref{masscomp} for the mass spectrum of closed strings compactified on a circle allows us to make a very interesting observation. In the limit $R\to \infty$, we see that the Kaluza--Klein modes become light, while the winding modes become infinitely massive and decouple from the theory, as we would expect from a decompactification limit. However, the new thing is that something similar happens also in the opposite limit $R\to 0$, where the Kaluza--Klein modes decouple, but now the spectrum of winding modes approaches a continuum! In fact, \eqref{masscomp} is invariant under:
\begin{equation}\label{RtoTR}
	n \leftrightarrow w\,,\qquad
	R \to \hat{R} = \frac{\ls^2}{R}\,,
\end{equation}
and this makes clear that, as far as the bosonic spectrum is concerned, the $R\to0$ and $R\to\infty$ limits are physically identical: the string spectrum looks like the one of an uncompactified theory. This symmetry of the bosonic spectrum is known as \emph{T-duality}. Notice that exchanging winding and momentum is equivalent to the following action on the bosonic zero-modes:
\begin{equation}
	\alpha_0 \to \alpha_0\,,\qquad
	\talpha_0 \to -\talpha_0\,,
\end{equation}
and this operation is extended to the bosonic non-zero modes as well, so that we can summarize the action of this symmetry on the coordinate field $X$ as follows. Rearrange the expansion of the original coordinate field $X$ corresponding to the compact direction as:
\begin{equation}
	X (\tau,\sigma) = X_L (\tau-\sigma) + X_R (\tau+\sigma) \,,
\end{equation}
where:
\begin{equation}\label{XLR}
\begin{aligned}
	X_L (\tau-\sigma) &= \frac{x^\mu}{2} + \frac{\ls}{\sqrt{2}} \alpha_0^\mu (\tau-\sigma)
		+ i \frac{\ls}{\sqrt{2}} \sum_{n\neq0} \frac{\alpha_n^\mu}{n} e^{-i(\tau-\sigma)}\,,\\ 
	X_R (\tau+\sigma) &= \frac{x^\mu}{2} + \frac{\ls}{\sqrt{2}} \alpha_0^\mu (\tau+\sigma)
		+ i \frac{\ls}{\sqrt{2}} \sum_{n\neq0} \frac{\alpha_n^\mu}{n} e^{-i(\tau+\sigma)}\,.
\end{aligned}
\end{equation}
From the conditions we derived above, we see that the coordinate $\hat{X}$ to be used in the T-dual description must satisfy:
\begin{equation}\label{NtoD}
	\partial_\tau \hat{X} = - \partial_\sigma X\,,\qquad
	\partial_\sigma \hat{X} = - \partial_\tau X.
\end{equation}
This fact identifies the T-dual coordinate to be:
\begin{equation}\label{TdX}
	\hat{X} (\tau,\sigma) =  X_L (\tau-\sigma) - X_R (\tau+\sigma)\,,
\end{equation}
in terms of the expressions~\eqref{XLR}.

So far we have shown that T-duality is a symmetry of the bosonic part of the spectrum, but we have to see how it acts on the fermionic part of a type II theory. Its action can be fixed by requiring superconformal invariance on the world-sheet, which implies:
\begin{equation}
	\psi_+ \to \psi_+\,,\qquad
	\psi_- \to - \psi_-\,,
\end{equation}
or, in terms of oscillators:
\begin{equation}
	\psi_r \to \psi_r\,,\qquad
	\tpsi_r \to -\tpsi_r\,.
\end{equation}
The important thing to notice is that these transformations reverse the chirality of the ground state in the right-moving R-R sector. This means that, since the bosonic spectrum is unchanged under T-duality, if we start with type IIA theory on a circle of radius $R$, we end up with type IIB theory on a circle of radius $\tfrac{\ls^2}{R}$, and vice-versa. This easily extends to the case of taking a T-duality in several dimensions. If we start with a given type II theory, we reach the other type II theory with an odd number of T-duality transformations, while with an even number we land back on the original theory. We can then say that the type II theories are decompactification limits of a single space of compact theories. In fact, the relation~\eqref{RtoTR} implies that in a compactified string theory we can limit ourselves to consider the region $R\ge \ls$, and this is the reason why $\ls$ is often referred to as the minimal length of the theory.

Since T-duality transforms type IIA into type IIB theory and vice-versa, the whole spectra of the two theories must map onto each other. This can in fact be seen to be true from a string theory perspective, but we just limit ourselves to see how the massless fields, namely the ones appearing in the supergravity actions~\eqref{IIBst} and~\eqref{IIAst} are transformed into each other~\cite{Bergshoeff:1995as}. In particular, we find that the odd rank R-R potentials are mapped onto even rank potentials and vice-versa. We limit ourselves to the case of a T-duality along a single direction, say $X^9$, while the indices $\mu$, $\nu$ span the untouched directions. Denoting the fields of the theory which lies at the end of the T-duality transformation with a hat, the formulae read:
\begin{equation}\label{Tduality}
\begin{gathered}
	\hat{G}_{99} = \frac{1}{G_{99}}\,,\qquad
	e^{2\hat{\Phi}} = \frac{e^{2\Phi}}{G_{99}}\,,\qquad
	\hat{G}_{\mu 9} = \frac{B_{\mu 9}}{G_{99}}\,,\qquad
	\hat{B}_{\mu 9} = \frac{G_{\mu 9}}{G_{99}}\,,\\
	\hat{G}_{\mu\nu} = G_{\mu\nu} - \frac{G_{\mu 9}G_{\nu 9}-B_{\mu 9}B_{\nu 9}}{G_{99}}\,,\qquad
	\hat{B}_{\mu\nu} = B_{\mu\nu} - \frac{B_{\mu 9}G_{\nu 9}-G_{\mu 9}B_{\nu 9}}{G_{99}}\,,\\
\begin{aligned}
	(\hat{C}_{p})_{\mu\cdots\nu\alpha9} &= (C_{p-1})_{\mu\cdots\nu\alpha}
		- (p-1) \frac{(C_{p-1})_{[\mu\cdots\nu|9}G_{|\alpha|9}}{G_{99}}\,,\\
	(\hat{C}_{p})_{\mu\cdots\nu\alpha\beta9} &= (C_{p+1})_{\mu\cdots\nu\alpha\beta9}
		+ p (C_{p-1})_{[\mu\cdots\nu\alpha}B_{\beta]9}
		+ p(p-1) \frac{(C_{p-1})_{[\mu\cdots\nu|9}B_{|\alpha|9}G_{|\beta]9}}{G_{99}}\,.
\end{aligned}	
\end{gathered}
\end{equation}

\subsection*{T-duality for open strings: D-branes}

It is natural to ask if and how T-duality extends to a theory with open strings too. Open strings do not satisfy any periodicity condition, and they clearly cannot wind around a periodic direction. Therefore, in their spectrum we will find the Kaluza--Klein modes but no winding modes, and this means that the $R \to 0$ limit looks physically very different from the decompactification limit $R\to\infty$.

However, this looks puzzling when we recall that theories with open strings automatically include closed strings too, since loop diagrams of open string theory such as the annulus can be also regarded as closed string diagrams (this is made precise via a modular transformation of the string amplitudes). How is then possible that, in the $R\to 0$ limit of a compactification, the closed strings ``feel'' all the directions as uncompactified, while the open strings effectively see a direction disappearing due to the decoupling of the Kaluza--Klein modes? After all, open and closed strings are mostly indistinguishable, at least in every point along their length, except the endpoints of the open string.

This puzzle can precisely be solved by implementing our latter argument. Since the difference between an open and a closed string lies in the endpoints of the former, we can think of the endpoints of the open strings to be confined to a hyperplane in space-time.

To clarify what we mean, consider open and closed strings living in $D$ spatial dimensions, and compactify one direction on a circle of radius $R$. In order to perform the $R\to0$ limit, we had better take a T-duality transformation, leading to a theory compactified on the dual circle $\hat{R}=\tfrac{\ls^2}{R}$, and then take the decompactification limit $\hat{R}\to\infty$. In the T-dual description, we have to use the coordinate $\hat{X}$ defined in~\eqref{TdX} (which is also valid for the open string provided we correctly identify the left and right-moving oscillators), and we see from~\eqref{NtoD} that a Neumann boundary condition for $X$ is transformed into a Dirichlet boundary condition for $\hat{X}$!

This exactly means that, in the T-dual theory, the open strings endpoint are confined to live on a $p$-dimensional hyperplane, where $p=D-1$. This hyperplane takes the name of Dirichlet $p$-brane, or \emph{D$p$-brane}, from the fact that the open strings attached to them satisfy Dirichlet boundary conditions on the directions which are transverse to the world-volume of the hyperplane~\cite{Polchinski:1995mt}. It is clear that, since a T-duality transformation exchanges Neumann and Dirichlet boundary conditions, an additional T-duality made on a longitudinal direction yields a D$(p-1)$-brane, while a T-duality in a transverse direction changes a D$p$-brane into a D$(p+1)$-brane.

We have introduced D-branes as rigid hyperplanes where open string endpoints lie, but the excitations of such open strings make D-branes fully dynamical objects in string theory. Among these excitations, the massless ones have the important peculiarity of not changing the D-brane energy, and they can therefore be seen as collective coordinates of the brane. Let us then consider the massless spectrum of the open strings attached to a single D$p$-brane. In the following (as reported in insert~\ref{i:notations} on page~\pageref{i:notations}), we will often represent a D-brane configuration by means of a table like the following one, where the symbols $-$ and $\cdot$ denote resepectively directions which are longitudinal and transverse to the world-volume of the D$p$-brane:
\begin{center}
\begin{tabular}{|c|ccc|ccc|}
\hline
&0&$\cdots$&$p$&$p$+1&$\cdots$&9\\
\hline
D$p$ &$-$&$-$&$-$&$\cdot$&$\cdot$&$\cdot$\\
\hline
\end{tabular}
\end{center}
We split the ten space-time coordinates as follows:
\begin{itemize}
\item $x^\alpha\,,\quad \alpha=0,\ldots,p\,:$ directions belonging to the world-volume of the D$p$-brane;
\item $x^i\,,\quad i=p+1,\ldots,9\,:$ directions transverse to the world-volume of the D$p$-brane. The brane is thought to be, for example, at the point $x^i=0$.
\end{itemize}
The open string states at the massless level are given by:
\begin{center}
\begin{tabular}{|c|l|}
\hline
\multicolumn{2}{|c|}{NS states} \\
\hline
$\psi_{-1/2}^\alpha \ket{0,k}$ &
$\to$ $1$ vector $A^\alpha$\\
$\psi_{-1/2}^i \ket{0,k}$ &
$\to$ $9-p$ real scalars $\Phi^i$\\
\hline
\end{tabular}
\qquad
\begin{tabular}{|c|l|}
\hline
\multicolumn{2}{|c|}{R states} \\
\hline
$\ket{s_0,s_1,s_2,s_3,s_4}$ &
$\to$ $16$ fermions\\
\hline
\end{tabular}
\end{center}
These states comprise precisely the vector multiplet of a $(p+1)$-dimensional gauge theory with gauge group $U(1)$ and sixteen preserved supercharges. This is one of the most important facts about D-branes: the low-energy dynamics of the lowest-lying open string states on a D$p$-brane describes a gauge theory living in a $(p+1)$-dimensional space-time. The scalars $\Phi^i$ describe the shape of the hyperplanes, and in fact they are precisely related to the embedding coordinate fields as:
\begin{equation}
	X^i = 2\pi\ls^2 \Phi^i\,.
\end{equation}
Notice that T-duality exchanges gauge field components with scalar fields. This can be explained in more detail by introducing appropriate Wilson lines in the toroidal compactification, see for example~\cite{Polchinski:1998}.

The analysis can be extended to the case of more than one D-brane. Consider for example two parallel D$p$-branes placed at the same point $x^i=y^i$ in the common tranverse space. In addition to the open strings starting and ending on the same brane, we will also have strings stretching between the two different branes, with the two possible orientations. This configuration can be represented by introducing a $2\times2$ Chan--Paton matrix $\lambda$ labeling the generic open string state:
\begin{equation}
	\lambda \otimes \text{``oscillators''}\ \ket{0,k}\,,\qquad
	\lambda = \begin{pmatrix}
		\text{D}p-\text{D}p & \text{D}p-\text{D}p' \\
		\text{D}p'-\text{D}p & \text{D}p'-\text{D}p' \\
		\end{pmatrix} \,,
\end{equation}
in which the matrix elements refer to the open strings attached to the first brane or to the second brane (denoted with a prime). Repeating the analysis above, we can easily see that the massless open string states fill precisely the vector multiplet of a $U(2)$ gauge theory in $p+1$ dimensions. In fact, if the two D$p$-branes were separated in transverse space, the states with an off-diagonal Chan--Paton matrix $\lambda$ would represent massive W-bosons, and the gauge symmetry would be broken down to $U(1)\times U(1)$. When the two branes coincide, the W-bosons become massless and a full $U(2)$ gauge symmetry is recovered.

This construction can of course be generalized to $N$ D-branes. Summarizing, the low-energy theory on a stack of $N$ coincident D$p$-branes is $(p+1)$-dimensional Super Yang--Mill theory with sixteen supercharges and gauge group $U(N)$.

Moving a brane off the stack, say from $x^i=0$ to a generic point $x^i$ in transverse space, corresponds to giving a non vanishing expectation value to the scalars $\Phi^i$ in the vector multiplet of the gauge theory, $x^i = 2\pi\ls^2 \Phi^i$, and this breaks the gauge symmetry to $U(N-1)\times U(1)$ via the Higgs mechanism.

%:Figure: D-brane stack
\begin{figure}
\begin{center}
\includegraphics[scale=.7]{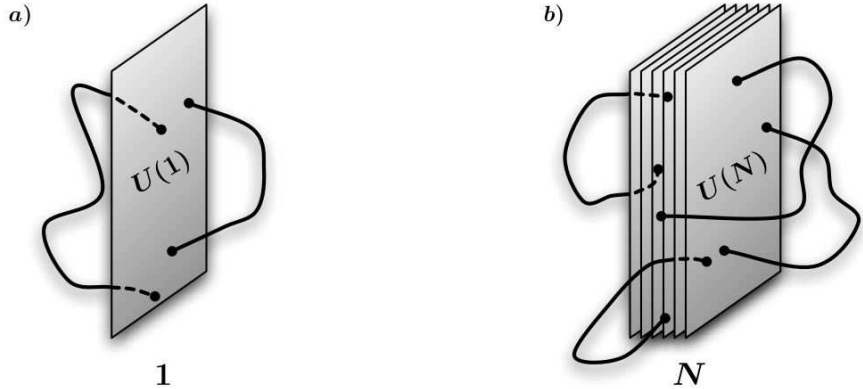}
\caption{{\small D-branes are dynamical hyperplanes where open strings end. \emph{a)} The low-energy dynamics of the massless open string states on a single D-brane gives a $U(1)$ gauge theory; \emph{b)} The theory living on a stack of $N$ coincident D-branes has gauge group $U(N)$.}}
\label{f:stack}
\end{center}
\end{figure}

\subsection*{D-branes and Ramond-Ramond charges}

We have seen that D-branes are physical objects of string theory, described by the fact that open strings are attached to them, and that the fluctuations of the open strings on the branes at low energy give rise to supersymmetric gauge theories.

In fact, the study of the gauge theory spectrum is an easy way to point out the remarkable fact that D-branes preserve \emph{half of the supersymmetries} of flat space, namely 16 supercharges out of 32. This can be confirmed at all levels by a careful analysis. This means that D-branes are \emph{BPS states}. An important property of BPS states in supersymmetric theories is that they must carry conserved charges, and in fact there is one (and only one) set of charges which is appropriate for coupling to D-branes, namely the antisymmetric Ramond-Ramond charges~\cite{Polchinski:1995mt}. In fact, the $(p+1)$-dimensional world-volume $\mathcal{M}_{p+1}$ of a D$p$-brane naturally couples to a $(p+1)$-form potential $C_{p+1}$ in the R-R sector via the following minimal coupling:
\begin{equation}\label{mincoup}
	 \mu_p \int_{\mathcal{M}_{p+1}} C_{p+1}\,,
\end{equation}
where $\mu_p$ is a coupling to be determined in terms of string quantities. We mentioned in section~\ref{s:strings} that strings are not the fundamental sources of R-R fields, and now we have seen that string theory also contains objects, the D-branes, which play this role. Notice that the fact that R-R potentials $C_p$ do not exist for any $p$ in each type II theory restricts the possible spectrum of D-branes in type IIA and type IIB superstring theory, as summarized in table~\ref{t:branes}.

In the next section, we will study in more detail the relationship between the fields living on a D-brane and the closed string fields, by deriving the D-brane world-volume action.
%:Table: D-brane spectrum
\begin{table}
\begin{center}
\begin{tabular}{|l|ccccc|ccccc|}
\hline
&\multicolumn{5}{|c|}{Type IIA}&\multicolumn{5}{|c|}{Type IIB}\\
\hline
D-brane & D0 & D2 & D4 & D6 & D8 & D$(-1)$ & D1 & D3 & D5 & D7 \\
R-R potential & $C_1$ & $C_3$ & $C_5$ & $C_7$ & $C_9$ 
& $C_0$ & $C_2$ & $C_4$ & $C_6$ & $C_8$ \\
\hline
\end{tabular}
\end{center}
\caption{{\small The spectrum of D-branes in type II theories. Notice that the D9-brane is not present in the standard type IIB theory since it would imply charge anomalies, but can exist in orientifolds thereof. The IIB D$(-1)$-brane, coupled to the R-R scalar $C_0$, is a ``D-instanton'', since it is a localized configuration in (euclidean) time.}}
\label{t:branes}
\end{table}

\section{The D-brane world-volume action}\label{s:wv}

We have seen that the massless string states on a D$p$-brane form the vector multiplet of a $(p+1)$-dimensional gauge theory with 16 supercharges. It is natural to ask how one deals with the dynamics on the world-volume of a D-brane at low-energy, and in particular how the theory on the brane ``feels'' the geometry of the bulk of space-time.

There are several ways to derive the effective action for the low-energy dynamics of the theory living on a D-brane. One can compute string disk amplitudes and insert closed string vertex operators, thus extracting the coupling between open and closed strings. Alternatively, one can derive those couplings by means of the boundary state formalism that we will shortly mention in the next section. Here, for the sake of brevity, we will instead derive the action somewhat heuristically by means of T-duality as in~\cite{Polchinski:1998}.

Let us start with the coupling to the metric and dilaton. A D$p$-brane has a $(p+1)$-dimensional world-volume, and the action will come from low-energy open string disk amplitudes which go as $\gs^{-1}$. We therefore expect a coupling of the form:
\begin{equation}\label{wv1}
	S_{\text{D}p} = - \tau_p \int_{\mathcal{M}} d^{p+1}\xi\ e^{-\Phi} \sqrt{-\det \hat{G}_{ab}} + \ldots\,,
\end{equation}
where $\xi^a$, $a=0,\ldots,p$ parameterize the world-volume directions, $\tau_p$ is the tension of the brane (on which we will have some more to say later on) including a factor of $\gs^{-1}$, and $\hat{G}_{ab}$ is the pull-back of the space-time metric onto the world-volume of the brane:
\begin{equation}\label{pullback}
	\hat{G}_{ab} = \frac{\partial x^\mu}{\partial \xi^a} \frac{\partial x^\nu}{\partial \xi^b} G_{\mu\nu}\,.
\end{equation}

The next step is to consider the coupling to the $B$-field and to the world-volume gauge field $A_a$. In order to find it, we consider the specific example of a D2-brane extended along the $x^1$ and $x^2$ directions, and turn on a constant field strength $F_{12}$ on the world-volume. Let us also choose a gauge in which $A_2 = x^1 F_{12}$. A T-duality along the direction $x^2$ will give a D1-brane with:
\begin{equation}
	x^2 = 2\pi\ls^2 A_2 = 2\pi\ls^2 x^1 F_{12}\,.
\end{equation}
We can interpret this result by saying that the D1-brane is tilted at an angle:
\begin{equation}
	\theta = \tan^{-1} \left(2\pi\ls^2 F_{12}\right)\,,
\end{equation}
with respect to the $X^2$ direction. This means that we have the following geometrical piece coming inside the world-volume action of the D1-brane from~\eqref{pullback}:
\begin{equation}
	\int dx^1 \sqrt{1-(\partial_1 x^2)^2} = \int dx^1 \sqrt{1 + (2\pi\ls^2 F_{12})^2}\,.
\end{equation}
This result may be generalized by boosting a D-brane to be aligned with the coordinate axes and then bringing $F_{ab}$ to a block-diagonal form, obtaining a product of factors similar to the one we just found. We therefore expect a term such as:
\begin{equation}\label{wv2}
	\int d^{p+1}x \sqrt{\det (\eta_{ab} + 2\pi\ls^2 F_{ab})}\,.
\end{equation}

Now, on the one hand we have to covariantize the term~\eqref{wv2} in order to incorporate the coupling~\eqref{wv1}. On the other hand, we have to be careful about preserving space-time gauge invariance, and this requires that $F$ and $B_2$ appear in a specific combination in the action. In fact, a space-time gauge transformation of the form $\delta B_2 = d\zeta$, where $\zeta$ is an arbitrary one-form, will give a surface term in the sigma-model action describing the string in the background. This has to be cancelled by an appropriate gauge transformation of $A$, that can be seen to be $\delta A = - \zeta / 2\pi\ls^2$. The gauge-invariant combination we have to consider is then $B + 2\pi\ls^2 F$. All these ingredients suggest the following form for the action:
\begin{equation}\label{DBI}
	S_{\text{DBI}} = -\tau_p \int_{\mathcal{M}_{p+1}} d^{p+1}\xi\ e^{-\Phi}
		\sqrt{\det (\hat{G}_{ab} + \hat{B}_{ab} + 2\pi\ls^2 F_{ab})}\,,
\end{equation}
where hats denote pull-backs of bulk fields onto the world-volume of the brane, as in~\eqref{pullback}, and we have introduced generic coordinates $\xi$ on the world-volume, not necessarily tied to the coordinate axes in space-time. The above action is known as \emph{Dirac--Born--Infeld} action, and describes the low-energy dynamics on the world-volume of a D$p$-brane for arbitrary values of the background fields in the NS-NS sector.

We are not done yet, since we have to consider the couplings in the R-R sector. As we already anticipated, the fundamental coupling of a D$p$-brane, as a $p$-dimensional object, is the following one to a $(p+1)$-rank potential~\eqref{mincoup}:
\begin{equation}\label{wv3}
	 \mu_p \int_{\mathcal{M}_{p+1}} \hat{C}_{p+1}\,,
\end{equation}
where $\mu_p$ is the R-R charge of the D-brane in appropriate units, and the hat again denotes the pull-back of the $C$-field onto the world-volume. The term~\eqref{wv3} does not give the complete answer, however, and we can compute the additional terms again by means of T-duality. Let us consider a D1-brane tilted in the $(X^1,X^2)$-plane. The computation of the pull-back of $C_2$ will give the following expression for the coupling~\eqref{wv3}:
\begin{equation}
	\int dx^0 dx^1 \left( (C_2)_{01} +\partial_1 x^2 (C_2)_{02} \right)\,,
\end{equation}
that under a T-duality in the direction $x^2$, using~\eqref{Tduality}, becomes a term like:
\begin{equation}
	\int dx^0 dx^1 dx^2 \left( (C_3)_{012} + 2\pi\ls^2 F_{12} (C_1)_{0} \right)\,.
\end{equation}
For higher dimensional branes, this procedure can be generalized to lower rank potentials. Recalling the gauge-invariant combination of $B_2$ and $F$ that we have to use, we can guess the final \emph{Wess--Zumino} term of the action~\cite{Li:1996pq,Douglas:1995bn}:
\begin{equation}\label{WZ}
	S_{\text{WZ}} = \mu_p \int_{\mathcal{M}_{p+1}} \sum_q \hat{C}_q \wedge e^{\hat{B}+2\pi\ls^2 F}\,,
\end{equation}
where the sum is meant to pick up the terms in the expansion corresponding to $(p+1)$-forms, which give a non-vanishing result to the integral over the world-volume.

The world-volume action for a D$p$-brane is then given by the sum of the Dirac--Born--Infeld part~\eqref{DBI} and the Wess--Zumino part~\eqref{WZ}:
\begin{equation}\label{Dpwv}
\begin{split}
	S_{\text{D}p} = &- \tau_p \int_{\mathcal{M}_{p+1}} d^{p+1}\xi\ e^{-\Phi}
		\sqrt{- \det \left(\hat{G}_{ab} + \hat{B}_{ab} + 2\pi\ls^2 F_{ab}\right) }\\
		 &+ \mu_p \int_{\mathcal{M}_{p+1}} \sum_q \hat{C}_q \wedge e^{\hat{B}+2\pi\ls^2 F}\,,
\end{split}
\end{equation}
plus the supersymmetric completion describing the coupling to the fermions, that we will not discuss here. Since we will use it in both versions, let us also write the equivalent expression in the Einstein frame:
\begin{equation}\label{wvE}
\begin{split}
	S_{\text{D}p}^{\text{(E)}} = &- \tau_p \int_{\mathcal{M}_{p+1}} d^{p+1}\xi\ e^{\frac{p-3}{4}\Phi}
		\sqrt{- \det \left[\hat{G}_{ab} + e^{-\Phi/2}\left(\hat{B}_{ab} + 2\pi\ls^2 F_{ab}\right)\right] }\\
		 &+ \mu_p \int_{\mathcal{M}_{p+1}} \sum_q \hat{C}_q \wedge e^{\hat{B}+2\pi\ls^2 F}\,.
\end{split}
\end{equation}

In addition to the terms that we just uncovered, the world-volume action may contain additional parts. In particular, since we recovered anomalous gauge couplings in~\eqref{Dpwv}, we might expect that there are also anomalous couplings associated to higher corrections in the curvature $R$. These can in fact be shown to be present \cite{Green:1997dd,Cheung:1998az}, but we will not discuss them since they will be of no relevance to the examples we are interested in.

So far, everything we considered was for a single D$p$-brane. What happens if we put $N$ D-branes on top of each other? The precise answer to this question, as far as the world-volume action is concerned, is not known at the moment of writing. The world-volume fields $A_a$ and $X^i$ representing the collective motions of the branes become matrices valued in the adjoint representation of $U(N)$. In particular, this means that the transverse coordinates are really to be regarded as $N\times N$ matrices, and this has a series of non-trivial consequences, such that, for example, the pull-backs of the bulk fields should be done using the covariant derivative, or that the action acquires terms of order $[X^i,X^j]$ and $[A_a,X^i]$. Another problem is the precise prescription one has to use in order to perform the trace that is needed for getting a gauge invariant quantity to use in the action. A possible prescription is given by the use of the ``symmetrized trace''~\cite{Tseytlin:1997cs}, but it has been shown that this procedure gives rise to some ambiguities.

Without entering into the details of the non-abelian extensions of~\eqref{Dpwv}, there is one important observation that we can make, and that follows for example unambiguously from the symmetrized trace prescription. By expanding any non-abelian extension of~\eqref{Dpwv} in a flat space background up to second order in the gauge fields, one sees that the leading order reproduces the bosonic action of $(p+1)$-dimensional Super Yang--Mills theory:
\begin{equation}
	S_{\text{SYM}} = -\frac{1}{\gym} \int d^{p+1}\xi\ \tr
		\left[ \frac{1}{2} F_{ab}F_{ab} + D_a \Phi^i D_a \Phi^i + \frac{1}{2}[\Phi^i,\Phi^j]^2 \right]\,.
\end{equation}
We will not be precise here about this derivation and the relation of the coupling constant to string theory quantities, because we prefer to face this issue from a slightly different prespective which will shed some light on what we mean by gauge/gravity correspondence. This will be done in section~\ref{s:probe}.

\subsection*{The D-brane tension and charge}

It remains to determine the coefficients $\tau_p$ and $\mu_p$ appearing in the world-volume action~\eqref{Dpwv}, namely the tension and the R-R charge of a D$p$-brane. In order to derive them, we need to perform the computation of the string vacuum amplitude between D-branes~\cite{Polchinski:1995mt}. Let us then consider two parallel D$p$-branes, respectively located at $x^i=0$ and $x^i=y^i$. The one-loop planar free energy for an open string with $9-p$ Dirichlet boundary conditions is given by the Coleman--Weinberg formula:
\begin{equation}
	F = -2 \times \frac{1}{2}\tr \ln (L_0-a)
		= \int_0^{\infty} \frac{dt}{t}\tr_{\text{NS}-\text{R}}\left[\frac{1+(-1)^F}{2}e^{-2\pi t (L_0-a)}\right]\,,
\end{equation}
where we have taken into account the GSO projection and the two possible orientations of the open strings. $L_0$ contains an additional term due to the energy of the strings stretched between the two branes:
\begin{equation}
	L_0 = \ls^2 p^2 + \frac{y^2}{(2\pi\ls)^2} + \sum_{n=1}^{\infty} \alpha_{-n} \cdot \alpha_n
		+ \sum_{r>0} r \psi_{-r} \cdot \psi_r\,,
\end{equation}
where now $p$ only denotes the momentum in the longitudinal directions. The explicit computation of the trace in the various sectors gives the following final result:
\begin{equation}\label{polchinski}
	F = V_{p+1} \int_{0}^{\infty} \frac{dt}{2t}\ (8\pi^2\ls^2t)^{-\frac{p+1}{2}} e^{-\frac{y^2 t}{2\pi\ls^2}}\
		\frac{f_3(q)^8 - f_4(q)^8 - f_2(q)^8}{f_1(q)^8}\,,
\end{equation}
where $V_{p+1}$ is the (infinite) volume of the brane, $q=e^{-2\pi t}$ and:
\begin{equation}
\begin{aligned}
	f_1(q) &= q^{1/24}\prod_{n=1}^{\infty} ( 1 - q^n )\,,&
	f_2(q) &= \sqrt{2}q^{1/24}\prod_{n=1}^{\infty} ( 1 + q^n )\,,\\
	f_3(q) &= q^{-1/48}\prod_{n=1}^{\infty} ( 1 + q^{n-1/2} )\,,&
	f_4(q) &= q^{-1/48}\prod_{n=1}^{\infty} ( 1 - q^{n-1/2} )\,.
\end{aligned}
\end{equation}

The amplitude~\eqref{polchinski} vanishes due to the ``abstruse identity'' satisfied by the functions $f_i(q)$, and this means that there is no force between two parallel D-branes of equal dimension, as it should be because of supersymmetry. However, we can extract information about tension and charge by noting that the one-loop amplitude we just computed is an annulus diagram, which can also be regarded as a closed string tree-level amplitude, with the topology of a cylinder and representing the emission and successive reabsorption of a closed strings by a D-brane. This is made explicit by implementing the modular transformation $t\to s=1/t$ to the closed string channel. The vanishing of the amplitude~\eqref{polchinski} is then interpreted as the mutual cancellation of the attractive forces (of gravitational type) due to the fields in the NS-NS sector of the closed string and the repulsive forces (of electromagnetic type) due to the fields in the R-R sector:
\begin{equation}
	F = \mathcal{A}_{\text{NS-NS}}+\mathcal{A}_{\text{R-R}}=0\,.
\end{equation}
In the limit $s\to\infty$, the amplitude is dominated by the lightest closed string fields, namely the massless ones. We can therefore expand the result~\eqref{polchinski} (after the modular transformation) obtaining:
\begin{equation}\label{pole}
	\mathcal{A}_{\text{NS-NS}}=-\mathcal{A}_{\text{R-R}}
		\simeq V_{p+1}\ 2\pi(4\pi^2\ls^2)^{3-p}\ G_{9-p}(y)\,,
\end{equation}
where $G_d(y) = \tfrac{\pi^2}{4}\Gamma\left(\tfrac{d}{2}-1\right)\tfrac{1}{y^{d-2}}$ is the scalar Green's function in $d$ dimensions.

In order to obtain the tension and charge of a D-brane, we can compare the result~\eqref{pole} with a field theory computation, namely the amplitude for the exchange of a graviton, dilaton and R-R field between two D-branes. The two ingredients we need are the propagators and coupling, and we will obtain the former from the supergravity actions~\eqref{IIBE} and~\eqref{IIAE} in the Einstein frame, and the latter from the world-volume action~\eqref{wvE}. After some calculations, one gets the following propagators for the graviton $h_{\mu\nu}$ and dilaton $\Phi$ with momentum $k$:
\begin{equation}
	\langle h_{\mu\nu} h_{\rho\sigma} \rangle
		= -\frac{2i\kappa^2}{k^2}\left[\eta_{\mu\rho}\eta_{\nu\sigma}
		+\eta_{\mu\sigma}\eta_{\nu\rho} - \frac{1}{4} \eta_{\mu\nu}\eta_{\rho\sigma}\right]\,,\qquad
	\langle \Phi \Phi \rangle
		= - \frac{2i\kappa^2}{k^2}\,.
\end{equation}
The two Feynman diagrams corresponding to the exchange of graviton and dilaton sum up to the following expression in position space:
\begin{equation}
	\mathcal{A} = 2 V_{p+1}\ \tau_p^2 \kappa^2\ G_{9-p}(y)\,.
\end{equation}
The comparison with $\mathcal{A}_{\text{NS-NS}}$ then gives the following value for $\tau_p$:
\begin{equation}\label{taup}
	\tau_p = \frac{\sqrt{\pi}(2\pi\ls)^{3-p}}{\kappa} = \frac{1}{(2\pi)^p \gs \ls^{p+1}}\,.
\end{equation}

We can perform the same comparison for the exchange of the R-R field $C_{p+1}$, and with the normalizations that we use in the world-volume action~\eqref{wvE} we finally get:
\begin{equation}\label{mup}
	\mu_p = \tau_p\,.
\end{equation}
This completes the determination of the world-volume action, since we have expressed all parameters in terms of the string quantities $\gs$ and $\ls$.

\section{The geometry of D-branes}\label{s:classol}

In the previous sections, we have reviewed the fundamental fact that the new objects of string theory called D-branes admit a perturbative description in terms of the open strings which are attached to them. 

However, the fundamental importance of D-branes resides mainly in the fact that they have a twofold interpretation. We have already seen that they couple to closed strings, being in fact the fundamental objects charged under the R-R gauge potentials, and that the one-loop vacuum energy between two D-branes can be also interpreted as the tree-level exchange of a closed string between the branes. This is the basis of what is called \emph{open/closed string duality}, that implies that D-branes can be also described as \emph{boundaries} of the conformal field theory of closed strings. This approach associates to every D-brane a state in the theory known as the \emph{boundary state}, which is very useful for analyzing various properties of D-branes, such as deriving in an alternative way their world-volume action, and also for a different approach to what we will treat in the present section~\cite{DiVecchia:1997pr,DiVecchia:1999uf}. Here, we do not have the space of describing this formalism, which is nicely reviewed for instance in~\cite{DiVecchia:1999rh,DiVecchia:1999fx}.

\begin{figure}
\begin{center}
\includegraphics[scale=.7]{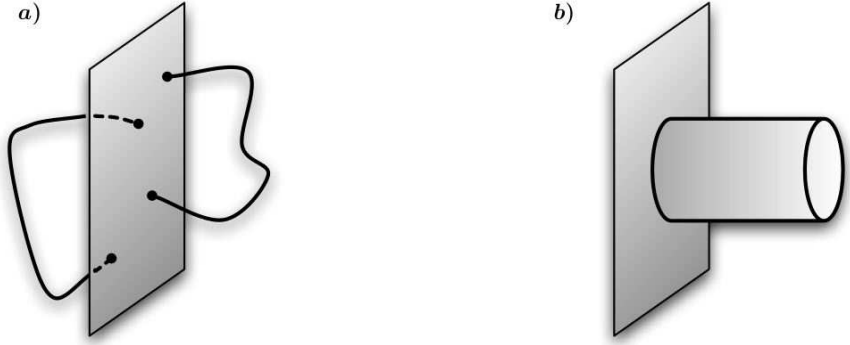}
\caption{{\small The two-fold interpretation of D-branes. \emph{a)} D-branes are hyperplanes where open strings end; \emph{b)} D-branes are boundaries of the superconformal theory of closed strings, and solitonic solutions of the low-energy effective action of closed string theory, charged under the Ramond-Ramond fields.}}
\label{f:openclosed}
\end{center}
\end{figure}

One of the main consequences of the closed string description of D-branes is that they can be regarded as classical solitons of the low-energy supergravity theories. In fact, the discovery of classical solutions of the low-energy string equations of motion describing extended $p$-dimensional solitonic objects charged under the R-R fields, called $p$-branes~\cite{Horowitz:1991cd,Duff:1995an}, preceded Polchinski's fundamental observation about the description of D-branes via open strings. One of the major advances of~\cite{Polchinski:1995mt} was precisely the identification of these solitonic solutions with the D-branes required by T-duality. This twofold interpretation, which we stress again will be crucial in the following, is depicted in figure~\ref{f:openclosed}.
%:Figure: Open/closed duality

Let us then look for solutions of the low-energy type II supergravity actions~\eqref{IIBE} and \eqref{IIAE} with these properties. Here we will not treat the problem of brane-like solution in much generality, but just derive the solution corresponding to extremal $p$-branes. A very nice reference about brane solutions in string and M-theory is~\cite{Argurio:1998cp}.

Our starting point is given by a consistent truncation of one of the two type II supergravity actions (in the Einstein frame), containing only one $(p+2)$-form field-strength $F_{p+2}$ (coming either from the NS-NS or the R-R sector), the metric and the dilaton:
\begin{equation}\label{II}
	S_{\text{II}} = \frac{1}{2\kappa^2} \Bigg\{ \int d^{10}x \sqrt{-\det G}\ R
		- \frac{1}{2} \int \Big[ d\Phi \wedge \hd{} d\Phi + e^{-a\Phi} F_{p+2} \wedge \hd{}F_{p+2}				\Big] \Bigg\}\,,
\end{equation}
where $a$ is an appropriate constant related to the rank of the form, as in~\eqref{IIBE}, \eqref{IIAE}. Since we are looking for a solution describing an object extended in the directions $x^0,\ldots,x^p$, we divide the coordinates in two groups and make some assumptions:
\begin{itemize}
\item $x^\alpha$, $\alpha=0,\ldots,p$, are the coordinates along the brane world-volume. We require that the solution preserves Poincar\'e symmetry in this $(p+1)$-dimensional space;
\item $x^i$, $i=p+1,\ldots,9$, are the coordinates transverse to the brane world-volume. We require that the solution preserves rotational $SO(9-p)$ symmetry in this $(9-p)$-dimensional space. The brane appears as a point in transverse space.
\end{itemize}
An ansatz compatible with the above hypotheses is:
\begin{equation}\label{braneans}
	ds^2 = e^{2A(r)}\ \eta_{\alpha\beta}dx^\alpha dx^\beta
		+ e^{2B(r)}\ \delta_{ij}dx^i dx^j\,,\qquad
	\Phi = \Phi(r)\,,
\end{equation}
where $r=(x^ix^i)^{1/2}$ is the radial coordinate in the rotationally symmetric transverse space. For the $(p+2)$-form we can choose either an electric ansatz, thus realizing a coupling like~\eqref{mincoup} or a magnetic ansatz, namely an electric ansatz for the Hodge dual field strength $\hd{}F_{p+2} = (\hd{}F)_{8-p}$. This corresponds to looking either for the solution of a $p$-brane or for the one of a $(6-p)$-brane, which is its electric-magnetic dual object. The simplest form for the electric ansatz is:
\begin{equation}
	C_{p+1} = (e^{C(r)}-1)\ dx^0 \wedge \dotsm \wedge dx^p\,.
\end{equation}

One then substitutes the above ansatz into the equations of motion. The final important ingredient is given by the fact that the correct normalizations are obtained by using the world-volume action~\eqref{Dpwv} as a source term added to the supergravity action~\eqref{II}:
\begin{equation}
	S = S_{\text{II}} + S_{\text{D}p}\,,
\end{equation}
where $S_{\text{D}p}$ is written as an integral over the ten-dimensional space-time by means of suitable delta functions in the transverse directions. This has the effect of fixing the functions $A(r)$, $B(r)$ and $C(r)$ in terms string parameters, via the quantity $\tau_p$ appearing in the world-volume action.

The solution of the equations of motion, when $F_{p+2}$ is interpreted as a R-R field strength, yields the following field configuration of type II supergravity:
\begin{equation}\label{DpsolE}
\begin{aligned}
	ds^2 &= H_p(r)^{-\frac{7-p}{8}}\ \eta_{\alpha\beta} dx^\alpha dx^\beta
		+ H_p(r)^{\frac{p+1}{8}}\ \delta_{ij} dx^i dx^j\,,\\
	e^{\Phi} &= H_p(r)^{\frac{3-p}{4}}\,,\\
	C_{p+1} &= (H_p(r)^{-1} - 1)\ dx^0 \wedge \dotsm \wedge dx^p\,,
\end{aligned}
\end{equation}
where $H_p(r)$ is a harmonic function in the transverse space given by:
\begin{equation}\label{H}
	H_p(r) = 1 + \frac{Q_p}{r^{7-p}}\,,\qquad
	Q_p = \frac{2\kappa^2\tau_p}{(7-p)\Omega_{8-p}}\,,
\end{equation}
where $\Omega_{q} = (2\pi)^{(q+1)/2} / \Gamma ((q+1)/2)$ is the volume of a unit $q$-sphere and the above expression is valid for $p < 7$. Notice that by ``factorizing the Newton constant'' in $Q_p$ one can easily read the tension of the $p$-brane, which is equal to $\tau_p$, namely proportional to $\gs^{-1}$. This again nicely fits with the fact that D$p$-branes are non-perturbative configurations of string theory that, given the dependence on the coupling, have no direct analogue in field theory. In fact, this result agrees perfectly with the appearance of $\tau_p$ as a coefficient of the Dirac--Born--Infeld action, as well as with the interpretation of D-branes in open string theory. Finally, the R-R charge of the brane can be extracted by computing:
\begin{equation}
	\mathcal{Q}_p = \frac{1}{\sqrt{2}\kappa} \int_{S^{8-p}} \hd{} dC_{p+1}
		= \sqrt{2}\kappa\tau_p\,.
\end{equation}
The fact that the charge is equal to the tension (in appropriate units) is a manifestation of the BPS property of D$p$-branes. We also see that $\mathcal{Q}_p\mathcal{Q}_{6-p} = 2\pi$, which is the correct Dirac quantization relation for electric-magnetic dual objects such as a D$p$ and a D$(6-p)$-brane.

Notice also that the metric of the solution~\eqref{DpsolE} can be written in the string frame in an even simpler form:
\begin{equation}\label{Dpsolst}
	ds^2 = H_p(r)^{-1/2}\ \eta_{\alpha\beta} dx^\alpha dx^\beta
		+ H_p(r)^{1/2}\ \delta_{ij} dx^i dx^j\,.
\end{equation}

Let us make some observation on these D-brane geometries. First of all, we can generalize the solution~\eqref{DpsolE} to the case of the geometry generated by $N$ coincident BPS $p$-branes easily, by changing $Q_p\to NQ_p$ in~\eqref{H}, and this will be useful later on. Next, we notice that all these solutions (for $p\neq3$) present a horizon at $r=0$, which is in fact a singular place of zero area. Instead, in the case of $N$ D3-branes one can see that the inverse quartic power of $r$ appearing in~\eqref{H} yields a cancellation between the vanishing of the horizon size and the divergence of the metric, leaving a horizon of finite size $r_{H} = \ls (4\pi\gs N)^{1/4}$. We will see some of the relevance of this fact in section~\ref{s:adscft}. As a final remark, notice that the expression~\eqref{DpsolE} is not accurate for the case of D3-branes since it does not take into account the self-duality of $\tilde{F}_5$. We have therefore to rewrite the solution in the following form:
\begin{equation}\label{D3E}
\begin{aligned}
	ds^2 &= H_3(r)^{-1/2}\ \eta_{\alpha\beta} dx^\alpha dx^\beta
		+ H_3(r)^{1/2}\ \delta_{ij} dx^i dx^j\,,\\
	e^{\Phi} &= 1\,,\\
	F_5 &= d H_3(r)^{-1} \wedge dx^0 \wedge \dotsm \wedge dx^3
		+ \hd{} ( d H_3(r)^{-1} \wedge dx^0 \wedge \dotsm \wedge dx^3 )\,.
\end{aligned}
\end{equation} 

\subsection*{F-strings and NS5-branes}

Since we started looking for classical supergravity solutions describing $p$-dimensional charged objects, we may wonder if there are any more of them. In fact, we did not consider the possibility of objects charged under the NS-NS two-form potential $B_2$. For instance, we know that the perturbative string of section~\ref{s:strings} is the fundamental electric source $B_2$, and we could look for its magnetic dual, namely a five-dimensional object electrically charged under an NS-NS six-form potential $B_6$. In fact, both solutions can be found, and we will summarize them here together with some observations.

Imposing an electric ansatz for $B_2$, one finds the following classical solution for the \emph{F-string} (that could also be called a ``NS1-brane'', namely a one-brane charged under the NS-NS field $B_2$) in the Einstein frame:
\begin{equation}\label{F1}
\begin{aligned}
	ds^2 &= Z_1(r)^{-3/4}\ \eta_{\alpha\beta} dx^\alpha dx^\beta
		+ Z_1(r)^{1/4}\ \delta_{ij} dx^i dx^j\,,\\
	e^{\Phi} &= Z_1(r)^{-1/2}\,,\\
	B_2 &= (Z_1(r)^{-1} - 1)\ dx^0 \wedge dx^1\,,
\end{aligned}
\end{equation}
where $Z_1(r)$ is again a harmonic function in transverse space given by:
\begin{equation}\label{Z1}
	Z_1(r) = 1 + \frac{K_1}{r^6}\,,\qquad
	Q_p = \frac{2\kappa^2 T}{6\Omega_7}\,,
\end{equation}
where $T= \tfrac{1}{2\pi\ls^2}$ is both the tension of the perturbative string and of the F-string. In fact, they can be seen as complementary descriptions of the same object (but notice the fact that the classical solution describes a string of infinite length).

The magnetic dual of the F-string is the \emph{NS5-brane}, whose solution in the Einstein frame reads:
\begin{equation}\label{NS5}
\begin{aligned}
	ds^2 &= Z_5(r)^{-1/4}\ \eta_{\alpha\beta} dx^\alpha dx^\beta
		+ Z_5(r)^{3/4}\ \delta_{ij} dx^i dx^j\,,\\
	e^{\Phi} &= Z_5(r)^{1/2}\,,\\
	B_6 &= (Z_5(r)^{-1} - 1)\ dx^0 \wedge \dotsm \wedge dx^5\,,
\end{aligned}
\end{equation}
where:
\begin{equation}\label{Z5}
	Z_1(r) = 1 + \frac{\ls^2}{r^2}\,.
\end{equation}
Notice that the solutions~\eqref{F1} and~\eqref{NS5} are present in both type II theories. One can see that the tension of the NS5-branes goes like $\gs^{-2}$, which means that it is also a non-perturbative object of string theory, and, if compared to the D-branes, has a more standard behavior of its mass, typical of field theory solitons. However, this is also an indication of the fact that NS5-branes do not have an interpretation in terms of open strings as D-branes have, and in fact it is difficult to study such objects beyond the regime of the classical solution.

\subsection*{M-branes}

Low-energy M-theory, namely eleven-dimensional supergravity, has also brane-like solutions. Since the only gauge potential in the action~\eqref{11dsugra} is the three-form $A_3$, we will have an M2-brane and its magnetic dual, the M5-brane. The M2-brane solution reads:
\begin{equation}\label{M2}
\begin{aligned}
	ds^2 &= f_2(r)^{-2/3}\ \eta_{\alpha\beta} dx^\alpha dx^\beta
		+ f_2(r)^{1/3}\ \delta_{ij} dx^i dx^j\,,\\
	A_3 &= f_2(r)^{-1} dx^0 \wedge dx^1 \wedge dx^2\,,
\end{aligned}
\end{equation}
where:
\begin{equation}
	f_2 (r) = 1 + \frac{\pi {l_{\text{P}}}^3}{r^3}\,.
\end{equation}
The M5-brane solution reads instead:
\begin{equation}\label{M5}
\begin{aligned}
	ds^2 &= f_5(r)^{-1/3}\ \eta_{\alpha\beta} dx^\alpha dx^\beta
		+ f_5(r)^{2/3}\ \delta_{ij} dx^i dx^j\,,\\
	A_6 &= f_5(r)^{-1} dx^0 \wedge dx^1 \wedge dx^2\,,
\end{aligned}
\end{equation}
where
\begin{equation}
	f_5 (r) = 1 + \frac{32\pi ^2 {l_{\text{P}}}^6}{r^6}\,.
\end{equation}

It is interesting to trace the eleven-dimensional origin of the branes of type IIA theory, when eleven-dimensional supergravity is compactified along the eleventh direction $x^{10}$. Let us start from the M2-brane. If one of its world-volume directions is along $x^{10}$, we will be left with the F-strings of type IIA, while if the M2-brane is transverse to $x^{10}$ we will obtain a type IIA D2-brane (one can check that this is indeed the result of the dimensional reduction of the solutions given above). Analogously, an M5-brane extended along $x^{10}$ will give rise to a D4-brane, while an M5-brane transverse to $x^{10}$ will result in a type IIA NS5-brane.

What about the D0 and D6-branes? Notice that the tension~\eqref{taup} of a D0-brane is $\tau_0 = \tfrac{1}{\gs\ls}$. Since D0-branes are BPS, the tension of $N$ D0-branes will be given by $N\tau_0 = \tfrac{N}{R_{10}}$, where $R_{10} = \gs\ls$. Moreover, when we go to the strong coupling regime of type IIA theory $\gs\to\infty$, we see that the mass spectrum of the D0-branes approaches a continuum. This is precisely the behavior appropriate for Kaluza--Klein modes, so we are uncovering an additional dimension of radius $R_{10}$! This is a hint of the type IIA/M-theory duality, that we do not have the space of describing in detail. However, we see that D0-branes have the eleven-dimensional interpretation of Kaluza--Klein states of the compactification. Their magnetic dual the D6-branes are instead interpreted as eleven-dimensional ``Kaluza--Klein monopoles''.

\section{Gauge theory from gravity: A first example}\label{s:probe}

Let us summarize one more time what we have found (and depicted in figure~\ref{f:openclosed}) about D-branes, and in particular their low-energy interpretations:
\begin{itemize}
\item On the one hand, the low-energy dynamics of massless open string states on a D$p$-brane gives rise to a $(p+1)$-dimensional supersymmetric gauge theory;
\item On the other hand, a D$p$-brane is described in the closed string channel as a classical solution of the low-energy supergravity equations of motion charged under the R-R field $C_{p+1}$.
\end{itemize}
A natural question to ask is if how we can use the interplay between these two interpretations in order to compute gauge theory quantities from classical supergravity, and vice-versa. The big amount of concepts, techniques and methods exploring this interplay can collectively be called \emph{gauge/gravity correspondence}, and will be our main focus from now on.

Let us then start with a first simple example on how quantum information about the gauge theory living on a D-brane is encoded in the corresponding supergravity solution and can be extracted from it. We have already derived the two main ingredients, that we rewrite here for convenience. The first is the classical solution~\eqref{DpsolE}-\eqref{Dpsolst} of type II supergravity describing the geometry generated by $N$ D$p$-branes placed at the origin of flat space (in the string frame):
\begin{equation}\label{Dpsol2}
\begin{aligned}
	ds^2 &= H_p^{-1/2}\ \eta_{\alpha\beta} dx^\alpha dx^\beta
		+ H_p^{1/2}\ \delta_{ij} dx^i dx^j\,,\\
	e^{\Phi} &= H_p^{\frac{3-p}{4}}\,,\\
	C_{p+1} &= (H_p^{-1} - 1)\ dx^0 \wedge \dotsm \wedge dx^p\,,
\end{aligned}
\end{equation}
where the harmonic function $H(r)$, $r= (x^i x^i)^{1/2}$, given in~\eqref{H}, reads:
\begin{equation}
	H_p(r) = 1 + \frac{Q_p}{r^{7-p}}\,,\qquad
	Q_p = \frac{2\kappa^2\tau_pN}{(7-p)\Omega_{8-p}}\,.
\end{equation}

The second ingredient is the world-volume action describing the low-energy dynamics of the theory living on a single D$p$-brane, whose bosonic part~\eqref{Dpwv} in the string frame reads:
\begin{equation}\label{Dpwv2}
\begin{split}
	S_{\text{D}p} = &- \tau_p \int d^{p+1}\xi\ e^{-\Phi}
		\sqrt{- \det \left(\hat{G}_{ab} + \hat{B}_{ab} + 2\pi\ls^2 F_{ab}\right) }\\
		 &+ \tau_p \int_{\mathcal{M}_{p+1}} \sum_q \hat{C}_q \wedge e^{\hat{B}+2\pi\ls^2 F}\,.
\end{split}
\end{equation}

\begin{figure}
\begin{center}
\includegraphics[scale=.7]{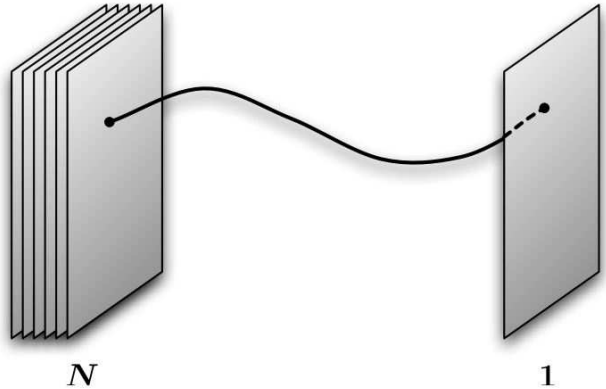}
\caption{{\small Probing the geometry generated by $N$ D-branes with a single moving D-brane.}}
\label{f:probe}
\end{center}
\end{figure}

The procedure we are going to follow in order to find the gauge coupling constant consists in substituting the fields of the solution~\eqref{Dpsol2} into the action~\eqref{Dpwv2}. Why is this supposed to work? Let us present two different interpretations of such a procedure, which will both be useful in later chapters.
\begin{itemize}
\item The first interpretation is geometrical. We can think of forming the D-brane configuration generating the supergravity solution as a step-by-step procedure, by taking the branes one at a time from infinity. This is of course a supersymmetric operation since the branes are BPS objects and there is no force between branes of equal dimension. We can therefore think of adding another brane to the system, a \emph{probe brane}, and of moving it slowly in the transverse space away from the other branes placed at the origin, as in figure~\ref{f:probe}. Neglecting the back-reaction of the probe, we can then use its effective world-volume action to explore the geometry. In particular, its moduli space will tell us information about the Coulomb branch of the gauge theory, where the gauge group is broken as $U(N+1) \to U(N)\times U(1)$.
\item The second interpretation is field-theoretic. We can simply think of studying the low-energy dynamics of the open strings living on the $N$ D-branes by using the world-volume action and carefully taking the $\ls\to0$ limit. Of course, a complete study would require the use of a fully non-abelian version of the action~\eqref{Dpwv2}, describing the low-energy dynamics of the theory living on $N$ branes. Such a non-abelian action is not completely known at the moment of writing, even if many prescriptions exist, with various levels of validity. However, we already mentioned in section~\ref{s:wv} that it is known that this non-abelian action reduces at low energies to the action of non-abelian Super Yang--Mills theory. Therefore, in order to extract information on the gauge theory living on the branes, we can think of using the action~\eqref{Dpwv2}, taking the $\ls\to0$ limit and then ``promote'' the resulting fields to non-abelian ones.
\end{itemize}
%:Figure: Probe
Both the described approaches are able to give relevant information on the low-energy gauge theory, and we will use them alternatively. Whenever they can be used at the same time, they can be shown to yield the same results for the gauge theory. However, each of them has advantages and disadvantages. For instance, the probe approach has the good point of being precisely related to a particular phase of the gauge theory, so that more information can be extracted on it (like the metric on the moduli space), but on the other hand, in order to be implemented, it needs the presence at least of a part of transverse space where the branes can be freely and supersymmetrically moved, a situation that is not always possible to obtain. The second approach does not have neither the above advantage nor the above disadvantage, and it is more difficult to establish a precise relation between gauge theory scales and supergravity coordinates. All these aspects are not very relevant in the highly supersymmetric setup we are considering in this section (recall that we have 16 unbroken supercharges), but will be extremely important in later chapters, when studying the application of these methods to less supersymmetric gauge theories.

In any case, let us proceed in our analysis. Our starting point is taking the \emph{static gauge}:
\begin{equation}
	x^a = \xi^a\,,\qquad
	x^i = x^i (\xi^a)\,,
\end{equation}
which can be chosen by implementing diffeomorphism invariance on the world-volume and simplifies the analysis considerably. The square root of the determinant in~\eqref{Dpwv2} can be expanded as follows:
\begin{multline}
	\sqrt{- \det \left(\hat{G}_{ab} + \hat{B}_{ab} + 2\pi\ls^2 F_{ab}\right) }\\
		= \sqrt{- \det G_{ab} } \left\{
		1 + \frac{1}{2} G^{ab}G_{ij} \partial_a x^i \partial_b x^j
		+ \frac{(2\pi\ls^2)^2}{4} G^{ac}G^{bd}F_{ab}F_{cd}\right\}\,,
\end{multline}
which can be interpreted, depending on the perspective, as a $\ls\to 0$ expansion, or (in the probe approach) as an expansion for slowly-varying world-volume fields. Using the explicit solution~\eqref{Dpsol2}, the action~\eqref{Dpwv2} now becomes:
\begin{equation}
	S_{\text{D}p} \simeq - \tau_p \int d^{p+1}x\
		\left\{ H^{-1} + \frac{1}{2} \partial_a x^i \partial_a x^i + \frac{(2\pi\ls^2)^2}{4}
		F_{ab}F_{ab}\right\}\
		 + \tau_p \int d^{p+1}\xi\ ( H^{-1} - 1 )\,.
\end{equation}
We see that all position-dependent terms cancel, as had to be expected because of the no-force condition which is a consequence of supersymmetry. Neglecting the unimportant constant term in the above action, and recalling that the coordinate fields $x^i$ are related to the gauge theory scalars as:
\begin{equation}
	x^i = 2\pi\ls^2 \Phi^i\,,
\end{equation}
we can rewrite the above result as:
\begin{equation}
	S_{\text{D}p} = - \tau_p \frac{(2\pi\ls^2)^2}{2}\int d^{p+1}x\
		\left\{ \frac{1}{2} \partial_a \Phi^i \partial_a \Phi^i + \frac{1}{4} F_{ab}F_{ab}\right\}\,,
\end{equation}
where we have included an additional factor of $\tfrac{1}{2}$ due to the normalization of the generators of the gauge group once we promote everything to the non-abelian level. The resulting action is precisely the kinetic bosonic part of the action of Super Yang--Mills theory in $p+1$ dimensions with 16 supercharges:
\begin{equation}\label{Dpprobe}
	S_{\text{SYM}} = - \frac{1}{g^2_{\text{D}p}} \int d^{p+1}x\
		\left\{ \frac{1}{2} \partial_a \Phi^i \partial_a \Phi^i + \frac{1}{4} F_{ab}F_{ab}\right\}\,,
\end{equation}
once we identify:
\begin{equation}\label{Dpcoupling}
	g^2_{\text{D}p} = \frac{2}{(2\pi\ls^2)^2 \tau_p} = 2 (2\pi)^{p-2} \gs \ls^{p-3}\,.
\end{equation}
This result gives the gauge coupling constant of the Super Yang--Mills theory living on $N$ D$p$-branes in flat space at low energies, and is the first piece of quantum information on a gauge theory that we recover from a classical supergravity solution. This result tells us that the coupling constant of Super Yang--Mills theory in $p+1$ dimensions (an example of which, that we will often consider in the following, is \Ne{4} Super Yang--Mills in four space-time dimensions, for which $g^2_{D3} = 4\pi\gs$) is simply a constant and does not run with the scale, apart the trivial dependence on the string scale $\ls$ which is necessary for dimensional reasons. Of course, at this level one could argue that we have no direct means of verifying that this is a quantum result and not simply the (identical) classical one, but we will later see in many less trivial examples that this method is indeed able to give us a lot of quantum information on gauge theories from gravity.

We can make some additional observations. Notice first that in the above action the scalars and gauge field kinetic terms come with the same coefficient, and this is another consequence of supersymmetry which is correctly taken into account. Second, if we interpret the above computation as a probe analysis, thus exploring the Coulomb branch of the theory when the gauge group is broken with the pattern $U(N+1)\to U(N)\times U(1)$, we can interpret the coefficient in front of the $9-p$ scalars in the effective action~\eqref{Dpprobe} as the metric on the moduli space of the theory, which in this case is simply:
\begin{equation}
	ds^2_{\mathcal{M}} = \frac{1}{g^2_{\text{D}p}} \delta_{ij} d\Phi^i d\Phi^j\,,
\end{equation}
as it should be due to the high amount of supersymmetry which does not allow additional structure: the moduli space is flat.

Having found the metric on the moduli space, let us now make another observation. The D$p$-brane solution~\eqref{Dpsol2} has only the R-R field $C_{p+1}$ turned on, so the expansion of the Wess-Zumino part of the world-volume action~\eqref{Dpwv2} just contains the standard D-brane minimal coupling. However, from a gauge theory point of view, it is interesting to see that the expansion of the Wess-Zumino action (for $p\ge3$) also contains the following ``theta-term'':
\begin{equation}\label{p-4inst}
	\frac{\tau_{p-4}}{8\pi^2}\int_{\mathcal{M}_{p+1}} C_{p-3}\wedge F \wedge F\,,
\end{equation}
where we used the explicit expression~\eqref{mup} to relate D-brane charges. In the case at hand, as we said, the solution~\eqref{Dpsol2} has vanishing $C_{p-3}$, so we can only conclude that we are in a theta-vacuum with $\tym=0$, and, as expected, no quantum anomaly to modify this result.

However, what is interesting about the expression~\eqref{p-4inst}, once we promote it to the non-abelian level, is that if we excite an instanton configuration in the gauge theory living on a D$p$-brane, namely a (euclidean) configuration with integer topological charge $k = \frac{1}{8\pi^2}\int \tr (F\wedge F)$, this precisely corresponds to turn on $k$ units of D$(p-4)$-brane charge. In fact, since \eqref{p-4inst} gives the theta-term in the Yang--Mills action of the $(p+1)$-dimensional gauge theory,we can guess that a D$(p-4)$-brane behaves as an instanton configuration of the theory on a D$p$-brane.

We may then think of recovering some information about instantons by probing the D$p$-brane solution with a D$(p-4)$-brane, instead of using a brane of the same dimension. In fact, as we will summarize in section~\ref{s:dictionary}, in more complicated setups a computation like~\eqref{p-4inst}, as well as a D$(p-4)$-probe analysis, will give us some interesting pieces of information about anomalies and instantons of the gauge theory.

\section{The AdS/CFT correspondence}\label{s:adscft}

In the previous section, we have seen a first example about how gauge theories arise from classical solution describing the geometry of $N$ D$p$-branes. Now let us concentrate on the case $p=3$. The world-volume theory on $N$ D3-branes at low energy is four-dimensional \Ne{4} Super Yang--Mills theory (sixteen preserved supercharges) with $U(N)$ gauge group. This theory is classically and quantum-mechanically scale-invariant, so the Poincar\'e symmetry combines with dilatations and special conformal transformations to give the conformal group $SO(2,4)\simeq SU(2,2)$. In addition, there is an R-symmetry group $SU(4)_R\simeq SO(6)_R$. The combination of all these symmetries with \Ne{4} supersymmetry requires the addition of sixteen conserved conformal supersymmetries and enlarges the whole global symmetry group to the supergroup $SU(2,2|4)$.

After this brief survey of the symmetries of the gauge theory living on the D3-branes, let us turn to the corresponding geometry. A stack of $N$ D3-branes in the string frame is described by the classical solution~\eqref{D3E}:
\begin{equation}\label{D3}
\begin{aligned}
	ds^2 &= H_3^{-1/2}\ \eta_{\alpha\beta} dx^\alpha dx^\beta
		+ H_3^{1/2}\ \delta_{ij} dx^i dx^j\,,\\
	e^{\Phi} &= 1\,,\\
	F_5 &= d H_3^{-1} \wedge dx^0 \wedge \dotsm \wedge dx^3
		+ \hd{} ( d H_3^{-1} \wedge dx^0 \wedge \dotsm \wedge dx^3 )\,,
\end{aligned}
\end{equation} 
where we recall the expression of the warp factor $H_3(r)$:
\begin{equation}
	H_3(r) = 1 + \frac{R^4}{r^4}\,,\qquad R^4 \equiv Q_3 = 4\pi\gs N \ls^4\,.
\end{equation}

We want to study the asymptotic regions of this geometry. Far from the sources, namely for $r\gg R$, the solution~\eqref{D3} approaches Minkowski space, since $H_3\sim 1$. It is more interesting to consider the ``near-horizon'' limit, namely the region $r \ll R$. In this case we can approximate $H_3 \sim \tfrac{R^4}{r^4}$, and the metric in~\eqref{D3} becomes:
\begin{equation}
	ds^2 = \left( \frac{r^2}{R^2}\ \eta_{\alpha\beta} dx^\alpha dx^\beta 
		+ \frac{R^2}{r^2} dr^2 \right) + R^2 d\Omega_5^2\,,
\end{equation}
where we have introduced spherical coordinates on the transverse space, and where $d\Omega_5^2$ is the metric on a round five-sphere. We see that the metric is naturally decomposed into two terms, one of which represents a five-sphere of radius $R$. What does the other term represent? With the change of coordinates $z=\tfrac{R^2}{r}$, the above metric becomes:
\begin{equation}\label{ads}
	ds^2_{\ads} = \frac{R^2}{z^2} \left( \eta_{\alpha\beta} dx^\alpha dx^\beta 
		+ dz^2 \right) + R^2 d\Omega_5^2\,,
\end{equation}
and we recognize the metric of five-dimensional Anti-de Sitter space written in the so-called ``Poincar\'e patch'' of coordinates. We have then shown that the geometry resulting from the near-horizon limit of the background generated by $N$ D3-branes is the space \ads~\cite{Maldacena:1998re}, where the radii of the two spaces are both given by $R$. In some sense, the D3-brane geometry~\eqref{D3} can be thought of as an interpolating geometry between Minkowski space and the \ads{} geometry~\eqref{ads}.

A first interesting observation we can make is that the isometry group of the solution~\eqref{ads} is $SO(2,4)\times SO(6)$, the two factors coming respectively from Anti-de Sitter space and from the sphere. These symmetries are exactly the conformal group and R-symmetry group of \Ne{4} Super Yang--Mills theory! Notice however that in the near-horizon limit that we have taken, there are no D3-branes left, but rather the geometry is a purely closed string background. In fact, one of the most interesting part of the story is that in the limit the open string dynamics (and then the gauge theory) and the closed string dynamics completely decouple, since one is taking $\ls\to0$ while keeping fixed $\gs$, $N$ as well as all physical length scales. And in fact, it is only in the near-horizon limit that one is able to show that also in the gravitational solution the full global symmetry is enlarged to $SU(2,2|4)$, precisely matching the one of the gauge theory.

These and other observations, about symmetries and the decoupling limit, suggest a bold conjecture about the equivalence between two drastically different theories. In fact, the \emph{Maldacena} or \emph{AdS/CFT conjecture}~ \cite{Maldacena:1998re,Gubser:1998bc,Witten:1998qj} (for a standard review see~\cite{Aharony:1999ti}) states that the two following theories are \emph{equivalent}:
\begin{itemize}
\item Type IIB superstring theory on \ads with string coupling $\gs$, where both $AdS_5$ and $S^5$ have radius $R$ and there are $N$ units of five-form flux $\tilde{F}_5$ through the $S^5$;
\item Superconformal \Ne{4} Super Yang--Mills theory with gauge group $U(N)$ and gauge coupling constant $g_{\text{YM}}$\,,
\end{itemize}
with the following identification of parameters:
\begin{equation}\label{adsparam}
	\gym = 4\pi\gs\,,\qquad R^4 = 4\pi\gs N\ls^4\,.
\end{equation}
By equivalence or \emph{duality} it is meant that there is a precise map between states and correlators in the two theories.

We stated the AdS/CFT duality in its stronger form, which unfortunately turns out to be almost untractable, because of profound difficulties in the quantization of string theory on a curved space such as \ads. Let us now present some limits of the above correspondence that are far more tractable, yet still highly non-trivial.

First, we can think of taking 't Hooft large $N$ limit on the gauge theory side~\cite{Hooft:1974jz}:
\begin{equation}
	N\to\infty\,,\qquad \lambda \equiv \gym N \text{ fixed.}
\end{equation}
In this limit, the diagrammatic expansion of the theory rearranges into a topological expansion of Feynman diagrams. Given~\eqref{adsparam}, we see that this corresponds to taking the weak coupling limit $\gs\to0$ in the string theory. Therefore, \Ne{4} Super Yang--Mills theory in the 't Hooft limit is dual to \emph{classical type IIB strings} on \ads.

Starting from the 't Hooft limit, we can think of additionally taking the strong coupling limit $\lambda\to\infty$. Remarkably, on the string theory side this corresponds to taking $\ls\to0$, and we are left with low-energy supergravity on \ads, namely a very tractable classical limit! We summarize the three levels of the correspondence in table~\ref{t:ads}.
%:Table: AdS/CFT
\begin{table}
\begin{center}
\begin{tabular}{|ccc|}
\hline
Type IIB string theory&&\Ne{4} $U(N)$\\
on \ads&&Super Yang--Mills theory\\
\hline
$\gs\,,\quad R/\ls$
& $\longleftrightarrow$
& $g_{\text{YM}}\,,\quad N$\\
&$\gym = 4\pi\gs$&\\
&$(R/\ls)^4 = 4\pi\gs N$&\\
\hline
Classical strings
& $\longleftrightarrow$
& 't Hooft limit \\
$\gs\to 0\,,\quad R/\ls$ fixed.
& & $N\to\infty\,,\quad \lambda=\gym N$ fixed.\\
\hline
Classical supergravity
& $\longleftrightarrow$
& Large 't Hooft coupling limit \\
$\gs\to0\,,\quad R/\ls\to\infty\,.$
& & $N\to\infty\,,\quad \lambda\to\infty\,.$\\
\hline
\end{tabular}
\end{center}
\caption{{\small Three forms of the AdS/CFT conjecture, in order of decreasing strength.}}
\label{t:ads}
\end{table}

The AdS/CFT duality we just introduced is probably the most powerful and best understood example of gauge/gravity correspondence. Let us stress again that it intends to be an \emph{exact} duality between a gauge theory in four dimensions and a theory without gauge degrees of freedom in a higher dimension. In some sense, then, all the information on the bulk higher-dimensional theory can be thought to be encoded in the four-dimensional one, as a sort of hologram. This gives rise to the concept of \emph{holography}~\cite{Hooft:1993gx,Susskind:1995vu}.
% "Engineering Gauge Theories"

\chapter{Engineering Gauge Theories}\label{c:engineering}

If we think that gauge theories arising from string theory should finally have something to do with the physical world, we might be interested in seeing if string and D-brane constructions can give rise to gauge theories with more ``realistic'' properties with respect to the ones we have encountered in the previous chapter.

What do we precisely mean by ``realistic''? Let us concentrate for a while on the case of four space-time dimensions. The theory living on a stack of D3-branes in flat space is \Ne{4} Super Yang--Mills, which is a theory with a high amount of preserved supersymmetry and superconformal invariance. Since the Standard Model of particle physics has neither supersymmetry nor conformal invariance, we would like to act in these two directions and try, by using string theory, to engineer gauge theories with less than sixteen supercharges and broken conformal symmetry. More generally, looking also at theories in dimensions different from four, whose dimensionful coupling makes them in any case non-conformal, we would like to build theories exhibiting a scale anomaly.

In this chapter, we will present some of the methods by which scale-anomalous gauge theories with reduced supersymmetry can be engineered in string theory via appropriate configurations of D-brane systems. Although we will not be able to reach non-supersymmetric theories, the analysis of this chapter will be the basis for the more concrete examples of gauge/gravity correspondence to be presented in chapters~\ref{c:8susy} and~\ref{c:4susy}.

\section{How to get realistic theories?}\label{s:realistic}

There are several useful methods for breaking the sixteen supersymmetries preserved by the theories living on D-branes in flat space. Among the most fruitful ones, there is the possibility of deforming highly supersymmetric theories by adding supersymmetry breaking masses/superpotential terms into the action, and then exploring what are the consequences of the deformation on the dual geometry. This is a method we will not pursue here, and we just refer the reader for instance to the review~\cite{Aharony:2002up} and its references.

We will instead follow a more direct approach, consisting in engineering D-brane configurations of type II string theory which, by means of open string computations or other considerations, can be seen to describe the gauge theories we are interested in. It is important to notice that our engineering procedure usually requires that we follow these two steps:
\begin{enumerate}
\item \emph{Reduce supersymmetry}. This is usually achieved by considering type II theory on an appropriate~\emph{closed string background} known to preserve a specific fraction of the supersymmetries of flat space;
\item \emph{Acquire scale anomaly}. This is usually achieved by engineering particular \emph{configurations of D-branes} (thus this step involves \emph{open} rather than closed strings) in the above closed string backgrounds.
\end{enumerate}

For instance, closed string backgrounds one can use in order to reduce the amount of supersymmetry of the ambient space include:
\begin{itemize}
\item \emph{Calabi--Yau manifolds}. In particular, Calabi--Yau twofolds preserve one-half of the 32 supercharges of flat space, while Calabi--Yau threefolds preserve one-fourth of them. We will mainly be interested in \emph{non-compact} Calabi--Yau manifolds. Let us also note that many useful examples are given by \emph{singular} manifolds, such as the \emph{conifold};
\item \emph{Orbifolds} of type II theories, namely quotients of a part of space-time by a discrete symmetry. These spaces can be seen as geometrically singular points in the moduli space of Calabi--Yau manifolds, which however admit a sensible conformal field theory description in terms of perturbative strings. In particular, $\mathbb{C}^2/\Gamma$ orbifolds, where $\Gamma$ is a discrete subgroup of $SU(2)$, preserve one-half of the supersymmetries of flat space, while one-fourth of the supercharges are preserved by $\mathbb{C}^3/\Gamma$ orbifolds, where $\Gamma\subset SU(3)$;
\item \emph{NS5-brane} backgrounds. Parallel NS5-branes in flat space are BPS configurations breaking 16 out of 32 supersymmetries.
\end{itemize}

Usually (we will be more precise later in specific examples), the standard ``bulk'' or ``regular'' D-branes in the above spaces host theories with reduced supersymmetry, according to the chosen closed string background. However, it can be seen that the gauge theories living on such branes have a matter content that does not make them scale-anomalous (for examples, four-dimensional theories will still have unbroken conformal symmetry). In order to overcome this obstruction, one has to engineer configurations of D-branes whose world-volume is, roughly speaking, topologically non-trivial. For instance one can consider:
\begin{itemize}
\item \emph{D-branes wrapped on supersymmetric cycles} of Calabi--Yau manifolds. We will see in the following that the non-trivial topology of the world-volume allows getting a scale-anomalous theory living at low energies on the flat part of the brane world-volume;
\item \emph{Fractional D-branes} on orbifold or conifold backgrounds. These branes can be thought of as D-branes wrapped on cycles that, in the limit in which the Calabi--Yau manifold degenerates into a metrically singular space, result to be wrapped on shrinking cycles, effectively losing some world-volume directions and being stuck at the singularity of the background;
\item \emph{Stretched D-branes} between NS5-branes. D-brane world-volumes can end on NS5-branes, and this causes the freezing of some of the moduli of the theory. As a result, the gauge theory living at low energies on the intersection of the D-branes and NS5-branes can acquire a scale anomaly.
\end{itemize}

Of course, up to now we just gave a general overview of some of the many available tools to build our more ``realistic'' gauge theories. The rest of this chapter is devoted to study them (and in particular the wrapped and fractional D-branes) in some detail, while also sketching some connections among the different approaches. We will mainly proceed by means of the study of simple examples, from which we can subsequently abstract some general properties of the relevant D-brane systems and the corresponding gauge theories. After the overview presented in this chapter, we will be ready to see the gauge/gravity correspondence at work in some interesting cases in chapters~\ref{c:8susy} and~\ref{c:4susy}.

\section{About D-branes on curved manifolds}\label{s:wrapped}

Let us start by considering type II string theory on a Calabi--Yau manifold. We will mainly be interested in non-compact Calabi--Yau twofolds and threefolds, which preserve respectively one-half and one-fourth of the 32 supersymmetries of flat space. The study of strings and D-branes on curved manifolds, and Calabi--Yau spaces in particular, has been one of the most studied topics in string theory since its first days, so we will just be able to sketch few aspects that we will need for our goal.

To be specific, we will only consider some of the consequences that arise when a D-brane has part of its world-volume extended on a curved space. We will be interested in D-branes wrapped on supersymmetric cycles inside Calabi--Yau manifolds. Notice that what we mean by supersymmetric cycle is precisely that the world-volume theory of the wrapped D-brane preserves supersymmetry.

At first sight, a D-brane whose world-volume extends along a curved manifold will not preserve any supersymmetry, since the condition for finding a covariantly constant spinor will include the spin connection $\omega_\mu^{ab}$ on the curved space and will formally look like:
\begin{equation}\label{nosusy}
	\left(\partial_\mu + \omega_\mu\right)\epsilon = 0\,,
\end{equation}
and this equation does not generally admit any solution. This means that supersymmetry cannot be realized in the world-volume of the curved brane in a conventional manner.

Rather, the low-energy field theory living on the world-volume of the brane has to be partially \emph{topologically twisted}~\cite{Bershadsky:1996qy}. The reason of this fact can be clarified as follows. Consider a D$p$-brane living in a ten-dimensional space-time $M$. There are $9-p$ scalars corresponding to the directions transverse to its world-volume $\mathcal{M}$, which are interpreted as collective coordinates. In more complicated setups, it is not accurate to consider them just as functions on $\mathcal{M}$. Denoting as $TM$ the tangent bundle of $M$, we can decompose it as:
\begin{equation}
	TM = T\mathcal{M} + N_{\mathcal{M}}\,,
\end{equation}
where $N_{\mathcal{M}}$ is the normal bundle of $\mathcal{M}$. Its dimension is precisely $9-p$ and we can then regard our transverse scalars to be actually sections of this bundle, and interpret the R-symmetry group $SO(9-p)_R$ of the world-volume gauge theory as the structure group of the normal bundle $N_{\mathcal{M}}$.

Consider now the theory on a (possibly non-compact) Calabi--Yau $n$-fold, and denote a real $d$-dimensional cycle inside it as $\Sigma_d$. A D$(p+d)$-brane wrapped on $\Sigma_d$ can be represented by the following table, where the symbols $-$, $\bigcirc$ and $\cdot$ respectively represent flat world-volume directions, wrapped world-volume directions and transverse directions:
\begin{center}
\begin{tabular}{|c|c|c|c|c|}
\multicolumn{2}{c}{ }&
\multicolumn{2}{c}{$\overbrace{\phantom{\qquad\qquad\qquad\quad}}^{\text{CY}_n}$}&
\multicolumn{1}{c}{ }\\
\hline
&$\,\,\,\,\mathbb{R}^{1,p}\,\,\,\,$
&$\quad\Sigma_{d}\quad$
&$\quad N_{\Sigma}\quad$
&$\mathbb{R}^{9-2n-p}$\\
\hline
D$(p+d)$ & $-$ & $\bigcirc$ & $\cdot$ & $\cdot$ \\
\hline
\end{tabular}
\end{center}
The normal bundle $N_{\mathcal{M}}$ is now naturally decomposed in two parts, corresponding to the non-trivial part inside the Calabi--Yau (denoted as $N_{\Sigma}$ in the table) and the trivial transverse flat part. Correspondingly, the full Lorentz group results broken as:
\begin{equation}\label{Lbreak}
	SO(1,9) \to SO(1,p) \times SO(d)_\Sigma \times SO(2n-d)_R \times SO(9-2n-p)_R\,.
\end{equation}
Now, the fact that the directions in $N_\Sigma$ are non-trivially fibered over the cycle $\Sigma_d$ means that there is a relation between the connection on the normal bundle $N_{\Sigma}$ and the (spin) connection on $\Sigma$. We have to pick a subgroup $SO(d)_R\subset SO(2n-d)_R$ and implement the following identification:
\begin{equation}\label{toptwist}
	SO(d)_\Sigma = SO(d)_R\,.
\end{equation}
This is a topological twist, since the behavior of all fields under Lorentz transformations, namely their spin, gets changed by this identification imposed by the embedding geometry. Now~\eqref{nosusy} is modified by the presence of the additional external ``gauge'' field $A_\mu$ coupled to the R-symmetry, and becomes:
\begin{equation}\label{twistsusy}
	\left(\partial_\mu + \omega_\mu - A_\mu\right)\epsilon = 0\,.
\end{equation}
The twist~\eqref{toptwist} schematically imposes $\omega_\mu = A_\mu$, so that~\eqref{twistsusy} can be simply satisfied by a constant spinor and supersymmetry can be preserved.

When we dimensionally reduce along the cycle $\Sigma_d$ for studying the low-energy theory living on the flat $\mathbb{R}^{1,p}$ part of the world-volume, the resulting field content will be a direct consequence of the twist~\eqref{toptwist}. More precisely, all fields with charges such that the condition $\omega_\mu = A_\mu$ is satisfied will result in massless fields of the $(p+1)$-dimensional gauge theory we are interested in.

The crucial property of this procedure is that the resulting low-energy theory on $\mathbb{R}^{1,p}$ is an \emph{ordinary} gauge theory, with no topological twist at all. The topological twist of the $(p+d+1)$-dimensional world-volume theory of the D$(p+d)$-brane is precisely what is needed in order to obtain an ordinary field theory living on the un-wrapped directions.

\subsection*{Example: D4-brane wrapped on $S^2$ inside a Calabi--Yau twofold}

Let us then see how the topological twist works in a specific example. We will consider a D4-brane wrapped on a two-sphere inside a Calabi--Yau twofold. As we already mentioned, the closed string background preserves 16 supercharges, and we therefore expect eight supercharges on the world-volume of the D-brane. The configuration is summarized in the following table:
\begin{center}
\begin{tabular}{|c|c|c|c|c|c|c|c|c|c|c|}
\multicolumn{4}{c}{ }&
\multicolumn{4}{c}{$\overbrace{\phantom{\qquad\qquad\qquad}}^{\text{CY}_2}$}
&\multicolumn{3}{c}{ }\\
\hline
&\multicolumn{3}{|c|}{$\mathbb{R}^{1,2}$}
&\multicolumn{2}{|c|}{$S^2$}
&\multicolumn{2}{|c|}{$N_2$}
&\multicolumn{3}{|c|}{$\mathbb{R}^3$}\\
\hline
D$4$ &$-$&$-$&$-$&$\bigcirc$&$\bigcirc$&$\cdot$&$\cdot$&$\cdot$&$\cdot$&$\cdot$\\
\hline
\end{tabular}
\end{center}

In flat space, the presence of the D4-brane breaks Lorentz invariance as $SO(1,9)\to SO(1,4)\times SO(5)_R$. The fact that the D4-brane is wrapped on $S^2$ introduces an additional breaking of $SO(1,4)$ into $SO(1,2) \times SO(2)_{S^2}$. The twist is then introduced by breaking the $R$-symmetry group $SO(5)_R$ into $SO(2)_R \times SO(3)$ and by identifying $SO(2)_R$ with $SO(2)_{S^2}$. In conclusion the pattern of Lorentz symmetry breaking is given by (compare with~\eqref{Lbreak}):
\begin{equation}
	SO(1,9) \to SO(1,4) \times SO(5)_R
		\to SO(1,2) \times SO(2)_{S^2} \times SO(2)_R \times SO(3)\,,
\end{equation}
with the two $SO(2)$ factors identified. The representations of the above groups in which the fields of the gauge theory living on the wrapped D4-branes transform are given by:
\begin{center}
\begin{tabular}{|c|ccc|ccc|}
\hline
&$SO(1,4)$&$\longrightarrow$&$SO(1,2) \times SO(2)_{S^2}$
&$SO(5)_R$&$\longrightarrow$&$SO(2)_R \times SO(3)$\\
\hline
Vector&$\mathbf{5}$&$\longrightarrow$&$(\mathbf{3},\mathbf{1}) \oplus (\mathbf{1},\mathbf{2})$
&$\mathbf{1}$&$\longrightarrow$&$(\mathbf{1},\mathbf{1})$\\
\hline
Scalars&$\mathbf{1}$&$\longrightarrow$&$(\mathbf{1},\mathbf{1})$
&$\mathbf{5}$&$\longrightarrow$&$(\mathbf{1},\mathbf{3}) \oplus (\mathbf{2},\mathbf{1})$\\
\hline
Fermions&$\mathbf{4}$&$\longrightarrow$&$(\mathbf{2},\mathbf{+}) \oplus (\mathbf{2},\mathbf{-})$
&$\mathbf{4}$&$\longrightarrow$&$(\mathbf{+},\mathbf{2}) \oplus (\mathbf{-},\mathbf{2})$\\
\hline
\end{tabular}
\end{center}

The fact that we are interested in the three-dimensional theory living on the flat part of the world-volume at very low energies implies that we must keep only the massless states, which are the ones with appropriate charges under the identified $SO(2)_D\equiv(SO(2)_{S^2}\times SO(2)_R)_{\text{diag}}\,$:
\begin{center}
\begin{tabular}{|c|c|}
\hline
&$SO(1,2) \times SO(2)_D \times SO(3)$\\
\hline
Vector&$(\mathbf{3},\mathbf{1},\mathbf{1})$\\
\hline
Scalars&$(\mathbf{1},\mathbf{1},\mathbf{3})$\\
\hline
Fermions&$(\mathbf{2},+,\mathbf{2})\oplus(\mathbf{2},-,\mathbf{2})$\\
\hline
\end{tabular}
\end{center}
These states form exactly the vector multiplet of three-dimensional Super Yang--Mills theory with eight supercharges. This means that, if we now take a configuration with $N$ D4-branes, the theory living on the flat part of their world-volume is \emph{pure} $U(N)$ \Ne{4} Super Yang--Mills theory in three space-time dimensions, which has a scale anomaly, as one can see for example by computing the coefficient of the one-loop effective coupling constant from insert~\ref{i:running} on page~\pageref{i:running}.

We have therefore shown in an explicit case that in this wrapped brane scenario, because of the topological twist that is enforced by the embedding geometry, the gauge theory living on the flat part of the world-volume of D-branes has the property of being a supersymmetric gauge theory exhibiting a scale anomaly.

In chapter~\ref{c:8susy}, we will use the example of wrapped D4-branes that we just introduced as our first explicit case of application of the gauge/gravity correspondence to a scale-anomalous gauge theory with less than 16 preserved supercharges.

%:Insert: Running coupling
\begin{Insertf}{One-loop running coupling constant}\label{i:running}
The one-loop effective action for a $D$-dimensional field theory expanded around a background which is a solution of the classical field equations can be expressed as:\\
\begin{multline*}
	S_{\text{eff}}=\frac{1}{4g_{\text{YM}}^2}\int d^Dx\Bigg\{
		\bar{F}^a_{\mu\nu}\bar{F}^a_{\mu\nu}+\frac{1}{2}\tr\log\Delta_1
		+\left(\frac{N_s}{2}-1\right)\tr\log\Delta_0\\
		-\frac{N_f}{2}\tr\log\Delta_{1/2}\Bigg\}\,, 
\end{multline*}
where a bar on an operator indicates that the operator is evaluated at the background value $\bar{A}_\mu^a$ of the gauge field, $N_s$ and $N_f$ are respectively the number of scalars and Weyl fermions and where:
\begin{equation*}
	\left(\Delta_1\right)^{ab}_{\mu\nu}=-\left(\bar{D}^2\right)^{ab}\delta_{\mu\nu}
		+2f^{acb}\bar{F}^c_{\mu\nu}\,,\quad
		\left(\Delta_0\right)^{ab}=-\left(\bar{D}^2\right)^{ab}\,,\quad
		\Delta_{1/2}=i\bar{\,/\hspace{-0.27cm}D}\,,
\end{equation*}
$D_{\mu}$ being the covariant derivative and $f^{abc}$ the gauge group structure constants. The part of the above determinants which is quadratic in the gauge fields can be extracted obtaining:
\begin{equation*}
	S_{\text{eff}}=\frac{1}{4}\int d^Dx F^2\left\{\frac{1}{g^2_{\text{YM}}}+I\right\}\,,
\end{equation*}
where:
\begin{equation*}
	I=\frac{1}{(4\pi)^{D/2}}\int_0^\infty \frac{ds}{s^{D/2-1}}e^{-\mu^2s}\ b\,,
\end{equation*}
$\mu$ being the regulating ``mass'' of the fields, and
\begin{equation*}
	b=\frac{D-26}{6}c_v+\frac{2^{[D/2]}N_f}{6}c_f+\frac{N_s}{6}c_s\,,
\end{equation*}
where $[D/2]=D/2$ if $D$ is even and $[D/2]=\tfrac{D-1}{2}$ if $D$ is odd, and where the constants $c$ set the normalization of the generators of the gauge group as $\tr_r ( T^a T^b ) = c_r \delta^{ab}$ in the representation $r$ under which the vector, the fermions and the scalars respectively transform.

In the case $D=3$, we get:
\begin{equation*}
	I_{D=3}=\frac{1}{(4\pi)^{3/2}}\int_0^\infty\frac{ds}{s^{1/2}}e^{-\mu^2s}\ b
		=\frac{b}{8\pi\mu}\,,
\end{equation*}
with
\begin{equation*}
	b_{D=3}=-\frac{23}{6}c_v+\frac{N_f}{3}c_f+\frac{N_s}{6}c_s\,.
\end{equation*}
\end{Insertf}
\begin{Insertl}
In the case $D=4$, we get:
\begin{equation*}
	I_{D=4}=\frac{1}{(4\pi)^{2}}\int_0^\infty\frac{ds}{s}e^{-\mu^2s}\ b\
		\Bigg\lvert_{\text{regular part}}
		=\frac{b}{8\pi^2}\ln\frac{\mu}{\epsilon}
\end{equation*}
where $\epsilon$ is a regulator and:
\begin{equation*}
	b_{D=4}=-\frac{11}{3}c_v+\frac{2N_f}{3}c_f+\frac{N_s}{6}c_s\,.
\end{equation*}
\end{Insertl}

\subsection*{Twists of the D5-brane theory}

Let us consider another example which will be very useful in chapters~\ref{c:8susy} and~\ref{c:4susy}, namely the one of a D5-brane wrapped on a two-sphere. In this case, we will explicitly show two inequivalent topological twists, giving rise to different four-dimensional gauge theories on the flat part of the world-volume of the D5-branes.

In general, in this case the pattern of Lorentz symmetry breaking~\eqref{Lbreak} will read:
\begin{equation}
	SO(1,9) \to SO(1,5) \times SO(4)_R
		\to SO(1,3) \times SO(2)_{S^2} \times SO(4)_R\,,
\end{equation}
and the topological twist will imply the identification of $SO(2)_{S^2}$ with a suitable $SO(2)\subset SO(4)_R$. Let us then write the R-symmetry group as:
\begin{equation}\label{susu}
	SO(4)_R = SU(2)_1 \times SU(2)_2\,,
\end{equation}
and consider two distinct possibilities:
\begin{itemize}
\item A D5-brane wrapped on a two-cycle inside a Calabi--Yau \emph{twofold}, as shown in the following diagram:
\begin{center}
\begin{tabular}{|c|c|c|c|c|c|c|c|c|c|c|}
\multicolumn{5}{c}{ }&
\multicolumn{4}{c}{$\overbrace{\phantom{\qquad\qquad\qquad}}^{\text{CY}_2}$}
&\multicolumn{2}{c}{ }\\
\hline
&\multicolumn{4}{|c|}{$\mathbb{R}^{1,3}$}
&\multicolumn{2}{|c|}{$S^2$}
&\multicolumn{2}{|c|}{$N_2$}
&\multicolumn{2}{|c|}{$\mathbb{R}^2$}\\
\hline
D$5$ &$-$&$-$&$-$&$-$&$\bigcirc$&$\bigcirc$&$\cdot$&$\cdot$&$\cdot$&$\cdot$\\
\hline
\end{tabular}
\end{center}
This corresponds to considering the diagonal subgroup in~\eqref{susu}, $SU(2)_D = (SU(2)_1\times SU(2)_2)_{\text{diag}}$, and identifying the $SO(2)_{S^2}$ with $SO(2)_D\subset SU(2)_D$. By an analysis perfectly analogous to the one performed in the previous subsection, one sees that the following fields survive the topological twist and therefore live at low energies on the four-dimensional flat part of the brane world-volume:
\begin{center}
\begin{tabular}{|c|c|}
\hline
&$SO(1,3) \times SO(2)_D$\\
\hline
Vector&$(\mathbf{4},\mathbf{1})$\\
\hline
Scalars&$2\times(\mathbf{1},\mathbf{1})$\\
\hline
Fermions&$(\mathbf{2}\oplus\mathbf{\bar{2}},+)\oplus
(\mathbf{2}\oplus\mathbf{\bar{2}},-)$\\
\hline
\end{tabular}
\end{center}
The matter content of the low-energy theory is then the one of pure \Ne{2} Super Yang--Mills theory in four space-time dimensions, thus as expected we are left with eight preserved supercharges (and a non-conformal four-dimensional gauge theory!).

\item A D5-brane wrapped on a two-cycle inside a Calabi--Yau \emph{threefold}, as shown in the following diagram:
\begin{center}
\begin{tabular}{|c|c|c|c|c|c|c|c|c|c|c|}
\multicolumn{5}{c}{ }&
\multicolumn{6}{c}{$\overbrace{\phantom{\qquad\qquad\qquad\qquad}}^{\text{CY}_3}$}\\
\hline
&\multicolumn{4}{|c|}{$\mathbb{R}^{1,3}$}
&\multicolumn{2}{|c|}{$S^2$}
&\multicolumn{4}{|c|}{$N_4$}\\
\hline
D$5$ &$-$&$-$&$-$&$-$&$\bigcirc$&$\bigcirc$&$\cdot$&$\cdot$&$\cdot$&$\cdot$\\
\hline
\end{tabular}
\end{center}
In this case, the identification is performed between $SO(2)_{S^2}$ and $SO(2)_1\subset SU(2)_1$, where $SU(2)_1$ is the first factor appearing in~\eqref{susu}. The fields surviving the topological twist (call the identified group again $SO(2)_D$) are:
\begin{center}
\begin{tabular}{|c|c|}
\hline
&$SO(1,3) \times SO(2)_D$\\
\hline
Vector&$(\mathbf{4},\mathbf{1})$\\
\hline
Fermions&$(\mathbf{2},+)\oplus(\mathbf{\bar{2}},-)$\\
\hline
\end{tabular}
\end{center}
Now we are clearly left with four supercharges, and the low-energy theory is pure, non-conformal \Ne{1} four-dimensional Super Yang--Mills theory, whose vector multiplet comprises a vector and a Majorana spinor.
\end{itemize}

\section{A simple $\mathbb{Z}_2$ orbifold}\label{s:Z2}

Let us now leave apart for a while D-branes wrapped on smooth manifolds, and let us instead pass to considering another very interesting situation mentioned in section~\ref{s:realistic}, namely branes in singular spaces. As an explicit example of branes at singularities, we consider D$p$-branes of type II string theory on the orbifold $\mathbb{R}^{1,5}\times$\cz. This is an example we will work out in detail, since it is probably the simplest setting in which scale-anomalous gauge theories on D-branes can be achieved. In an orbifold we can benefit of having full control over perturbative string theory, and choosing a particularly simple one allows to develop our understanding without hindering it with technicalities. We will uncover here many interesting features of D-branes on orbifolds and the corresponding gauge theories, and then formalize them a little more generally in the next section.

The \cz{} orbifold is obtained by identifying the last four coordinates of $\mathbb{R}^{1,9}$ under a reflection:
\begin{equation}\label{Z2}
	x^r \simeq - x^r\,,\qquad r=6,7,8,9,
\end{equation}
which is the action of the unique non-trivial generator $g$ of the $\mathbb{Z}_2$ group. It will also be useful to introduce complex coordinates:
\begin{equation}\label{Z2c}
\begin{aligned}
	z^1 &= x^6 + i x^7 \,,&\qquad z^2 &= x^8 + i x^9 \,.\\
	z^1 &\simeq - z^1\,,& z^2 &\simeq - z^2\,.
\end{aligned}
\end{equation}
Notice that the Lorentz group is broken by the orbifold into $SO(1,9) \to SO(1,5) \times SO(4)$.

The $\mathbb{Z}_2$ action~\eqref{Z2} translates on the bosonic and fermionic oscillators as follows:
\begin{equation}\label{Z2osc}
	\alpha_n^r \to - \alpha_n^r\,,\qquad \talpha_n^r \to - \talpha_n^r\,,\qquad 
	\psi_n^r \to - \psi_n^r\,,\qquad \tpsi_n^r \to - \tpsi_n^r\,,\qquad r=6,7,8,9
\end{equation}
(where tilded oscillators are of course defined for closed strings only).

\subsection*{D-branes on \cz}

We want to introduce in the setting a D$p$-brane transverse to the orbifolded directions, as in the following table:
\begin{center}
\begin{tabular}{|c|ccc|ccc|c|c|c|c|}
\multicolumn{7}{c}{ }&
\multicolumn{4}{c}{$\overbrace{\phantom{\qquad\qquad\quad}}^{\cz}$}\\
\hline
&0&$\cdots$&$p$&$p$+1&$\cdots$&5&6&7&8&9\\
\hline
D$p$ &$-$&$-$&$-$&$\cdot$&$\cdot$&$\cdot$&$\cdot$&$\cdot$&$\cdot$&$\cdot$\\
\hline
\end{tabular}
\end{center}
To summarize, we have splitted the ten space-time coordinates as follows:
\begin{itemize}
\item $x^\alpha\,,\quad \alpha=0,\ldots,p\,:$ directions belonging to the world-volume of the D$p$-brane;
\item $x^i\,,\quad i=p+1,\ldots,5\,:$ non-orbifolded coordinates transverse to the D$p$-brane;
\item $x^A = \{x^\alpha,x^i\}\,,\quad A=0,\ldots,5\,:$ all the non-orbifolded coordinates;
\item $x^r\,,\quad r=6,\ldots,9$ (also denoted as $z^m\,,\quad m=1,2$): orbifolded coordinates (which are all transverse to the D$p$-brane).
\end{itemize}

\begin{figure}
\begin{center}
\includegraphics[scale=.7]{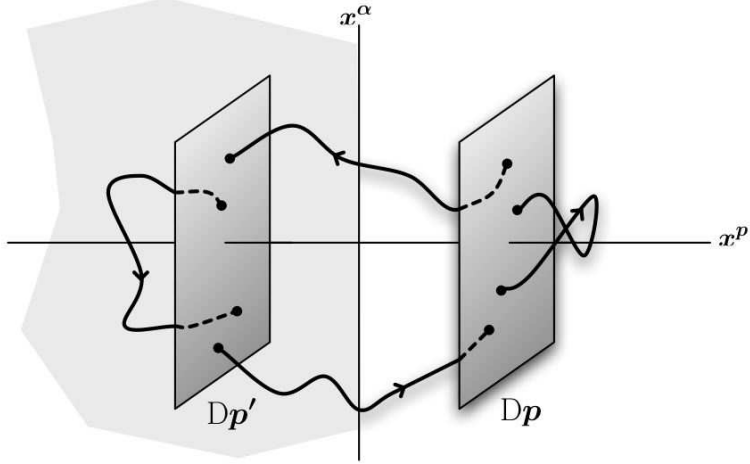}
\caption{{\small A \emph{regular} D$p$-brane in an orbifold has images on the covering space.}}
\label{f:regbranes}
\end{center}
\end{figure}

There is an important observation we can immediately make. Working in the covering space, a single D-brane at an arbitrary point in transverse space is not an invariant configuration under the orbifold action~\eqref{Z2}. This means that, in order to build an invariant configuration, a D$p$-brane at a point $x^r=y^r$ must necessarily have an \emph{image} located at $x^r=-y^r$, as shown in figure~\ref{f:regbranes}.
%:Figure: Regular brane with image
In order to describe the open string spectrum, then, we have to treat four different kinds of strings, namely strings with both endpoints on one of the two branes, or with one endpoint on each. This means we will need a $2\times 2$ Chan--Paton matrix $\lambda$ to identify open string states of the form:
\begin{equation}\label{genregstate}
	\lambda \otimes \text{``oscillators''}\ \ket{0,k}\,,\qquad
	\lambda = \begin{pmatrix}
		\text{D}p-\text{D}p & \text{D}p-\text{D}p' \\
		\text{D}p'-\text{D}p & \text{D}p'-\text{D}p' \\
		\end{pmatrix} \,.
\end{equation}
The action of the $\mathbb{Z}_2$ orbifold on a string state will therefore comprise two parts, the action~\eqref{Z2osc} on the oscillators and an action on the Chan--Paton factors:
\begin{equation}\label{CPaction}
	\lambda\ \to\ \gamma(g)\ \lambda\ \gamma(g)^{-1}\,,
\end{equation}
where $\gamma(g)$ is an appropriate representation of $\mathbb{Z}_2$. How can we find it in the present example? Looking at figure~\ref{f:regbranes}, we see that the action of the element $g$ exchanges the brane with its image, and therefore the action on the general state~\eqref{genregstate} can be chosen as:
\begin{equation}
	\gamma(g) = \sigma^1.
\end{equation}
This is the regular representation of the orbifold group. If we choose for $\lambda$ the basis given by the Pauli matrices (plus the $2\times 2$ identity matrix), $\un$ and $\sigma^1$ are even, while $\sigma^2$ and $\sigma^3$ are odd under~\eqref{CPaction}.

We now have all the ingredients for studying the first level of the open string spectrum. The surviving states are the ones which are even under the combined action of~\eqref{Z2osc} and~\eqref{CPaction}. In the NS sector we are left with:
\begin{center}
\begin{tabular}{|cc|l|}
\hline
\multicolumn{3}{|c|}{NS states} \\
\hline
$\tfrac{\un+\sigma^1}{2} \otimes \psi_{-1/2}^\alpha \ket{0,k}$, &
$\tfrac{\un-\sigma^1}{2} \otimes \psi_{-1/2}^\alpha \ket{0,k}$ &
$\to$ $2$ vectors\\
$\tfrac{\un+\sigma^1}{2} \otimes \psi_{-1/2}^i \ket{0,k}$, &
$\tfrac{\un-\sigma^1}{2} \otimes \psi_{-1/2}^i \ket{0,k}$ &
$\to$ $2(5-p)$ real scalars\\
$\tfrac{\sigma^3+i\sigma^2}{2} \otimes \psi_{-1/2}^r \ket{0,k}$, &
$\tfrac{\sigma^3+i\sigma^2}{2} \otimes \psi_{-1/2}^r \ket{0,k}$ &
$\to$ $4$ complex scalars\\
\hline
\end{tabular}
\end{center}
On the ground state of the R sector, we can choose the action of $g$ to be:
\begin{equation}
	\ket{s_0,s_1,s_2,s_3,s_4} \to (-1)^{s_3+s_4+1} \ket{s_0,s_1,s_2,s_3,s_4}\,,
\end{equation}
so that the surviving states are:
\begin{center}
\begin{tabular}{|cc|l|}
\hline
\multicolumn{3}{|c|}{R states} \\
\hline
$\tfrac{\un+\sigma^1}{2} \otimes \ket{s_0,s_1,s_2,s_3,s_4=s_3}$, &
$\tfrac{\un-\sigma^1}{2} \otimes \ket{s_0,s_1,s_2,s_3,s_4=s_3}$ &
$\to$ $16$ fermions\\
$\tfrac{\sigma^3+i\sigma^2}{2} \otimes \ket{s_0,s_1,s_2,s_3,s_4=-s_3}$, &
$\tfrac{\sigma^3+i\sigma^2}{2} \otimes \ket{s_0,s_1,s_2,s_3,s_4=-s_3}$ &
$\to$ $16$ fermions\\
\hline
\end{tabular}
\end{center}
(where in the counting we have taken into account the fact that the GSO projection selects states for which $\sum s_i = \text{even}$).

The massless open string spectrum is the one of a supersymmetric gauge theory in $p+1$ dimensions, with 8 real supercharges and gauge group $U(1)\times U(1)$. In fact, one vector, $5-p$ real scalars and 8 fermions make up a vector multiplet, so we have two of them. The remaining fields form two matter multiplets that transform in the ``bifundamental'' charge representations $(+1,-1)$ and $(-1,+1)$ of the gauge group (one can easily compute the charges by commuting the relevant Chan--Paton $\sigma^i$ matrices). The matter content of the theory can be summarized in the following \emph{quiver diagram}:
\begin{center}
\includegraphics[scale=.7]{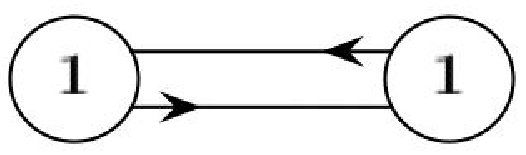}
\end{center}
Quiver gauge theories and quiver diagrams, that we will encounter many times from now on, are described in some more generality in insert~\ref{i:quiver} on page~\pageref{i:quiver}. A very important observation is that the brane configuration has left only one quarter of the original supersymmetries unbroken - half of the supersymmetries of a D-brane in flat space. This is in agreement with the fact that a $\mathbb{C}^2/\Gamma$ orbifold, with $\Gamma \subset SU(2)$, breaks half of the space-time supersymmetry.

If we create a stack of $N$ such \emph{regular} D-branes, the gauge theory will become a $U(N) \times U(N)$ quiver theory with 8 supercharges and two bifundamental matter multiplets transforming in the \hyp{N}{\bar{N}}, \hyp{\bar{N}}{N} representations, namely represented by the following diagram:
\begin{center}
\includegraphics[scale=.7]{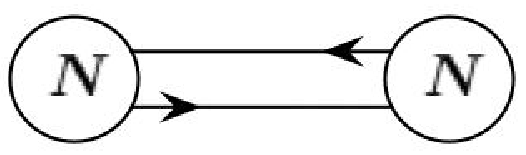}
\end{center}
Notice that the resulting gauge theory has only 8 supersymmetries, but still does not exhibit any scale anomaly due to the matter content. In fact, one can compute the coefficient of the one-loop running coupling constant (see insert~\ref{i:running} on page~\pageref{i:running}) and see that it vanishes.

%:Insert: Quiver gauge theories
\begin{Insertf}{Quiver gauge theories}\label{i:quiver}
A quiver gauge theory can be defined, for all our purposes, as a supersymmetric gauge theory whose field content is encoded in a so-called \emph{quiver diagram}~\cite{Douglas:1996sw}. A quiver diagram consists of \emph{nodes} and \emph{arrows}. The simplest quiver diagram is drawn in the following figure:
\begin{center}
\includegraphics[scale=0.5]{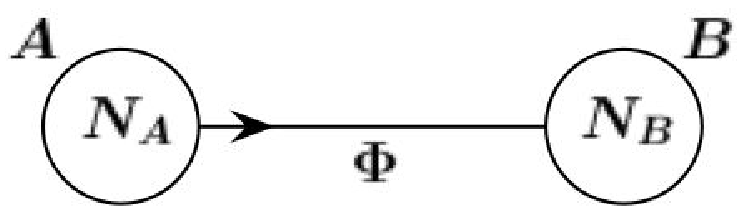}
\end{center}
Each node $i$ represents a gauge group $U(N_i)$ and the corresponding vector multiplet in the adjoint representation, while each arrow represents a matter multiplet transforming in the bifundamental representation of the two gauge groups whose nodes the arrow connects. This means that the above diagram represents a $U(N_A)\times U(N_B)$ gauge theory with a matter multiplet $\Phi$ transforming in the \hyp{N_A}{\bar{N}_B} representation.

Quiver theories can also contain matter multiplets in the adjoint representation, naturally denoted by an arrow beginning and ending at the same node, as in the following figure:
\begin{center}
\includegraphics[scale=0.5]{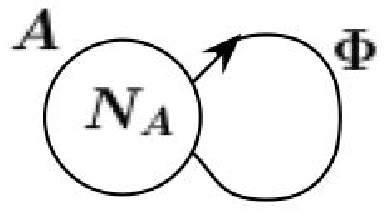}
\end{center}

Notice that the same quiver diagram may have a different meaning depending on the amount of supersymmetry, since, of course, vector and matter multiplets have different field content with different supersymmetry.

In the case of \Ne{1} theories (in four-dimensional language), the quiver diagram also encodes the tree-level cubic superpotential. For each closed triangle in the diagram there is an associated cubic term in the superpotential. The following figure is a prototype example of a tree-level superpotential $W = \tr \Phi_1 \Phi_2 \Phi_3$:
\begin{center}
\includegraphics[scale=0.5]{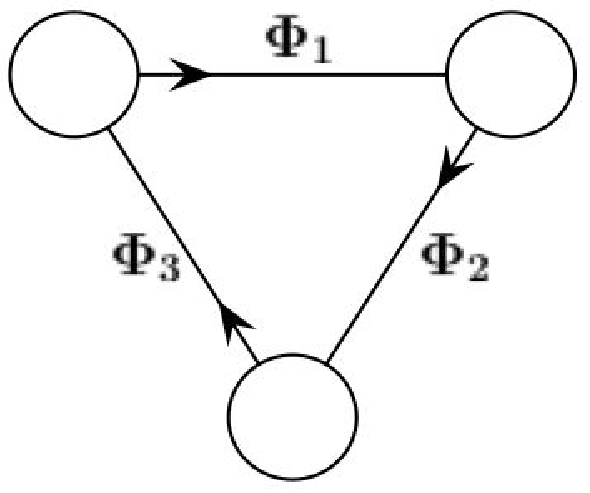}
\end{center}
\end{Insertf}
\begin{Insertl}
Quiver gauge theories arise for instance on D-branes at orbifold singularities. For example, a regular D3-brane at an $A_{N-1}$ singularity, namely transverse to a \cZ{N} orbifold, supports on its world-volume a gauge theory represented by the following \Ne{2} quiver diagram:
\begin{center}
\includegraphics[scale=0.5]{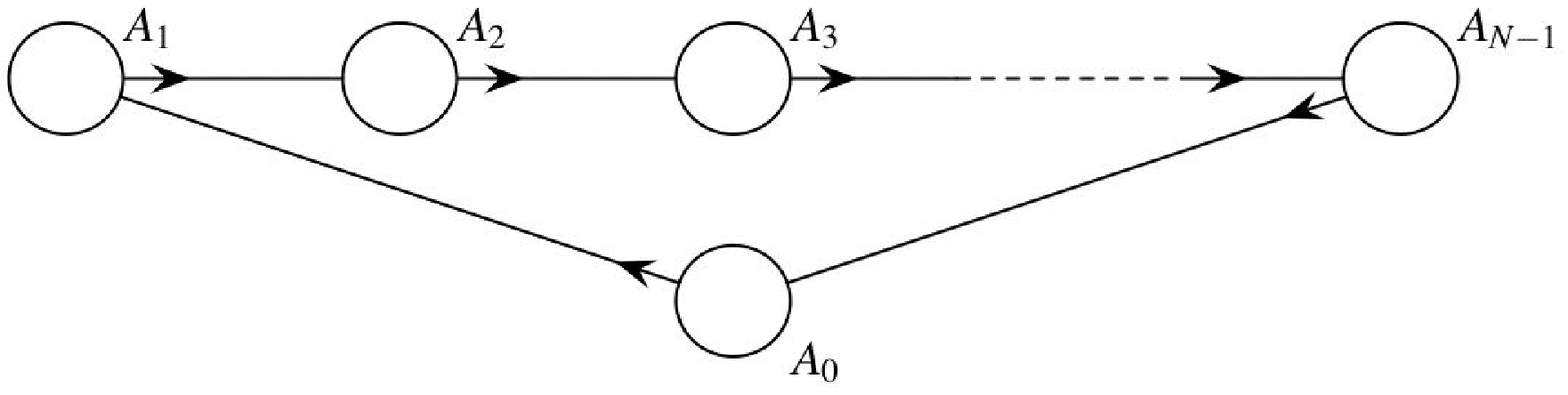}
\end{center}
This is exactly the extended (affine) Dynkin diagram of the $A_{N-1}$ group. This is a general result: D-branes on $\mathbb{C}^2/\Gamma$ orbifolds, where $\Gamma$ is a discrete subgroup of $SU(2)$, support \Ne{2} quiver gauge theories whose quiver diagram is given by the extended A-D-E Dynkin diagram of $\Gamma$.
\end{Insertl}

\subsection*{Fractional branes}

\begin{figure}
\begin{center}
\includegraphics[scale=.7]{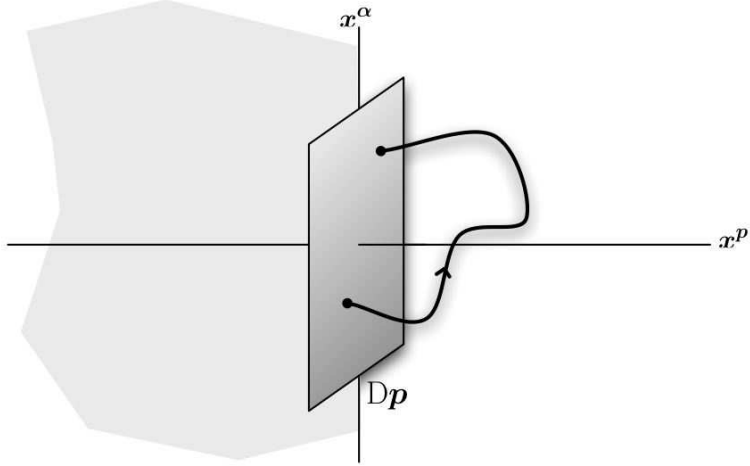}
\caption{{\small A \emph{fractional} D$p$-brane is charged under the twisted closed string fields, and is stuck at the orbifold fixed point.}}
\label{f:fracbranes}
\end{center}
\end{figure}

Are the D-branes we have studied up to now elementary configurations in our \cz{} orbifold? The fact that the gauge group is generically $U(1)\times U(1)$ points to the fact that this is not the case. In fact, when the D-brane sits at the origin, in principle we no longer need any brane image in the covering space, as in figure~\ref{f:fracbranes}. However, it is clear that if we want the brane to be able to move in the $x^r$ direction, we need to include its image even when it is placed at the orbifold fixed point (indeed this is what we have done before). In other words, a brane without images is constrained to stay at the orbifold fixed point $x^r=0$, and cannot move off.
%:Figure: Fractional brane

We will call such an object \emph{fractional brane}, for a reason to be understood in a while. Now, we do not need Chan--Paton factors to distinguish between the brane and its image, and the matrix $\lambda$ is simply a real number. The representation $\gamma(g)$ of the orbifold group can be chosen as one of the two one-dimensional irreducible representations of $\mathbb{Z}_2$, namely:
\begin{equation}\label{CPfrac}
	\gamma(g) = +1\qquad \text{or} \qquad \gamma(g) = -1\,.
\end{equation}
In either case, the relevant action of the orbifold group on the open string states is just the one on the oscillators. This means that the orbifold projection retains the following states at the massless level:
\begin{center}
\begin{tabular}{|c|l|}
\hline
\multicolumn{2}{|c|}{NS states} \\
\hline
$\psi_{-1/2}^\alpha \ket{0,k}$ &
$\to$ $1$ vector\\
$\psi_{-1/2}^i \ket{0,k}$ &
$\to$ $5-p$ real scalars\\
\hline
\end{tabular}
\qquad
\begin{tabular}{|c|l|}
\hline
\multicolumn{2}{|c|}{R states} \\
\hline
$\ket{s_0,s_1,s_2,s_3,s_4=s_3}$ &
$\to$ $8$ fermions\\
\hline
\end{tabular}
\end{center}
These states comprise a single vector multiplet. The gauge theory living on a fractional brane is just \emph{pure} $U(1)$ Super Yang--Mills theory in $p+1$ dimensions with 8 supercharges. Fractional branes are the elementary brane configurations living in the orbifold.

Notice that we have two (equivalent) types of fractional branes, call them A and B, corresponding to the two irreducible representations in~\eqref{CPfrac}. If we put the two types of branes together, the computation of the open string spectrum is perfectly analogous to the one we did for regular branes, but we can now have a different number of branes of the two types, thus getting the following quiver diagram:
\begin{center}
\includegraphics[scale=.7]{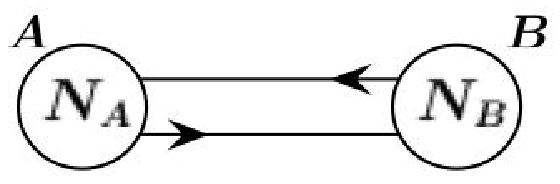}
\end{center}

We note that when $N_A=N_B=N$ the quiver diagram is identical to the one representing the gauge theory living on $N$ regular branes. This is a hint of the fact that a regular brane can be thought of as a bound state of two ``fractions'', the fractional branes of different types (hence the name ``fractional branes'').

However, let us first make an important observation. Beyond having reduced supersymmetry, we can see that the gauge theory living on a generic configuration of fractional branes (with $N_A\neq N_B$) has a gauge coupling constant that runs. We thus see explicitly that fractional branes are among the easiest setups to get scale-anomalous gauge theories with less than 16 supersymmetries on the world-volume of D-branes.

\subsection*{Closed strings and orbifold resolution}

In order to elaborate more on the nature of fractional branes, we now turn to a closed string point of view, and consider the spectrum of the closed strings on our \cz{} orbifold. Since we limit ourselves to the massless states, the fields in the untwisted sector of the theory can be obtained by direct dimensional reduction of the fields of ten dimensional type II supergravity along the forms invariant under the orbifold identification~\eqref{Z2c}. It is immediate to see that the surviving forms are $1$, $dz^1\wedge dz^2$, $dz^m\wedge d\bar{z}^n$, $d\bar{z}^1\wedge d\bar{z}^2$, and $dz^1\wedge dz^2\wedge d\bar{z}^1\wedge d\bar{z}^2$, and we can summarize them in the following Hodge diamond:%
\footnote{Hodge diamonds are briefly discussed in insert~\ref{i:hodge} on page~\pageref{i:hodge}.}
\begin{equation}\label{zh1}
	\ghodgesm\ \ = \hodgesm{1}{0}{1}{4}\,.
\end{equation}

Turning to the single twisted sector, the explicit analysis of the string spectrum gives the result summarized in table~\ref{t:twisted} for the twisted fields, which have a six-dimensional dynamics.
%:Table: Twisted sector spectrum
\begin{table}
\begin{center}
\begin{tabular}{|c|c|c|}
\hline
& NS-NS twisted sector & R-R twisted sector \\
\hline
Type IIB & $b$,\, $\xi^1$, $\xi^2$, $\xi^3$ & $A_0$, $A_2$ \\
Type IIA & $b$,\, $\xi^1$, $\xi^2$, $\xi^3$ & $A_1$, $A_3$ \\
\hline
\end{tabular}
\end{center}
\caption{{\small Bosonic massless spectrum of the twisted sector of type II superstring theories on \cz.}}
\label{t:twisted}
\end{table}
The fields of table~\ref{t:twisted} can be thought of as giving rise to a Hodge diamond too, since they can be seen as the zero-mode reduction of the fields of type II supergravity along an ``exceptional'' two-cycle $\Sigma$. From this point of view, the NS-NS scalar moduli $b$ and $\xi^i$ can be thought of as coming respectively from the ten-dimensional $B$-field and metric on a single two cycle, as summarized in the Hodge diamond:
\begin{equation}\label{zh2}
	\hodgesm{0}{0}{0}{1}\,.
\end{equation}

But where does the exceptional cycle come from? The idea is that the moduli $\xi^i$ of the string spectrum are ``desingularization'' moduli. Turning on a non-vanishing background value of these fields smooths out the singularity of the orbifold space, replacing it with a two-sphere. More precisely, let us define:
\begin{itemize}
\item $\xi = \xi^1+i\xi^2$ is a complex field which acts as modulus for the deformation of the complex structure of the space.
 \item $r = \xi^3$ is the modulus related to the deformation of the K\"ahler structure of the space. The desingularization procedure implemented by giving a non vanishing expectation value to $r$ is known as \emph{resolution} of the orbifold.
\end{itemize}
Notice that there is an $SU(2)$ symmmetry relating the $\xi^i$ fields which makes our choice above just a matter of definition. However, the reason why we introduced it is to point out that usually there are two inequivalent ways to smooth out singular spaces such as orbifolds, even if in the case of \cz{} we are considering they are in fact equivalent. We will consider the complex structure deformation in more detail in section~\ref{s:geom}.

Here, let us consider the resolution of the orbifold obtained by turning on the K\"ahler modulus $r$. The resulting four-dimensional space is known as an ``asymptotically locally euclidean'', or ALE, space. In the case we are considering, we have the simplest ALE space, the Eguchi--Hanson space~\cite{Eguchi:1979gw}, which corresponds to the blow-up of \cz. Its metric can be cast in the form:
\begin{equation}\label{EHm}
	ds^2 = \left(1-\frac{a^4}{r^4}\right)^{-1} dr^2
		+ r^2 \left(1-\frac{a^4}{r^4}\right) (\sigma^3)^2
		+ r^2 \left((\sigma^1)^2 + (\sigma^2)^2\right)\,,
\end{equation}
where the one-forms $\sigma^i$ parameterize a round three-sphere.%
\footnote{In insert~\ref{i:S3} on page~\pageref{i:S3} we summarize some parameterizations of $S^3$ which will be useful in many places in the following. The particular parameterization we are using in~\eqref{EHm} is the third one in insert~\ref{i:S3}.}

%:Insert: Hodge diamonds
\begin{Insert}{Few basic elements of complex (co)homology}\label{i:hodge}
On a \emph{complex $n$-fold} $M$ we can introduce complex coordinates $z^i$ and $\bar{z}^{\bar{\imath}}$, $i,\bar{\imath}=1,\ldots,n$. We can define $(p,q)$-forms as having $p$ antisymmetric holomorphic indices and $q$ antisymmetric anti-holomorphic indices:
\begin{equation*}
	\omega^{(p,q)} = \omega_{i_1\cdots i_p\bar{\jmath}_1\cdots\bar{\jmath}_q}\
		dz^{i_1} \wedge dz^{i_p}
		\wedge d\bar{z}^{\bar{\jmath}_1} \wedge d\bar{z}^{\bar{\jmath}_q}\,.
\end{equation*}
The standard exterior derivative naturally decomposes into the sum of the nilpotent operators $\partial = dz^i \partial_i$ and $\bar{\partial} = d\bar{z}^{\bar{\imath}} \partial_{\bar{\imath}}$. The \emph{Dolbeault cohomology} is defined as:
\begin{equation*}
	H^{p,q}_{\bar{\partial}}(M) = \frac{\bar{\partial}\text{-closed $(p,q)$-forms}}
		{\bar{\partial}\text{-exact $(p,q)$-forms}}\,.
\end{equation*}
The dimension of $H^{p,q}_{\bar{\partial}}$ is the \emph{Hodge number} $h^{p,q}$. By defining the Laplacian operator, one also sees that the harmonic $(p,q)$-forms are in one to one correspondence with $H^{p,q}_{\bar{\partial}}$.

The full set of Hodge numbers is conventionally displayed as a \emph{Hodge diamond}. In particular, the Hodge diamonds of a complex 2-fold and 3-fold are respectively:
\begin{equation*}
	\ghodgesm\ \ ,\qquad\quad \ghodge\ \ .
\end{equation*}

\emph{Calabi--Yau $n$-folds} are complex manifolds with $SU(n)$ holonomy. Due to their particular properties, the Hodge diamonds of the unique compact Calabi--Yau 2-fold (the K3 manifold, which can be obtained from a blow-up of a $T^4/\mathbb{Z}_N$ orbifold via a resolution procedure analogous to the one described in the text) and of a generic Calabi--Yau 3-fold are respectively constrained to be:
\begin{equation*}
	\hodgeK\ \ ,\qquad\quad
		\hodgeCY{$h^{1,1}\hspace{-0.3cm}$}{$h^{2,1}\hspace{-0.3cm}$}\ \ .
\end{equation*}

Because of the duality between homology and cohomology, from the Hodge diamond of a Calabi--Yau manifold we can also read off the number of its independent $(p,q)$-cycles, defined as equivalence classes of chains.
\end{Insert}

%:Insert: S^3 parameterizations
\begin{Insertf}{Some parameterizations of $S^3$}\label{i:S3}
We describe some parameterizations of a round three-sphere that are useful in many places in the text. For each parameterization we give the explicit embedding in $\mathbb{R}^4$ with the ranges of the chosen coordinates, the left-invariant one-forms and the resulting metric.\\
\begin{center}
\begin{tabular}{|l|l|}
\hline
\multicolumn{2}{|c|}{\bf{Parameterization 1}}\\
\hline
Embedding & $\begin{cases}
	x^1 = \cos\psi \cos\theta\\
	x^2 = \cos\psi \sin\theta\\
	x^3 = \sin\psi \cos\phi\\
	x^4 = \sin\psi \sin\phi
	\end{cases}$ \\
\hline
Range & $0\leq\psi\leq\tfrac{\pi}{2}\quad 0\leq\theta,\phi\leq2\pi$\\
\hline
One-forms & $\begin{cases}
	\sigma^1 = -\sin(\theta+\phi)d\psi + \frac{\sin2\psi}{2}\cos(\theta+\phi)(d\theta-d\phi)\\
	\sigma^2 = \cos(\theta+\phi)d\psi + \frac{\sin2\psi}{2}\sin(\theta+\phi)(d\theta-d\phi)\\
	\sigma^3 = -\cos^2\psi d\theta - \sin^2\psi d\phi
	\end{cases}$ \\
\hline
Metric & $ds^2 = d\psi^2 + \cos^2\psi d\theta^2 + \sin^2\psi d\phi^2$\\
\hline
\multicolumn{2}{c}{}\\
\hline
\multicolumn{2}{|c|}{\bf{Parameterization 2}}\\
\hline
Embedding & $\begin{cases}
	x^1 = \cos\phi \sin\theta \sin\psi\\
	x^2 = \sin\phi \sin\theta \sin\psi\\
	x^3 = \cos\theta \sin\psi\\
	x^4 = \cos\psi
	\end{cases}$ \\
\hline
Range & $0\leq\psi,\theta\leq\pi\quad 0\leq\phi\leq2\pi$\\
\hline
One-forms & $\begin{cases}
	\sigma^1 = \sin\theta\cos\phi d\psi + \sin\psi(\sin\psi\sin\phi+\cos\psi\cos\theta\cos\phi)d\theta\\
		\qquad + \sin\psi\sin\theta(-\cos\psi\sin\phi+\sin\psi\cos\theta\cos\phi)d\phi\\
	\sigma^2 = \sin\theta\sin\phi d\psi + \sin\psi(\sin\psi\cos\phi+\cos\psi\cos\theta\sin\phi)d\theta\\
		\qquad + \sin\psi\sin\theta(\cos\psi\cos\phi+\sin\psi\cos\theta\sin\phi)d\phi\\
	\sigma^3 = \cos\theta d\psi - \sin\psi\cos\psi\sin\theta d\theta - \sin^2\psi\sin^2\theta d\phi
	\end{cases}$ \\
\hline
Metric & $ds^2 = d\psi^2 + \sin^2\psi (d\theta^2 + \sin^2 \theta d\phi^2 )$\\
\hline
\end{tabular}
\end{center}
\end{Insertf}
\begin{Insertl}
\begin{center}
\begin{tabular}{|l|l|}
\hline
\multicolumn{2}{|c|}{\bf{Parameterization 3 (Euler angles)}}\\
\hline
Embedding & $\begin{cases}
	x^1 = \cos\left(\tfrac{\theta}{2}\right) \sin\left(\tfrac{\psi+\phi}{2}\right)\\
	x^2 = \cos\left(\tfrac{\theta}{2}\right) \cos\left(\tfrac{\psi+\phi}{2}\right)\\
	x^3 = -\sin\left(\tfrac{\theta}{2}\right) \cos\left(\tfrac{\psi-\phi}{2}\right)\\
	x^4 = \sin\left(\tfrac{\theta}{2}\right) \sin\left(\tfrac{\psi-\phi}{2}\right)
	\end{cases}$ \\
\hline
Range & $0\leq\psi\leq4\pi \quad 0\leq\theta\leq\pi \quad 0\leq\phi\leq2\pi$\\
\hline
One-forms & $\begin{cases}
	\sigma^1 = \tfrac{1}{2}(\cos\psi d\theta + \sin\theta \sin\psi d\phi)\\
	\sigma^2 = \tfrac{1}{2}(-\sin\psi d\theta + \sin\theta \cos\psi d\phi)\\
	\sigma^3 = \tfrac{1}{2}(d\psi + \cos\theta d\phi)
	\qquad\qquad\qquad\qquad\qquad\qquad\qquad\quad
	\end{cases}$ \\
\hline
Metric & $ds^2 = \tfrac{1}{4}(d\psi^2 + d\theta^2 + d\phi^2 + 2\cos\theta d\psi d\phi)$\\
\hline
\end{tabular}
\end{center}
\end{Insertl}

This ALE space has two very important properties. First, the point $r=a$ is a singularity unless the periodicity of $\psi$ (see insert~\ref{i:S3} on page~\pageref{i:S3}) is taken to be $2\pi$. Near there, the space looks topologically like $\mathbb{R}^2\times S^2$, and the $S^2$ is precisely the blown-up two-cycle of finite size we were looking for. Moreover, the unusual periodicity of $\psi$ has the consequence that the space at infinity looks asymptotically like $S^3/\mathbb{Z}_2$, and therefore has the same asymptotics of the \cz{} orbifold we started with.

\subsection*{Fractional branes as wrapped branes}

The above considerations suggest that regular branes on the orbifold can be described, when the cycle is blown up and the orbifold resolved, as simply being transverse to the resulting ALE space.

What about fractional branes? The first thing to notice is that, if we analyze the closed string fields that couple to a fractional D$p$-brane, the latter results to be charged also under the fields in the twisted sector. This is very natural and had to be expected: these twisted couplings are the closed string counterpart of the property of fractional branes of being stuck at the orbifold fixed point. In particular, among the twisted fields of table~\ref{t:twisted}, a fractional D$p$-brane turns out to be charged under the NS-NS scalar $b$ and the R-R form $A_{p+1}$.

However, we said in the previous subsection that the twisted fields can be seen as reductions of the type IIB fields along an exceptional cycle $\Sigma$, so that we can write:
\begin{equation}
	b = \int_\Sigma B\,,\qquad
	A_{p+1} = \int_\Sigma C_{p+3}\,.
\end{equation}
The latter coupling suggests an intriguing interpretation, since it is the minimal coupling appropriate for a D$(p+2)$-brane! The interpretation is the following: a fractional D$p$-brane can be seen as a \emph{D$(p+2)$-brane wrapped on the exceptional two-cycle $\Sigma$}, in the orbifold limit where the geometrical volume of the cycle vanishes. However, the tension of the fractional brane remains finite even in the limit, due to its charge under the twisted field $b$, whose flux through $\Sigma$ should necessarily be non-vanishing for this interpretation to make sense.

This interpretation is confirmed by a careful mathematical construction due to McKay and Kronheimer~\cite{McKay:1980??,Kronheimer:1989zs}, in which the crucial ingredient is the one-to-one correspondence between irreducible representations of the orbifold group (that, as discussed above, define fractional branes) and blown-up two-cycles in the resolved geometry.

Let us see this from another point of view. Another hint of the validity of the description of fractional branes as wrapped branes comes from the following short computation~\cite{Dasgupta:1999wx}. Consider a D$(p+2)$-brane with some non-zero amount of $B+2\pi\ls^2 F$ on it, that we can think of as having some D$p$-brane dissolved into the world-volume. Let us also consider another D$(p+2)$-brane with opposite charge. The total world-volume interaction coming from the Wess--Zumino action will be:
\begin{equation}
	\tau_{p+2} \int C_{p+1} \wedge \left\{ (B+2\pi\ls^2 F)-(B+2\pi\ls^2 \tilde{F})\right\}\,,
\end{equation}
where we indicate with $\tilde{F}$ the world-volume gauge field of the second D$(p+2)$-brane. How can we obtain from this configuration a vanishing D$(p+2)$-brane charge, but a unit D$p$-brane charge? Let us denote as $\Sigma$ a particular two-dimensional subspace of the world-volume and choose:
\begin{equation}\label{choice}
	2\pi\ls^2 \int_\Sigma (F-\tilde{F}) = \frac{\tau_{p}}{\tau_{p+2}}\,,\qquad
	\frac{\tau_{p+2}}{\tau_p} \int_{\Sigma} B = \frac{1}{2}\,.
\end{equation}
This will give us vanishing D$(p+2)$-brane charge, and a total D$p$-brane charge of $\tfrac{1}{2}+\tfrac{1}{2}=1$. We then may identify the two halves as fractional D$p$-branes. In this description, however, they are completely delocalized in the world-volume of the higher dimensional D-$\bar{\text{D}}$ system. In order to make them localized, it is natural to identify $\Sigma$ with the two-cycle of the $A_1$ ALE space which is the blow-up of the \cz{} orbifold. In the limit where the cycle shrinks to zero size, the D$p$-branes are completely localized in the orbifolded directions, but are of course free to move supersymmetrically in the remaining transverse space, in agreement with our previous analysis.

From this short check, we have also recovered a very important piece of information on our geometry. Notice that, in the limiting process where the two-cycle shrinks to zero geometrical volume, the background flux of the $B$-field through it remains given by~\eqref{choice}:
\begin{equation}\label{bbkgd}
	\frac{1}{(2\pi\ls)^2} \int_{\Sigma} B = \frac{1}{2}\,.
\end{equation}
In fact, though the way we have derived it might seem not very convincing, this is precisely the correct value of $B$-flux that is needed in order to have a sensible conformal field theory description of strings propagating on a \cz{} orbifold~\cite{Aspinwall:1995zi,Douglas:1997xg}, and will turn out to be very important in the following.

Let us make another observation. Up to now, we have considered only a single type of fractional brane, but in the \cz{} orbifold we have two different types of them, the second one corresponding to the ``extra'' node in the extended Dynkin diagram which is the quiver diagram of the theory. This fractional brane can be seen as wrapping a non-independent cycle $\Sigma_0 = -\Sigma$, with an additional background value of the world-volume gauge field turned on along the cycle, that ensures the fact that the fractional brane of this type has positive untwisted charge.

\subsection*{The world-volume action of fractional branes}

Having introduced the interpretation of fractional D$p$-branes on orbifolds as D$(p+2)$-branes wrapped on shrinking two-cycles of the blown-up ALE space, we are now in a good position to compute their world-volume action, which will be very useful in the gauge/gravity examples of the next chapters.

The strategy we will follow is simple. The world-volume action of a fractional D$p$-brane is given by the action of a wrapped D$(p+2)$-brane when we correctly take into account the properties of the vanishing cycles.

We still limit ourselves to the specific example of a \cz{} orbifold we have been considering in this section. To be precise, let us start by considering a fractional D$p$-brane of type A transverse to \cz. Recall that the action~\eqref{Dpwv} of a D$(p+2)$ brane in the string frame is given by:
\begin{equation}\label{Dp+2}
\begin{split}
	S_{\text{D}(p+2)} = &- \tau_{p+2} \int d^{p+3}\xi\ e^{-\Phi}
		\sqrt{- \det \left(\hat{G}_{ab} + \hat{B}_{ab} + 2\pi\ls^2 F_{ab}\right) }\\
		 &+ \tau_{p+2} \int_{\mathcal{M}_{p+3}} \sum_q \hat{C}_q \wedge e^{\hat{B}+2\pi\ls^2 F}\,.
\end{split}
\end{equation}
From now on, in this subsection we will leave the pull-backs understood to avoid clutter. Let us start by considering the Dirac--Born--Infeld part of the action. Denote with $\xi^\alpha$, $\alpha=0,\ldots,p$, the coordinates along the unwrapped world-volume and suppose that the only non-vanishing part of $B$ is given by:
\begin{equation}
	B = b\ \omega_2\,, \qquad \frac{b}{(2\pi\ls^2)} = \frac{1}{2} + \frac{\tilde{b}}{(2\pi\ls^2)}\,,
\end{equation}
where $\omega_2$ is the anti-self dual two-form which is dual to the two-cycle $S^2$, and $\tilde{b}$ is the fluctuation of the ``twisted'' field $b$ around its background value given in~\eqref{bbkgd}. The two-form $\omega_2$ satisfies the following important properties:
\begin{equation}\label{omegaprop}
	\omega_2=-\hd{4} \omega_2\,, \qquad \int_{S^2}\omega_2=1\,,
	\qquad \int_{\text{ALE}}\omega_2\wedge\omega_2=-\frac{1}{2}\,.
\end{equation}
Now, we assume that the world-volume gauge-field is turned on only in the directions $x^\alpha$. Supposing that the metric has no support on $S^2$, we then get the following expression for the Dirac--Born--Infeld part of the action:
\begin{equation}\label{fracDBI}
\begin{split}
	S_{\text{DBI}}^{\text{f}} &= - \tau_{p+2} \int d^{p+1}\xi\ e^{-\Phi}
		\sqrt{- \det \left(G_{\alpha\beta} + 2\pi\ls^2 F_{\alpha\beta}\right)} \ 
		\int_{S^2} b\ \omega_2 \\
		&= -\frac{\tau_p}{2} \int d^{p+1}\xi\ e^{-\Phi}
		\sqrt{- \det \left({G}_{\alpha\beta} + 2\pi\ls^2 F_{\alpha\beta}\right)} \ 
		\left(1+\frac{\tilde{b}}{2\pi^2\ls^2}\right)\,,
\end{split}
\end{equation}
where we have used~\eqref{omegaprop}. From the coefficient in front of the action, we immediately see that the tension of the fractional brane results to be half of the tension $\tau_p$ of a bulk brane. This fits nicely with what we know about fractional branes.

Turning to the Wess--Zumino part, we have to decompose the R-R potentials in two components, respectively outside and along the cycle, as:
\begin{equation}
   C_q=\bar{C}_q+A_{q-2}\wedge\omega_2\,.
\end{equation}
Let us start by considering the decomposition of $C_{p+3}$, where in order to be able to integrate on the $(p+1)$-dimensional world-volume of the fractional brane we should only consider the ``twisted'' component:
\begin{multline}
	\tau_{p+2}\int_{\mathcal{M}_{p+3}}C_{p+3}
		=\tau_{p+2}\int_{\mathcal{M}_{p+3}}A_{p+1}\wedge\omega_2\\
		=\tau_{p+2}\int_{\mathcal{M}_{p+1}}A_{p+1}\ \int_{S^2}\omega_2
		=\frac{\tau_p}{2}\int_{\mathcal{M}_{p+1}}\frac{A_{p+1}}{2\pi^2\ls^2}\,.
\end{multline}
Considering now $C_{p+1}$, one starts by getting the following term:
\begin{equation}
	\tau_{p+2}\int_{\mathcal{M}_{p+3}}C_{p+1}\wedge B
		\to\tau_{p+2}\int_{\mathcal{M}_{p+3}}\bar{C}_{p+1} b\wedge\omega_2
      		=\frac{\tau_p}{2}\int_{\mathcal{M}_{p+1}}
      		\bar{C}_{p+1}\left(1+\frac{\tilde{b}}{2\pi^2\ls^2}\right)\,.
\end{equation}
Therefore the first term of the expansion of the Wess--Zumino action becomes:
\begin{equation}
	\frac{\tau_p}{2}\int_{\mathcal{M}_{p+1}}\left[
      		\bar{C}_{p+1}\left(1+\frac{\tilde{b}}{2\pi^2\ls^2}\right)
      		+\frac{A_{p+1}}{2\pi^2\ls^2}\right]\,.
\end{equation}
If we consider any other lower rank potential, we see that the relevant contributions to the Wess--Zumino action always involve the following combinations of fields:
\begin{equation}
	\mathcal{C}_q=
		\bar{C}_q\left(1+\frac{\tilde{b}}{2\pi^2\ls^2}\right)
		+\frac{A_q}{2\pi^2\ls^2}\,,
\end{equation}
Recalling~\eqref{fracDBI}, the world-volume action for a fractional D$p$-brane of type A on a \cz{} orbifold can be cast in the following simple form:
\begin{equation}\label{fracwv}
\begin{split}
	S_{\text{D}p}^{\text{f}} = &-\frac{\tau_p}{2} \int d^{p+1}\xi\ e^{-\Phi}
		\sqrt{- \det \left({G}_{\alpha\beta} + 2\pi\ls^2 F_{\alpha\beta}\right)} \
		\left(1+\frac{\tilde{b}}{2\pi^2\ls^2}\right)\\
		&+ \frac{\tau_{p}}{2} \int_{\mathcal{M}_{p+1}} \sum_q \mathcal{C}_q \wedge e^{2\pi\ls^2 F}\,.
\end{split}
\end{equation}
This expression for the action of a fractional D$p$-brane is also confirmed by the couplings of the brane to the bulk fields, computed with the boundary state formalism~\cite{Bertolini:2000dk} and with explicit computations of string scattering amplitudes on a disk~\cite{Merlatti:2000ne}.

\section{D-branes on orbifolds: a summary}\label{s:orbsumm}

In the previous section, we have studied in detail the explicit example of D$p$-branes on a $\mathbb{R}^{1,5}\times\cz$ orbifold, and we have uncovered a lot of features which are characteristic of branes on orbifolds in general.

In this section we will try summarizing the properties of D-branes on orbifold spaces, without relying on a particular example. We are mainly interested on type II string theory on two different general kinds of orbifolds:
\begin{itemize}
\item $\mathbb{R}^{1,5}\times \mathbb{C}^2/\Gamma$ orbifolds, with $\Gamma$ a discrete subgroup of $SU(2)$. These orbifolds break one half of the supersymmetries of flat space, so the theory living on a D-brane transverse to the orbifold preserves 8 real supercharges. Orbifolds of this kind arise as singular limits of Calabi--Yau twofolds (in the compact case, the K3 manifold).
\item $\mathbb{R}^{1,3}\times \mathbb{C}^3/\Gamma$ orbifolds, with $\Gamma$ a discrete subgroup of $SU(3)$. These orbifolds break three quarters of the supersymmetries of flat space, so the theory living on a D-brane transverse to the orbifold preserves 4 real supercharges. Orbifolds of this kind arise as singular limits of Calabi--Yau threefolds.
\end{itemize}

Let us introduce the regular representation $\mathcal{R}$ of the orbifold group, which is a reducible representation of dimension $\abs{\Gamma}$, the order of the group. This can be decomposed into irreducible representations $\mathcal{D}_I$ of dimension $n_I$, $I=1,\ldots,k-1$, as:
\begin{equation}
	\mathcal{R} = \bigoplus_I n_I \mathcal{D}_I\,,\qquad
	\sum_{I=0}^{k-1} n_I = \abs{\Gamma}\,.
\end{equation}

We distinguish two kinds of D-branes on orbifolds:
\begin{itemize}
\item \emph{Regular} D-branes are defined by the fact that the Chan--Paton factors of the open strings attached to them transform in the regular representation $\mathcal{R}$ of the orbifold group $\Gamma$.

The gauge theory living on the world-volume of $N$ regular D$p$-branes is a $(p+1)$-dimensional quiver gauge theory with gauge group $\prod_{I=0}^{m-1} U(n_I N)$ and matter in the bifundamental representation of adjacent nodes in the quiver diagram. Due to the matter content, the gauge theory does not exhibit a scale anomaly.

In the case of $\mathbb{C}^2/\Gamma$ orbifolds, $\Gamma \subset SU(2)$, we can find a general recipe for deriving the quiver diagram of the theory. Start by the following decomposition of representations of $\Gamma$:
\begin{equation}
	\mathcal{Q} \otimes \mathcal{D}_I = \bigoplus_J a_{IJ} \mathcal{D}_J\,,
\end{equation}
where $\mathcal{Q}$ is the defining two-dimensional representation. One finds that $a_{IJ}$ is the adjacency matrix of the extended (affine) Dynkin diagrams of the A-D-E series, which allows to classify the discrete subgroups of $SU(2)$ in an A-D-E classification. This fact, known as McKay correspondence, has even more surprising consequences, since one finds that the matrix $a_{IJ}$ also precisely gives the connections of the quiver diagram of the corresponding gauge theory arising on D-branes in the orbifold! For example, in insert~\ref{i:quiver} on page~\pageref{i:quiver} we show the diagram corresponding to $\Gamma = A_{N-1} = \mathbb{Z}_N$, an abelian group whose $N$ irreducible representations are one-dimensional.

\item \emph{Fractional} D-branes of type $I$ ($I=0,\ldots,k-1$) are defined by the fact that the Chan--Paton factors of the open strings attached to them transform in the irreducible $I$-th representation $\mathcal{D}_I$ of the orbifold group $\Gamma$.

The gauge theory living on the world-volume of $N$ fractional D$p$-branes of type $I$ is \emph{pure} $(p+1)$-dimensional Super Yang--Mills theory with gauge group $U(n_I N)$.

Fractional D$p$-branes can generally be interpreted as D$(p+2)$-branes wrapped on a two-cycle of the ALE space arising as the blow-up of the orbifold, in the limit where this cycle shrinks to zero size. This gives a natural interpretation to many of their properties, and allows to obtain in a simple way important results such as their world-volume action and their tension, which is related to the one of the corresponding regular brane by $\tau_p^{\text{f}} = \tfrac{\tau_p}{\abs{\Gamma}}$.

The blown-up ALE space is obtained by a resolution procedure analogous to the one described in section~\ref{s:Z2}. In the case of the $A_{N-1}$ series, relevant to the blow-up of $\cZ{N}$ orbifolds, the metric of the ALE space can be written explicitly as the multi-centered Gibbons--Hawking metric~\cite{Gibbons:1979xm}:
\begin{equation}\label{GH}
\begin{gathered}
	ds^2 = V^{-1}(dz - \mathbf{A}\cdot \mathbf{y})^2 + V d\mathbf{y}\cdot d\mathbf{y}\,,\\
	V = \sum_{i=0}^{N-1}\frac{1}{\abs{\mathbf{y}-\mathbf{y}_i}}\,,\qquad
	\mathbf{\nabla} V = \mathbf{\nabla} \times \mathbf{A}\,,
\end{gathered}
\end{equation}
where for simplicity we omitted some dimensionful constants. Through a change of coordinates, the Eguchi--Hanson metric~\eqref{EHm} can be written as the $N=2$ case of~\eqref{GH}. The ALE space~\eqref{GH} has $N-1$ blown-up two-spheres on which we can wrap a D$(p+2)$-brane in order to obtain $N-1$ different types of fractional D$p$-branes when the cycle shrinks. Each two-cycle is associated to a node of the extended Dynkin diagram of the $A$-series, except the trivial node. The $N$th type of fractional brane is obtained by wrapping a D$(p+2)$-brane on the non-independent cycle corresponding to the trivial node of the extended diagram, with a suitable world-volume gauge field turned on, as we saw in the specific example of \cz.
\end{itemize}

\subsection*{Orbifolds of AdS/CFT}

A brief comment is in order about the possibility of realizing in this less supersymmetric context an exact gauge/gravity duality as the one described in section~\ref{s:adscft}. In fact this turns out to be possible, but only when the four-dimensional gauge theory involved in the duality is conformal, as in the original AdS/CFT duality.

To be a little more specific, consider $N$ regular D3-branes transverse to a \cZ{M}{} orbifold. The gauge theory living on their world-volume, as we have learnt, is a superconformal quiver gauge theory with \Ne{2} supersymmetry whose quiver diagram is given in insert~\ref{i:quiver} on page~\pageref{i:quiver}.

One can then take a near-horizon limit on the D3-brane geometry, exactly as the one performed in section~\ref{s:adscft} for D3-branes in flat space. The resulting geometry turns out to be $\ads/\mathbb{Z}_M$, where the $\mathbb{Z}_M$ factor acts only on the five-sphere leaving a great circle $S^1$ fixed. Since the $AdS_5$ part is untouched, we still expect the same matching of symmetries we found for the \Ne{4} theory, and conclude that type IIB string theory on $\ads/\mathbb{Z}_M$ is dual to the \Ne{2} superconformal quiver theory~\cite{Kachru:1998ys}. Notice that a similar construction exists also for \Ne{1} theories.

\section{Stretched branes and a duality web}\label{s:HW}

In this section we will introduce some useful dualities that will allow us to understand better the connection between fractional and wrapped branes, and between them and a new scenario consisting of branes stretched between branes~\cite{Hanany:1997ie}.%
\footnote{For a review on this approach see for instance~\cite{Giveon:1998sr}}

Let us start from the Gibbons--Hawking metric~\eqref{GH} that, as we have explained, describes the $A_{N-1}$ ALE space obtained by resolving a $\cZ{N}$ orbifold. By changing variables in an opportune way, we can pick the vector $\mathbf{y}$ in ~\eqref{GH} to be parameterized by $\mathbf{y}=(x^7,x^8,x^9)$, while identifying $z=x^6$. We now perform a T-duality transformation along the compact direction $x^6$ using the rules~\eqref{Tduality}. The resulting new background reads:
\begin{equation}\label{deloc}
\begin{aligned}
	ds^2 &= \eta_{\alpha\beta}dx^\alpha dx^\beta + V(\mathbf{y}) ((dx^6)^2 + \delta_{ij} dx^i dx^j)\,,\\
	e^{2\Phi} &= V(y) = \sum_{i=0}^{N-1}\frac{\ls}{\abs{\mathbf{y}-\mathbf{y}_i}}\,,
\end{aligned}
\end{equation}
where $\alpha,\beta=0,\ldots,5$ and $i,j=7,8,9$ and we reinstated correct dimensionful units. The background~\eqref{deloc} has also to include a NS-NS two-form field $B_{6i}$ (coming from the fact that the ALE background had non-vanishing $G_{6i}$) which is a vector satisfying $\nabla V = \nabla \times \mathbf{B}$.

As a necessary result of applying the T-duality rules, we have reached a solution describing an object which is completely delocalized along $x^6$. However, in the ALE space we started with, we have particular points along the compact direction $x^6$ where two-cycles shrink, so we expect that winding states behave in some special way, implying a localized structure for the resulting new background. The simplest possibility for a solution to be completely localized along $\mathbf{x}=(x^6,\ldots,x^9)$ is being harmonic there. Therefore, we replace $V(y)$ with:
\begin{equation}\label{NSV}
	V(x) = 1 + \sum_{i=0}^{N-1}\frac{\ls^2}{(\mathbf{x}-\mathbf{x}_i)^2}\,.
\end{equation}

We can now easily identify the string frame background~\eqref{deloc}, together with the function~\eqref{NSV}, as a solution describing a system of $N$ NS5-branes~\eqref{NS5} arranged on a circle on $x^6$! We have then discovered that systems of parallel NS5-branes on a circle are T-dual to ALE spaces.
%:Figure: Stretched branes
\begin{figure}
\begin{center}
\includegraphics[scale=.7]{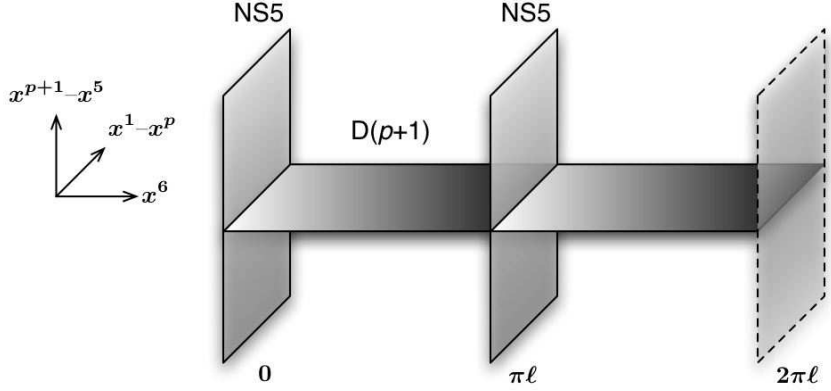}
\caption{{\small The scenario of branes stretched between branes. A single D$(p+1)$ brane is suspended between two parallel NS5-branes along the compact direction $x^6$ of circumference $2\pi\ell$. The two D$(p+1)$-brane segments connecting neighboring NS5-branes have the T-dual interpretation of fractional D$p$-branes on a \cz{} orbifold.}}
\label{f:stretch}
\end{center}
\end{figure}

\begin{figure}
\begin{center}
\includegraphics[scale=.6]{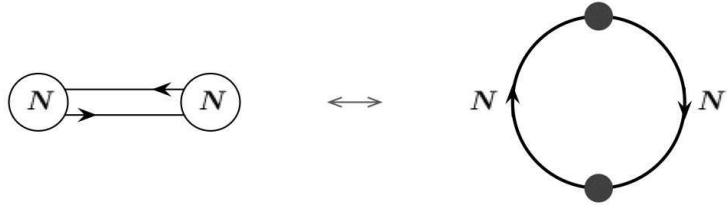}
\caption{{\small A quiver diagram (here the simple $A_1$ diagram) has the dual interpretation of a diagram representing D-branes stretched between NS5-branes along a compact direction. Notice that the role of nodes and arrows is reversed.}}
\label{f:NSquiver}
\end{center}
\end{figure}

Let us now consider for simplicity a blown-down $A_1$ ALE space, namely a \cz{} orbifold. What happens to the dual background~\eqref{deloc} once we consider regular or fractional D$p$-branes in the orbifold? A T-duality along $x^6$ transforms a D$p$-brane into a D$(p+1)$-brane, so in the T-dual setup we have a D$(p+1)$-brane stretched along $x^6$ between the two parallel NS5-branes, as in the following table where $\abs{-}$ denotes a longitudinal direction in which a brane may have a finite extent:
\begin{center}
\begin{tabular}{|c|ccc|ccc|c|c|c|c|}
\hline
&0&$\cdots$&$p$&$p+1$&$\cdots$&5&6&7&8&9\\
\hline
NS5 &$-$&$-$&$-$&$-$&$-$&$-$&$\cdot$&$\cdot$&$\cdot$&$\cdot$\\
\hline
D$p$ &$-$&$-$&$-$&$\cdot$&$\cdot$&$\cdot$&$\abs{-}$&$\cdot$&$\cdot$&$\cdot$\\
\hline
\end{tabular}
\end{center}
The general setup of branes suspended between branes which is T-dual to a \cz{} orbifold is depicted in figure~\ref{f:stretch}. It preserves eight supersymmetries, since each type of brane breaks half of the supersymmetries of flat space, and this matches the supersymmetry of a D-brane in the orbifold. Notice also that the quiver diagram describing the gauge theory living on a regular D-brane on \cz{} has a nice dual interpretation in terms of this configuration, as shown in figure~\ref{f:NSquiver}.
%:Figure: Dual quiver

In this singular ALE space, we know that fractional branes are tensionful because of the non-vanishing flux~\eqref{bbkgd} of $B$ along the shrinking two-cycle $\Sigma$. The T-duality operation maps this flux onto the distance $L$ between the NS5-branes in the $x^6$ direction, so that the two segments in figure~\ref{f:stretch} have lengths proportional to $2\pi\ell \int_{\Sigma} B$ and $2\pi\ell (1-\int_{\Sigma} B)$, respectively. Due to~\eqref{bbkgd}, we see that in the background T-dual to the orbifold the two NS5-branes are equally spaced along $x^6$.

This fact leads us to the interpretation of fractional branes in this dual setup. The two types of fractional D$p$-branes on \cz{} are mapped to the two consecutive D$(p+1)$-brane segments in figure~\ref{f:stretch}, which together make a full regular brane. In this scenario, the interpretation of fractional branes as ``fractions'' of a regular brane is literal and straightforward. Notice that a brane segment can exist independently only if the two NS5-branes are located at the same point in the space $(x^7,x^8,x^9)$, and this is the manifestation of the fact that fractional branes are stuck at the orbifold fixed point. Notice finally that all the reasoning above can easily be extended to the full $A_{N-1}$ ALE series.

The stretched brane scenario also allows us to describe how fractional branes on orbifolds are related to the wrapped branes we introduced in section~\ref{s:wrapped}. In fact, we can think of performing a new T-duality on the system with the NS5-branes, this time along one of the directions transverse to both the NS5 and D-branes. The NS5-brane background will again transform into an ALE space, but the fate of the stretched D$(p+1)$-brane segment is rather different, since it becomes a D$(p+2)$-brane wrapped on a two-cycle of the ALE space, which now has a finite geometrical volume proportional to the distance of the NS5-branes in the original setup! This configuration is precisely the one of a D$(p+2)$-brane wrapped on a two-cycle inside a non-compact Calabi--Yau twofold that we studied in section~\ref{s:wrapped}, and we have now seen how it is related to fractional branes via an interesting chain of dualities.

\section{Breaking more supersymmetry}\label{s:geom}

Up to now, we have mainly considered setups with eight preserved supercharges on the world-volume of D-branes. Now we will be specifically interested in breaking additional supersymmetry. This could be obtained simply by using tools we already presented, such as branes wrapped on cycles inside Calabi--Yau threefolds, or $\mathbb{C}^3/\Gamma$ orbifolds, where $\Gamma$ is a discrete subgroup of $SU(3)$. In fact, we will explicitly use both these classes of examples in chapter~\ref{c:4susy}, but here we would like to follow a different route, that in our opinion is useful to shed some light on the importance of the underlying geometry of the systems we are considering, and in particular on the deep relation existing between gauge theories and geometry.

\subsection*{Discovering the conifold via gauge superpotentials}

Our starting point will again be the example we treated most, the \cz{} orbifold, that we will consider here from a more geometrical point of view. Specifically, we consider fractional D3-branes in the orbifold, thus concentrating on the case of four-dimensional gauge theories.

Recall from section~\ref{s:Z2} that the \cz{} orbifold is defined by the identifications~\eqref {Z2c} of two complex coordinates:
\begin{equation}\label{Z2again}
	z_1 \to -z_1\,,\qquad z_2 \to -z_2\,.
\end{equation}
We can introduce three complex variables that are invariant under~\eqref{Z2again}:
\begin{equation}
	u = z_1^2\,,\qquad v = z_2^2\,,\qquad z = i z_1 z_2\,.
\end{equation}
Thus, we can naturally regard our \cz{} orbifold as a geometrical hypersurface embedded in $\mathbb{C}^3$, defined in terms of the above variables by the following algebraic equation:
\begin{equation}\label{geomZ2}
	uv + z^2 = 0\,.
\end{equation}
To obtain another useful form of the above equation, let us change variables according to $u = x + iy$, $v= x - iy$, so that it reads:
\begin{equation}\label{geomZ22}
	x^2 + y^2 + z^2 = 0\,.
\end{equation}
In this description, it is immediate to point out the singularity at the origin $x=y=z=0$, and to identify the shrinking two-sphere we have introduced in section~\ref{s:Z2} by essentially taking the real part of~\eqref{geomZ22}.

In section~\ref{s:Z2}, we have discussed a way to remove the singularity of this orbifold by turning on the K\"ahler deformation parameter $r$. We also observed that the NS-NS twisted spectrum of the string also contains an additional complex scalar $\xi$, whose vacuum expectation value parameterizes the deformation of the complex structure of the space encoded in~\eqref{geomZ22}. In fact, one can see that turning on the modulus $\xi$ modifies~\eqref{geomZ22} as follows:
\begin{equation}\label{deforb}
	x^2 + y^2 + z^2 + \varepsilon^2 = 0\,,
\end{equation}
where we have defined $\varepsilon = (2\pi\ls^2)^2 \xi$. Thanks to this deformation, we see that the singularity at the origin has been removed and the two-sphere has acquired a finite radius. This finite radius of $S^2$ was also a property of the resolution of the orbifold we studied in section~\ref{s:Z2}, and this is an indication that we are performing an equivalent operation, as one can easily understand by recalling the symmetry which relates in this case the K\"ahler and complex deformation parameters. However, the two ways one may use in order to remove the singularity (the resolution by blowing up two-spheres and the deformation by changing the defining equation) are not always equivalent, as we will see.

It is interesting to study the effect that the deformation~\eqref{deforb} has on the low-energy gauge theory living on the world-volume of fractional D3-branes. When studying the spectrum of the open strings attached to a fractional D3-brane, we limited ourselves to the standard on-shell massless states, which gave rise to pure \Ne{2} Super Yang--Mills theory. However, it is possible to construct in string theory also the auxiliary fields that would be necessary for a superfield description of the low-energy gauge theory~\cite{Billo:2002hm}. In particular, define three auxiliary fields $D^c$, $c=1,2,3$ as:
\begin{equation}
	D^c \sim \bar{\eta}^c_{ab} \psi^a_{-1/2} \psi^b_{-1/2} \ket{0,k}\,,
\end{equation}
where $\bar{\eta}^c_{ab}$ are the anti-self dual 't Hooft symbols. If we define $D=D^1+iD^2$, it is not too difficult to show that the string disk amplitude $\langle \mathcal{V}_\xi \mathcal{V}_D \rangle \sim \xi D$ is non-vanishing. This means that the low-energy non-abelian gauge theory effective action acquires a term of the form:
\begin{equation}\label{FI}
	S_{\text{FI}} \sim \int d^4x \left( \xi \tr D + c.c. \right)\,,
\end{equation}
that can be also completed to a full \Ne{2} supersymmetric Fayet--Iliopoulos term by considering the $r$ modulus in addition to $\xi$~\cite{Douglas:1996sw}.

What is the gauge theory origin of this Fayet--Iliopoulos term? Using an \Ne{1} language, we can think of it as arising from a tree-level superpotential for the chiral superfield $\Phi$ in the adjoint representation of the $U(N)$ gauge group, given by:
\begin{equation}\label{Wtree}
	W(\Phi) = \xi \tr \Phi\,.
\end{equation}
To translate~\eqref{Wtree} into geometrical quantities, we define the complex variable $t = 2\pi\ls^2 \tr\Phi$, which naturally parameterizes the plane in which the fractional branes are free to move supersymmetrically. In terms of $t$ and of the deformation parameter appearing in~\eqref{deforb}, we then have:
\begin{equation}\label{Wgeom}
	W(t) = (2\pi\ls^2)^3\ W(\Phi) = \varepsilon t\,,
\end{equation}
so that we can write the defining equation of the deformed geometry as:
\begin{equation}\label{defW}
	x^2 + y^2 + z^2 + W'(t)^2 = 0\,.
\end{equation}
Notice that, in this case, the linear superpotential has the only consequence of adding a constant term to the action, which then remains \Ne{2} supersymmetric. In geometrical terms, we have a trivial fibration of the four-dimensional ALE space over the complex plane parameterized by $t$. The non-trivial part of the ambient geometry then remains the one of a Calabi--Yau twofold, and we can understand the reason of the preserved supersymmetry.

However, from the latter observation, it seems natural to ask what happens when we consider a non-trivial fibration of the ALE space over the $t$-plane. This corresponds to taking a general superpotential in~\eqref{defW}. In particular, we can take the simplest superpotential which is known to break the supersymmetry of the gauge theory from \Ne{2} down to \Ne{1}, namely a mass term for the adjoint superfield:
\begin{equation}
	W(\Phi) = \frac{1}{2} m \tr \Phi^2\,.
\end{equation}
Using~\eqref{Wgeom}, we see that~\eqref{defW} becomes the following equation in $\mathbb{C}^4$:
\begin{equation}\label{Wcon}
	x^2 + y^2 + z^2 + w^2 = 0\,.
\end{equation}
where we have introduced the new complex coordinate $w = 2\pi\ls^2 m t$. We can decouple the massive adjoint multiplet by taking $m\to\infty$, but in doing so we keep $\ls^2 m$ finite so that the geometry~\eqref{Wcon} still makes sense. A stack of fractional D3-branes in the geometry~\eqref{Wcon} then supports pure \Ne{1} Super Yang--Mills theory! The non-trivial fibration of the ALE space over the $t$-plane has led us to the geometry of a Calabi--Yau threefold, which breaks additional supersymmetry.

In fact, the the equation~\eqref{Wcon} defines a very well known non-compact Calabi--Yau threefold, the \emph{conifold}~\cite{Romans:1985an,Candelas:1990js}, which, as we have just seen, is a natural setup in order to study geometric duals of four-dimensional \Ne{1} gauge theories. We will elaborate more on this setup in chapter~\ref{c:4susy}.

However, let us now continue with the analysis of this space. The first observation we can make is that it is evident from~\eqref{Wcon} that the conifold is a \emph{singular} space. The conifold can be described as a cone over the space $T^{1,1} = \frac{SU(2)\times SU(2)}{U(1)}$, which can be topologically thought of as $S^3\times S^2$. At the tip of the cone, the volume of both spheres vanishes, and we have the singularity.%
\footnote{The conifold and the $T^{1,1}$ space will be studied in more detail in section~\ref{s:conifold} and in insert~\ref{i:T11} on page~\pageref{i:T11}.}
The singularity present at the origin of the space is in a certain sense ``worse'' that the singularity of the orbifold we started with, because a conformal field theory (such as perturbative string theory) on the conifold is not well defined.

\begin{figure}
\begin{center}
\includegraphics[scale=.5]{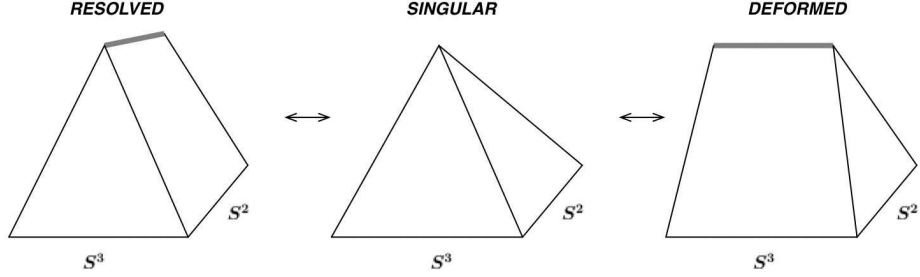}
\caption{{\small The singularity of the conifold can be removed in two ways. Deformation of the complex structure blows up the $S^3$ yielding the \emph{deformed conifold}, while deformation of the K\"ahler structure blows up the $S^2$ and gives the \emph{resolved conifold}.}}
\label{f:conifold}
\end{center}
\end{figure}

Nevertheless, also in the case of the conifold we have the possibility of smoothing out the space by removing the singularity. We can act in two ways that, differently from what happened in the \cz{} orbifold, are now inequivalent:
\begin{itemize}
\item \emph{Resolution}. The deformation of the K\"ahler structure leaves the defining equation~\eqref{Wcon} invariant, but removes the singularity by blowing up the two-sphere of $T^{1,1}$, which acquires a finite volume at any radius.
\item \emph{Deformation}. The deformation of the complex structure modifies~\eqref{Wcon} as:
\begin{equation}\label{defconifold}
	x^2 + y^2 + z^2 + w^2 = \varepsilon^2\,,
\end{equation}
where $\varepsilon$ is a new deformation parameter. It is easy to see that the resulting space, the deformed conifold, has a three-sphere of finite volume at any radius.
\end{itemize}
The singular, resolved and deformed conifold spaces are depicted in figure~\ref{f:conifold}.
%:Figure: Conifold

Notice that, in perfect analogy to our interpretation of fractional branes on orbifolds as wrapped branes (see sections~\ref{s:Z2}-\ref{s:orbsumm}), the fractional D3-branes stuck at the tip of the cone of the singular conifold can be thought of as D5-branes wrapped on the blown-up $S^2$ of the resolved conifold. It is also very interesting to follow the fate of the branes once we deform the conifold, and this will be the subject of the next subsection.

Before proceeding, let us complete the picture of D-branes on the conifold by considering what happens to the regular D-branes on the \cz{} orbifold once the geometry flows to the conifold via the construction we just described. Recall from section~\ref{s:Z2} that the gauge theory living on $N$ regular D3-branes on \cz{} is a $U(N)\times U(N)$ quiver theory described by the following diagram:
\begin{center}
\includegraphics[scale=.7]{Z2quiver2}
\end{center}
In \Ne{1} notation, the matter content consists of two adjoint chiral superfields $\Phi$ and $\tilde{\Phi}$, and four chiral superfields, $A_i$ and $B_i$, $i=1,2$, transforming respectively in the \hyp{N}{\bar{N}}, \hyp{\bar{N}}{N}{} representations. There is also a tree-level superpotential of the form:
\begin{equation}
	W \sim \tr \Phi\ (A_1 B_1 - A_2 B_2) + \tr \tilde{\Phi}\ (B_1 A_1 - B_2 A_2)\,. 
\end{equation}
We have seen that in order to recover the conifold geometry, we have to add to the tree-level superpotential a supersymmetry-breaking mass term for the adjoint chiral superfields:
\begin{equation}
	\frac{1}{2} m\ ( \tr \Phi^2 - \tr \tilde{\Phi}^2 )\,.
\end{equation}
Integrating out the adjoint superfields, we get the quartic superpotential:
\begin{equation}\label{Wreg}
	W \sim \tr (A_1 B_1 A_2 B_2) - \tr (B_1 A_1 B_2 A_2)\,. 
\end{equation}
We can then conclude that the theory living at low energies on the world-volume of $M$ regular D3-branes transverse to the conifold is a $SU(M)\times SU(M)$ \Ne{1} gauge theory~\cite{Klebanov:1998hh}, with matter fields $A_i$, $B_i$ and tree-level superpotential given by~\eqref{Wreg}. The amount of matter makes this gauge theory conformal, in agreement with our general considerations. We can also generalize the configuration by considering $M$ regular and $N$-fractional D3-branes on the conifold. The resulting \Ne{1} gauge theory will have gauge group $SU(M+N)\times SU(M)$, plus the usual matter in the bifundamental representations, and will be a non-conformal theory.

\subsection*{Geometric transitions}

Summarizing our findings, $N$ fractional D3-branes at the tip of the conifold, or better $N$ D5-branes wrapped on the blown-up two-sphere of the resolved conifold, support on their world-volume pure \Ne{1} Super Yang--Mills theory with $SU(N)$ gauge group.

What happens to the branes if we deform the complex structure and give finite volume to the three-sphere of $T^{1,1}$, thus reaching the deformed conifold? The idea put forward in~\cite{Gopakumar:1998ki,Vafa:2000wi}, is that in the transition of the topology D-branes disappear and are replaced in the deformed geometry by \emph{fluxes} of the NS-NS and R-R three-form field strengths. Thus we get a pure closed string background, which is interpreted (at large $N$) as the infrared geometric dual of the \Ne{1} gauge theory on the D5-branes. The transition from the resolved to the deformed geometry has the dual gauge theory intepretation of \emph{gaugino condensation}, and in fact the deformation parameter can essentially be seen as the expectation value of the gaugino condensate.

We will not insist here on the original motivations and derivation of this duality, which came from (a mirror description of) the embedding in the superstring of a duality between topological strings and Chern--Simons theory. Instead, we will study it from a low-energy supergravity perspective in section~\ref{s:conifold}, where the link between the deformation of the geometry and gaugino condensation will appear clearly.

%:Figure: Geometric transition
\begin{figure}
\begin{center}
\includegraphics[scale=.6]{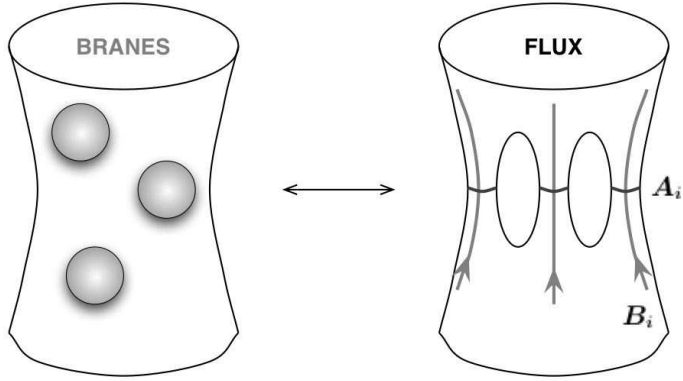}
\caption{{\small Geometric transitions exchange D-branes on resolved manifolds with fluxes on deformed manifolds.}}
\label{f:flop}
\end{center}
\end{figure}
What we just sketched for the conifold is a particular case of the more general framework of ``geometric transitions'', initiated in~\cite{Gopakumar:1998ki,Vafa:2000wi},%
\footnote{See also for instance~\cite{Cachazo:2001jy,Cachazo:2001gh,Cachazo:2001sg,Cachazo:2002pr}.} 
in which dualities are interpreted as transitions in the geometry. In this framework, four-dimensional gauge theories are engineered via configurations of D5-branes wrapped on two-spheres of resolved manifolds. The resolved Calabi--Yau threefold where the D5-branes are wrapped is obtained, as in the previous subsection, as a non-trivial fibration of an ALE space over a complex plane, dictated by the tree-level superpotential for an adjoint superfield which breaks the \Ne{2} supersymmetry down to \Ne{1}:
\begin{equation}
	x^2 + y^2 + z^2 + W'(t)^2 = 0\,.
\end{equation}
By this construction, one can consider a very large class of \Ne{1} quiver gauge theories. In all cases, the infrared dynamics is captured by the dual deformed geometry, defined by:
\begin{equation}\label{defCY}
	x^2 + y^2 + z^2 + W'(t)^2 + f_{n-1}(t)= 0
\end{equation}
(with $f_{n-1}$ a polynomial of degree $n-1$ in $t$, $n$ being the degree of $W'(t)$), where branes have been replaced by fluxes through (compact and non-compact) three-cycles, as depicted in figure~\ref{f:flop}. We will not enter into more details of this framework here, but just introduce an important ingredient which we will use in the following chapters. This piece of information on \Ne{1} gauge theories that is possible to extract from this framework is the \emph{effective superpotential} which is non-perturbatively generated in the infrared. It is given in terms of the type IIB complex three-form $G_3 = dC_2 + (C_0 + i e^{-\Phi}) dB_2$ and of the holomorphic $(3,0)$-form of the non-compact Calabi--Yau threefold:
\begin{equation}\label{omega30}
	\Omega = \frac{1}{2\pi i}\oint_{F=0} \frac{dx\wedge dy\wedge dz\wedge dt}{F}\,,
\end{equation}
where $F(x,y,z,t)=0$ is the defining algebraic equation~\eqref{defCY} of the manifold in $\mathbb{C}^4$. The remarkable proposal of~\cite{Taylor:1999ii,Vafa:2000wi} is that the effective superpotential of the gauge theory is given by the superpotential of the type IIB string on the Calabi--Yau:
\begin{equation}\label{VafaW}
	W_{\text{eff}} = \sum_{i}\
		\left[\ \int_{A_i} G_3 \int_{B_i} \Omega
		- \int_{A_i} \Omega \int_{B_i} G_3\ \right]\,,
\end{equation}
where we have chosen a standard basis of orthogonal cycles on the non-compact Calabi-Yau, $A_i$ being compact and $B_i$ being non-compact three-cycles, as in figure~\ref{f:flop}. The formula~\eqref{VafaW}, which was originally obtained by considering tensions of domain walls in the gauge theory, will be used in chapter~\ref{c:4susy} to obtain the effective superpotential of some \Ne{1} gauge theories of interests, starting from supergravity solutions and geometrical considerations.

% "Gauge/Gravity with Eight Supercharges"

\chapter{Gauge/Gravity with Eight Supercharges}\label{c:8susy}

In this chapter we will examine some explicit examples of gauge/gravity correspondence for scale-anomalous gauge theories, in the case the theory preserves eight real supercharges. We start from two specific examples of classical solutions dual to a three-dimensional gauge theory, allowing us to concretely expose the main issues one encounters when looking for such solutions and trying to extract from them relevant gauge theory information. After studying the two examples, we continue with some general discussion on the gauge/gravity correspondence for the considered class of gauge theories and end the chapter with a study of \Ne{2} four-dimensional Super Yang--Mills theory from supergravity.

\section{An example with wrapped D-branes}\label{s:wD4}

In this section we analyze our first explicit example of gauge/gravity correspondence in a case where the gauge theory has a scale anomaly and less than 16 supersymmetries. In fact, as we already did several times in the previous chapters, instead of giving a general recipe for finding supergravity solutions and interpreting them in gauge theory terms, we prefer concentrating on the illustration of a simple example, and then to extract general properties as a subsequent step.

The example we are going to study consists of a system of $N$ D4-branes wrapped on a two-cycle inside a non-compact Calabi--Yau twofold~\cite{Maldacena:2000mw,DiVecchia:2001uc}. This is exactly the first configuration we used to illustrate the topological twist in section~\ref{s:wrapped}, so we already know that the gauge theory living on the $(2+1)$-dimensional ``flat'' part of the world-volume of the D4-branes is pure Super Yang--Mills theory in 2+1 dimensions with 8 supercharges (which gives \Ne{4} supersymmetry in three-dimensional language).

The configuration under study is schematically shown in the following table:
\begin{center}
\begin{tabular}{|c|c|c|c|c|c|c|c|c|c|c|}
\multicolumn{7}{c}{ }&
\multicolumn{4}{c}{$\overbrace{\phantom{\qquad\qquad\qquad}}^{\text{CY}_2}$}\\
\hline
&\multicolumn{3}{|c|}{$\mathbb{R}^{1,2}$}
&\multicolumn{3}{|c|}{$\mathbb{R}^3$}
&\multicolumn{2}{|c|}{$S^2$}
&\multicolumn{2}{|c|}{$N_2$}\\
\hline
D$4$ &$-$&$-$&$-$&$\cdot$&$\cdot$&$\cdot$&$\bigcirc$&$\bigcirc$&$\cdot$&$\cdot$\\
\hline
\end{tabular}
\end{center}

\subsection*{Finding the supergravity solution}

The goal is to find a classical description of our wrapped D4-branes, similarly to what we did for D-branes in flat space in chapter~\ref{c:dbranes}. This subsection will be rather technical, so the reader who is not interested in the details of the derivation can directly proceed to the solution~\eqref{newsol} and to the next subsection, where the dual gauge theory is discussed.

One could in principle work in ten-dimensional type IIA supergravity, write a suitable ansatz for the system, then solve the equations of motion and find the solution. This is, however, not a simple task because it is not easy to implement directly in ten dimensions the topological twist that we have seen to be necessary from the point of view of the gauge theory living  on the brane.

We therefore proceed in a longer way that has been introduced in~\cite{Maldacena:2000mw}. We take inspiration from the fact that the near-horizon $AdS_5$ geometry generated by D3-branes in flat space can also be thought of as a domain-wall solution of the five-dimensional gauged supergravity theory which is obtained by compactifying type IIB supergravity on $S^5$. In such case, however, things are simple and none of the scalars or gauge fields of gauged supergravity are turned on. However, the gauge fields of gauged supergravity can be used as the main tool for implementing the topological twist at the level of classical solutions, thus allowing us to construct solutions describing (the near-horizon geometry of) D$p$-branes wrapped on supersymmetric cycles.

Let us then consider the generic case of a $p$-brane in a $D$-dimensional theory (where of course $D=10$ for the case of a D-brane or NS5-brane and $D=11$ for the case of an M-brane). Analogously to the case of D3-branes, we can now compactify the relevant supergravity on the sphere $S^{D-p-2}$, obtaining a $(p+2)$-dimensional gauged supergravity theory. The maximal gauge group of this gauged supergravity will be given by the isometry group $SO(D-p-1)$ of the sphere we compactified on.

In this theory, we look for a domain wall solution that preserves the desired amount of supersymmetry, but now we also turn on appropriate gauge fields to implement the topological twist. How does this work? In gauged supergravity the supersymmetry preserving condition contains also the gauge fields and can schematically be written as $(\partial_\mu+\omega_\mu-A_\mu)\ \epsilon =0$. As discussed in section~\ref{s:wrapped}, the twist corresponds to the identification of some of the gauge fields with the spin connection of the manifold around which the brane is wrapped, $A_\mu=\omega_\mu\,$, so that the request of finding covariantly constant spinors is equivalent to that of just finding constant spinors.

\begin{figure}
\begin{center}
\includegraphics[scale=.7]{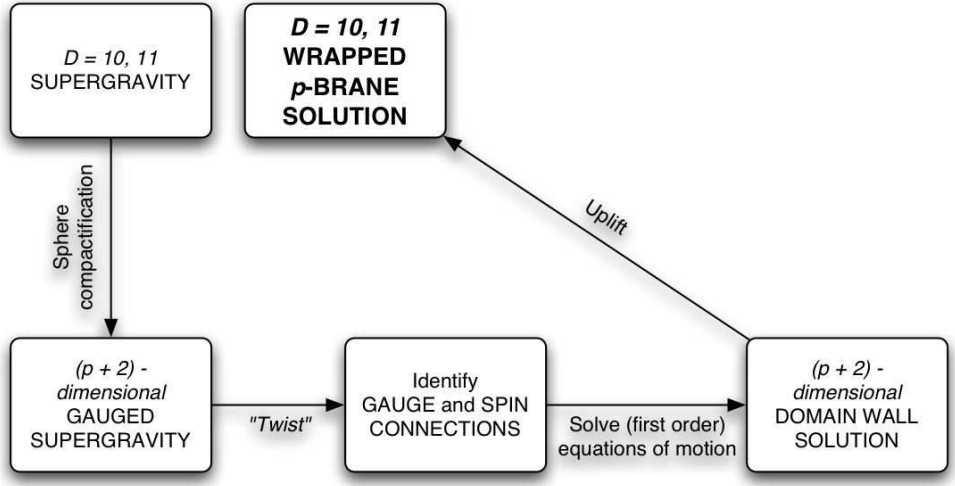}
\caption{{\small Finding the supergravity solution describing a wrapped $p$-brane.}}
\label{f:wsolution}
\end{center}
\end{figure}

Once the solution with the correct properties has been found, the last step is to uplift it to $D$ dimensions by using the formulae given in~\cite{Cvetic:1999xp,Cvetic:2000dm}. The whole procedure is summarized in figure~\ref{f:wsolution}.
%:Figure: Supergravity solution of wrapped branes

In fact, for the specific example we treat in this section we will follow a slightly longer way. Instead of starting from a six-dimensional gauged supergravity as it would be natural for a D4-brane, we look for an eleven-dimensional solution describing an M5-brane wrapped on a two-sphere inside a Calabi--Yau two-fold. The D4-brane solution we are interested in will then be obtained by direct dimensional reduction from eleven to ten dimensions. This choice is mainly due to the fact that we do not explicitly know the relevant six-dimensional gauged supergravity, but on the other hand it allows us to have a non-trivial eleven-dimensional M-brane solution as a by-product.

We therefore start looking for a domain wall solution in seven-dimensional gauged supergravity with the required properties. The starting point is the gauged supergravity considered in~\cite{Liu:1999ai,Maldacena:2000mw}. It is a $U(1)\times U(1)$ consistent truncation of the $SO(5)$ gauged supergravity arising when one compactifies eleven dimensional supergravity on a four-sphere $S^4$. The bosonic field content of the truncated theory consists of the metric, two $U(1)$ gauge fields $A^{(1,2)}$ and two scalar fields $\lambda_{1,2}$.

By setting to zero the supersymmetry variations, one finds the following domain wall solution:
\begin{subequations}\label{7dimsolution}
\begin{align}
	ds_{(7)}^2&=\left(\frac{R_A}{R_0}\right)^2e^{2\rho}e^{\lambda}\, 
		\eta_{ij}d\xi^id\xi^j \nonumber\\
		&\qquad+R_A^2\left(e^\lambda(e^{2\rho}
		-{\textstyle\frac{1}{4}}) (d\ttheta^2+\sin^2\ttheta d\tphi^2)+
 		e^{-4\lambda}d\rho^2\right)\,, \\
	A^{(1)}&=\frac{R_A}{4}\cos\ttheta d\tphi\, , \ \ A^{(2)}=0\,, \\
	\lambda&\equiv\lambda_2\, , \ \ 2\lambda_1+3\lambda_2=0\,, \\ 
	e^{5\lambda}&=\frac{e^{2\rho}+ke^{-2\rho}-\frac{1}{2}} {e^{2\rho}-\frac{1}{4}}\,, 
\end{align}
\end{subequations}
where $\ttheta$ and $\tphi$ are the coordinates along the two-sphere that supports the wrapping, $R_A=2(\pi N)^{1/3}{l_\text{P}}$ is the radius of the $AdS_7$ space appearing in the near horizon limit of the usual ``flat'' M5-brane solution~\eqref{M5}, $R_0$ is an arbitrary integration constant with dimension of a length and $k$ is a (dimensionless) integration constant. Notice that all the coordinates entering in the above solution are dimensionless, except those spanning the unwrapped part of the world-volume of the wall, $\xi^i$, $i=0,\ldots,3$, which have dimensions of a length.

The seven-dimensional solution can be lifted to eleven dimensions with the help of the formuae given in~\cite{Cvetic:1999xp}, which we rewrite here:
\begin{subequations}\label{uplift}
\begin{align}
	d\hat{s}^2&=\tilde{\Delta}^{1/3}ds_{(7)}^2+g^{-2}\tilde{\Delta}^{-2/3}
		\left(X_0^{-1}d\mu_0^2+
		\sum_{i=1}^{2}
		X_i^{-1}\left(d\mu_i^2+\mu_i^2\left(d\phi_i+g{\cal A}^{(i)}
		\right)^2\right)\right)\, , \\ 
	\hde d\hat{C}_3&=
		2g\sum_{\alpha=0}^{2}\left(X_\alpha^2\mu_\alpha^2-
		\tilde{\Delta}X_\alpha\right)\varepsilon_{(7)}+
		g\tilde{\Delta}X_0\, \varepsilon_{(7)}+
		\frac{1}{2g}\sum_{\alpha=0}^{2}X_\alpha^{-1}\ \hd{7}dX_\alpha
		\wedge d(\mu_\alpha^2) \nonumber \\
		&\ \ +\frac{1}{2g^2}\sum_{i=1}^{2}
		X_i^{-2}d(\mu_i^2)\wedge\left(d\phi_i+g{\cal A}^{(i)}\right)
		\wedge\hd{7}{\cal F}^{(i)}\, .
\end{align}
\end{subequations}
Here and below, hats will always refer to eleven-dimensional quantities (recall that our notations are summarized in insert~\ref{i:notations} on page~\pageref{i:notations}). The above formulae are written in the notation of~\cite{Cvetic:1999xp}: $g$ is the seven  dimensional gauged supergravity coupling constant, $\varepsilon_{(7)}$ is the seven dimensional volume form, ${\cal A}^{(1,2)}$ are the two $U(1)$ gauge fields, the $X_{\alpha}$ are a suitable parameterization of the 2 scalars present in the theory and $\tilde{\Delta}$ is given by:
\begin{equation}
	\tilde{\Delta}\equiv\sum_{\alpha=0}^{2}X_\alpha\mu_\alpha^2\, , 
\end{equation}
where $\mu_\alpha$ parameterize a two-sphere: $\mu_0^2+\mu_1^2+\mu_2^2=1$. The quantities appearing in the uplift formulae are given in terms of those appearing in~\eqref{7dimsolution} by the following expressions:
\begin{equation}\label{translation}
\begin{gathered}
	\frac{1}{g^2}=\left(\frac{R_A}{2}\right)^2\,, \qquad
	X_0=X_1=e^{2\lambda}\,, \qquad
	X_2=e^{-3\lambda}\,, \qquad
	{\cal A}^{(1,2)}=2A^{(2,1)}\,, \\ 
	\Delta \equiv e^{3\lambda}\tilde{\Delta}=e^{5\lambda}\cos^2\chi+\sin^2\chi\,,\qquad 
	\varepsilon_{(7)}=-\sqrt{-\det G_{(7)}}\, 
	d\xi_0\wedge \cdots d\xi_3\wedge d\ttheta\wedge d\tphi\wedge d\rho\,, 
\end{gathered} 
\end{equation}
where we have chosen the following parameterization for $\mu_i$: 
\begin{equation}
	\mu_0=\cos\chi\cos\theta\, , \qquad
	\mu_1=\cos\chi\sin\theta\, , \qquad
	\mu_2=\sin\chi\,. 
\end{equation}
By using the above expressions we are now ready to write the full solution in eleven dimensions:
\begin{subequations}\label{11dimsolution}
\begin{align}
	d\hat{s}^2&=\Delta^{1/3}\left(\left(\frac{R_A}{R_0}\right)^2
		e^{2\rho}\, \eta_{ij}d\xi^id\xi^j + 
		R_A^2(e^{2\rho}-{\textstyle\frac{1}{4}})
		(d\ttheta^2+\sin^2\ttheta d\tphi^2)\right) \nonumber \\ 
		&\ \ \ +\Delta^{-2/3}\left(\frac{R_A}{2}\right)^2\Bigg(
		\frac{4\Delta}{e^{5\lambda}}d\rho^2+\Delta d\chi^2+
		\cos^2\chi(d\theta^2+\sin^2\theta d\varphi^2) \nonumber \\
		&\ \ \ + e^{5\lambda}\sin^2\chi\left(d\psi+\cos\ttheta d\tphi\right)^2 \Bigg)\, , \\ 
	\hde d\hat{C}_3&=\frac{R_A^6}{R_0^4}\, 
		e^{4\rho}(e^{2\rho}-\tfrac{1}{4})\sin\ttheta\Bigg(
		2(\Delta+2)\, d\xi^0\wedge\cdots\wedge d\xi^3
		\wedge d\ttheta\wedge d\tphi\wedge d\rho  \nonumber \\
		&\ \ \ +\frac{1}{4}\partial_\rho(e^{5\lambda})\sin(2\chi)\, 
		d\xi^0\wedge\cdots\wedge d\xi^3
		\wedge d\ttheta\wedge d\tphi\wedge d\chi \nonumber \\ 
		&\ \ \ +\frac{1}{16}\frac{e^{5\lambda}\sin(2\chi)}
		{(e^{2\rho}-\frac{1}{4})^2\sin\ttheta}\, 
		d\xi^0\wedge\cdots\wedge d\xi^3
		\wedge d\rho \wedge d\chi \wedge \left(d\psi+\cos\ttheta d\tphi\right)\Bigg)\,  
		\label{11dimsolutionb} 
\end{align}
\end{subequations}
where we have also relabeled the angles appearing in~\eqref{uplift}: $\phi_1=\varphi$, $\phi_2=\psi$. This solution describes the near-horizon geometry of an M5-brane wrapped on a two-sphere. The unwrapped world-volume coordinates are $\xi^0,\ldots, \xi^3$, the wrapped ones are $\ttheta$ and $\tphi$, and the remaining coordinates are transverse to the brane. From~\eqref{11dimsolutionb} we can compute the three-form potential:
\begin{equation}\label{C3}
\begin{aligned}
	\hat{C}_3=&\frac{R_A^3}{8}\, \frac{e^{5\lambda}\cos^3\chi\cos\theta\sin\ttheta}
		{\Delta}\, d\ttheta\wedge d\tphi\wedge d\varphi  \\ 
		&+\frac{R_A^3}{8}\, \frac{e^{5\lambda}
		(\Delta+2)\cos^2\chi\sin\chi\cos\theta}{\Delta^2}\, 
		d\chi\wedge d\varphi\wedge\left(d\psi+\cos\ttheta d\tphi\right) \\
		&+\frac{R_A^3}{8}\, \frac{\partial_\rho(e^{5\lambda})
		\cos^3\chi\sin^2\chi\cos\theta}{\Delta^2}\, 
		d\rho\wedge d\varphi\wedge\left(d\psi+\cos\ttheta d\tphi\right)\, . \\ 
\end{aligned}
\end{equation}
The last step consists of compactifying to ten dimensions the M5-brane solution just obtained along one of its non-wrapped world-volume coordinates (that we choose to be $\xi^3$), in order to get the solution describing the geometry of $N$ wrapped D4-branes. The compactification is obtained by means of the standard expressions in the ten dimensional string frame:
\begin{subequations}\label{compactification}
\begin{align}
	(G_{\text{st}})_{\mu\nu}&=(\hat{G}_{33})^{1/2}\hat{G}_{\mu\nu}\, , \\
		e^{2\Phi}&=(\hat{G}_{33})^{3/2}\, , \\
	(C_3)_{\mu\nu\rho}&=(\hat{C}_3)_{\mu\nu\rho}\,,  
\end{align}
\end{subequations}
where we have split eleven dimensional indices in $\hat{\mu}=\left\{\mu, \xi^3\right\}\,$. This is all we need to get the final expression for the wrapped D4-brane solution~\cite{DiVecchia:2001uc}:
\begin{subequations}\label{10dimsolution}
\begin{align}
	ds^2&=\left(\frac{R_A}{R_0}\right)^3\Delta^{1/2}e^{3\rho}\, 
		\eta_{\alpha\beta}d\xi^\alpha d\xi^\beta+
		\frac{R_A^3}{R_0}\Delta^{1/2}e^{\rho}
		(e^{2\rho}-{\textstyle\frac{1}{4}})
		(d\ttheta^2+\sin^2\ttheta d\tphi^2)  \nonumber \\ 
		&\ \ \ +\frac{R_A^3}{4R_0}\Delta^{-1/2}e^\rho\Bigg(
		\frac{4\Delta}{e^{5\lambda}}d\rho^2+\Delta d\chi^2+
		\cos^2\chi(d\theta^2+\sin^2\theta d\varphi^2) \nonumber \\
		&\ \ \ + e^{5\lambda}\sin^2\chi\left(d\psi+\cos\ttheta d\tphi\right)^2  
		\Bigg)\, , \\ 
	e^{2\Phi}&=\left(\frac{R_A}{R_0}\right)^3\Delta^{1/2}e^{3\rho}\, ,  \\ 
	C_3&=\frac{R_A^3}{8}\, \frac{e^{5\lambda}\cos^3\chi\cos\theta\sin\ttheta}
		{\Delta}\, d\ttheta\wedge d\tphi\wedge d\varphi  \nonumber \\ 
		&\ \ \ +\frac{R_A^3}{8}\, \frac{e^{5\lambda}
		(\Delta+2)\cos^2\chi\sin\chi\cos\theta}{\Delta^2}\, 
		d\chi\wedge d\varphi\wedge\left(d\psi+\cos\ttheta d\tphi\right) \nonumber \\
		&\ \ \ +\frac{R_A^3}{8}\, \frac{\partial_\rho(e^{5\lambda})
		\cos^3\chi\sin^2\chi\cos\theta}{\Delta^2}\, 
		d\rho\wedge d\varphi\wedge\left(d\psi+\cos\ttheta d\tphi\right)\, ,  
\end{align}
\end{subequations}
where the functions $e^{5\lambda}$ and $\Delta$ entering the solution are given by:
\begin{subequations}\label{e5lambdaDelta}
\begin{align}
	e^{5\lambda}&=\frac{e^{2\rho}+ke^{-2\rho}-\frac{1}{2}}
		{e^{2\rho}-\frac{1}{4}}\, , \\
	\Delta&=e^{5\lambda}\cos^2\chi+\sin^2\chi\, . 
\end{align}
\end{subequations}
Before proceeding let us briefly summarize the roles of the various coordinates and constants appearing in~\eqref{10dimsolution}-\eqref{e5lambdaDelta}:
\begin{itemize}
\item $\xi^{\alpha,\beta}$ ($\alpha,\beta=0,1,2$) are the coordinates along the unwrapped brane world-volume;
\item $\ttheta\,,\tphi$ are the coordinates along the wrapped world-volume;
\item $\rho$ is a radial coordinate transverse to the brane;
\item $\chi,\theta,\varphi,\psi$ parameterize the ``twisted'' four-sphere transverse to the brane;
\item $R_A$ is the radius of the $AdS_7$ space appearing in the near horizon geometry of the usual ``flat'' M5-brane solution, which is given in terms of ten dimensional quantities by $R_A=2 \ls (\pi \gs N)^{1/3}\,$;
\item $R_0$ is an arbitrary integration constant with dimension of a length that we will show to set the scale of the radius of the $S^2$ on which the D4-branes are wrapped;
\item $k$ is a dimensionless integration constant.
\end{itemize}
All coordinates are dimensionless except $\xi^\alpha$ which have dimension of a length.

A D4-brane is coupled naturally to a $5$-form potential while the solution given above contains a R-R $3$-form potential. However, the latter is related to $C_5$ by the duality relation $dC_5=\hd{} dC_3$ (in the string frame). By using it we get:
\begin{equation}
\begin{split}
	C_5=&\ \frac{R_A^6}{R_0^4}\, \Delta e^{4\rho}
		\left(e^{2\rho}-\tfrac{1}{4}\right) \sin\ttheta\, 
		d\xi^0\wedge d\xi^1\wedge d\xi^2\wedge d\ttheta\wedge d\tphi  \\ 
		&-\frac{R_A^6}{R_0^4}\, \frac{1}{2}e^{4\rho}\sin^2\chi 
		\ d\xi^0\wedge d\xi^1\wedge d\xi^2\wedge d\rho
		\wedge\left(d\psi+\cos\ttheta d\tphi\right)  \\
		&-\frac{R_A^6}{R_0^4}\, \frac{1}{8}e^{4\rho}e^{5\lambda}\sin(2\chi)
		\ d\xi^0\wedge d\xi^1\wedge d\xi^2\wedge d\chi
		\wedge\left(d\psi+\cos\ttheta d\tphi\right)\, .
\end{split}
\end{equation}

The supergravity solution for the D4-branes wrapped on $S^2$ as given in~\eqref{10dimsolution} is written in a way in which  the role of the different coordinates and factors is not immediately clear. The first thing that we can do in order to clarify the role of the various terms appearing in the solution is to extract the warp factors for the longitudinal and transverse part of the metric in the string frame. They are given in terms of a function $H$ that for a D4-brane is related to the dilaton through the following relation:
\begin{equation}\label{myH}
	H=e^{-4\Phi}=\left(\frac{R_0}{R_A}\right)^6\Delta^{-1}e^{-6\rho}\,.
\end{equation}
Using the previous definition of $H$, one can immediately see that the dependence on $H$ of the four longitudinal unwrapped directions of the metric is the one corresponding to four ``flat'' world-volume directions: $H^{-1/2}\eta_{\alpha\beta}d\xi^\alpha d\xi^\beta\,$, as expected. We also expect three transverse directions ($\theta\,$, $\varphi$ and a suitable combination of $\rho$ and $\chi\,$) to be flat, apart from the usual warp factor $H^{1/2}\,$. This can be seen to be correct by using instead of the coordinates $\rho$ and $\chi$ the following new coordinates:
\begin{equation}\label{mycoord}
\begin{cases}
	r=\frac{R_A^3}{2R_0^2}e^{2\rho}\cos \chi \\ \\
	\sigma=\frac{R_A^3}{2R_0^2}\left[e^{2\rho}\left(e^{2\rho}-\frac{1}{4}\right)
		e^{5\lambda}\right]^{1/2}\sin \chi
\end{cases}
\end{equation}
which have dimensions of a length. In terms of the new coordinates~\eqref{mycoord}, the solution for the metric, dilaton and R-R 5-form becomes:
\begin{subequations}\label{newsol}
\begin{align}
	ds^2&=H^{-1/2}\left[\eta_{\alpha\beta}d\xi^\alpha d\xi^\beta+\mathcal{Z}R_0^2
		(d\ttheta^2+\sin^2\ttheta d\tphi^2)\right]\label{newsolm}\\
		&\qquad\quad+H^{1/2}\left[dr^2+r^2\left(d\theta^2+\sin^2\theta d\varphi^2\right)
		+\frac{1}{\mathcal{Z}}\left(d\sigma^2+\sigma^2\left(d\psi+\cos\ttheta 		d\tphi\right)^2\right)\right]\nonumber\,,\\
	e^\Phi&=H^{-1/4}\,,\label{newsold}\\
	C_5&=d\xi^0\wedge d\xi^1\wedge d\xi^2\wedge
		\left[\frac{1}{H}\mathcal{Z}R_0^2\sin\ttheta\ d\ttheta\wedge d\tphi
		-\frac{1}{\mathcal{Z}}\sigma d\sigma\wedge\left(d\psi+\cos\ttheta 		d\tphi\right)\right]\,,\label{newsolC5}
\end{align}
\end{subequations}
where the functions $H$ and $\mathcal{Z}$ are (implicitly) defined as:
\begin{equation}
	H(r,\sigma)=\left(\frac{R_0}{R_A}\right)^6\Delta^{-1}(r,\sigma)e^{-6\rho(r,\sigma)}\,,\qquad
	\mathcal{Z}(r,\sigma)=e^{-2\rho(r,\sigma)}\left(e^{2\rho(r,\sigma)}-\frac{1}{4}\right)\,.
\end{equation}
The structure of the solution in the form~\eqref{newsol} is much clearer. First of all one can distinguish the trivial ``flat'' part of the solution from the nontrivial part coming from the internal directions of the four-dimensional Calabi--Yau space.  In this sense, the coordinates $r$ and $\sigma$ that we have introduced represent two radial directions, respectively in the ``flat'' transverse space and in the space $N_2$ transverse to the brane but nontrivially fibered on the two-cycle on which the brane is wrapped. Moreover, the function $\mathcal{Z}$ represents the ``running volume'' of the two-cycle, with the constant $R_0$ being the radius of the $S^2$ when $\mathcal{Z}=1\,$ (see insert~\ref{i:wEH} on page~\pageref{i:wEH} for another interesting interpretation), while in the part of the metric containing $\sigma$ and $\psi$ we can easily see the twist which is required for having a supersymmetric gauge theory living on the brane. Finally, also the R-R potential has a quite standard part ($H^{-1}$ times the volume form of the longitudinal space), plus an additional part due to the twist.

%:Insert: Wrapped branes and Eguchi-Hanson
\begin{Insert}{Wrapped branes and the Eguchi--Hanson geometry}\label{i:wEH}
The supergravity solutions of branes wrapped on cycles described in the text can also be seen as realization of ``warped'' Eguchi--Hanson spaces.

For example, starting from the wrapped D4-brane solution~\eqref{newsol}, we can define a new coordinate $z$ and a function $\widetilde{\mathcal{Z}}$ as:
\begin{equation*}
	z = R_0 \left(1+\frac{\sigma^2}{R_0^2}\right)^{1/4}\,,\qquad
	\widetilde{\mathcal{Z}} = \mathcal{Z} \left(1+\frac{\sigma^2}{R_0^2}\right)^{-1/2}\,,
\end{equation*}
in terms of which the metric~\eqref{newsolm} becomes:
\begin{equation}\label{ALE}
\begin{aligned}
   ds_{\text{st}}^2&=H^{-1/2}\left\{\eta_{\alpha\beta}d\xi^\alpha 
        d\xi^\beta+\widetilde{\mathcal{Z}}\ z^2(d\ttheta^2+\sin^2\ttheta d\tphi^2)\right\}\\
        &\qquad+H^{1/2}\Bigg\{dr^2+r^2\left(d\theta^2+\sin^2\theta d\varphi^2\right)\\
        &\qquad+\frac{1}{\widetilde{\mathcal{Z}}}\left[4\left(1-\frac{R_0^4}{z^4}\right)^{-1}dz^2
        +z^2\left(1-\frac{R_0^4}{z^4}\right)
        \left(d\psi+\cos\ttheta d\tphi\right)^2\right]\Bigg\}\,.
\end{aligned}
\end{equation}
The metric we have obtained on the four-dimensional space spanned by the coordinates $\{\ttheta,\tphi,z,\psi\}$ is that of a ``warped'' Eguchi--Hanson space~\eqref{EHm}. This fact provides additional evidence of the geometrical structure of the background: the D4-branes are wrapped on the two-sphere, of radius $R_0\,$, inside the simplest ALE space (which corresponds to the blow-up of a \cz{} orbifold).
\end{Insert}

\subsection*{The dual gauge theory}

We are now ready to show how the classical solution we have just found is capable of giving relevant information on the gauge theory living on the D-branes. In this specific example, we will be able to get the perturbative running coupling constant and metric on the moduli space of the three-dimensional gauge theory under consideration. We will also use this section to illustrate in full detail some techniques we will use many times in all further examples.

Recall how we extracted the (constant) gauge coupling of the gauge theory living on the world-volume of a D$p$-brane in flat space in chapter~\ref{c:dbranes}. The fields of the supergravity solution were substituted into the world-volume action of a D$p$-brane, and this could have a twofold interpretation:
\begin{itemize}
\item a geometric interpretation, namely the one of moving a single D-brane along a flat direction, away from the remaining $N$ branes generating the geometry. The single D-brane acts then as a \emph{probe} in the geometry (neglecting the back-reaction of the single brane on the background), as was shown in figure~\ref{f:probe}. The expansion of the world-volume action of the probe for small velocities then reads the metric on the moduli space of the gauge theory in the Coulomb phase $U(N+1) \to U(N)\times U(1)$.
\item a field-theoretic interpretation, namely the one of studying the open string dynamics on the stack of D-branes at low energies. The full non-abelian world-volume action is not known, but one uses the fact that it reduces to the Super Yang--Mills action in the field theory limit $\ls\to 0$ to promote the fields to non-abelian fields.
\end{itemize}
The two interpretations were equivalent in the case considered in chapter~\ref{c:dbranes}, and they are still equivalent now, since we have three flat directions in which the wrapped D4-branes can move. For now, we will adopt the probe point of view, in order to comment on the moduli space metric.

Let us therefore study the dynamics of a probe D4-brane wrapped on $S^2$ in the geometry generated by the solution~\eqref{newsol}. As in chapter~\ref{c:dbranes}, we exploit diffeomorphism invariance to choose the ``static gauge'', namely we align the world-volume coordinates $\xi^0,\ldots,\xi^3$ of the probe brane with the space-time coordinated given by the solution. The world-volume action for a single D4-brane in the string frame in the static gauge, in the case in which $B$ is vanishing, is given by~\eqref{Dpwv}:
\begin{equation}\label{wprobe}
	S_{\text{D}4}=-\tau_4
		\int{d^{3}\xi d\ttheta d\tphi\ e^{-\Phi}
		\sqrt{-\det \left[\hat{G}_{ab}+
		2\pi \ls^2 F_{ab}\right]}}
		+\tau_4\int_{{\cal M}_5}{\big(
		\hat{C}_5+2\pi\ls^2 \hat{C}_3\wedge F\big)}\,, 
\end{equation}
where $a,b=\{0,1,2,\ttheta,\tphi\}$ and as usual hats denote pull-backs onto the brane world-volume.

Where are we able to move our probe? We expect that it is free to move only in the three-dimensional space outside the Calabi--Yau twofold, spanned by the coordinates $r,\theta,\phi$. In fact, let us compute the static potential between the probe and the stack of $N$ D4-branes, simply by substituting the solution \eqref{newsol} into~\eqref{wprobe}. The contribution of the Dirac--Born--Infeld part is given by: 
\begin{equation}
	e^{-\Phi}\sqrt{-\det G_{ab}}=\sin\ttheta\ \frac{\mathcal{Z}R_0^2}{H}
		\left(1+\frac{\sigma^2H}{\mathcal{Z}^2R_0^2}\right)^{1/2}\,.
\end{equation}
Adding to it the Wess--Zumino part, whose contribution is computed using the expression~\eqref{newsolC5} of the R-R 5-form, we get the following expression for the static potential:
\begin{equation}\label{potential}
	S_{\text{pot}}=-\tau_4
		\int d^{3}\xi d\ttheta d\tphi\ \sin\ttheta
		\frac{\mathcal{Z}R_0^2}{H}
		\left[\left(1+\frac{\sigma^2H}{\mathcal{Z}^2R_0^2}\right)^{1/2}-1\right]\,.
\end{equation}
We see that in general there is a force between the branes, and when this happens the configuration is not supersymmetric. This is because, as we said, the brane is not allowed to move supersymmetrically inside the Calabi--Yau space. We then have to allow the probe brane to move only in the ``flat'' part of the transverse space, keeping it fixed at the locus $\sigma=0$ in the ``internal'' transverse space. This makes the potential \eqref{potential} vanish, yielding a supersymmetric configuration. Therefore in the following we will always work at the ``supersymmetric locus'' $\sigma=0$.

In order to study the dynamics of the probe brane, we will allow the transverse coordinates $x^i=\{r,\theta,\phi\}$ to depend on the ``flat'' world-volume coordinates $\xi^\alpha$ but not on the ``wrapped'' ones $\ttheta$ and $\tphi$. Moreover, the gauge field $F_{\alpha\beta}$ is defined to be non-vanishing only on the ``flat'' part of the world-volume. Let us start from the Dirac--Born--Infeld part of the action in~\eqref{wprobe}. By expanding the determinant, we find:
\begin{multline}
	S_{\text{DBI}}\simeq-\tau_4\int d^{3}\xi d\ttheta d\tphi\
		e^{-\Phi}\sqrt{-\det G_{\alpha\beta}}\\
		\times\left\{1+\frac{1}{2}G^{\alpha\beta}G_{ij}\partial_\alpha X^i\partial_\beta X^j
		+\frac{(2\pi\ls^2)^2}{4}G^{\alpha\gamma}G^{\beta\delta}
		F_{\alpha\beta}F_{\gamma\delta}\right\}\,.
\end{multline}
Inserting the expressions \eqref{newsolm} and \eqref{newsold} for the metric and dilaton we get:
\begin{multline}
	S_{\text{DBI}}=-\frac{\tau_4}{2}\int d^{3}\xi d\ttheta d\tphi\ \sin\ttheta
		\frac{\mathcal{Z}R_0^2}{H}\\
		\times \left\{1+\frac{1}{2}H 
		\big[(\partial r)^2+r^2\left((\partial\theta)^2+\sin^2\theta(\partial\varphi)^2\right)\big]
		+\frac{(2\pi \ls^2)^2}{4}H F^2\right\}\,,
\end{multline}
where we have included an additional factor of $1/2$ due to the normalization of the generators of the gauge group.

Before proceeding, let us notice a fact which will be of importance in the following. At a distance given by:
\begin{equation}
	r_{\text{e}}=\frac{R_A^3}{8R_0^2}=\frac{\pi g_s \ls^3 N}{R_0^2}\,,
\end{equation}
the function $\mathcal{Z}$ vanishes, which means the probe becomes effectively tensionless! This fact implies that we have problems in thinking of the geometry as being generated by the subsequent stacking of the $N$ branes, as we did in chapter~\ref{c:dbranes}. Could this be responsible for the singular nature of the solution? The vanishing of the probe tension is a signal of the fact that at that scale there are new light degrees of freedom entering into play, and that they are not accounted for by the supergravity approximation, which therefore cannot be trusted for $r < r_e$. The singularity is then just to be interpreted as a limitation of supergravity, and we will ignore it and show that (perturbative) gauge theory information is present anyway. All this situation, that we will see reappearing in many cases and will study a bit more in depth in a while, is known as \emph{enhan\c con} mechanism~\cite{Johnson:1999qt}.

As in chapter~\ref{c:dbranes}, we interpret the transverse scalars as Higgs fields for the gauge theory living on the brane: $x^i=2\pi\ls^2\Phi^i$. Then, defining the scale $\mu$ as $r=2\pi\ls^2 \mu$ and integrating over the volume of the two-sphere on which the brane is wrapped, we obtain the final expression for the Dirac--Born--Infeld part:
\begin{multline}\label{wdbi}
	S_{\text{DBI}}=-\frac{4\pi \tau_4}{2}
		\int d^{3}\xi\
		\frac{\mathcal{Z}R_0^2}{H}\\
		\times\left\{1+\frac{(2\pi\ls^2)^2}{2}H 
		\big[(\partial\mu)^2+\mu^2\left((\partial\theta)^2
		+\sin^2\theta(\partial\varphi)^2\right)\big]
		+\frac{(2\pi\ls^2)^2}{4}H F^2\right\}\,.
\end{multline}
Turning now to the Wess--Zumino part, the pullback of $C_3$ is given by:
\begin{equation}
	\hat{C}_3=\frac{1}{8}R_A^3\cos\theta\sin\ttheta\ \partial_\alpha \varphi\ d\xi^\alpha\wedge 		d\ttheta\wedge d\tphi \,. 
\end{equation}
Then from~\eqref{wprobe} we get:
\begin{equation}\label{wwz}
\begin{split}
	S_{\text{WZ}}&=\tau_4\int d^3\xi d\ttheta d\tphi\ \sin\ttheta\left\{
		\frac{\mathcal{Z}R_0^2}{H}
		+\frac{2\pi\ls^2 R_A^3}{16}\cos\theta\varepsilon^{\alpha\beta\gamma}
		\partial_\alpha \varphi F_{\beta\gamma}\right\}\\
		&= 4\pi \tau_4 \int d^3\xi \left\{
		\frac{\mathcal{Z}R_0^2}{H}
		+\frac{2\pi\ls^2 R_A^3}{16}\cos\theta\varepsilon^{\alpha\beta\gamma}
		\partial_\alpha \varphi F_{\beta\gamma}\right\}\\
\end{split}
\end{equation}
Putting~\eqref{wdbi} and~\eqref{wwz} together and substituting the expressions for $\tau_4$, $R_A$ and for the function $\mathcal{Z}$, the probe action finally becomes:
\begin{multline}\label{wfinprobe}
	S_{\text{D}4}=-\frac{R_0^2}{2\pi \gs \ls}
		\int d^{3}\xi \left(1-\frac{\gs \ls N}{2R_0^2\mu}\right)
		\Bigg\{\frac{1}{2} 
		\big[(\partial\mu)^2+\mu^2\left((\partial\theta)^2
		+\sin^2\theta(\partial\varphi)^2\right)\big]
		+\frac{1}{4} F^2\Bigg\}\\
		+\frac{N}{8\pi}\int d^3\xi \cos\theta
		\varepsilon^{\alpha\beta\gamma}\partial_\alpha\varphi F_{\beta\gamma}\,.
\end{multline}

Notice that, unlike the case of D-branes in flat space, the coefficient of $F^2$ in the effective action is not constant, and this means that the gauge coupling constant is running. What is striking is that we can get precise quantitative information. In fact, we can read the perturbative running gauge coupling constant of the three-dimensional gauge theory as a function of the scale $\mu$:
\begin{equation}\label{wrunning}
	\frac{1}{g^2_{\text{YM}}(\mu)}=
		\frac{1}{g^2_{\text{YM}}}\left(1-\frac{g^2_{\text{YM}}N}{4\pi\mu}\right)\,, 
\end{equation}
where we have defined the bare coupling as:
\begin{equation}\label{wbare}
	g^2_{\text{YM}}=\frac{2\pi \gs \ls}{R_0^2}\,,
\end{equation}
The resulting running is in perfect agreement with the gauge theory expectations, computable for instance from insert~\ref{i:running} on page~\pageref{i:running}. We have then just given our first example of extraction of a relevant piece of quantum information on a supersymmetric scale-anomalous gauge theory from a classical supergravity solution.

Let us proceed and analyze the metric on the moduli space of the theory, which, as discussed before, is seen by the effective action of the probe. Our result~\eqref{wfinprobe} does not give explicitly the full metric on the moduli space of $\mathcal{N}=4\,$, $D=2+1$ Super Yang--Mills theory yet. In fact such a metric must be hyperK\"ahler \cite{Alvarez-Gaume:1981hm} and in~\eqref{wfinprobe} we have only three moduli, and not four as it should be in a hyperK\"ahler metric. We need an extra modulus that can be obtained by dualizing the vector field. In order to do that, we regard the original action in~\eqref{wfinprobe} as a function of $F_{\alpha\beta}$ and we add to it a term:
\begin{equation}\label{addterm}
	- \int\ \Sigma\ dF\,,
\end{equation}
so that the equation of motion for the auxiliary field $\Sigma$ enforces the Bianchi identity for $F$ on shell. By partially integrating the additional term in~\eqref{addterm}, we are left with the following action:
\begin{multline}\label{dual1}
	S_{\text{D}4}=-
		\int d^3 \xi\ \frac{1}{g^2_{\text{YM}}(\mu)}
		\Bigg\{\frac{1}{2} 
		\big[(\partial\mu)^2+\mu^2\left((\partial\theta)^2
		+\sin^2\theta(\partial\varphi)^2\right)\big]
		+\frac{1}{4} F^2\Bigg\} \\
		+\frac{N}{8\pi}\int d^3 \xi \cos\theta 
		\varepsilon^{\alpha\beta\gamma}\ \partial_\alpha\varphi\ F_{\beta\gamma}
		+\frac{1}{2}\int d^3 \xi 
		\varepsilon^{\alpha\beta\gamma}\ \partial_\alpha\Sigma\ F_{\beta\gamma}\,.
\end{multline} 
We can then eliminate $F$ by means of its equation of motion that follows from~\eqref{dual1}:
\begin{equation}
	F_{\beta\gamma}=g^2_{\text{YM}}(\mu)\varepsilon^{\alpha\beta\gamma}
		\left[\frac{N}{4\pi}\cos\theta\partial_\alpha\varphi
		+\partial_\alpha\Sigma\right]\,,
\end{equation}
and we arrive at an action that contains four moduli, given by:
\begin{multline}
	S_{\text{D}4}= -\frac{1}{2}
		\int d^3 \xi \Bigg\{
		\frac{1}{g^2_{\text{YM}}(\mu)} 
		\big[(\partial\mu)^2+\mu^2\left((\partial\theta)^2
		+\sin^2\theta(\partial\varphi)^2\right)\big]\\
		+g^2_{\text{YM}}(\mu)\left(\frac{N\cos\theta}{4\pi}\partial\varphi
		+\partial\Sigma\right)^2\Bigg\}\,.
\end{multline} 
The complete metric on the moduli space $\mathcal{M}$ of the gauge theory, in terms of the 4 scalars $\mu$, $\theta$, $\phi$ and $\Sigma$ is finally given by:
\begin{equation}\label{wMetricMS}
	ds^2_{\mathcal{M}}=
		\frac{1}{g^2_{\text{YM}}(\mu)}
		\left(d\mu^2+\mu^2d\Omega^2\right)
		+g^2_{\text{YM}}(\mu)
		\left(d\Sigma+
		\frac{N\cos\theta}{4\pi}d\varphi\right)^2\,,
\end{equation}
where $d\Omega^2=d\theta^2+\sin^2\theta d\varphi^2\,$. The metric in~\eqref{wMetricMS} is indeed hyperK\"ahler since it has precisely the form of the Taub-NUT metric~\cite{Hawking:1977jb}. However, because of the form given in~\eqref{wrunning} of the function $g_{\text{YM}}(\mu)$ our metric has a ``negative mass'' and thus is singular. This is due
to the fact that with our probe analysis we are only able to reproduce the perturbative behaviour of the gauge theory. As discussed in~\cite{Johnson:1999qt}, the complete metric should also include the instanton contribution, becoming a completely nonsingular generalization of the Atiyah-Hitchin metric.

\section{An example with fractional D-branes}\label{s:fD2}

The second example we will consider is a system of fractional D-branes on a \cz{} orbifold, where the $\mathbb{Z}_2$ action reverses the last four coordinates of $\mathbb{R}^{1,9}$:
\begin{equation}
	x^r \to -x^r\,,\qquad r=6,7,8,9.
\end{equation}
We extensively studied this setup in section~\ref{s:Z2}, where we discovered that a fractional D$p$-brane at the fixed point of a \cz{} orbifold supports on its world-volume a $(p+1)$-dimensional gauge theory with 8 supercharges.

Therefore, the gauge theory living on $N$ fractional D2-branes on $\mathbb{R}^{1,5}\times\cz$ is the same $(2+1)$-dimensional \Ne{4} Super Yang--Mills theory that we considered in section~\ref{s:wD4}, where we engineered it by means of wrapped D4-branes. We will now study it from another perspective. After studying the fractional D2-branes, we will modify the setup by adding to the system D6-branes extendend along the orbifolded directions, and we will see that this allows the description of matter multiplets in the fundamental representation of the gauge group.

Our starting point is then a configuration of type IIA string theory made of $N$ fractional D2-branes extended along $x^0,x^1,x^2$~\cite{DiVecchia:2001uc}, as shown schematically in the following table:
\begin{center}
\begin{tabular}{|c|c|c|c|c|c|c|c|c|c|c|}
\multicolumn{7}{c}{ }&
\multicolumn{4}{c}{$\overbrace{\phantom{\qquad\qquad\qquad}}^{\cz}$}\\
\hline
&0&1&2&3&4&5&6&7&8&9\\
\hline
D$2$ &$-$&$-$&$-$&$\cdot$&$\cdot$&$\cdot$&$\cdot$&$\cdot$&$\cdot$&$\cdot$\\
\hline
\end{tabular}
\end{center}

\subsection*{Finding the supergravity solution}

Let us now describe the derivation of the classical solution. The procedure is again described in some detail, so the reader interested in the dual gauge theory might skip to the next subsection, where only the final solution~\eqref{solution} is needed.

In order to describe the above system by means of a supergravity solution, we have to recall how the low-energy fields which appear in the effective action behave in the orbifold background. As we learned in chapter~\ref{c:engineering}, our background is characterized by the presence of a 2-form $\omega_2$, dual to the exceptional 2-cycle $\Sigma$ of the ALE space which is obtained by the resolution of the orbifold, that satisfies the properties~\eqref{omegaprop}. In the orbifold limit, the volume of $\Sigma$ vanishes, but (as we have seen in~\eqref{bbkgd}) the background value of the integral of $B_2$ on it has to remain finite in order to define a sensible CFT~\cite{Aspinwall:1995zi,Douglas:1997xg}:
\begin{equation}\label{asp}
	\int_{\Sigma}B_2=\frac{\left(2\pi \ls\right)^2}{2}\,.
\end{equation}
The supergravity fields can have components along the vanishing cycle, so the following decompositions hold for the NS-NS two-form and the R-R three-form:
\begin{equation}\label{decomp}
	B_2=\bar{B}_2+b\,\omega_2\,,\qquad
	C_3=\bar{C}_3+A_1\wedge\omega_2\,.
\end{equation}
Since we will be looking for supergravity solutions which represent branes without a $B_2$ field in their world-volume, in the following we will put $\bar{B}_2=0$, so we simply have:
\begin{equation}\label{fieldform}
	B_2=b\,\omega_2\,,\qquad
	C_3=\bar{C}_3+A_1\wedge\omega_2\,,
\end{equation}
where, because of eq.~\eqref{asp}:
\begin{equation}\label{asp2}
	b=\frac{\left(2\pi \ls \right)^2}{2}+\tilde{b}\,,
\end{equation}
and $\tilde{b}$ represents the fluctuation around the background value of $b$.  The fields $b$ and $A_1$ in~\eqref{fieldform} are the supergravity counterparts of the twisted fields that we found in chapter~\ref{c:engineering} by studying the massless spectrum of closed strings on the orbifold.

We then proceed looking for the classical solution. In this section we will always work in the Einstein frame for convenience. As we discussed in section~\ref{s:classol}, in order to find a D-brane solution we need to add to the type IIA effective action the world-volume action~\eqref{fracwv} of a fractional D2-brane as a boundary term, so that the full action is $S_{\text{IIA}}+S_{\text{D}2}^{\text{f}}$.

The first step is the substitution of the form~\eqref{fieldform} of the fields into the type IIA supergravity action~\eqref{IIAE}, and we get:
\begin{multline}\label{6dimaction}
	S^\prime_{\text{IIA}}=
		\frac{1}{2\kappa^2}\Bigg\{
		\int d^{10}x\sqrt{-G}\, R
		-\frac{1}{2}\int\Big(d\Phi\wedge\hd{}d\Phi
		- e^{3\Phi/2}dC_1\wedge\hd{} dC_1
		- e^{\Phi/2}d\bar{C}_3\wedge\hd{}d\bar{C}_3
		\Big)\\
		-\frac{1}{4}\int_{\mathbb{R}^{1,5}}\Big(e^{-\Phi}db\wedge\hd{6}db
		-e^{\Phi/2}G_2\wedge\hd{6}G_2
		-2b\wedge d\bar{C}_3\wedge dA_1
		\Big)\Bigg\}\,,   
\end{multline}
where we have introduced the quantity: 
\begin{equation}
	G_2\equiv dA_1-C_1\wedge db\, .
\end{equation}
By varying the previous action one finds the equations of motion for the fields $C_1\,$, $\bar{C}_3\,$, $A_1\,$, $b$ and $\Phi$ respectively:
\begin{subequations}\label{orbeqs}
\begin{align} 
	d\left(e^{3\Phi/2}\ \hd{} dC_1\right)
		-\frac{1}{2}e^{\Phi/2}db\wedge\hd{6}G_2\wedge\Omega_4
		+2\kappa^2\frac{\delta \mathcal{L}_b}{\delta C_1}&=0\,,\label{orbeqC1}\\
	d\left(e^{\Phi/2}\ \hd{} d\bar{C}_3\right)
		+\frac{1}{2}db\wedge dA_1\wedge\Omega_4
		+2\kappa^2\frac{\delta \mathcal{L}_b}
		{\delta \bar{C}_3}&=0\,,\label{orbeqC3}\displaybreak[1]\\
	d\left(e^{\Phi/2}\ \hd{6} G_2\right)
		+db\wedge d\bar{C}_3+4\kappa^2\frac{\delta \mathcal{L}_b}{\delta
		A_1}&=0\,,\label{orbeqA1}\displaybreak[1]\\
	d\left(e^{-\Phi}\ \hd{6}db-e^{\Phi/2}C_1\wedge \hd{6} G_2\right)
		+d\bar{C}_3\wedge dA_1+4\kappa^2\frac{\delta \mathcal{L}_b}
		{\delta b}&=0\,,\label{orbeqb}\\
	d\hd{}d\Phi+\frac{3}{4}e^{3\Phi/2}dC_1\wedge\hd{}dC_1
		+\frac{1}{4}e^{\Phi/2}
		d\bar{C}_3\wedge\hd{}d\bar{C}_3\qquad\qquad\qquad\nonumber&\\
		+\frac{1}{4}\Big[e^{-\Phi}\ db\wedge\hd{6}db
		+\frac{1}{2}
		e^{\Phi/2}\  G_2 \wedge \hd{6} G_2
		\Big]\wedge\Omega_4+2\kappa^2\frac{\delta \mathcal{L}_b}
		{\delta\Phi}&=0\label{orbeqphi}\,,   
\end{align}
\end{subequations}
where we have defined $\Omega_4=\delta(x^6)\cdots\delta(x^9)\ dx^6\wedge\cdots\wedge
dx^9\,$ and $\mathcal{L}_b$ denotes the boundary lagrangian coming from $S_{\text{D}2}^{\text{f}}$. Of course one also has the Einstein equations, that we will not write here for simplicity (see~\cite{DiVecchia:2001uc} for details).

The next step is finding an appropriate ansatz. For the ``untwisted'' fields we consider the following standard D2-brane ansatz (compare with~\eqref{DpsolE}):
\begin{subequations}\label{ansatz}
\begin{align}
	ds^2&=H_2^{-5/8}\ \eta_{\alpha\beta} dx^\alpha dx^\beta
		+H_2^{3/8}\left(\delta_{ij}dx^idx^j + \delta_{rs}dx^r dx^s\right)\,,\\
	e^\Phi&=H_2^{1/4}\,,\\
	\bar{C}_3&=\left(H_2^{-1}-1\right)dx^0\wedge dx^1\wedge dx^2\,,\label{ansC3}
\end{align}
\end{subequations}
where we have divided the coordinates in three groups: $x^{\alpha,\beta}=\left\{x^0,x^1,x^2\right\}$ denote the coordinates along the world-volume of the branes, $x^{i,j}=\left\{x^3,x^4,x^5\right\}$ denote the unorbifolded transverse coordinates, while we recall that $x^{r,s}=\left\{x^6,x^7,x^8,x^9\right\}$ are the coordinates along the orbifold. The function $H_2$ depends on the radial coordinate $\rho=[(x^3)^2+\ldots+(x^9)^2]^{1/2}$ of the space transverse to the D2-branes. All ``untwisted'' fields not appearing in~\eqref{ansatz} are taken to vanish.

In order to find a sensible ansatz for $A_1$ we need to take a more careful look at the contributions coming from the boundary action describing the world-volume theory of the branes. Let us recall the world-volume action~\eqref{fracwv}, whose relevant terms in the Einstein frame read:
\begin{equation}\label{D2act}
\begin{split}
	S_{\text{D}2}^{\text{f}} \to &-\frac{\tau_2}{2}\int{d^{3}x\ e^{-\Phi/4}
		\sqrt{-\det G_{\alpha\beta}}
		\left(1+\frac{\tilde{b}}{2\pi^2 \ls^2}\right)}\\
		&+\frac{\tau_2}{2}\int_{{\cal M}_3} \left[
		\bar{C}_3\left(1+\frac{\tilde{b}}{2\pi^2\ls^2}\right)+
		\frac{A_3}{2\pi^2\ls^2}\right]\,.
\end{split}
\end{equation}
Since the above action does not depend on $A_1$, \eqref{orbeqA1} simply becomes:
\begin{equation}
	d\left(e^{\Phi/2}\ \hd{6} G_2\right)
		+db\wedge d\bar{C}_3=0\,.
\end{equation}
Taking into account the expression~\eqref{ansC3} for $\bar{C}_3\,$, we see that this equation is easily satisfied by imposing: 
\begin{equation}\label{duansA1}
	e^{\Phi/2}\ \hd{6} G_2=H_2^{-1} db\wedge dx^0 \wedge dx^1 \wedge dx^2\,,
\end{equation}
so that we obtain:
\begin{equation}
	dA_1= \frac{1}{2}\varepsilon_{ijk} \partial_ib dx^j\wedge dx^k\,,\\
\end{equation}
where $\varepsilon_{ijk}$ is such that $\varepsilon_{345}=\varepsilon^{345}=+1\,$. 

We are now ready to find the complete solution. Inserting the ans\"atze into the equations of motion~\eqref{orbeqs} and computing all the relevant contributions coming from the boundary action after some algebra we get:
\begin{subequations}
\begin{equation}\label{H2}
	\left(\delta^{ij}\partial_i\partial_j 
		+\delta^{rs}\partial_r\partial_s\right) H_2
		+\frac{1}{2}\delta^{ij}\partial_i b\partial_j b\delta(x^6)\cdots\delta(x^9)
		+\kappa^2\tau_2 N\delta(x^3)\cdots\delta(x^9)=0\,,
\end{equation}
from~\eqref{orbeqC3},
\begin{equation}\label{2eq}
	H_2^{-1}\delta^{ij}\partial_i \partial_j b
		-\frac{\kappa^2 \tau_2 N}{\pi^2\ls^2}\delta(x^3)\cdots\delta(x^5)=0\,,   
\end{equation}
from~\eqref{orbeqb} and:
\begin{multline}\label{3eq}
	\frac{1}{4}H_2^{-1}
		\Big(\left(\delta^{ij}\partial_i\partial_j
		+\delta^{rs}\partial_r\partial_s\right) H_2
		+\frac{1}{2}\delta^{ij}\partial_i b\partial_j b\delta(x^6)\cdots\delta(x^9)\Big)\\
		+\frac{\kappa^2 \tau_2 N}{4}\delta(x^3)\cdots\delta(x^9)=0\,,
\end{multline}
from~\eqref{orbeqphi}.
\end{subequations}
The above equations are satisfied by the following solution describing $N$ fractional D2-branes transverse to a \cz{} orbifold:
\begin{subequations}\label{solution}
\begin{align}
	ds^2&=H_2^{-5/8}\ \eta_{\alpha\beta}dx^\alpha dx^\beta
		+H_2^{3/8} \left(\delta_{ij}dx^idx^j
		+ \delta_{rs}dx^r dx^s\right)\,,\\
	e^\Phi&=H_2^{1/4}\,,\\
	\bar{C}_3&=\left(H_2^{-1}-1\right)dx^0\wedge dx^1\wedge dx^2\,,\\
	A_1&=-4\gs\pi^2\ls^3 N \cos\theta d\varphi\,,\\
	b&=\frac{\left(2\pi \ls \right)^2}{2}
		\left(1-\frac{2 \gs \ls N}{r}\right)\,,
 \end{align}
\end{subequations} 
where $H_2$ is the solution of~\eqref{H2} and we have introduced spherical coordinates $(r,\theta,\varphi)$ in the space $\{x^3,x^4,x^5\}$. One can also show after some computation that with our ansatz the Einstein equations are also satisfied provided that~\eqref{H2} holds.

From the solution~\eqref{solution} we can also compute the expressions for the field $A_3$ which is more natural from a string theory perspective. Using the duality relation valid in the Einstein frame,	$dA_3 = e^{\Phi/2}\ \hd{6} G_2 - db \wedge \bar{C}_3$, we find:
\begin{equation}
	A_3 = \tilde{b} \  dx^0\wedge dx^1\wedge dx^2\,.
\end{equation}

A careful study of the solution~\eqref{solution} shows that there is a naked singularity for the value of $r$ that makes $H_2$ vanish. We will see in the next subsection that, similarly to what happened with the wrapped D4-branes in section~\ref{s:wD4}, this singularity is excised by an enhan\c con mechanism, and gauge theory information can therefore be extracted from the supergravity configuration.

\subsection*{The dual gauge theory}

We can now pass to the gauge theory. We will proceed as in the previous section, and show how it is possible to extract the perturbative gauge coupling constant and metric on moduli space of the \Ne{4}, three-dimensional gauge theory living on the branes by a probe analysis.

We will then study the world-volume theory of a probe fractional D2-brane placed in the background~\eqref{solution} at some finite distance $r$ in the transverse space $\left\{x^3,x^4,x^5\right\}$. As we noticed, in the probe approach we are supposed to recover information on the Coulomb branch $U(N+1)\to U(N)\times U(1)$ of the gauge theory.

Let us start from the world-volume action for a single fractional D2-brane, which, as we recall from chapter~\ref{c:engineering}, in the Einstein frame is given by:
\begin{multline}\label{probact}
	S_{\text{D}2}^{\text{f}}=-\frac{\tau_p}{2}
		\int{d^{3}x\  e^{-\Phi/4}
		\sqrt{-\det \left [G_{\alpha\beta}+e^{-\Phi/2}
		2\pi\ls^2 F_{\alpha\beta}\right]} 
		\left(1+\frac{\tilde{b}}{2\pi^2\ls^2}\right)}  \\
		+\frac{\tau_p}{2}\int_{{\cal M}_3}{\big(
		{\cal C}_3+2\pi\ls^2{\cal C}_1\wedge F\big)}\,, 
\end{multline}
where we have chosen the static gauge, hats denote pullbacks onto the brane world-volume and the fields $\mathcal{C}_3$ and $\mathcal{C}_1$ are given by:
\begin{subequations}\label{curlyc}
\begin{align}
	\mathcal{C}_3&=\bar{C}_3\left(1+\frac{\tilde{b}}{2\pi^2\ls^2}\right)+
		\frac{A_3}{2\pi^2\ls^2}=\frac{b}{2\pi^2\ls^2}-1\,,\\
	\mathcal{C}_1&={C}_1\left(1+\frac{\tilde{b}}{2\pi^2\ls^2}\right)+
		\frac{A_1}{2\pi^2\ls^2}=-2 \gs \ls N \cos\theta d\varphi\,.
\end{align}
\end{subequations}

The computation is analogous to the one performed in the previous section. We regard the coordinates $\{x^3,x^4,x^5\}$ transverse to the probe brane as Higgs fields of the dual gauge theory: $x^i=2\pi\ls^2\Phi^i$. We also define polar coordinates $(\mu,\theta,\varphi)$ in the moduli space of the $\Phi^i$, so that the resulting energy/radius relation is given by $r=2\pi\ls^2 \mu$.

Expanding the world-volume action for slowly varying world-volume fields and keeping only up to quadratic terms in their derivatives we easily see that position dependent terms cancel, as is should be since the system is BPS, and we are left with the following effective action:
\begin{multline}\label{notsofinalprobe}
	S_{\text{D}2}^{\text{f}}\simeq -\frac{\ls}{4\gs}
		\int d^3 x \frac{b}{2\pi^2\ls^2}
		\left\{\frac{1}{2}
		\big[(\partial\mu)^2+\mu^2\left((\partial\theta)^2
		+\sin^2\theta(\partial\varphi)^2\right)\big]
		+\frac{1}{4}F^2\right\} \\
		-\frac{N}{8\pi}\int d^3 x \cos\theta 
		\varepsilon^{\alpha\beta\gamma}\partial_\alpha\varphi F_{\beta\gamma}\,.
\end{multline} 
When $b=0$, the effective tension of the probe vanishes and this means that also in this case an enhan\c con mechanism is taking place at the radius:
\begin{equation}
	r_{\text{e}}= 2 \gs \ls N\,.
\end{equation}
Substituting in~\eqref{notsofinalprobe} the expression of $b$ in terms of $\mu$, we obtain:
\begin{multline}\label{finalprobe}
	S_{\text{D}2}^{\text{f}}= -\frac{\ls}{4\gs}
		\int d^3 x \left[1-\frac{g_sN}{\pi\ls \mu}\right]
		\left\{\frac{1}{2}
		\big[(\partial\mu)^2+\mu^2\left((\partial\theta)^2
		+\sin^2\theta(\partial\varphi)^2\right)\big]
		+\frac{1}{4}F^2\right\} \\
		-\frac{N}{8\pi}\int d^3 x \cos\theta 
		\varepsilon^{\alpha\beta\gamma}\partial_\alpha\varphi F_{\beta\gamma}\,.
\end{multline} 
From the coefficient of the gauge field kinetic term in the previous action we can read the running coupling constant:
\begin{equation}\label{running}
	\frac{1}{\gym(\mu)}=
		\frac{1}{\gym}\left(1-\frac{\gym N}{4\pi\mu}\right)\,,
\end{equation}
where we have defined the bare coupling as:
\begin{equation}\label{bare}
	\gym=\frac{4\gs}{\ls}\,.
\end{equation}
The result~\eqref{running} agrees with the one obtained from the solution describing wrapped D4-branes, and again is exactly what expected for the gauge theory under consideration, as one can verify from insert~\ref{i:running}.

Precisely as in the case of the wrapped branes described in section \ref{s:wD4}, \eqref{finalprobe} does not give explicitly the full hyperK\"ahler metric on the moduli space of the gauge theory.
We can obtain the needed extra modulus by dualising the vector field into a scalar, using exactly the same procedure which brought us from~\eqref{wfinprobe} to.~\eqref{wMetricMS} in section~\ref{s:wD4}.%
\footnote{In this case, the dualisation procedure can also be done directly in the original three-dimensional world-volume action, as in~\cite{Schmidhuber:1996fy,Johnson:1999qt}, and one gets the same result.} The resulting moduli space metric is of course identical to the one found in~\eqref{wMetricMS} by probing the geometry of $N$ D4-branes wrapped on $S^2$. Again, we obtain the hyperK\"ahler Taub-NUT metric, but with a ``negative mass'' which makes it singular, which means we are only able to recover the perturbative behavior of the gauge theory.

\subsection*{Adding D6-branes}

Now we introduce another ingredient, that allows us to add matter multiplets to the gauge theory. In fact, if we consider adding $M$ D6-branes extended along the world-volume of the D2-branes, as well as wrapping the whole orbifold space, the strings stretching between the D2 and D7-branes make up $M$ matter multiplets in the fundamental representation of the $U(N)$ gauge group of the theory living on the D2-branes.

We will therefore expand the analysis above by studying the following configuration of D-branes on \cz~\cite{DiVecchia:2001uc}:
\begin{center}
\begin{tabular}{|c|c|c|c|c|c|c|c|c|c|c|}
\multicolumn{7}{c}{ }&
\multicolumn{4}{c}{$\overbrace{\phantom{\qquad\qquad\qquad}}^{\cz}$}\\
\hline
&0&1&2&3&4&5&6&7&8&9\\
\hline
D$2$ &$-$&$-$&$-$&$\cdot$&$\cdot$&$\cdot$&$\cdot$&$\cdot$&$\cdot$&$\cdot$\\
\hline
D$6$ &$-$&$-$&$-$&$\cdot$&$\cdot$&$\cdot$&$-$&$-$&$-$&$-$\\
\hline
\end{tabular}
\end{center}

Again, the first step is finding the supergravity solution, which has to be a generalization of~\eqref{solution}. The ans\"atze for metric and dilaton that we will use are the standard ones for a configuration of intersecting branes~\cite{Papadopoulos:1996uq}, so for the case of a D2/D6 brane system in the Einstein frame we have:
\begin{subequations}\label{ansatz26}
\begin{align}
	ds^2&=H_2^{-5/8}H_6^{-1/8}\eta_{\alpha\beta}dx^\alpha dx^\beta
		+H_2^{3/8}H_6^{7/8}\delta_{ij}dx^idx^j
	+H_2^{3/8}H_6^{-1/8}\delta_{rs}dx^r dx^s\,,\\
	e^\Phi&=H_2^{1/4}H_6^{-3/4}\,,
\end{align}
\end{subequations}
while $\bar{C}_3$ retains its form~\eqref{ansC3}.

However, there are additional fields turned on, since a D6-brane is magnetically charged under $C_1$, and we also have additional source terms coming from the world-volume action of the D6-branes. The relevant terms in such an action can be showed (for instance with the boundary state approach~\cite{Bertolini:2001qa}) to be given by:
\begin{multline}\label{D6act}
	S_{\text{D}6}\to\tau_6\left\{
		-\int d^{7}x\ e^{\frac{3}{4}\Phi}\sqrt{-\det G_{\rho\sigma}}
		+\int_{\mathcal{M}_{7}}C_{7}\right\}\\
		+\frac{\tau_2}{2}\frac{1}{2(2\pi\ls)^2}\left\{
		\int d^{3}\xi \sqrt{-\det G_{\alpha\beta}}\ \tilde{b}
		-\int_{\mathcal{M}_{3}}A_{3}\right\}+\ldots\,,
\end{multline}
where the indices $\alpha,\beta,\ldots$ run along the common world-volume $\mathcal{M}_{3}$ and $\rho,\sigma,\ldots$ along the whole D6-brane world-volume $\mathcal{M}_{7}\,$, and dots denote higher order terms that are irrelevant as source terms in the equations of motion.

Since the full boundary action does not depend on the fields $C_1$ and $A_1$, \eqref{orbeqC1} and~\eqref{orbeqA1} read:
\begin{subequations}
\begin{align} 
	d\left(e^{3\Phi/2}\ \hd{} dC_1\right)
		-\frac{1}{2}e^{\Phi/2}db\wedge\hd{6}G_2\wedge\Omega_4&=0\,,
		\label{ansC1}\\
	d\left(e^{\Phi/2}\ \hd{6} G_2\right)
		+db\wedge d\bar{C}_3&=0\,.
\end{align}
\end{subequations}
One solves the above equations by again imposing~\eqref{duansA1} and:
\begin{equation}\label{duansC1}
	e^{3\Phi/2}\ \hd{} dC_1 = -d\left(H_6^{-1}\right)
		dx^0\wedge dx^1\wedge dx^2\wedge dx^6\wedge\cdots\wedge dx^9\,.
\end{equation}
One therefore arrives to the following expressions for $A_1$ and $C_1\,$:
\begin{subequations}\label{ansatz226}
\begin{align}
	dA_1&=C_1\wedge db+\frac{1}{2}\varepsilon_{ijk}H_6\partial_ib dx^j\wedge dx^k\,,\\
	dC_1&=\frac{1}{2}\varepsilon_{ijk}\partial_iH_6 dx^j\wedge dx^k\,.
\end{align}
\end{subequations}

Inserting the ans\"atze \eqref{ansatz26}-\eqref{ansatz226} into the equations of motion~\eqref{orbeqs}, after manipulations that are slightly more complicated than in the case with only D2-branes, one finally finds the following solution corresponding to our fractional D2/D6 system:
\begin{subequations}\label{solution26}
\begin{align}
	ds^2&=H_2^{-5/8}H_6^{-1/8}\eta_{\alpha\beta}dx^\alpha dx^\beta
		+H_2^{3/8}H_6^{7/8}\delta_{ij}dx^idx^j
		+H_2^{3/8}H_6^{-1/8}\delta_{rs}dx^r dx^s\,,\\
	e^\Phi&=H_2^{1/4}H_6^{-3/4}\,,\\
	\bar{C}_3&=\left(H_2^{-1}-1\right)dx^0\wedge dx^1\wedge dx^2\,,\\
	C_1&=\frac{\gs \ls M}{2}\cos\theta d\varphi\,,\\
	A_1&=-\pi^2\ls^2 \frac{\gs\ls (4N-M)}
		{H_6} \cos\theta d\varphi\,,\\
	b&=\frac{Z}{H_6}\,,
 \end{align}
\end{subequations} 
where:
\begin{equation}
	H_6(r)=1+\frac{\gs\ls M}{2r}\,,\qquad
	Z(r)=\frac{\left(2\pi \ls \right)^2}{2}
		\left(1-\frac{\gs \ls (2N-M)}{r}\right)\,,
\end{equation}
and where $H_2$ is now the solution of:
\begin{equation}\label{H226}
	\left(\delta^{ij}\partial_i\partial_j 
		+H_6\delta^{rs}\partial_r\partial_s\right) H_2
		+\frac{1}{2}H_6\delta^{ij}\partial_i b\partial_j b\delta(x^6)\cdots\delta(x^9)
		+\kappa^2 \tau_2 N\delta(x^3)\cdots\delta(x^9)=0\,,
\end{equation}

From the solution given in eq.~\eqref{solution26} we can also compute the expressions for the field $C_7$ which appear naturally in the string theory. Given that $dC_7 = -e^{3\Phi/2} \hd{} dC_1$ we get the standard expression:
\begin{equation}
	C_7 = (H_6^{-1}-1) \ dx^0\wedge dx^1\wedge dx^2\wedge dx^6\wedge\cdots\wedge dx^9\,.
\end{equation}

The moduli space of the gauge theory can be explored by means of a probe fractional D2-brane as before. One notices that the effective tension of the probe vanishes when $Z=0$, which means that the enhan\c con radius is now $r_{\text{e}}=\gs \ls (2N-M)$. The resulting effective action~\eqref{finalprobe} gets modified as follows:
\begin{equation}\label{finalprobe26}
\begin{split}
	S_{\text{probe}}&= -\frac{\ls}{4\gs}
		\int d^3 x \left[1-\frac{g_s(2N-M)}{2\pi\ls \mu}\right]\\
		&\qquad\times\left\{\frac{1}{2}
		\big[(\partial\mu)^2+\mu^2\left((\partial\theta)^2
		+\sin^2\theta(\partial\varphi)^2\right)\big]
		+\frac{1}{4}F^2\right\} \\
		&\qquad-\frac{1}{16\pi}\int d^3 x (2N-M)\cos\theta 
		\varepsilon^{\alpha\beta\gamma}\partial_\alpha\varphi F_{\beta\gamma}\,.
\end{split}
\end{equation} 
From the coefficient of $F^2$ we can again read the effective running coupling:
\begin{equation}\label{running26}
	\frac{1}{\gym(\mu)}=
		\frac{1}{\gym}
		\left(1-\gym\frac{2N-M}{8\pi\mu}\right)\,,
\end{equation}
which perfectly agrees with gauge theory expectations, as one can verify from insert~\ref{i:running} on page~\pageref{i:running}.

In order to extract the metric on the moduli space, we have to dualize again the vector into a scalar $\Sigma$, by repeating the procedure used in the previous subsection. The final result is:
\begin{equation}\label{MetricMS26}
	ds^2_{\mathcal{M}}=
		\frac{1}{g^2_{\text{YM}}(\mu)}
		\left(d\mu^2+\mu^2d\Omega^2\right)
		+g^2_{\text{YM}}(\mu)
		\left(d\Sigma+
		\frac{(2N-M)\cos\theta}{8\pi}d\varphi\right)^2\,,
\end{equation}
where we recall that $d\Omega^2=d\theta^2+\sin^2\theta d\varphi^2\,$. In this case, we see that we have to distinguish two cases. If $M<2N$, the situation is similar to the case $M=0$ studied beforehand, namely we obtain a singular Taub--Nut metric, which is a signal that non-perturbative contribution should be taken into account in order to smooth out the moduli space metric. In the case $M>2N$, instead, the metric is regular, and this is expected from the gauge theory side: with this amount of matter, the full metric on the moduli space coincides with the perturbative one. Notice also that in this case no enhan\c con mechanism takes place.

This ends our analysis of the first simple examples of scale-anomalous and less supersymmetric gauge/gravity correspondence. In the next section, we will try to make the point about what we have learned from these examples.

\section{Summary of the gauge/gravity dictionary}\label{s:dictionary}

The explicit examples studied in the two previous sections gave us a flavor about how the gauge/gravity correspondence works when the gauge theory involved is scale-anomalous and preserves less than 16 supersymmetries. This section is devoted to a little more general analysis of the gauge/gravity dictionary, as well as to some comments and discussions on the approaches we have encountered.

\subsection*{Gauge theory from gravity}

Let us start from some general formulae. We have seen that the first relevant piece of information we can recover is the one-loop running coupling constant of the gauge theory living on a particular configuration of D-branes.

In particular, the examples studied in sections~\ref{s:wD4} and~\ref{s:fD2}, as well as the more general considerations of chapter~\ref{c:engineering}, showed that one efficient way to engineer a $(p+1)$-dimensional gauge theory with 8 supercharges is to consider D$(p+2)$-branes wrapped on two-cycles. This is of course the essence of the wrapped brane setup, and is also correct for the fractional D$p$-branes, which have the interpretation of D$(p+2)$-branes wrapped on a vanishing two-cycle. In order to get a general formula for the coupling constant, then, we can start from the Dirac--Born--Infeld part of the world-volume action of a D$(p+2)$-brane, and consider what happens when we consider two longitudinal directions to be wrapped on a non-trivial two-cycle. This is of course closely related to the derivation of the world-volume action of fractional branes in chapter~\ref{c:engineering}.

We then recall one more time the expression Dirac--Born--Infeld action for a D$(p+2)$-brane in the string frame:
\begin{equation}\label{Sp+2}
	S_{\text{DBI}} = -\tau_{p+2} \int d^{p+3} \xi\ e^{-\Phi}
		\sqrt{-\det \left( \hat{G}_{ab} + \hat{B}_{ab} + 2\pi\ls^2 F_{ab} \right)}\,.
\end{equation}
Now, divide the coordinates in two groups:
\begin{itemize}
\item $\xi^\alpha$, $\alpha=0,\ldots,p$ are the coordinates along the $(p+1)$-dimensional ``flat'' part of the world-volume. The only non-vanishing components of the world-volume gauge field $F$ are along these directions, while we assume that the metric elements are diagonal, $G_{\alpha\beta} = H^{-1/2} \delta_{\alpha\beta}$ (where for now we just define $H$ to be a function independent of $x^\alpha$), and $B_{\alpha\beta}=0$.
\item $\xi^A$, $A=p+1,p+2$ parameterize the two-cycle on which the branes are wrapped. Along these coordinates, there may be a non-trivial metric $G_{AB}$ and $B$-field $B_{AB}$.
\end{itemize}
With these assumptions, we can expand the determinant in~\eqref{Sp+2} and extract the quadratic part in  the gauge field-strength:
\begin{equation}\label{probegen}
\begin{split}
	&-\tau_{p+2}\ \frac{(2\pi\ls^2)^2}{8} \int d^{p+3}\xi\ e^{-\Phi}
		\sqrt{-\det G_{\alpha\beta}}\ G^{\alpha\beta} G^{\gamma\delta}
		F_{\alpha\gamma} F_{\beta\delta}
		\sqrt{\det \left( G_{AB} + B_{AB}\right)}\\
	= &-\frac{1}{g^2_{\text{D}p}}\ \frac{1}{(2\pi\ls)^2} \int d^{p+3}\xi\ e^{-\Phi}
		H^{\frac{3-p}{4}}
		\sqrt{\det \left( G_{AB} + B_{AB}\right)}\
		\frac{1}{4}F_{\alpha\beta} F_{\alpha\beta}\,,
\end{split}
\end{equation}
where we have inserted an additional factor of  $\tfrac{1}{2}$ due to the standard normalization of the generators of the non-abelian gauge group, and where $g^2_{\text{D}p}=2 (2\pi)^{p-2} \gs \ls^{p-3}$ is the coupling constant~\eqref{Dpcoupling} of the gauge theory living on a D$p$-brane in flat space. We can already point out the important fact that the gauge coupling constant, namely the coefficient of $-\tfrac{1}{4} F_{\alpha\beta}F_{\alpha\beta}$, depends essentially on the ``stringy volume'' of the two-cycle $S^2$, namely on $\int d^2\xi\ \sqrt{\det ( G_{AB} + B_{AB})}$:
\begin{equation}\label{gymgen}
	\frac{1}{\gym} = \frac{1}{g^2_{\text{D}p}}\ \frac{1}{(2\pi\ls)^2} \int d^2\xi\ e^{-\Phi}
		H^{\frac{3-p}{4}}
		\sqrt{\det \left( G_{AB} + B_{AB}\right)}\,,
\end{equation}
where now the integral runs over the directions $\xi^A$ of the two-cycle. With no additional assumptions on $p$ and on the dependence of the dilaton on the warp factor, it is difficult to additionally simplify~\eqref{gymgen}. This cannot easily be done in general, since the dilaton dependence on $H$ is different in the case of a ``real'' wrapped D$(p+2)$-brane with respect to the case of a fractional D$p$-brane. We are thus forced to consider the two cases separately:
\begin{subequations}
\begin{align}
		e^{\Phi} &= H^{\frac{1-p}{4}}\,, &
		G_{AB} &= H^{-1/2} g_{AB}\,, &
		B_{AB} &= 0 &
		&\text{(wrapped D$(p+2)$-branes)} \\
		e^{\Phi} &= H^{\frac{3-p}{4}}\,, &
		G_{AB} &= 0\,, &
		B_{AB} &\neq 0 &
		 &\text{(fractional D$p$-branes)}
\end{align}
\end{subequations}
Therefore one sees that in both cases all warp factors cancel, and the coupling constant of the $(p+1)$-dimensional gauge theory is finally given by:
\begin{equation}\label{gymwrapfrac}
	\frac{1}{\gym} = \begin{cases}
		\frac{1}{g^2_{\text{D}p}}\ \frac{1}{(2\pi\ls)^2}
			\int d^2\xi\ \sqrt{\det g_{AB}}\qquad &\text{(wrapped D$(p+2)$-branes)} \\
		\frac{1}{g^2_{\text{D}p}}\ \frac{1}{(2\pi\ls)^2}
			\int d^2\xi\ \sqrt{\det B_{AB}}\qquad &\text{(fractional D$p$-branes)}
	\end{cases}
\end{equation}
One can easily verify that the above formulae indeed reproduce the results we obtained in sections~\ref{s:wD4} and~\ref{s:fD2} for wrapped D4-branes and fractional D2-branes respectively. An important observation is that we derived~\eqref{gymwrapfrac} for configurations made of a single type of branes. This means that when dealing with systems of branes of different dimensions, such as the fractional D2/D6 system we studied in section~\ref{s:fD2}, one cannot directly use~\eqref{gymwrapfrac}, but should rather use~\eqref{gymgen} as a starting point. As an additional check, notice that if $g_{AB}$ and $B_{AB}$ in~\eqref{gymwrapfrac} are trivial, so that the configuration is simply made of D$(p+2)$-branes in flat space, one correctly gets $\gym = g^2_{\text{D}(p+2)}$.

The coupling constant is not the only piece of information one can extract from supergravity. For example, in the previous sections of this chapter we used two supergravity solutions to get information on the moduli space of three-dimensional \Ne{4} Super Yang--Mills, and the Wess-Zumino part of the world-colume action entered the analysis crucially. Now, we would like to concentrate for a while on the case of D-brane configurations yielding four-dimensional gauge theories, since this will be the topic of the remaining part of this chapter and of chapter~\ref{c:4susy}. We first notice that in this case, where $p=3$, \eqref{gymgen} simplifies~\cite{DiVecchia:2003ne}:
\begin{equation}\label{gym4}
	\frac{1}{\gym} = \frac{1}{4\pi\gs}\ \frac{1}{(2\pi\ls)^2} \int d^2\xi\ e^{-\Phi}
		\sqrt{\det \left( G_{AB} + B_{AB}\right)}\,.
\end{equation}

The additional piece of information we can extract on a four-dimensional gauge theory is the theta-angle $\tym$, namely the coefficient of the topological term $\tfrac{1}{32\pi^2}F_{\alpha\beta} (\hd{}F)_{\alpha\beta}$ in the action. This term comes from the Wess-Zumino part of the world-volume action of a wrapped D5-brane:
\begin{equation}
	S_{\text{WZ}} = \tau_5 \int \sum_q C_q \wedge e^{B+2\pi\ls^2 F}\,,
\end{equation}
from whose expansion we can derive:
\begin{equation}\label{tym4}
	\tym = \frac{1}{2\pi\gs\ls^2} \int_{S^2} \left( C_2 + C_0\ B_2\right)\,.
\end{equation}
The relations~\eqref{gym4} and~\eqref{tym4} will be rather useful in the following.

Let us now pause for a very important observation. All the relations we have derived in this section up to now have the property of giving a gauge theory quantity \emph{in terms of functions of supergravity coordinates}. But this is not the end of the story because in general the gauge coupling constant~\eqref{gymgen} will be given by an appropriate function of a coordinate $r$ of the geometry:
\begin{equation}\label{gymr}
	\frac{1}{\gym} = F(r)\,,
\end{equation} 
as we can see also from the examples in sections~\ref{s:wD4} and~\ref{s:fD2}. This means that in order to establish the gauge/gravity dictionary we have to link the coordinate $r$ with quantities of the dual gauge theory. This is the subject of the next subsection.

\subsection*{The energy/radius relation}

As just discussed, we then need an expression relating the relevant coordinate $r$ appearing in~\eqref{gymr} to gauge theory parameters:
\begin{equation}\label{er}
	r = G(\mu)\,,
\end{equation}
where $\mu$ is the scale at which the gauge theory is defined. We will call~\eqref{er} \emph{energy/radius} or \emph{gauge/gravity} relation. Once we have at our disposal explicit expressions for~\eqref{gymr} and~\eqref{er}, we can obtain pure gauge theory quantities in terms of the appropriate parameters.

In the case of D-branes in flat space, where the involved gauge theories have no scale anomaly (namely are conformal in the case in which the world-volume is four-dimensional), things are easy. We can simply use the relation derived in chapter~\ref{c:dbranes} between the transverse coordinates and the expectation values of the scalar fields of the vector multiplet:
\begin{equation}
	x^i = 2\pi\ls^2\ \Phi^i\,,
\end{equation}
to derive a natural energy/radius relation between the radial coordinate $r=(x^i x^i)^{1/2}$ in the transverse space and the energy scale $\mu = (\Phi^i\Phi^i)^{1/2}$ of the gauge theory:
\begin{equation}\label{stst}
	r = 2\pi\ls^2\ \mu\,.
\end{equation}
In fact, we already used~\eqref{stst} in all cases treated in this chapter. We can call this relation ``stretched string'' energy/radius relation, since it has the equivalent interpretation that the energy scale is given by the energy of a string (read from the Nambu--Goto action) which is stretched between a stack of D-branes and a probe D-brane at distance $r$ from the stack, as in figure~\ref{f:probe}:
\begin{equation}
	E \sim \int^r dr \sqrt{-G_{00}G_{rr}} \sim r\,.
\end{equation}

However, even if we successfully used~\eqref{stst} also in cases in which we dealt with scale-anomalous theories, in principle things are not unambiguous in such cases~\cite{Peet:1998wn}, and $r$ and $\mu$ could be related differently. What we can say is that, as long as a brane can move supersymmetrically in a part of the transverse space (and we can therefore rely on the probe approach, as described in chapter~\ref{c:dbranes}), the identification of the radius of such space with the scale $\mu$ gives accurate one-loop gauge theory results. This is precisely the case we have dealt with up to now.

However, in the following we will deal with cases in which there is no space for freely moving a D-brane off the stack generating the geometry, and the probe mechanism fails. In these cases, one should follow a different strategy. The general idea is that one must identify a combination or function of supergravity coordinates which on general grounds is dual to a specific \emph{protected} quantity of the dual gauge theory. This can therefore fix the energy/radius relation, and we will see the precise way this works in some specific examples treated in the next chapter. Of course, the relation~\eqref{stst} can also be seen as a particular result of this strategy, where the gauge theory operators are the scalar fields of the vector multiplet of the gauge theory, which have protected dimension one.

\subsection*{Origins of the correspondence and the enhan\c con mechanism}

We have seen in some examples that the procedures we introduced \emph{work}, in the sense that the gauge theory information that we recover from them is relevant and correct. But \emph{why} does all this work? This apparently naive question opens the way to a quite wide range of different issues and considerations, and, at least to our knowledge, a complete and satisfactory answer is not known yet.

The main point is that, unlike the case of the original AdS/CFT correspondence involving a superconformal gauge theory, that we briefly treated in section~\ref{s:adscft}, it is quite difficult to define unambiguously an appropriate limit in which the gauge theory dynamics completely decouples from the dynamics in the bulk, so that an exact duality can be established.

In fact, the minimal point of view on this problem is that an exact duality, at least at the supergravity level, \emph{cannot} be established. In this subsection, we will try to discuss a bit this and some related 
issues, such as the enhan\c con mechanism~\cite{Johnson:1999qt} already mentioned in sections~\ref{s:wD4} and~\ref{s:fD2}.

Let us then start by considering the enhan\c con more carefully. The easiest setup in order to study it is the one considered in section~\ref{s:fD2}, describing $N$ fractional D2-branes on a \cz{} orbifold. We have seen the enhan\c con appearing as the point where the effective tension of a probe fractional D2-brane vanishes, because the fluctuations of the twisted field $b$ cancel its background value. On the other hand, in the case of the wrapped D4-branes of section~\ref{s:wD4} the enhan\c con locus is the point where the effective volume of the two-cycle the branes are wrapped on goes to zero, so that, also in this case, the effective tension of a probe wrapped D4-brane vanishes. Recalling then the general expression~\eqref{probegen}, we can say that the enhan\c con mechanism takes place, for both fractional and wrapped branes, where the ``stringy volume'' of the relevant two-cycle vanishes due to the fluctuations of the supergravity fields:
\begin{equation}\label{enhanconradius}
	\int d^2\xi\ \sqrt{\det{G_{AB}+B_{AB}}} = 0 \qquad
	\text{at}\quad r=r_e\,.
\end{equation}
It is not difficult to see that the above result implies that the effective tension of \emph{any} fractional or wrapped D$p$-brane probe (and not only the tension of a probe of the same dimension of the branes generating the geometry) vanishes at the enhan\c con radius:
\begin{equation}
	\tau_p^{\text{eff}} (r_e) = 0\,.
\end{equation}
What does all this mean? Let us first try to look at this from a geometrical perspective, and subsequently we will think of the consequences on the gauge theory side. Roughly speaking, the meaning of the above result is that our supergravity solutions for fractional and wrapped branes have something wrong, as in fact one could expect from the fact that they present a naked singularity surrounded by a \emph{repulson} region in which gravity becomes repulsive.

In fact, we could think of our supergravity solution to be formed as a step-by-step procedure, by bringing the branes one after the other from infinity to the origin, since they are BPS objects which do not feel any mutual force. But at the enhan\c con radius, before reaching the origin, all of them become tensionless. We can think of the branes as making up a spherical ``shell'' around the region of the repulson geometry, so that the singularity is cloaked and the classical solution makes sense only up to the enhan\c con radius, the last point where, so to say, the branes can be thought of as localized sources.

However, when the supergravity description fails, it usually means that we are missing some new light degrees of freedom in our analysis, and this is precisely what is happening here. For definiteness, consider our fractional D2-branes of section~\ref{s:fD2}. Looking at the solution~\eqref{solution}, it is clear that they can be seen as $U(1)$ monopoles for the R-R twisted field $A_1$. At the enhan\c con radius, such monopoles become massless, but in addition, as we have seen, \emph{all} probes have a vanishing effective tension. In particular, we can consider the two types of  fractional D0-branes, which are \emph{electrically} charged under $A_1$. They have precisely the right quantum numbers to behave as $W$-bosons of the gauge symmetry we are considering and, when they become massless at the enhan\c con radius, there is an enhancement of the symmetry from $U(1)$ to $SU(2)$. These fractional D0-branes are then the new light degrees of freedom appearing at the enhan\c con that we were looking for. This also explains the reason of the name that was given to this mechanism, due to the enhanced gauge symmetry which is realized.

The upshot is that the geometry in the interior of the enhan\c con locus should be described by an $SU(2)$ monopole, and in fact this fits nicely with the comment about the desingularization of the moduli space metric via the Atiyah--Hitchin manifold that we made in sections~\ref{s:wD4} and~\ref{s:fD2}. There, we also observed that this desingularization is a consequence of non-perturbative phenomena in the dual gauge theory.

In fact, the meaning of the enhan\c con on the gauge theory side is straighforward once we recall our general result~\eqref{gymgen} for the running coupling constant. The enhan\c con radius is precisely the point where the gauge coupling diverges or, in other words, where the non-perturbative contributions become relevant, spoiling the perturbative result. This fact explains why our supergravity solutions were only able to give us information on the \emph{perturbative} regime of the considered gauge theories: the non-perturbative physics is ``hidden'' inside the enhan\c con. A full understanding of gauge theory dynamics at the non-perturbative level would at least require the inclusion of the new light degrees of freedom in the analysis, and this have not been completely analyzed at the moment of writing. An attempt based on the duality with the heterotic string was made in~\cite{Wijnholt:2001us}.

Some more light on the meaning of the enhan\c con mechanism comes from the dual scenario of stretched branes we described in section~\ref{s:HW}, so let us see in some detail how the features we discovered are translated in this context. We continue working with our three-dimensional gauge theory. Recall from section~\ref{s:HW} that both the systems we studied, fractional D2-branes on \cz{} and D4-branes wrapped on a two-cycle inside a Calabi--Yau twofold, are mapped by T-duality into the same configuration of type IIB string theory, in which a D3 brane of finite length is stretched between two parallel NS5-branes, as in the following table:
\begin{center}
\begin{tabular}{|c|c|c|c|c|c|c|c|c|c|c|}
\hline
&0&1&2&3&4&5&6&7&8&9\\
\hline
NS5 &$-$&$-$&$-$&$-$&$-$&$-$&$\cdot$&$\cdot$&$\cdot$&$\cdot$\\
\hline
D3 &$-$&$-$&$-$&$\cdot$&$\cdot$&$\cdot$&$\abs{-}$&$\cdot$&$\cdot$&$\cdot$\\
\hline
\end{tabular}
\end{center}
where we recall that $\abs{-}$ denotes the finite extension of a brane along a longitudinal dimension. This system of branes preserves eight supercharges, and the three-dimensional gauge theory we have been studying lives on the intersection of the D3 and NS5-branes. Notice also that the twisted fields of the string in the orbifold are mapped to world-volume fields of the NS5-branes, and this is a natural way to see that their dynamics is six-dimensional.

\begin{figure}
\begin{center}
\includegraphics[scale=.6]{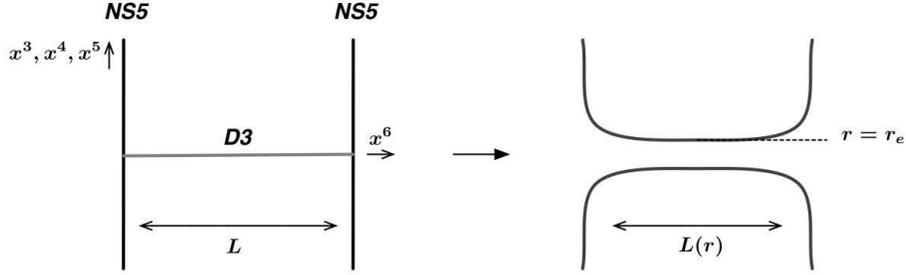}
\caption{{\small The enhan\c con mechanism for three-dimensional Super Yang--Mills theory in the T-dual scenario of branes suspended between branes. Quantum mechanically, the distance $L$ between the two NS5-branes becomes a function of $r=[(x^3)^2+(x^4)^2+(x^5)^2]^{1/2}$. At the enhan\c con radius, $L(r_e)=0$ and new light degrees of freedom come into play because the NS5-branes touch.}}
\label{f:enhancon}
\end{center}
\end{figure}

The length $L$ of the stretched D3-branes along $x^6$ is the T-dual of the ``stringy volume'' of the two-cycle which is present in the fractional and wrapped brane setups. Quantum mechanically, we know that the stringy volume runs as a function of the radius of a three-dimensional space, and this has the gauge theory interpretation of the running of the coupling constant. Here, we then see that quantum mechanically the D3-branes exert a force on the NS5-branes and tend to bend them, while $L$ becomes a function of $r=[(x^3)^2+(x^4)^2+(x^5)^2]^{1/2}$. The enhan\c con radius is the locus where $L(r_e)$ vanishes and the two NS5-branes touch, as shown in figure~\ref{f:enhancon}. In this setup, it is therefore very easy to understand qualitatively what are the new light degrees of freedom coming into play: they are world-volume fields of the NS5-branes that become massless when the branes touch. These states come from D1-branes, and one can see that when the NS5-branes touch the gauge symmetry of the theory is in fact enhanced from $U(1)$ to $SU(2)$.
%:Figure: The enhancon and stretched branes

Notice that the enhan\c con mechanism we just described is not at all limited to the case of a three-dimensional gauge theory as the one we just considered, but has general validity. For instance, in the next section we will see that it takes place also for four-dimensional theories arising on fractional D3-branes and wrapped D5-branes, in which case the enhan\c con locus is a ring instead of being a two-sphere. Also in this case, the new relevant light degrees of freedom are the ones which have electric charge under the bulk field with respect to which the branes generating the solution are magnetically charged, and are respectively fractional D-instantons and wrapped D1-branes. Now, however, the quantum numbers of the states do not allow the interpretation of having an enhanced gauge symmetry, but rather we have  a two-form potential carried by tensionless strings. This is in agreement with the fact that the dual scenario of D4-branes suspended between NS5-branes now involves NS5-branes of type IIA string theory, whose world-volume theory has a self-dual two-form as fundamental field.

To summarize, then, we can say that the enhan\c con locus for a fractional D$p$-brane (or a wrapped D$(p+2)$-brane) is $S^{4-p}$, and that for even $p$ the theory in the interior has an enhanced $SU(2)$ gauge symmetry, while for odd $p$ there is the $A_1$ two-form gauge theory of tensionless strings. In any case, one should consider these more ``exotic'' theories, beyond the usual supergravity, in order to study  the non-perturbative regime of the dual gauge theories. We can also rephrase this by saying that the enhan\c con prevents us to take a proper decoupling limit, as we were instead able to do in the case of \Ne{4} Super Yang--Mills. This explains why an exact duality at the supergravity level is out of reach in the present context of less supersymmetric and scale-anomalous gauge theories.

Before passing to our first example dealing with a four-dimensional non-conformal gauge theory, let us make another observation on the interpretation of our ability of extracting information on gauge theory from classical supergravity fields. As we mentioned several times, this fact can be thought of as a consequence of open/closed string duality. This is however rather surprising, since such a duality is known to mix string states of different mass levels, while we are only considering massless states in both the open and closed string channels.

However, it has been shown via a careful string theory computation that, at least for a class of fractional D-branes on orbifolds, the divergences of the massless open string channel, which are related to the gauge theory quantities we are used to consider such as the running of the gauge coupling constant, are precisely captured by only the massless states in the closed string channel, and therefore by supergravity~\cite{DiVecchia:2003ae}. This fact gives a firm string theory basis to the analysis we have been performing throughout this report.

\section{\Ne{2}, $d=4$ Super Yang--Mills from D-branes}\label{s:Ne2}

In this section we pass to the study of another gauge theory with eight preserved supercharges, \Ne{2} Super Yang--Mills theory in four space-time dimensions. Similarly to what we did in sections~\ref{s:wD4} and~\ref{s:fD2}, we will engineer it via two different configurations of D-branes. A stack of $N$ D5-branes wrapped on a two-cycle inside a Calabi--Yau twofold will give us information on the pure $U(N)$ theory, while a system of fractional D3/D7 branes on a \cz{} orbifold will allow us to couple the theory to matter hypermultiplets in the fundamental representation of the gauge group.

We will recover from both the setups correct results for the perturbative running coupling constant and chiral anomaly, while we will be unable to enter the non-perturbative regime described by the exact solution by Seiberg and Witten~\cite{Seiberg:1994rs}, since also in this case an enhan\c con mechanism takes place preventing us to analyze the gauge theory at strong coupling.

\subsection*{Wrapped D5-branes}

Our first example of brane configuration able to give information on a four-dimensional gauge theory is made of $N$ D5-branes wrapped on a supersymmetric two-cycle inside a Calabi--Yau twofold~\cite{Gauntlett:2001ps,Bigazzi:2001aj}:
\begin{center}
\begin{tabular}{|c|c|c|c|c|c|c|c|c|c|c|}
\multicolumn{7}{c}{ }&
\multicolumn{4}{c}{$\overbrace{\phantom{\qquad\qquad\qquad}}^{\text{CY}_2}$}\\
\hline
&\multicolumn{4}{|c|}{$\mathbb{R}^{1,3}$}
&\multicolumn{2}{|c|}{$\mathbb{R}^2$}
&\multicolumn{2}{|c|}{$S^2$}
&\multicolumn{2}{|c|}{$N_2$}\\
\hline
D$5$ &$-$&$-$&$-$&$-$&$\cdot$&$\cdot$&$\bigcirc$&$\bigcirc$&$\cdot$&$\cdot$\\
\hline
\end{tabular}
\end{center}

This system is T-dual to the system of D4-branes we considered in section~\ref{s:wD4}, and in fact the supergravity solution can be found in much the same way. Since the Calabi--Yau twofold preserves 16 of the 32 supercharges of flat space, we expect that the theory living on the D5-branes will preserve, similarly to the T-dual D4-branes, 8 supercharges, or \Ne{2} in the language appropriate to the theory living on the four-dimensional ``flat'' part of the world-volume. In fact, we already confirmed these expectations in section~\ref{s:wrapped}, by means of a group theory analysis of the topological twist. 

The procedure for finding the supergravity solution is exactly the one outlined in section~\ref{s:wD4} and summarized in figure~\ref{f:wsolution}. One starts by imposing a suitable domain wall ansatz in seven-dimensional gauged supergravity, where a $U(1)$ gauge field is chosen to fulfill the requirements of the topological twist. The seven-dimensional solution is then uplifted to the ten-dimensional string frame by means of the formulae in~\cite{Cvetic:2000dm}. We will not present the details of the construction here, for which we refer to
\cite{Gauntlett:2001ps,Bigazzi:2001aj,DiVecchia:2002ks}. The final ten-dimensional solution in the string frame is given by:
\begin{equation}\label{wD5sol}
\begin{aligned}
	ds^2 &= e^{\Phi} \Bigg\{ \eta_{\alpha\beta}dx^\alpha dx^\beta + \gs N\ls^2 \Bigg[
		z \left( d\ttheta^2 + \sin^2\ttheta d\tphi^2 \right) + e^{2x} dz^2\\
		&\qquad+ d\theta^2 + \frac{e^{-x}}{\Omega}
		\cos^2\theta \left(d\phi_1^2 + \cos\ttheta d \tphi \right)^2
		+\frac{e^x}{\Omega} \sin^2\theta d\phi_2^2
		\Bigg] \Bigg\}\,,\\
	e^{2\Phi} &= e^{2z} \left[1-\sin^2\theta \frac{1+c e^{-2z}}{2z}\right]\,,\\
	C_2 &= \gs N \ls^2 \phi_2\ d\left[\frac{\sin^2\theta}{\Omega e^x}
		\left(d\phi_1^2 + \cos\ttheta d \tphi \right)\right]\,,
\end{aligned}
\end{equation}
where $c$ is an integration constant and:
\begin{equation}
	\Omega = e^x \cos^2\theta + e^{-x} \sin^2\theta\,.
\end{equation}
The above solution can again be showed to be singular, and one can see via a probe computation perfectly analogous to the one performed in section~\ref{s:wD4} that the singularity is again cloaked by an enhan\c con mechanism. In this situation, however, the enhan\c con locus has the shape of a ring instead of a two-sphere.

In order to clarify the role of the different coordinates and functions in~\eqref{wD5sol}, we proceed as in section~\ref{s:wD4} and introduce a change of coordinates analogous to~\eqref{mycoord}~\cite{DiVecchia:2002ks}:
\begin{equation}
	r = \sin\theta e^z\,,\qquad \sigma = \sqrt{z} \cos\theta e^{z-x}\,.
\end{equation}
In the new coordinate system the metric of the solution reads:
\begin{equation}\label{wD5newsol}
\begin{split}
	ds^2 &= H^{-1/2} \left[ \eta_{\alpha\beta}dx^\alpha dx^\beta
		+ \gs N \ls^2 z \left(d\ttheta^2 + \sin^2\ttheta d\tphi^2 \right)\right]\\
		&\qquad+ \gs N \ls^2 H^{1/2} \left\{ dr^2 + r^2 d\phi_2^2 
		+ \frac{1}{z}
		\left[ d\sigma^2 + \sigma^2 \left( d\phi_1+\cos\ttheta d\tphi\right)^2\right]\right\} \,,
\end{split}
\end{equation}
where we have defined $H = e^{-2\Phi}$ and $z$ is now meant as an implict function of $r$ and $\sigma$. Now it is easy to identify the transverse plane $(r,\phi_2)$ where the branes can move supersymmetrically, while $\sigma$ can be interpreted as a radial coordinate in the non-trivially fibered transverse two-dimensional space inside the Calabi--Yau twofold.

We can now use the solution~\eqref{wD5newsol} to extract gauge theory information on pure \Ne{2} $U(N)$ Super Yang--Mills theory. The relevant formulae are the ones given in section~\ref{s:dictionary}. As in section~\ref{s:wD4}, the geometrical features of the configuration force us to work at the ``supersymmetric locus'' $\sigma=0$, where the static potential between a D5-brane probe and the stack of branes generating the classical solution vanishes. The gauge coupling constant and theta-angle are given by~\eqref{gym4} and~\eqref{tym4}:
\begin{subequations}\label{gtwD5}
\begin{align}
	\frac{1}{\gym} &= \frac{1}{4\pi\gs}\ \frac{1}{(2\pi\ls)^2}
		\int_{S^2} d\ttheta d\tphi\ e^{-\Phi}\sqrt{\det{G_{AB}}}
		= \frac{N}{4\pi^2} \ln \frac{r}{r_0}\,,\\
	\tym &= \frac{1}{2\pi\gs\ls^2} \int_{S^2} C_2 = -2N\phi_2\,.\label{tymwD5}
\end{align}
\end{subequations}
where $r_0$ is a regulator and the integrals are made along the two-cycle on which the branes are wrapped, parameterized by $(\ttheta,\tphi)$.

The next crucial point is the implementation of the correct energy/radius relation. Since we have a part of transverse space in which the branes can move freely, we can implement the ``stretched string'' relation, which in this case implies:
\begin{equation}\label{erwD5}
	r = 2\pi\ls^2\ \mu\,,\qquad r_0 = 2\pi\ls^2\ \Lambda\,,
\end{equation}
where $\mu$ is the subtraction energy at which the theory is defined and $\Lambda$ is the dynamically generated scale. The coupling constant of the gauge theory is then given by:
\begin{equation}
	\frac{1}{\gym} = \frac{N}{4\pi^2}\ln \frac{\mu}{\Lambda}\,,
\end{equation}
and the $\beta$-function reads:
\begin{equation}
	\beta(g_{\text{YM}}) = -\frac{N}{8\pi^2} g_{\text{YM}}^3\,,
\end{equation}
which is the correct perturbative result for \Ne{2} Super Yang--Mills, whose $\beta$-function, apart from instanton corrections, is exact at one-loop.

Let us now pass to the chiral anomaly. The gauge theory has an anomalous $U(1)_R$ R-symmetry group. From~\eqref{tymwD5}, it is natural to identify this symmetry with shifts in the angular coordinate $\phi_2$. However, \eqref{tymwD5} also tells us that a generic shift in $\phi_2$, though being an isometry of the metric, is \emph{not} a symmetry of the full solution~\eqref{wD5sol}, since $C_2$ is not invariant.

Now, we know that an R-symmetry transformation of parameter $\epsilon$ in this \Ne{2} theory changes the theta-angle as:
\begin{equation}
	\tym \to \tym - 4 N \epsilon\,,
\end{equation}
which corresponds to a shift of $\phi_2$ of the form:
\begin{equation}
	\phi_2 \to \phi_2 + 2\epsilon\,,
\end{equation}
and the latter transformation, as we noticed, is not generically a symmetry of the solution. Are there nevertheless some values of $\epsilon$ for which the solution is invariant? We have to recall that the flux of $C_2$ through the cycle is not arbitrarily defined, but is instead a periodic variable of period $(2\pi\ls)^2\gs$. This means that transformations of the kind:
\begin{equation}
	\phi_2 \to \phi_2 + \frac{\pi k}{N}\,,
\end{equation}
with $k$ integer, modify the expression
\begin{equation}
	\frac{1}{(2\pi\ls)^2\gs} \int_{S^2} C_2 = -\frac{N\phi_2}{\pi}
\end{equation}
for an irrelevant integer number, thus being symmetries of the supergravity configuration. This means that the $U(1)$ group of the shifts in $\phi_2$, which we identified with the $U(1)_R$ symmetry of the gauge theory, is broken down to $\mathbb{Z}_{4N}$. This anomalous breaking is the correct result for the gauge theory, and we therefore see another example of gauge theory information which is very naturally encoded in a classical solution.

Now, recalling some of our considerations from section~\ref{s:probe}, we can think of extracting another piece of information about our four-dimensional \Ne{2} theory, namely the action of an instanton configuration. We learnt that, because of the expansion of the Wess--Zumino part of the world-volume action (from which in fact we just derived the theta-term of the gauge theory action), a D$(p-4)$-brane plays the role of an instanton configuration for the theory on a D$p$-brane. Since our geometry is generated by D5-branes wrapped on a two-cycle, it is natural to expect that an instanton configuration will be given by a (euclidean) D1-brane wrapping the same two-cycle.

This expectation is immediate to verify. The action of a euclidean D1-brane wrapped on $S^2$ in the background~\eqref{wD5sol} is given by:
\begin{equation}
	S_{\text{D}1} = \tau_1 \int_{S^2} d\theta d\tphi\ e^{-\Phi}\sqrt{\det{G_{AB}}} + \tau_1 \int_{S^2} C_2\,,
\end{equation}
which recalling the expressions computed in~\eqref{gtwD5}, becomes:
\begin{equation}\label{instwD5}
	S_{\text{D}1} = \frac{8\pi^2}{\gym}-i\tym\,,
\end{equation}
which is the correct result for the action of an instanton. A $k$-instanton configuration corresponds to considering $k$ of the above D1-branes, and the resulting action is simply $k$ times~\eqref{instwD5} due to the BPS nature of the branes. This agrees with what is expected from field theory.

\subsection*{Fractional D3/D7-branes}

Let us now present the last example of this chapter, which is again a system of fractional branes on the simple \cz{} orbifold. The gauge theory we want to describe is, as in the previous subsection, \Ne{2} Super Yang--Mills in four space-time dimensions, so we must consider D3-branes. In order to have a $U(N)$ gauge theory, and to also introduce $M$ hypermultiplets in the fundamental representation of the gauge group, we add $M$ D7-branes extended along the orbifold, analogously to what we did in section~\ref{s:fD2} with D2 and D6-branes. The configuration we are going to study is then the following one:
\begin{center}
\begin{tabular}{|c|c|c|c|c|c|c|c|c|c|c|}
\multicolumn{7}{c}{ }&
\multicolumn{4}{c}{$\overbrace{\phantom{\qquad\qquad\qquad}}^{\cz}$}\\
\hline
&0&1&2&3&4&5&6&7&8&9\\
\hline
D$3$ &$-$&$-$&$-$&$-$&$\cdot$&$\cdot$&$\cdot$&$\cdot$&$\cdot$&$\cdot$\\
\hline
D$7$ &$-$&$-$&$-$&$-$&$\cdot$&$\cdot$&$-$&$-$&$-$&$-$\\
\hline
\end{tabular}
\end{center}

The supergravity solution has been found in~\cite{Bertolini:2000dk,Bertolini:2001qa}, and additional aspects of the construction were outlined in~\cite{Polchinski:2000mx,Grana:2001xn,Billo:2001vg,Bertolini:2002xu}. The procedure one follows in order to find the solution is exactly the one we explained in section~\ref{s:fD2}. One splits the ten-dimensional type IIB supergravity action in two parts, one involving the ``untwisted'' fields and one (six-dimensional) involving the ``twisted'' fields that one obtains by decomposing the type IIB fields along the non-trivial anti-self-dual two-form $\omega_2$ defined in~\eqref{omegaprop}:
\begin{equation}
	B_2 = b\ \omega_2 \,,\qquad
	C_2 = A_0\ \omega_2 \,.
\end{equation}

By imposing a usual D3/D7 ansatz and solving the type IIB equations of motion, one finally arrives to the following supergravity configuration:
\begin{subequations}\label{fD3sol}
\begin{align}
	ds^2 &= H_3^{-1/2} \eta_{\alpha\beta} dx^\alpha dx^\beta
		+H_3^{1/2} ( \delta_{ij} dx^i dx^j + e^{-\Phi} \delta_{rs} dx^r dx^s )\,,\\
	\tilde{F}_5 &= d H_3^{-1} dx^0 \wedge \dotsm \wedge dx^3
		+ \hd{} (d H_3^{-1} dx^0 \wedge \dotsm \wedge dx^3)\,,\\
	\tau &= C_0 + i e^{-\Phi} = i \left( 1 - \frac{\gs M}{2\pi} \ln \frac{z}{z_0} \right)\,,\\
	e^{-\Phi}\ b &= \frac{(2\pi\ls)^2}{2} \left( 1 + \frac{2N-M}{\pi} \ln \frac{r}{r_0}\right)\,,\\
	A_0 + C_0\ b &= -2\pi\gs\ls^2 (2N-M)\ \theta\,,
\end{align}
\end{subequations}
where $z = x^4 + i x^5 = r e^{i\theta}$ spans the two-dimensional ``flat'' transverse space and $H_3$ is an appropriate function of $(x^4,\ldots,x^9)$. As we are now used to find, the solution~\eqref{fD3sol} presents a naked singularity of repulson type, which is again accounted for by the enhan\c con mechanism.

Let us now pass to recovering some gauge theory quantities from the solution~\eqref{fD3sol}. The relevant relations are still given by~\eqref{gym4} and~\eqref{tym4}:
\begin{subequations}
\begin{align}
	\frac{1}{\gym} &= \frac{1}{4\pi\gs}\ \frac{1}{(2\pi\ls)^2} \int_{S^2} e^{-\Phi} B_2
		= \frac{1}{8\pi\gs} + \frac{2N-M}{8\pi^2}\ln \frac{r}{r_0}\,,\\
	\tym &= \frac{1}{2\pi\gs\ls^2} \int_{S^2} C_2 = -(2N-M)\theta\,,
\end{align}
\end{subequations}
and the energy/radius relation we will implement is again given by~\eqref{erwD5}:
\begin{equation}
	r = 2\pi\ls^2\ \mu\,,\qquad r_0 = 2\pi\ls^2\ \Lambda\,,
\end{equation}
so that we obtain the following correct perturbative $\beta$-function of the gauge theory:
\begin{equation}
	\beta(g_{\text{YM}}) = - \frac{2N-M}{16\pi^2}\ g^3_{\text{YM}}\,.
\end{equation}

The analysis of the chiral anomaly is also analogous to the one we performed in the previous subsection. An R-symmetry transformation of parameter $\epsilon$ changes the theta-angle as follows:
\begin{equation}
	\tym \to \tym - 2\epsilon (2N-M)\,,
\end{equation}
so that the allowed transformations of $\theta$ in the solution, namely the ones respecting the periodicity of $\int_{S^2} C_2$, are only the one in the $\mathbb{Z}_{2(2N-M)}$ subgroup of $U(1)_R$. The supergravity solution then correctly accounts for the chiral symmetry breaking $U(1)_R \to \mathbb{Z}_{2(2N-M)}$ of the dual gauge theory.

As in the previous subsection, we can also extract the correct instanton action from our configuration. Since the gauge theory lives on fractional D3-branes, we need to probe the geometry with the action of a fractional D$(-1)$-brane, namely a fractional D-instanton. The computation is perfectly analogous to the one performed in the previous subsection, and one obtains the correct result:
\begin{equation}
	S_{\text{D}(-1)} = \frac{8\pi^2}{\gym}-i\tym\,.
\end{equation}

We have now concluded our analysis of some applications of the gauge/string correspondence to theories with eight preserved supercharges. In the next chapter, we will consider examples of theories with additionally reduced supersymmetry, concentrating in particular on brane configurations dual to \Ne{1} gauge theories in four dimensions.

% "Gauge/Gravity with Four Supercharges"

\chapter{Gauge/Gravity with Four Supercharges}\label{c:4susy}

This chapter ends our journey through non-conformal and less-supersymmetric cases of the gauge/gravity correspondence by analyzing some classical solutions dual to \Ne{1} Super Yang--Mills theory in four space-time dimensions. The importance of these studies relies in the fact that this is the supersymmetric theory which more closely resembles a phenomenologically realistic gauge theory, since it is believed that they share crucial properties such as asymptotic freedom in the ultraviolet and confinement, chiral symmetry breaking and generation of a mass gap in the infrared. In order to study \Ne{1} theories, we consider three specific examples: The Maldacena--N\'u\~nez solution, the warped deformed conifold and a system of fractional D3-branes on a \czz{} orbifold, and we show how one can extract information on gauge theory properties such as the chiral anomaly, gaugino condensation, the gauge coupling at all loops and non-perturbative superpotentials.

\section{The Maldacena--N\'u\~nez solution}\label{s:MN}

The first brane configuration we consider in order to engineer a non-conformal gauge theory with four supercharges is again a system of D5-branes wrapped on a supersymmetric cycle. Differently from the case of the wrapped D5-branes treated in section~\ref{s:Ne2}, now the particular embedding of the gauge in the spin connection will have the interpretation of wrapping the branes on a two-cycle in a Calabi--Yau threefold, thus breaking additional supersymmetry. The configuration under study is the following:
\begin{center}
\begin{tabular}{|c|c|c|c|c|c|c|c|c|c|c|}
\multicolumn{5}{c}{ }&
\multicolumn{6}{c}{$\overbrace{\phantom{\qquad\qquad\qquad\qquad}}^{\text{CY}_3}$}\\
\hline
&\multicolumn{4}{|c|}{$\mathbb{R}^{1,3}$}
&\multicolumn{2}{|c|}{$S^2$}
&\multicolumn{4}{|c|}{$N_4$}\\
\hline
D$5$ &$-$&$-$&$-$&$-$&$\bigcirc$&$\bigcirc$&$\cdot$&$\cdot$&$\cdot$&$\cdot$\\
\hline
\end{tabular}
\end{center}

We already presented the topological twist at hand in section~\ref{s:wrapped}, where we showed that the resulting theory on the four-dimensional ``flat'' part of the world-volume of the D5-branes is pure \Ne{1} Super Yang--Mills theory in four space-time dimensions.

\subsection*{The supergravity solution}

The construction of the supergravity solution proceeds in a way which is analogous to the one used in chapter~\ref{c:8susy}, namely by looking for an appropriate domain-wall solution of seven-dimensional gauged supergravity (since we are considering five-branes), with the needed twist. We do not give here the details of the construction~\cite{Chamseddine:1997nm,Maldacena:2000yy,Papadopoulos:2000gj,DiVecchia:2002ks}, but let us pause for an important observation. If we just turn on a $U(1)$ gauge field of the seven-dimensional gauged supergravity, the resulting domain wall solution can be shown to be singular, in perfect agreement with all the examples presented in the previous chapter.

However, unlike the other examples we presented so far, in the present case the singular nature of the solution can be cured~\cite{Chamseddine:1997nm,Maldacena:2000yy}. How? We said that we identified a $U(1) \subset SU(2)_L$ with the spin-connection of the two-sphere on which the wall is wrapped, with an ansatz like:
\begin{equation}
	A \sim \cos \ttheta d\tphi,,
\end{equation}
but we can also think of turning on the other components of the $SU(2)_L$ gauge fields:
\begin{equation}
	A^1 \sim a(r) d\ttheta\,,\qquad
	A^2 \sim a(r) \sin\ttheta d\tphi\,,\qquad
	A^3 \sim \cos \ttheta d\tphi,,
\end{equation}
where $a(r)$ is a function that approaches zero in the $r\to\infty$ limit. This means that the twist is the usual one for large $r$, as it should be since we are by now used of thinking of large distances as the ultraviolet region of a gauge theory, where the perturbative degrees of freedom used for discussing the topological twist are the relevant ones. However, we have introduced a deformation at short distances that makes the solution smooth everywhere. 

This is the first time we are able to achieve such a result, which seems a specific feature of cases with four supersymmetries, and will give us the possibility of exploring additional aspects of the gauge theory under consideration, since there is no longer an enhan\c con mechanism at work. In fact, such a resolution of the singularity (which takes place in the present case as well as in the case of the conifold we will consider in the next section) can be mapped to a phenomenon of \Ne{1} gauge theory which was not present in more supersymmetric cases, namely the chiral symmetry breaking due to gaugino condensation. Unlike the one-loop $U(1)_R$ anomaly, this additional breaking is a pure infrared phenomenon which manifests itself at the level of supergravity with the deformation and desingularization of the solution. We will comment more on this below, when we study the dual gauge theory in detail.

Let us now present the full ten-dimensional Maldacena--N\'u\~nez solution in the string frame~\cite{Maldacena:2000yy}:
\begin{subequations}\label{MN}
\begin{align}
	ds^2 &= e^{\Phi} \left[
		\eta_{\alpha\beta} dx^\alpha dx^\beta + \gs N \ls^2 \left(
		e^{2h} ( d\ttheta^2 + \sin^2 \ttheta d\tphi^2 )
		+ d\rho^2 + \sum_{a=1}^{3} (\sigma^a - A^a)^2\right)\right]\,,\label{MNmetric}\\
	e^{2\Phi} &= \frac{\sinh 2\rho}{2e^h}\,,\\
	F_3 &= \gs N \ls^2
		\left[ 2\ (\sigma^1 - A^1) \wedge (\sigma^2 - A^2) \wedge (\sigma^3 - A^3)
		- \sum_{a=1}^{3} F^a \wedge \sigma^a \right]\,,
\end{align}
\end{subequations}
where the indices $\alpha,\beta$ run as usual from 0 to 3, $\rho$ is a dimensionless coordinate which is related to the physical radial distance $r$ via $r = \sqrt{\gs N}\ \ls\ \rho$ and where:
\begin{equation}
\begin{gathered}
	A^1 = -\frac{1}{2} a(\rho) d\ttheta\,,\qquad
	A^2 = \frac{1}{2} a(\rho) \sin\ttheta d\tphi\,,\qquad
	A^3 = -\frac{1}{2} \cos \ttheta d\tphi\,,\\
	e^{2h} = \rho \coth 2\rho - \frac{\rho^2}{\sinh^2 2\rho}-\frac{1}{4}\,,\qquad
	a(\rho) = \frac{2\rho}{\sinh 2\rho}\,,
\end{gathered}
\end{equation}
and $F^a = dA^a + \epsilon^{abc} A^b \wedge A^c$. The $\sigma^a$ are left-invariant one-forms parameterizing the round three-sphere that was used to lift the solution from seven to ten dimensions (see the third parameterization in insert~\ref{i:S3} on page~\pageref{i:S3}):
\begin{equation}
\begin{gathered}
	\sigma^1 = \frac{1}{2} ( \cos \psi d\theta + \sin \psi \sin \theta d\phi )\,,\qquad
	\sigma^2 = -\frac{1}{2} ( \sin \psi d\theta - \cos \psi \sin \theta d\phi )\,,\\
	\sigma^3 = \frac{1}{2} ( d\psi + \cos \theta d\phi )\,.
\end{gathered}
\end{equation}

Since we will need it in the following, let us also compute the explicit expression of the R-R potential $C_2$:
\begin{multline}\label{MNC2}
	C_2 = \frac{\gs N \ls^2}{4} \left[ (\psi + \psi_0)
		(\sin\theta d\theta \wedge d\phi - \sin\ttheta d \ttheta\wedge d\tphi)
		+ \cos \ttheta \cos \theta d\tphi \wedge d\phi\right]\\
		+ \frac{\gs N \ls^2}{2} a(\rho)
		\left[ d\ttheta \wedge \sigma^1 - \sin\ttheta d\tphi \wedge \sigma^2\right]\,,
\end{multline}
where we have taken into account an arbitrary integration constant $\psi_0$. 

As we anticipated, thanks to the non-abelian ansatz in gauged supergravity, the solution~\ref{MN} is completely smooth, including the point $\rho=0$. In fact, one easily sees that for small $\rho$ the angular part of the six-dimensional part metric~\eqref{MNmetric} transverse to the four flat coordinates $x^\alpha$, reduces to:
\begin{multline}\label{MNsphere}
	ds^2_5 = \frac{1}{4} (\cos \psi \sin\theta d\phi - \sin\psi d\theta - \sin\ttheta d\tphi)^2\\
		+ \frac{1}{4} (\sin\psi \sin\theta d\phi + \cos\psi d\theta + d\ttheta)^2
		+ \frac{1}{4} (d\psi + \cos\ttheta d\tphi + \cos\theta d\phi)^2\,.
\end{multline}
Although it is quite difficult to see it, the above metric can be shown to be a redundant expression for the metric on a round three-sphere~\cite{Minasian:1999tt}, which of course has no singularity.

\subsection*{The dual gauge theory}

Let us now turn to the dual \Ne{1} $U(N)$ Super Yang--Mills theory. We already know how to extract from supergravity the running coupling constant and chiral anomaly of a four-dimensional gauge theory from our general discussion in section~\ref{s:dictionary} and specific \Ne{2} examples in section~\ref{s:Ne2}. We are going to use again the relations~\eqref{gym4}-\eqref{tym4}:
\begin{equation}\label{MNgauge}
	\frac{1}{\gym} = \frac{1}{4\pi\gs} \frac{1}{(2\pi\ls)^2}
		\int_{S^2} d^2\xi\ e^{-\Phi} \sqrt{\det G}\,,\qquad
	\tym = - \frac{1}{2\pi\gs\ls^2} \int_{S^2} C_2\,,
\end{equation}
where $S^2$ denotes the two-cycles on which the D5-branes are wrapped and $G$ and $C_2$ are meant to be space-time fields restricted to the cycle.

The problem is now the correct identification of the two-cycle. Naively, and driven by analogy with chapter~\ref{c:8susy}, we would say that it is simply the one parameterized by the angles $\ttheta$ and $\tphi$. However, things are not so simple~\cite{Bertolini:2002yr}. In order to clarify this issue, let us study the large $\rho$ limit of the solution. The five-dimensional angular part of the metric~\eqref{MNmetric} for large $\rho$ reduces to:
\begin{equation}
	ds^2_{5} \sim \rho\ (d\ttheta^2 + \sin^2\ttheta d\tphi^2)
		+ \frac{1}{4} (d\theta^2 + \sin^2\theta d\phi^2)
		+ \frac{1}{4} (d\psi + \cos\ttheta d\tphi + \cos\theta d\phi)^2\,,
\end{equation}
which can be recognized as the metric of $T^{1,1}$ (see insert~\ref{i:T11} on page~\pageref{i:T11}), although rescaled in a way it is no longer an Einstein space. The topologically non-trivial cycles can then be derived from insert~\ref{i:T11} on page~\pageref{i:T11}, and in particular the $S^2$ of minimal volume can be written as:%
\footnote{There is another choice of parameterization besides~\eqref{MNcycle} that we will not consider, but that gives fully equivalent results~\cite{Bertolini:2002yr}.}
\begin{equation}\label{MNcycle}
	\psi=0\text{ or }2\pi \,,\quad \ttheta = -\theta\,,\quad \tphi = - \phi\,.
\end{equation}
One then finds, integrating on the above cycle and using~\eqref{MN} and~\eqref{MNC2}, that the gauge theory quantities~\eqref{MNgauge} are given in terms of supergravity parameters as:
\begin{equation}\label{MNgauge2}
	\frac{1}{\gym} = \frac{N}{16\pi^2} \left(e^{2h(\rho)} + (a(\rho) -1)^2\right)
		= \frac{N}{4\pi^2}\ \rho \tanh \rho\,,\qquad
	\tym = - N \psi_0\,.
\end{equation}
%:Insert: T^1,1
\begin{Insert}{$T^{1,1}$ and its cycles}\label{i:T11}
The space $T^{1,1}$ appears ubiquitously in geometrical setups dual to \Ne{1} gauge theories in four dimensions. It is an Einstein space that can be defined as the coset space:
\begin{equation*}
	T^{1,1} = \frac{SU(2)\times SU(2)}{U(1)}\,.
\end{equation*}
The metric on $T^{1,1}$ can be cast in the form:
\begin{equation*}
	ds^2_{T^{1,1}} =
		\frac{1}{9} \bigg( d\psi + \sum_{i=1}^2 \cos\theta_i d\phi_i \bigg)^2
		+ \frac{1}{6} \sum_{i=1}^2 \left( d\theta_i^2 + \sin^2\theta_i d\phi_i^2\right)\,,
\end{equation*}
where $0\le \psi \le 4\pi$, $0\le \theta_i\le \pi$ and $0\le \phi_i\le 2\pi$. This form of the metric explicitly shows that $T^{1,1}$ is an $S^1$ bundle over $S^2\times S^2$. Using the metric, one can also compute the volume: 
\begin{equation*}
\vol(T^{1,1}) = \tfrac{16\pi^3}{27}\,.
\end{equation*}

Topologically, the $T^{1,1}$ manifold can be thought of as $S^2\times S^3$. The two cycles can be identified as:
\begin{equation*}
\begin{aligned}
	&S^2\,: \qquad & \psi &= 0\,,\quad \theta_1 = \theta_2\,,\quad \phi_1 = - \phi_2\,,\\
	&S^3\,: \qquad & \theta_1 &= \phi_1 = 0\,.
\end{aligned}
\end{equation*}
\end{Insert}
The next step is the identification of $\rho$ with gauge theory scales, via an appropriate energy/radius relation. However, in this case things are not so simple as in the previous chapter, and in order to find the correct gauge/gravity relation we first have to study chiral symmetry breaking in detail. Notice that, besides the observations we made in chapter~\ref{c:8susy}, the difficulties of establishing an energy/radius relation are due to the fact that the present setup has fluxes replacing branes, in the spirit of the geometric transitions discussed in section~\ref{s:geom}. In the absence of ``real'' branes, it is not at all easy, and in fact improper, to implement a relation such as the one coming from the ``stretched string'' argument.

Let us then turn to the chiral anomaly. From~\eqref{MNgauge2}, we see that $U(1)$ R-symmetry transformations in the gauge theory have to do with shifts in the angular variable $\psi$ in supergravity. On the other hand, it is also clear that shifts in $\psi$ are \emph{not} a symmetry of the full solution~\eqref{MN}. In fact, the only remnant of the symmetry is the choice of $\psi$ to be fixed at one of the values $0$ or $2\pi$, as one can see from~\eqref{MNcycle}, recalling that the period of $\psi$ is $4\pi$. This $\mathbb{Z}_2$ symmetry is all that is left of the R-symmetry, and this is the correct expected result, since in \Ne{1} Super Yang--Mills theory gaugino condensation breaks the $\mathbb{Z}_{2N}$ unbroken symmetry of the ultraviolet down to $\mathbb{Z}_2$.

We would like to analyze this phenomenon more closely, starting from the large $\rho$ ultraviolet region. In this region, we can identify the two-cycle on which the branes are wrapped as simply being the one parameterized by $(\ttheta,\tphi)$, and we get:
\begin{equation}
	\tym \sim \int_{S^2} C_2 = \frac{N}{2\pi} (\psi + \psi_0)\,.
\end{equation}
The above result is invariant for shifts of the form:
\begin{equation}\label{uvsymm}
	\psi \to \psi + \frac{2\pi k}{N}\,,
\end{equation}
which are thus symmetries of the solution for large $\rho$. Since an R-symmetry transformation of parameter $\epsilon$ changes the theta angle by $\tym \to \tym - 2 N\epsilon$, from~\eqref{MNgauge2} we see that this transformation translates in supergravity as:
\begin{equation}
	\psi \to \psi + 2\epsilon\,.
\end{equation}
This means that the remaining ultraviolet symmetries in~\eqref{uvsymm} correspond to the breaking of $U(1)_R$ down to $\mathbb{Z}_{2N}$, which is the correct R-symmetry breaking in the ultraviolet.

Therefore, we have seen that the supergravity solution perfectly accounts for the chiral symmetry breaking phenomenon in the gauge theory, both in the ultraviolet where a $\mathbb{Z}_{2N}$ subgroup of $U(1)_R$ is unbroken, and in the infrared where the addtional breaking $\mathbb{Z}_{2N}\to\mathbb{Z}_{2}$ takes place. The final step is to understand what is the reason of the latter breaking in supergravity, and this can be easily traced to the appearance of the function $a(\rho)$ in the solution~\eqref{MN}. It is $a(\rho)$ which is responsible for the deviation of the explicit form of $C_2$ from its asymptotic value, which in turn generates the infrared chiral symmetry breaking.

In gauge theory, we know that the $\mathbb{Z}_{2N}\to\mathbb{Z}_{2}$ breaking is a consequence of gaugino condensation, and it is therefore natural to identify this gauge theory quantity with the supergravity quantity $a(\rho)$~\cite{Apreda:2001qb,DiVecchia:2002ks}. This is crucial since it is the main ingredient that we need to determine the energy/radius relation. In fact, the gaugino condensate is a protected operator in the gauge theory, which is related to the dynamical scale via $\langle \tr \lambda\lambda \rangle = \Lambda^3$. This means that, recalling that the function $a(\rho)$ is dimensionless and thus introducing the subtraction scale $\mu$ of the gauge theory, we can identify:
\begin{equation}
	\mu^3\ a(\rho) = \Lambda^3\,,
\end{equation}
namely:
\begin{equation}\label{MNer}
	\frac{\Lambda^3}{\mu^3} = \frac{2\rho}{\sinh 2\rho}\,.
\end{equation}
This relation gives implicitly the energy/radius relation between supergravity coordinates and gauge theory scales, and will be used in the following to extract the $\beta$-function of the gauge theory. In fact, we can write:
\begin{equation}
	\beta(g_{\text{YM}}) = \frac{\partial g_{\text{YM}}}{\partial \ln \frac{\mu}{\Lambda}}
		= \frac{\partial g_{\text{YM}}}{\partial \rho}
		\frac{\partial \rho}{\partial \ln \frac{\mu}{\Lambda}}\,.
\end{equation}
Now, let us first use the asymptotic behavior of~\eqref{MNgauge2} for large $\rho$, neglecting subleading exponential corrections. This is given by:
\begin{equation}\label{MNgymas}
	\frac{1}{\gym} \sim \frac{N\rho}{4\pi^2}\,.
\end{equation}
We therefore get:
\begin{equation}\label{noexpc}
\begin{aligned}
	\frac{\partial g_{\text{YM}}}{\partial \rho} &= \frac{\pi}{\sqrt{N}} \rho^{-3/2}
		= - \frac{N g^3_{\text{YM}}}{8\pi^2}\,,\\
	\frac{\partial \rho}{\partial \ln \frac{\mu}{\Lambda}}
		&= \frac{3}{2} \left(1-\frac{1}{2\rho}\right)^{-1}
		= \frac{3}{2} \left(1- \frac{N\gym}{8\pi^2}\right)^{-1}\,,
\end{aligned}
\end{equation}
where in the latter equality we used the asymptotic relation~\eqref{MNgymas} to trade $\rho$ for $g_{\text{YM}}$. The final result we obtain is then:
\begin{equation}\label{MNNSVZ}
	\beta(g_{\text{YM}}) = -\frac{3Ng^3_{\text{YM}}}{16\pi^2}
		\left(1- \frac{N\gym}{8\pi^2}\right)^{-1}\,,
\end{equation}
which is the correct NSVZ $\beta$-function at all-loops computed in the Pauli--Villars renormalization scheme for \Ne{1} Super Yang--Mills theory~\cite{Novikov:1983uc}. Now, the identification~\eqref{MNer} between $a(\rho)$ and the gaugino condensate is an exact equation due to the fact that the operator is protected in the gauge theory. This means that we can think of taking into account the exponential corrections to~\eqref{noexpc} by using the full relation~\eqref{MNer} instead of just using its asymptotic value. This modifies~\eqref{noexpc} as follows:
\begin{equation}
	\frac{\partial \rho}{\partial \ln \frac{\mu}{\Lambda}}
		= \frac{3}{2} \left(1-\frac{1}{2\rho}
		+\frac{2e^{-4\rho}}{1-e^{-4\rho}}\right)^{-1}\,,
\end{equation}
and the $\beta$-function~\eqref{MNNSVZ} is then modified as:
\begin{equation}\label{MNbeta}
	\beta(g_{\text{YM}}) = -\frac{3Ng^3_{\text{YM}}}{16\pi^2}
		\left(1- \frac{N\gym}{8\pi^2}
		+\frac{2e^{-16\pi^2/Ng^2}}{1-e^{-16\pi^2/Ng^2}}\right)^{-1}\,.
\end{equation}

We would like to make some observations on this final result for the scale anomaly of \Ne{1} Super Yang--Mills engineered via wrapped D5-branes. It is already quite remarkable that a classical supergravity computation, together with physical input about the identification between supergravity coordinates and gauge theory energies, is able to give a closed and analytically correct result for a quantum scale anomaly at all loops as~\eqref{MNNSVZ}. It is maybe even more remarkable that the full result~\eqref{MNbeta} also includes contributions which look like non-perturbative modifications to the running of the coupling. In fact, the dependence of the corrections on $g_{\text{YM}}$ is such that it is tempting of interpreting them as non-perturbative contributions coming from ``fractional instantons'', namely instantons of fractional charge $\tfrac{2}{N}$. However, there is debate in the literature about this interpretation, since the full theory living on the D5-branes is six-dimensional (the flat four dimensional space-time times a two-sphere) and it is at present unclear how to decouple, even in some specific regime of energies, the four-dimensional dynamics from the Kaluza--Klein modes on $S^2$ within the region of validity of supergravity. Nevertheless, even in absence of a purely field-theoretic analysis (that would be of course extremely interesting to perform), the fact that the above corrections are indeed a property of the four dimensional gauge theory, instead of a spurious phenomenon due to modes coming from the ultraviolet completion of the theory, would receive an indirect confirmation if a result like~\eqref{MNbeta} could also be recovered by a brane setup with a different ultraviolet completion. We will see in the next section that this is the case, by performing an analogous study of the warped deformed conifold solution.

Before going on, let us also mention that, also in the \Ne{1} case at hand, the action of an instanton configuration can be correctly recovered by considering a euclidean D1-brane probe wrapped on the same two-cycle of the D5-branes. The computation is perfectly analogous to the one we did in section~\ref{s:Ne2} and gives the expected result~\eqref{instwD5}.

\subsection*{Confining strings}

Thanks to its non-singular nature, the Maldacena--N\'u\~nez solution~\eqref{MN} is able to give us another relevant piece of information on the dual \Ne{1} gauge theory at the non-perturbative level, namely the tension of the confining $q$-strings~\cite{Herzog:2001fq}.

Confining $q$-strings in a $SU(N)$ gauge theory ($q=1,\ldots,N-1$) can be thought of as tubes connecting $q$ probe quarks with $q$ probe antiquarks, and correspond to Wilson loops in tensor representations with $q$ indices. Of course, one has the symmetry $q \to N-q$ obtained by replacing quarks with antiquarks. An interesting question one could investigate is how the tension of the confining string scales with $q$. In particular, a behavior like:
\begin{equation}
	T_{q+q'} < T_q + T_{q'}
\end{equation}
would imply that the $q$-string does not decay into strings with smaller $q$.

What is a confining $q$-string in the background~\eqref{MN}? It should be described as $q$ coincident fundamental strings placed at $\rho=0$ (since it is in that region of distances that confinement takes place) and extending inside the space spanned by $x^\alpha$. However, we know that at small $\rho$ there is a non-vanishing flux of $F_3$ through a three-sphere, and this has the consequence of blowing up the strings into a three-brane wrapping an $S^2$ inside the $S^3$. In order to understand this phenomenon, it is easier to go into the S-dual description, where the D5-branes generating the geometry turn into NS5-branes and the confining strings into D1-branes which expand into D3-branes. We can therefore probe the S-dual solution for small $\rho$ with the world-volume action of an appropriately wrapped D3-brane with $q$ units of world-volume gauge field $F$ turned on. This computation was first done for a T-dual case in~\cite{Bachas:2000ik}.

The S-dual solution describing $N$ NS5-branes wrapped on a two-sphere inside a Calabi--Yau threefold assumes the following form for small $\rho$:
\begin{subequations}\label{MNstring}
\begin{align}
	ds^2 &\sim \eta_{\alpha\beta} dx^\alpha dx^\beta
		+ \gs N \ls^2 (d\psi^2 + \sin^2\psi (d\theta^2 + \sin^2\theta d\phi^2))\,,\\
	B_2 &\sim \gs N \ls^2 \left(\psi - \frac{\sin 2\psi}{2}\right) \sin \theta d\theta \wedge d\phi\,.
\end{align}
\end{subequations}
Notice that the coordinates appearing in~\eqref{MNstring} should not be confused with the one we have used for the Maldacena--N\'u\~nez solution up to now. In fact, here we have used the fact, noticed in~\eqref{MNsphere}, that the angular metric reduces for small $\rho$ to the metric on a round three-sphere in order to choose a much simpler parameterization of the metric (the second in insert~\ref{i:S3} on page~\pageref{i:S3}).

Let us then probe the geometry with a D3-brane wrapped on the two-sphere $(\theta,\phi)$ inside the three-sphere $(\psi,\theta,\phi)\,$. The following world-volume gauge field is turned on:
\begin{equation}
	F = -\frac{\gs q}{2}\sin\theta d\theta\wedge d\phi\,.
\end{equation} 
The relevant Dirac--Born--Infeld part of the action of the probe is as usual:
\begin{equation}
	S=-{\tau_3}\int d^2x\ d\theta\ d\phi\
	e^{-\Phi}\sqrt{-\det \left(\hat{G}_{ab}+\hat{B}_{ab}+2\pi\ls^2 F_{ab}\right)}\,.
\end{equation}
Using the solution~\eqref{MNstring}, the determinant turns out to be:
\begin{equation}\label{det}
	\sqrt{-\det \left(\hat{G}_{ab}+\hat{B}_{ab}+2\pi\ls^2 F_{ab}\right)}
		= g_s N\ls^2 \sin\theta
		\left[\sin^4\psi + \left(\psi-\frac{\sin 2\psi}{2}-\frac{\pi q}{N}\right)^2\right]^{1/2}\,,
\end{equation}
whose minimum value is reached when:
\begin{equation}
	\psi = \frac{\pi q}{N}\,.
\end{equation}
At this value of $\psi$, the determinant~\eqref{det} becomes:
\begin{equation}
	g_s N \ls^2 \sin\theta \sin \frac{\pi q}{N}\,,
\end{equation}
and from the probe action (after performing the trivial integration in $\theta,\phi$) one can read the following effective tension for the confining $q$-string:
\begin{equation}\label{tension}
	T_q = \frac{N}{2\pi^2\ls^2}\ \sin \frac{\pi q}{N}\,.
\end{equation}
This behavior of the tension of the confining $q$-string has been found also with other approaches~\cite{Douglas:1995nw,Hanany:1998hr} and seems also supported by lattice calculations in non-supersymmetric pure glue gauge theory~\cite{Lucini:2001nv,DelDebbio:2001kz}. Notice finally that the $N\rightarrow\infty$ limit of~\eqref{tension} gives for the tension of the confining one-string the one of the fundamental string, $T_1\rightarrow\tfrac{1}{2\pi\ls^2}$, as expected.

\section{The conifold and \Ne{1} gauge theories}\label{s:conifold}

The next case we are going to consider is the prototype example of the ``geometric transition'' framework we discussed in chapter~\ref{c:engineering}, the conifold~\cite{Romans:1985an,Candelas:1990js}. We will not have the space to explain all derivations in detail as far as the classical solutions are concerned, but we will limit ourselves to essentially following the lines of the nice reviews~\cite{Herzog:2001xk,Herzog:2002ih}.

In section~\ref{s:geom}, we showed how a relevant deformation of a \cz{} orbifold makes the geometry flow to the conifold, which breaks additional supersymmetry down to eight supercharges. The theory living on a stack of $M$ D3-branes on the conifold is then an \Ne{1} four-dimensional superconformal quiver theory with gauge group $SU(M)\times SU(M)$, matter in bifundamentals and a quartic superpotential.

We have also seen that adding to the configuration $N$ fractional D3-branes stuck at the singular point, which have the interpretation of D5-branes wrapping the vanishing two-cycle of the conifold geometry, changes the gauge group to $SU(M+N)\times SU(M)$ and makes the theory non-conformal. In the following, we will report on the construction of a supergravity solution describing this brane configuration and we will discuss as usual what gauge theory information we can recover from the classical solution.

\subsection*{Fractional D-branes on the conifold and the duality cascade}

Let us start by recalling that the conifold can be seen as a six-dimensional cone over the space $T^{1,1}$, so that the metric can be written as:
\begin{equation}
	ds^2_6 = dr^2 + r^2 ds^2_{T^{1,1}}\,,
\end{equation}
where the Einstein metric on $T^{1,1}$ is given in insert~\ref{i:T11} on page~\pageref{i:T11}. If we place $M$ D3-branes at the apex of the cone, we find the following metric in the standard way:
\begin{equation}
	ds^2 = H^{-1/2}\ \eta_{\alpha\beta} dx^\alpha dx^\beta
		+ H^{1/2} (dr^2 + r^2 ds^2_{T^{1,1}})\,,
\end{equation}
where the normalization of the fluxes requires~\cite{Gubser:1998vd}:
\begin{equation}
	H(r) = 1+ \frac{R^4}{r^4}\,,\qquad
	R^4 = \frac{27}{16}\ 4\pi\gs\ls^4 M\,.
\end{equation}

Let us now add $N$ fractional D3-branes, which act as sources of $F_3$ through the $S^3$ of $T^{1,1}$ (compare the similar behavior of the fractional D3-branes on \cz{} studied in section~\ref{s:Ne2}, where it was interpreted as charge under the twisted R-R field). We therefore want to find a solution with the following normalized fluxes:
\begin{equation}
	\frac{1}{(2\pi\ls)^2\gs} \int_{S^3} F_3 = N\,,\qquad
	\frac{1}{(2\pi\ls)^4\gs} \int_{T^{1,1}} F_5 = M\,.
\end{equation}
Such a solution was found in~\cite{Klebanov:2000nc}. It will be useful to introduce the following bases of one-forms:
\begin{equation}
\begin{gathered}
	e^1 = - \sin \theta_1 d\phi_1\,,\qquad
	e^2 = d\theta_1\,,\qquad
	e^3 = \cos\psi \sin\theta_2 d\phi_2 - \sin\psi d\theta_2\,,\\
	e^4 = \sin\psi \sin\theta_2 d\phi_2 + \cos\psi d\theta_2\,,\qquad
	e^5 = d\psi + \cos\theta_1 d\phi_1 + \cos\theta_2 d\phi_2\,,
\end{gathered}
\end{equation}
and:
\begin{equation}\label{gbasis}
	g^1 = \frac{e^1-e^3}{\sqrt{2}}\,,\qquad
	g^2 = \frac{e^2-e^4}{\sqrt{2}}\,,\qquad
	g^3 = \frac{e^1+e^3}{\sqrt{2}}\,,\qquad
	g^4 = \frac{e^2+e^4}{\sqrt{2}}\,,\qquad
	g^5 = e^5\,,
\end{equation}
in terms of which the metric on $T^{1,1}$ can be written as:
\begin{equation}
	ds^2_{T^{1,1}} = \frac{1}{9}(g^5)^2 + \frac{1}{6} \sum_{i=1}^4 (g^i)^2\,.
\end{equation}

The full solution of~\cite{Klebanov:2000nc} can then be cast in the following form:
\begin{subequations}\label{KT}
\begin{align}
	ds^2 &= h^{-1/2}(r)\ \eta_{\alpha\beta} dx^\alpha dx^\beta
		+ h^{1/2}(r) (dr^2 + r^2 ds^2_{T^{1,1}})\,,\\
	B_2 &= \frac{3\gs N \ls^2}{4}\ \ln \frac{r}{r_0}\ 
		(g^1\wedge g^2 + g^3 \wedge g^4)\,,\\
	F_3 &= \frac{\gs N\ls^2}{4}\ g^5\wedge
		(g^1\wedge g^2 + g^3 \wedge g^4)\,,\\
	\tilde{F}_5 &= \mathcal{F}_5 + \hd{} \mathcal{F}_5\,,\qquad
	\mathcal{F}_5 = 27\pi\gs\ls^4 M_{\text{eff}}(r) \vol(T^{1,1})\,,\label{KTF5}
\end{align}
\end{subequations}
where $r_0$ is a regulator and:
\begin{equation}
\begin{aligned}
	h(r) &= 27\pi\ls^4\ \frac{\gs M + \frac{3(\gs N)^2}{2\pi}
		\left( \ln\frac{r}{r_0}+\frac{1}{4}\right)}{4r^4}\,,\\
	M_{\text{eff}}(r) &= M + \frac{3 \gs N^2}{2\pi}\ \ln\frac{r}{r_0}\,.
\end{aligned}
\end{equation}

As one might have expected, this solution has a naked singularity at the point where $h(r)$ vanishes. However, in the case of the Maldacena-N\'u\~nez solution considered in the previous section we learned that, when constructing a solution dual to a gauge theory with four supercharges, it is sometimes possible to deform it and smooth out the singularity. Is there this possibility in the case of the conifold? Notice that the five-form flux, due to the expression~\eqref{KTF5}, is no longer quantized, and this fact can be traced to the supergravity equation of motion:
\begin{equation}
	d\tilde{F}_5 = H_3 \wedge F_3\,.
\end{equation}
Our identification of the rank of the gauge groups with the flux of $\tilde{F}_5$ has thus to be modified, and we could say that the gauge group is $SU(M_{\text{eff}}+N)\times SU(M_{\text{eff}})$, but only when $M_{\text{eff}}$ is integer at the special radii $r_k = r_0 \exp (-2\pi k/3\gs N)$. This means that the rank of the gauge group, when decreasing $r$, is reduced as $M_{\text{eff}} = M - kN$.

All this has a very nice meaning in the dual gauge theory. By studying the flux of $B_2$, that we know to be related to the running coupling constant, one discovers that there is a scale when the coupling of $SU(M+N)$ diverges. This forces us to make an \Ne{1} \emph{Seiberg duality} transformation~\cite{Seiberg:1994pq}. Since the $SU(M+N)$ gauge factor has $2M$ flavors in the fundamental representation, the Seiberg duality transformation changes the gauge group into $SU(2M - (M+N))= SU(M-N)$, and we finally obtain a $SU(M)\times SU(M-N)$ gauge theory closely resembling the theory we started with. This behavior precisely match the non-conservation of the flux of $\tilde{F}_5$ in supergravity. This phenomenon has been called a ``duality cascade''~\cite{Klebanov:2000hb}.

At some point, however, the duality cascade must stop, since having negative $M_{\text{eff}}$ makes no physical sense. This is mapped to reaching the singularity of the solution. Luckily, at this endpoint of the cascade the gauge group is simply reduced to $SU(N)$: we know that chiral symmetry breaking takes place, and this could modify the solution and remove the singularity as in section~\ref{s:MN}. The details of this mechanism are the subject of the next subsection.

\subsection*{The warped deformed conifold}

To remove the singularity of the solution~\eqref{KT}, the conifold must be replaced~\cite{Klebanov:2000hb} by the deformed conifold~\eqref{defconifold}, that we rewrite as follows:%
\footnote{Notice that, differently from chapter~\ref{c:engineering}, the coordinates in~\eqref{defcon2} have engineering dimensions $[x]=[y]=[z]=[t]=[\varepsilon]=L^{3/2}$, due to the form of the metric. This will be important later in this section.}
\begin{equation}\label{defcon2}
	x^2 + y^2 + z^2 + t^2 = \varepsilon^2\,.
\end{equation}
Recall that the singularity of the conifold is removed through the blow-up of the three-sphere $S^3$ of $T^{1,1}$. The metric on the deformed conifold can be given in terms of~\eqref{gbasis} as~\cite{Candelas:1990js,Minasian:1999tt}:
\begin{equation}
\begin{split}
	ds^2_6 &= \frac{\varepsilon^{4/3}}{2}K(\tau)
		\Bigg[ \frac{1}{3K^3(\tau)}(d\tau^2+(g^5)^2)\\
		&\qquad+ \cosh^2 \frac{\tau}{2} [(g^3)^2+(g^4)^2]
		+ \sinh^2 \frac{\tau}{2} [(g^1)^2+(g^2)^2]\Bigg]\,,
\end{split}
\end{equation}
where:
\begin{equation}
	K(\tau) = \frac{(\sinh 2\tau - 2\tau)^{1/3}}{2^{1/3}\sinh\tau}\,.
\end{equation}
Notice that we can introduce another radial coordinate $r$ for large $\tau$, in terms of which $ds^2_6 \to dr^2 + r^2 ds^2_{T^{1,1}}$:
\begin{equation}\label{rvstau}
	r^2 = \frac{3}{2^{5/3}}\varepsilon^{4/3}e^{2\tau/3}\,.
\end{equation}

By imposing suitable ans\"atze and solving the first order BPS equations, Klebanov and Strassler found the following solution of type IIB supergravity, called the \emph{warped deformed conifold}~\cite{Klebanov:2000hb}:
\begin{subequations}\label{KS}
\begin{align}
	ds^2 &= h^{-1/2}(\tau)\ \eta_{\alpha\beta} dx^\alpha dx^\beta
		+ h^{1/2}(\tau)\ ds^2_6\,,\\
	B_2 &= \frac{\gs N \ls^2}{2}\ 
		[ f(\tau) g^1\wedge g^2 + k(\tau) g^3 \wedge g^4 ]\,,\\
	F_3 &= \frac{\gs N\ls^2}{2}\
	\{ g^5\wedge g^3\wedge g^4 
	+ d[ F(\tau) (g^1\wedge g^3 + g^2 \wedge g^4)] \}\,,\\
	\tilde{F}_5 &= \mathcal{F}_5 + \hd{} \mathcal{F}_5\,,\qquad
		\mathcal{F}_5= \frac{\gs N^2 \ls^4}{4}
		\ell (\tau) g^1 \wedge g^2 \wedge g^3 \wedge g^4 \wedge g^5\,,
\end{align}
\end{subequations}
where:
\begin{equation}\label{KSfunctions}
\begin{aligned}
	F(\tau) &= \frac{\sinh\tau-\tau}{2\sinh\tau}\,,\\
	f(\tau) &= \frac{\tau \coth\tau-1}{2\sinh\tau}\ (\cosh\tau-1)\,,\\
	k(\tau) &= \frac{\tau \coth\tau-1}{2\sinh\tau}\ (\cosh\tau+1)\,,\\
	\ell(\tau) &= \frac{\tau \coth\tau-1}{4\sinh^2\tau}
		(\sinh 2\tau - 2\tau)\,,\\
	h(\tau) &= (\gs N \ls^2)^2 2^{2/3} \varepsilon^{-8/3}
		\int_{\tau}^{\infty} dx\ \frac{x\coth x-1}{\sinh^2 x}\
		(\sinh 2x - 2x)^{1/3}\,.
\end{aligned}
\end{equation}

The solution~\eqref{KS} is completely smooth. In fact, the geometry for small $\tau$ is very similar to the one we found for the Maldacena--N\'u\~nez solution in~\eqref{MNsphere}. The function $h(\tau)$ approaches the constant value $h(\tau)\sim (\gs N \ls^2)^2 2^{2/3} \varepsilon^{-8/3} a_0$ for $\tau\to 0$,  where one can evaluate numerically $a_0\sim 0.71805$, and the geometry becomes approximately $\mathbb{R}^{1,3}$ times the deformed conifold:
\begin{multline}
	ds^2 \to \frac{\varepsilon^{4/3}}{2^{1/3} a_0^{1/2}\gs N\ls^2}\ \eta_{\alpha\beta} dx^\alpha dx^\beta
		+ a_0^{1/2} 6^{-1/3} \gs N \ls^2\\
		\times \left\{ \frac{1}{2} d\tau^2 + \frac{1}{2} (g^5)^2
		+ (g^3)^2 + (g^4)^2 + \frac{1}{4}\tau^2 [ (g^1)^2 + (g^2)^2 ]\right\}\,,
\end{multline}
which degenerates to $\mathbb{R}^{1,3} \times S^3$ for $\tau=0$.

\subsection*{The dual gauge theory}

The supergravity solution~\eqref{KT} describing D-branes on the conifold is in general dual to a complicated gauge theory, with two gauge groups, matter in bifundamentals and a quartic superpotential. We could proceed for example by analyzing the running of the coupling constants~\cite{Herzog:2001xk}, but we prefer taking a different route since we would rather like to study pure \Ne{1} Super Yang--Mills theory.

In fact, pure $SU(N)$ Super Yang--Mills is the theory lying at the end of the duality cascade. On the other hand, the deformation of the conifold is dual to chiral supersymmetry breaking, which is a phenomenon taking place only in this simpler theory. Therefore, also in view of what we learned in chapter~\ref{c:engineering}, we expect that both solutions with $M=0$, namely with only $N$ fractional D3-branes, and in particular the non-singular solution~\eqref{KS} by Klebanov and Strassler which does not depend on $M$ at all, are able to give us information on pure \Ne{1} $SU(N)$ Super Yang--Mills theory. Our starting relations are the usual ones given in~\eqref{gym4}-\eqref{tym4}:
\begin{equation}\label{KSrelations}
	\frac{1}{\gym} = \frac{1}{4\pi\gs} \frac{1}{(2\pi\ls)^2}\int_{S^2} B_2\,,\qquad
	\tym = \frac{1}{2\pi\gs\ls^2} \int_{S_2} C_2\,,
\end{equation}
where the two-cycle $S^2$ (see insert~\ref{i:T11} on page~\pageref{i:T11}) is given by:
\begin{equation}
	\psi = 0 \,,\qquad \theta_1 = \theta_2\,,\qquad \phi_1 = - \phi_2\,.
\end{equation}

Recalling that in section~\ref{s:MN} we were able to derive the correct energy/radius relation by studying chiral symmetry beraking, we start by considering the chiral anomaly~\cite{Klebanov:2002gr}, which can be read off the conifold in a very similar manner.

In fact, the asymptotic metric~\eqref{KT} has a $U(1)_R$ symmetry associated with shifts in $\psi$. The theta-angle in this case is given by:
\begin{equation}\label{thetaKT}
	\tym = \frac{1}{2\pi\gs\ls^2} \int_{S_2} C_2 = N\psi
\end{equation}
(where we have written a local expression for $C_2$), and it is not invariant for arbitrary shifts of $\psi$. Exactly as in the case of the Maldacena--N\'u\~nez solution, the invariance of~\eqref{thetaKT} under $\psi \to \psi+\frac{2\pi k}{N}$ corresponds to the anomalous breaking of the $U(1)_R$ down to $\mathbb{Z}_{2N}$. This breaking is an ultraviolet phenomenon and in fact we are able to read it from the asymptotic Klebanov--Tseytlin solution~\eqref{KT}.

The full solution~\eqref{KS} by Klebanov and Strassler does not have this $\mathbb{Z}_{2N}$ invariance anymore. All the freedom we have, as in the previous section, is the choice $\psi=0,2\pi$. This is the remaining $\mathbb{Z}_2$ symmetry of the pure \Ne{1} theory which leaves the gaugino condensate invariant.

Let us analyze this fact a little bit more, since we would like to use it in order to extract a gauge /gravity relation to be used in the present case. Recalling from~\eqref{KSfunctions} that the function $F(\tau)$ which enters the expression of $F_3$ has the asymptotic value $F(\tau)\sim \tfrac{1}{2}$ for large $\tau$, and in fact, at this value of $F(\tau)$, $F_3$ reduces to the form given in the asymptotic solution~\eqref{KS}. We can therefore say that the additional anomalous breaking $\mathbb{Z}_{2N}\to \mathbb{Z}_2$ can be traced to the radial dependence of the difference of $C_2$ with respect to its asymptotic value~\cite{Imeroni:2002me}:
\begin{equation}
	\delta C_2 \sim \frac{1}{2} - F(\tau) = \frac{\tau}{2\sinh\tau}\,.
\end{equation}
This expression will play the same role as the function $a(\rho)$ in section~\ref{s:MN}. In fact, following the same reasoning which led us to~\eqref{MNer}, we will impose the following energy/radius relation:
\begin{equation}\label{KSer}
	\frac{\Lambda^3}{\mu^3} = \frac{\tau}{2\sinh\tau}\,.
\end{equation}

Let us then proceed by computing the running coupling constant. By substituting the solution in~\eqref{KSrelations}, we get the following expression:
\begin{equation}
	\frac{1}{\gym} = \frac{N}{4\pi^2}\ k(\tau) \sim \frac{N\tau}{8\pi^2}\,,
\end{equation}
where the last expression is the leading term for large $\tau$. From this expression, together with~\eqref{KSer}, we can compute the $\beta$-function of the gauge theory in exactly the same way as in section~\ref{s:MN}, and the final expression reads:
\begin{equation}\label{KSbeta}
	\beta(g_{\text{YM}}) = -\frac{3Ng^3_{\text{YM}}}{16\pi^2}
		\left(1- \frac{N\gym}{8\pi^2}
		+\frac{2e^{-16\pi^2/Ng^2}}{1-e^{-16\pi^2/Ng^2}}\right)^{-1}\,.
\end{equation}
The above result is precisely the same as the one we obtained in~\eqref{MNbeta} from the Maldacena--N\'u\~nez solution, namely it is the all-loop NSVZ $\beta$-function plus corrections which have the form of fractional instanton contributions of topological charge $\tfrac{1}{N}$. In our opinion (compare the discussion below~\eqref{MNbeta}), this fact strengthens the validity of this result for the pure \Ne{1} SU(N) gauge theory.

\subsection*{The Veneziano--Yankielowicz superpotential}

The gauge theory information we can extract from Klebanov and Strassler's warped deformed conifold does not end here. In fact, we will now analyze a completely new piece of information that we can obtain - the non-perturbative effective superpotential which is generated in the infrared of the gauge theory, as first conjectured by Veneziano and Yankielowicz on the basis of anomaly considerations~\cite{Veneziano:1982ah}. The computation of the effective superpotential that we will explicitly rederive here was sketched in~\cite{Cachazo:2001jy,Giddings:2001yu}.

In fact, in section~\ref{s:geom} we already mentioned Vafa's proposal for the identification between the superpotential of \Ne{1} supergravity and the effective superpotential of the gauge theory in the ``geometric transitions'' framework. Since we know that the deformed conifold is the prototype example of such a transition, let us see whether we can apply Vafa's methods here. The main ingredient we will implement is the formula for the effective superpotential in terms of the fluxes of the complex three-form $G_3$ of the classical type IIB solution and the periods of the holomorphic (3,0)-form $\Omega$ of the Calabi--Yau threefold we are considering. We already introduced the general formula~\eqref{VafaW}, which we repeat here:
\begin{equation}\label{Vafa}
	W \propto \sum_{i}\
		\left[\ \int_{A_i} G_3 \int_{B_i} \Omega
		- \int_{A_i} \Omega \int_{B_i} G_3\ \right]\,,
\end{equation}
where $A_i$ and $B_i$ are the standard orthogonal three-cycles of the Calabi--Yau, respectively compact and non-compact. 

We will here concentrate on the case of the pure \Ne{1} $SU(N)$ Super Yang--Mills theory, which one obtains when considering $N$ fractional D3-branes (namely, $N$ wrapped D5-branes) and no regular brane on the conifold (which is then deformed, as a dual phenomenon to chiral symmetry breaking in the gauge theory, as we have seen).

The first step is the identification of the three-cycles $A$ and $B$ (in the case at hand there is just a single cycle of each kind), which is easy once we recall the cycles of $T^{1,1}$ described in insert~\ref{i:T11} on page~\pageref{i:T11}. We have:
\begin{equation}\label{concycles}
\begin{aligned}
	A\text{ (compact)}\,&: \qquad & r&\text{ constant,} & \theta_1 &= 0\,, & \phi_1 &= 0\,,\\
	B\text{ (non-compact)}\,&: \qquad & \psi&=0\,, & \theta_1 &= \theta_2\,, & \phi_1 &= - \phi_2\,.
\end{aligned}
\end{equation}

Let us compute the fluxes of $G_3 = F_3 -i H_3$ on the cycles~\eqref{concycles}. It is instructive to start with the singular Klebanov--Tseytlin solution~\eqref{KT}, for which we already computed:
\begin{equation}\label{GfluxA}
	\frac{1}{(2\pi\ls)^2\gs} \int_A G_3 = N\,.
\end{equation}
Passing to the $B$-cycle, its non-compactness forces us to evaluate the needed integral over the radial variable $r$ up to a cut-off $r_c$. Moreover, since the solution is singular we cannot reach the value $r=0$, so we introduce an additional cut-off $r_0$ on short distances. With these assumptions (and taking for simplicity $\tym=0$), we find:
\begin{equation}
	\int_B F_3 = 0\,,\qquad
	\int_B H_3 = 6\pi \gs N\ls^2 \int_{r_0}^{r_c} \frac{dr}{r}
		= 6\pi \gs N\ls^2 \ln \frac{r_c}{r_0}\,,
\end{equation}
which implies:
\begin{equation}\label{GfluxBKT}
	\frac{1}{(2\pi\ls)^2\gs} \int_B G_3 = \frac{3N}{2\pi i}\ \ln \frac{r_c}{r_0}\,.
\end{equation}
From the above expression we easily recognize the correct one-loop running coupling of \Ne{1} Super Yang--Mills, if we implement the usual ``stretched string'' energy/radius relation we are used to from the previous chapter.

Consider now the smooth Klebanov--Strassler solution~\eqref{KS}. The flux of $G_3$ along $A$ is unchanged, and is still given by~\eqref{GfluxA}. As for the flux through $B$, we compute:
\begin{equation}
	\int_B F_3 = 0\,,\qquad
	\int_B H_3 = 4\pi \gs N\ls^2 \int_0^{\tau_c} d\tau k'
		= 4\pi \gs N\ls^2  k(\tau_c)\,,
\end{equation}
where we notice that now we can reach $\tau=0$ since the solution is non-singular, but we still need a cut-off $\tau_c$ due to the non-compactness of the cycle. Therefore we obtain:
\begin{equation}\label{GfluxB}
	\frac{1}{(2\pi\ls)^2\gs} \int_B G_3 = \frac{2 N k(\tau_c)}{2\pi i}\,.
\end{equation}
Since the two solutions~\eqref{KT} and~\eqref{KS} are supposed to describe the same geometry for large values of the radial coordinate, in that region we must identify the fluxes. Expanding~\eqref{GfluxB} for large $\tau_c$, since $k(\tau_c)\sim\tfrac{\tau_c}{2}$ we get:
\begin{equation}\label{GfluxBlargetau}
	\frac{1}{(2\pi\ls)^2\gs} \int_B G_3 \sim \frac{N \tau_c}{2\pi i}\,.
\end{equation}
Imposing the equality of~\eqref{GfluxBlargetau} and~\eqref{GfluxBKT}, we get the following relation between the two radial coordinates, valid for large $\tau$:
\begin{equation}
	\tau_c = 3 \ln \frac{r_c}{r_0}\,,
\end{equation}
which agrees with the relation $r^2=\frac{3}{2^{5/3}}\varepsilon^{4/3}e^{2\tau/3}$ given in~\eqref{rvstau}, once we identify $r_0=\tfrac{3^{1/2}\varepsilon^{2/3}}{2^{5/6}}$.

The next ingredient we need is given by the periods of the holomorphic $(3,0)$-form $\Omega$ along the two-cycles. Recall again that the deformed conifold can be described as the $F(x,y,z,t)=0$ hypersurface~\eqref{defcon2} in $\mathbb{C}^4$:
\begin{equation}\label{defeq}
	F = x^2 + y^2 + z^2 + t^2 - \varepsilon^2\,.
\end{equation}
The holomorphic $(3,0)$-form $\Omega$ is defined as in~\eqref{omega30}:
\begin{equation}
	\Omega = \frac{1}{2\pi i} \oint_{F=0} \frac{dx\wedge dy\wedge dz\wedge dt}{F}
		= \frac{dx\wedge dy\wedge dz}{2\sqrt{\varepsilon^2-x^2-y^2-z^2}}\,.
\end{equation}
In order to identify the cycles and compute the periods, let us first take a look to insert~\ref{i:K3} on page~\pageref{i:K3}, where a non-trivial two-cycle of a Calabi--Yau twofold is constructed. In the present case, the three-cycle $A$ is given by a fibration over the two-cycle of insert~\ref{i:K3}, where we take $x \in [-\varepsilon,\varepsilon]$, $y \in [-\sqrt{\varepsilon^2-x^2},\sqrt{\varepsilon^2-x^2}]$. We then compute:
\begin{equation}
	\int_A \Omega = \int_{-\varepsilon}^{\varepsilon} dx
		\int_{-\sqrt{\varepsilon^2-x^2}}^{\sqrt{\varepsilon^2-x^2}} dy
		\int_{\gamma_y} \frac{dz}{2\sqrt{(\varepsilon^2-x^2)-y^2-z^2}}
		= \int_{-\varepsilon}^{\varepsilon} dx\
		2\pi \sqrt{\varepsilon^2-x^2} = \pi^2\varepsilon^2\,,
\end{equation}
where we have used the result given in insert~\ref{i:K3} on page~\pageref{i:K3} for the integral of $\Omega^{(2,0)}$ on the two-cycle. On the noncompact three-cycle $B$, we again need a cutoff on distances. Recall that the complex coordinates in~\eqref{defeq} have dimension $L^{3/2}$, and the cutoff must have the same dimension. In order to use a cutoff with dimension of a length, we then use $r_c^{3/2}$, where $r_c$ is the one used in~\eqref{GfluxBKT}:
\begin{equation}\label{fullperiod}
\begin{split}
	\int_B \Omega & =  \int_{\varepsilon}^{r_c^{3/2}} dx\
		2\pi \sqrt{\varepsilon^2-x^2} =
		2\pi\varepsilon^2
		\int_{\pi/2}^{\arcsin\frac{r_c^{3/2}}{\varepsilon}} d\alpha \cos^2\alpha\\
		& = \pi r_c^{3/2} \sqrt{\varepsilon^2-r_c^3}
		+ \pi \varepsilon^2 \arcsin\frac{r_c^{3/2}}{\varepsilon} - \frac{1}{2}\pi^2\varepsilon^2\,.
\end{split}
\end{equation}
Expanding this result for large $r_c\,$ we get:
\begin{equation}
	\int_B \Omega \sim 2\pi i \left[
		\frac{r_c^3}{2}-\frac{\varepsilon^2}{4}
		+\frac{\varepsilon^2}{4}\ln\frac{\varepsilon^2}{4}
		- \frac{\varepsilon^2}{2}\ln r_c^{3/2}\right]\,.
\end{equation}
%:Insert: 2-cycle of K3
\begin{Insert}{The simplest two-cycle of K3}\label{i:K3}
As in section~\ref{s:geom}, the \cz{} orbifold can be defined as the hypersurface $F(u,v,w)=0$ in $\mathbb{C}^3$, where:
\begin{equation*}
F = u^2 + v^2 + w^2\,.
\end{equation*}
This space is clearly singular at the origin of $\mathbb{C}^3$. To remove the singularity, we can deform the complex structure by replacing the above expression with:
\begin{equation*}
F = u^2 + v^2 + w^2 - \varepsilon^2\,,
\end{equation*}
where we take $\varepsilon$ to be a real parameter. Let us define the unique holomorphic (2,0)-form $\Omega^{(2,0)}$ as:
\begin{equation*}
\Omega^{(2,0)} = \frac{1}{2\pi i}\oint_{F=0}
		\frac{du \wedge dv \wedge dw}{F}
		= \frac{du \wedge dv}
		{2\sqrt{\varepsilon^2 - u^2-v^2}}\,,
\end{equation*}
We can identify the non-contractible two-cycle in these coordinates by making $u$ run from $-\varepsilon$ to $\varepsilon$ and $v$ run around the branch cut of the denominator of $\Omega^{(2,0)}$ corresponding to $u$ in the above range, as in the following figure:
\begin{center}
\includegraphics[scale=0.5]{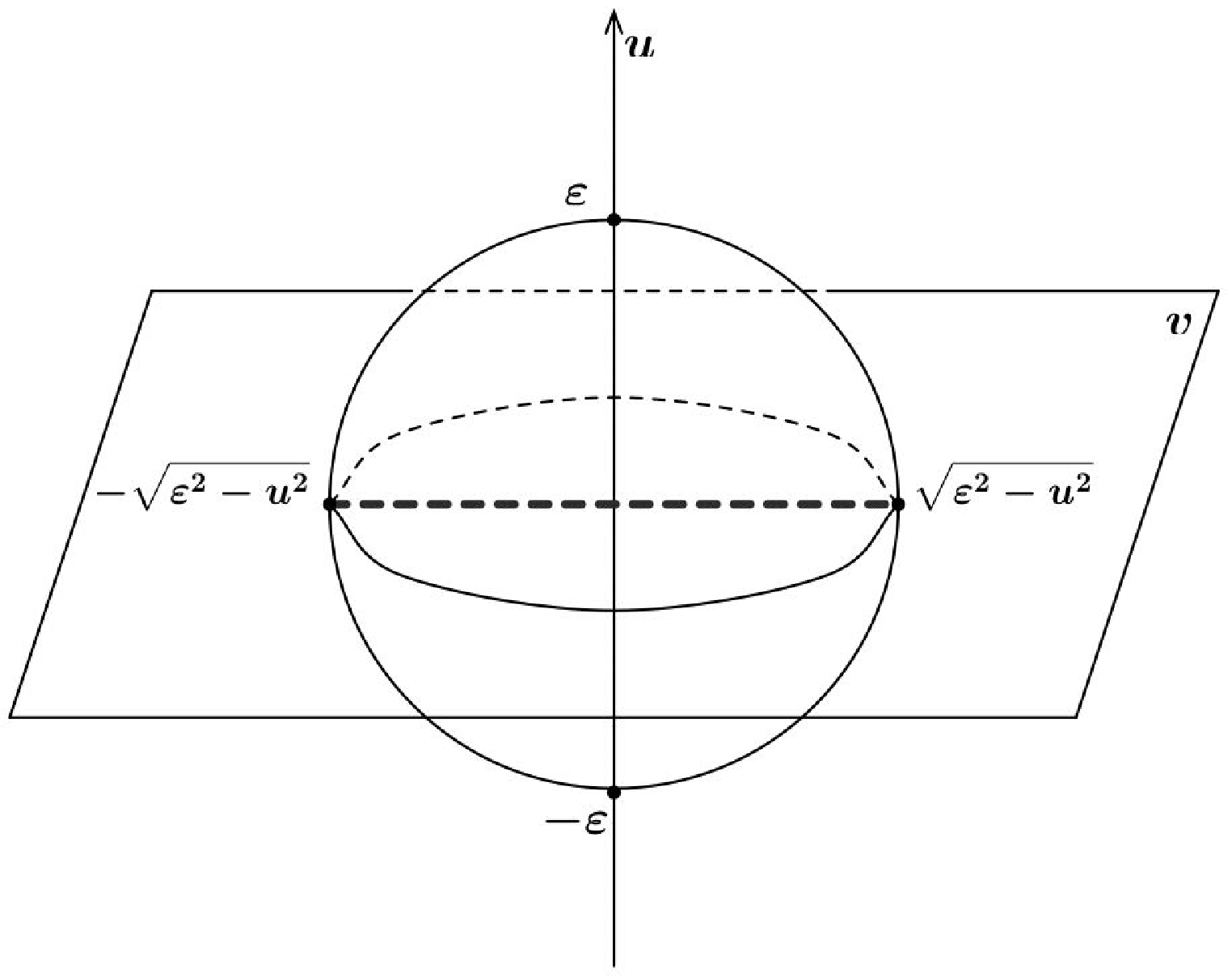}
\end{center}
The period of the holomorphic (2,0)-form along the cycle is then given by:
\begin{equation*}
	\int_{S^2}\ \Omega^{(2,0)}
		= \int_{-\varepsilon}^{\varepsilon} du
		\int_{\gamma_u} \frac{dv}{2\sqrt{\varepsilon^2-u^2-v^2}}
		= \int_{-\varepsilon}^{\varepsilon} du
		\int_{\gamma_\infty} \frac{dw}{2iw}
		= 2 \pi \varepsilon\,.
\end{equation*}
\end{Insert}

Let us then summarize all the ingredients we have found for the computation of the superpotential. The results of the periods along $B$ are given for large radius, and we will use the coordinate $r$, instead of $\tau\,$, since they are equivalent at large distances and it is $r$ which has a more direct gauge theory interpretation at one-loop.
\begin{equation}
\begin{aligned}
	\frac{1}{(2\pi\ls)^2\gs}\int_A G_3 &= N\,,&
	\int_A \Omega &= \pi^2 \varepsilon^2\,,\\
	\frac{1}{(2\pi\ls)^2\gs}\int_B G_3 &= \frac{1}{2\pi i} 3N \ln \frac{r_c}{r_0}\,,&
	\int_B \Omega &= 2\pi i \left[
		\frac{r_c^3}{2}-\frac{\varepsilon^2}{4}
		+\frac{\varepsilon^2}{4}\ln\frac{\varepsilon^2}{4r_c^3}\right]\,.
\end{aligned}
\end{equation}
The superpotential is given by:
\begin{equation}
\begin{split}\label{conW1}
	W_{\text{eff}} &= -\frac{1}{2\pi i}\ \frac{1}{(2\pi\ls)^2\gs}\ \frac{1}{(2\pi\ls^2)^3}\ 
		\left[\ \int_{A} G_3 \int_{B} \Omega
		- \int_{A} \Omega \int_{B} G_3\ \right]\\
	&= -\frac{N}{(2\pi\ls^2)^3}\ 
		\left[\ 
		3\ \frac{\varepsilon^2}{4}\ln \frac{r_c}{r_0}
		+ \frac{r_c^3}{2} - \frac{\varepsilon^2}{4}
		+ \frac{\varepsilon^2}{4}
		\ln \frac{\varepsilon^2}{4r_c^3}
		\ \right]\,.
\end{split}
\end{equation}
How to interpret the deformation parameter of the conifold, appearing always in the combination $\tfrac{\varepsilon^2}{4}$? The latter is a quantity of dimensions $L^3$ which we have seen to be related to the infrared chiral symmetry breaking of the theory, and therefore to the gaugino condensate $S$ of the gauge theory. It is therefore natural to expect that the ``stretched string'' gauge/gravity identifications will now read:
\begin{equation}
	r_c = 2\pi\ls^2\ \mu\,,\qquad
	r_0 = 2\pi\ls^2\ \Lambda\,,\qquad
	\frac{\varepsilon^2}{4} = (2\pi\ls^2)^3 S\,,
\end{equation}
so that, neglecting the constant term, the superpotential~\eqref{conW1} will look as follows:
\begin{equation}\label{VYsuperpot}
	W_{\text{eff}} = NS \left( 1- \ln \frac{S}{\Lambda^3} \right)\,,
\end{equation}
which is exactly the Veneziano--Yankielowicz effective superpotential for pure \Ne{1} Super Yang--Mills theory.%
\footnote{As an aside, notice that if we consider subleading corrections to the $B$-period of $\Omega$ coming from the full expression~\eqref{fullperiod}, we find fractional instanton contributions similar to the ones found in~\eqref{KSbeta} (we thank Alberto Lerda, Paolo Merlatti and Cumrun Vafa for discussions on this point). A direct comparison is however difficult because the renormalization scheme is different in the two cases.

Besides that, notice that one could also be interested in reproducing the superpotential~\eqref{VYsuperpot} from the Maldacena--N\'u\~nez solution. This is however not a straightforward problem, since~\eqref{Vafa} does not hold for backgrounds with varying dilaton (for related work, from a different perspective, see~\cite{Muck:2003zf}).}
By minimizing it, we find the value of the gaugino condensate $S$ in the $N$ supersymmetric vacua of the theory:
\begin{equation}
	S = \Lambda^3\ e^{2\pi ik/N}\,,
\end{equation}
where $k=1,\ldots,N$.

\section{SQCD from fractional D3-branes}\label{s:sqcd}

The study of our last example of gauge/gravity correspondence for a non-conformal gauge theory brings us back to an orbifold of type IIB string theory. Unlike the examples considered in the previous chapter, however, the orbifold under consideration will break three-quarters of the supersymmetry of flat space, so that a D-brane living in such a background will have a world-volume theory with 4 preserved supercharges.

\subsection*{Fractional branes on \czz}

We are going to consider a system of fractional D3-branes on $\mathbb{R}^{1,3}\times\czz$~\cite{Douglas:1997de}.%
\footnote{Notice that in what follows we will always consider the \czz{} orbifold \emph{without discrete torsion}. Discrete torsion could be introduced since there are non-trivial two-cocycles~\cite{Vafa:1986wx,Vafa:1995rv,Douglas:1998xa,Douglas:1999hq}, but we will not consider this case here since the theory living on D3-branes on the orbifold with discrete torsion has a matter content that makes it conformal~\cite{Douglas:1998xa}.}
As usual, we denote with $x^\alpha$, $\alpha=0,\ldots,3$ the coordinates transverse to the orbifold, while we introduce three complex coordinates in the orbifolded directions $x^r$, $r=4,\ldots,9$ as follows:
\begin{equation}
	z_1 = x^4 + i x^5\,,\qquad
	z_2 = x^6 + i x^7\,,\qquad
	z_3 = x^8 + i x^9\,.
\end{equation}

The two $\mathbb{Z}_2$ factors of the orbifold group have generators denoted with $g_1$ and $g_2$, whose action is:
\begin{equation}\label{zzact}
\begin{tabular}{c|ccc}
	& $z_1$ & $z_2$ & $z_3$ \\
	\hline
	$g_1$ & $z_1$ & $-z_2$ & $-z_3$ \\
	$g_2$ & $-z_1$ & $z_2$ & $-z_3$ \\
\end{tabular}
\end{equation}
The remaining two elements of the group are of course the identity $e$ and $g_3 = g_1 g_2$.

We want to introduce D3-branes transverse to the \czz{} orbifold, as in the following table:
\begin{center}
\begin{tabular}{|c|c|c|c|c|c|c|c|c|c|c|}
\multicolumn{5}{c}{ }&
\multicolumn{6}{c}{$\overbrace{\phantom{\qquad\qquad\qquad\qquad}}^{\czz}$}\\
\hline
&0&1&2&3&4&5&6&7&8&9\\
\hline
D$3$ &$-$&$-$&$-$&$-$&$\cdot$&$\cdot$&$\cdot$&$\cdot$&$\cdot$&$\cdot$\\
\hline
\end{tabular}
\end{center}

The necessary ingredients of the analysis are the same ones presented in section~\ref{s:Z2}. A fractional D-brane will be stuck at the orbifold fixed point $z_1=z_2=z_3=0$, and the Chan--Paton matrices $\lambda$ of the strings attached to it will transform according to~\eqref{CPaction}:
\begin{equation}
	\lambda\ \to\ \gamma(g)\ \lambda\ \gamma(g)^{-1}\,,
\end{equation}
where $\gamma(g)$ is one of the four irreducible one-dimensional representations of $\mathbb{Z}_2 \times \mathbb{Z}_2$:
\begin{equation}\label{zzirrep}
\begin{aligned}
	&\text{A}:\qquad & \gamma(e) &= +1\quad & \gamma(g_1) &= +1\quad
		& \gamma(g_2) &= +1\quad & \gamma(g_3) &= +1 \,,\\
	&\text{B}:\qquad & \gamma(e) &= +1\quad & \gamma(g_1) &= +1\quad
		& \gamma(g_2) &= -1\quad & \gamma(g_3) &= -1 \,,\\
	&\text{C}:\qquad & \gamma(e) &= +1\quad & \gamma(g_1) &= -1\quad
		& \gamma(g_2) &= +1\quad & \gamma(g_3) &= -1 \,,\\
	&\text{D}:\qquad & \gamma(e) &= +1\quad & \gamma(g_1) &= -1\quad
		& \gamma(g_2) &= -1\quad & \gamma(g_3) &= +1 \,,\\
\end{aligned}
\end{equation}
which means that there are four different types of fractional branes, that we denote with A, B, C and D. Let us then analyze the massless level of the spectrum of the open strings attached to a fractional D3-brane. One can easily see that the orbifold projection, which acts trivially on the Chan--Paton factors and according to~\eqref{zzact} on the oscillators, preserves the following massless states:
\begin{center}
\begin{tabular}{|c|l|}
\hline
\multicolumn{2}{|c|}{NS states} \\
\hline
$\psi_{-1/2}^\alpha \ket{0,k}$ &
$\to$ $1$ vector\\
\hline
\end{tabular}
\qquad
\begin{tabular}{|c|l|}
\hline
\multicolumn{2}{|c|}{R states} \\
\hline
$\ket{s_0,s_1,s_2,s_3=s_2,s_4=s_2}$ &
$\to$ $4$ fermions\\
\hline
\end{tabular}
\end{center}
These states comprise a single \Ne{1} vector multiplet. Therefore, the gauge theory living on a fractional D3-brane is \emph{pure} \Ne{1} four-dimensional $U(1)$ Super Yang--Mills theory. On a stack of $N$ fractional branes of the same type we will have pure $U(N)$ Super Yang--Mills.

What happens when we put fractional D3-branes of different types together? Each brane will carry an \Ne{1} vector multiplet, so we expect a $U(N_{\text{A}})\times U(N_{\text{B}})\times U(N_{\text{C}})\times U(N_{\text{D}})$ theory. In addition, from an analysis perfectly analogous to the one performed in section~\ref{s:Z2}, one finds that there are pairs of chiral multiplets transforming in the bifundamental representation of each adjacent gauge groups, so that the full quiver diagram of the theory living on $N_k$ branes of type $k$ is given by:
\begin{center}
\includegraphics[scale=.5]{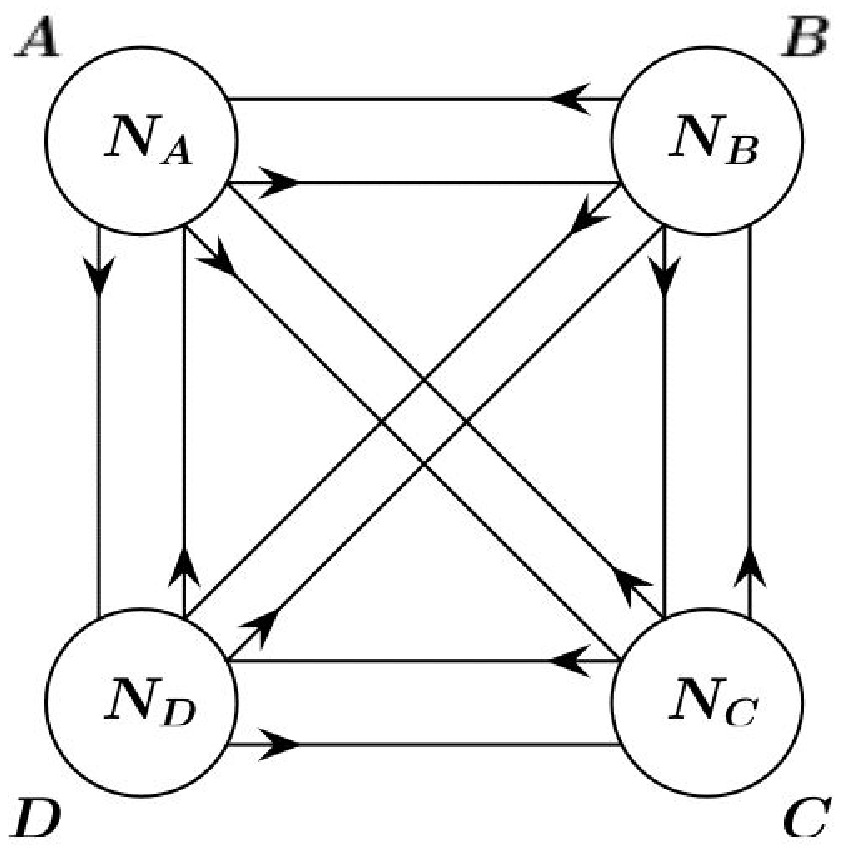}
\end{center}

Our goal is to use this orbifold configuration in order to engineer $U(N)$ \Ne{1} SQCD, which is a theory with $N_f$ flavors of ``quarks'' and ``antiquarks'' transforming in the fundamental representation of the gauge group. We could proceed as we did in section~\ref{s:Ne2}, by introducing D7-branes wrapping the orbifold directions, and this route was indeed followed in~\cite{Marotta:2002gc}.%
\footnote{See also the review~\cite{DiVecchia:2003ne}, and the recent work~\cite{Ouyang:2003df,Nunez:2003cf} for the related interesting issue of adding fundamental matter to the gauge theories dual to the Maldacena--N\'u\~nez and Klebanov--Strassler backgrounds that we described in the previous sections.}
However, there is an alternative way~\cite{Berenstein:2003fx}, which is probably even simpler and allows one to see the main features of the theory emerge in a very natural way.

This alternative strategy amounts simply to consider a configuration of $N$ fractional D3-branes of type A and $N_f$ fractional D3-branes of type B. The full $U(N)\times U(N_f)$ quiver theory will therefore be given by the following diagram:
\begin{center}
\includegraphics[scale=.5]{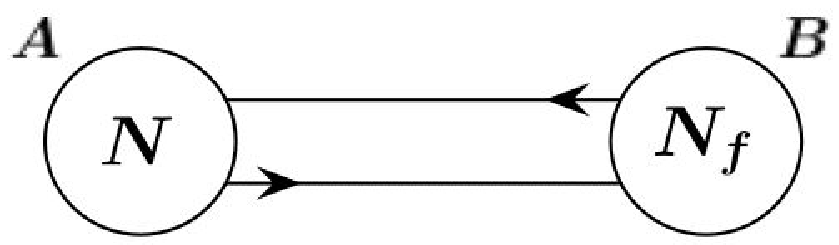}
\end{center}
Now, the point is that if we concentrate on the theory living on the branes of type A by a suitable selection of the open strings with the appropriate Chan--Paton factors, what we are left with is precisely $U(N)$ SQCD with $N_f$ chiral multiplets transforming in the fundamental and $N_f$ in the antifundamental representation of the gauge group. Of course, in the full theory there is the relevant difference that the flavor symmetry is gauged, but we will see that this system will be able to give us a big deal of information on SQCD via the gauge/gravity correspondence.

\subsection*{Closed strings and supergravity solution}

The closed type II string spectrum on \czz{} is similar to the one on \cz. The untwisted massless fields can be obtained via the direct reduction of the type II fields along the forms surviving the orbifold projection. They can easily be shown to give the following Hodge diamond:
\begin{equation}\label{zzh1}
	\ghodge\ \ =\hodgeCY{3}{3}\,.
\end{equation}
As for the three twisted sectors, the full quiver diagram presented in the previous subsection, with all four types of fractional branes, directly suggests the presence of three non-trivial two-cycles which are shrinking in the orbifold limit (one for each quiver node, except the trivial one). Corresponding to each cycle $\mathcal{C}_i$, $i=1,2,3$, there are two NS-NS twisted field, that we denote with $b^i$ and $\xi^i$. The former can be interpreted as the component of $B$ on the cycles, while the second are parameters of the geometrical blow-up, exactly as in the case of \cz. The contribution of each twisted sector to the Hodge diamond, and the total diamond, can therefore be seen to be:
\begin{equation}\label{zzh2}
\begin{tabular}{ccc}
One twisted sector\, & \qquad\qquad & Full Hodge diamond\, \\
\hodgeCY{1}{0}\,, & & \hodgeCY{6}{3}\,,\\
\end{tabular}
\end{equation}
which confirms the fact that we are dealing with a Calabi--Yau orbifold which breaks three-quarters of supersymmetry (see insert~\ref{i:hodge} on page~\pageref{i:hodge}). We obtained the above information in a somewhat indirect way, but it is easy to explicitly confirm it all with a careful analysis of the closed string spectrum.

Due to the evident similarities with the case of the \cz{} orbifold, we can expect that also the construction of the supergravity solution will proceed in a similar way. This is indeed the case, and the solution was constructed in~\cite{Bertolini:2001gg}. One introduces the following decomposition:
\begin{equation}
	B = b_i\ \omega_2^{(i)}\,, \qquad
	C_2 = (A_0)_i\ \omega_2^{(i)}\,,
\end{equation}
where the anti-self dual two-forms $\omega_2^{(i)}$, dual to the cycles $\mathcal{C}_i$, are defined and normalized as:
\begin{equation}\label{zzomeganorm}
	\int_{\mathcal{C}_i} \omega_2^{(j)} = \delta_i^j\,,\qquad
	\int \omega_2^{(i)} \wedge \omega_2^{(i)} = -\frac{1}{4}\,,\qquad
	\hd{4} \omega_2^{(i)} = - \omega_2^{(i)}\,.
\end{equation}

Proceeding as in sections~\ref{s:fD2} and~\ref{s:Ne2}, one finally finds the following solution of the equations of motion of type IIB supergravity, describing a system of $N_k$ fractional D3-branes of type $k$ placed at the origin $z_1=z_2=z_3=0$ of the orbifold space:
\begin{equation}\label{zzsol}
\begin{aligned}
	ds^2 &= H_3^{-1/2} \eta_{\alpha\beta} dx^\alpha dx^\beta
		+H_3^{1/2} \delta_{rs} dx^r dx^s\,,\\
	\tilde{F}_5 &= d H_3^{-1} dx^0 \wedge \ldots \wedge dx^3
		+ \hd{} ( d H_3^{-1} dx^0 \wedge \ldots \wedge dx^3)\,,\\
	G_3 &= d\gamma_i \wedge \omega_2^{(i)}\,,
\end{aligned}
\end{equation}
where we have defined the three-forms:
\begin{equation}
\begin{aligned}
	G_3 &= dC_2 + ( C_0 + i e^{-\phi} ) dB_2\,,\\
	\gamma_i &= (A_0)_i + i b_i\,.
\end{aligned}
\end{equation}
The explicit expression of $\gamma_i$ in the solution is given by:
\begin{equation}\label{zzgamma}
	\gamma_i = i K \left[ \frac{\pi}{2\gs} + f_i(N_k) \ln \frac{z_i}{\epsilon} \right]\,,
\end{equation}
where $\epsilon$ is a regulator, $K = 4\pi \gs \ls^2$ and the $f_i$ are functions of the numbers of the different types of fractional branes:
\begin{equation}\label{fNk}
\begin{aligned}
	f_1 (N_k) = N_{\text{A}} + N_{\text{B}} - N_{\text{C}} - N_{\text{D}}\,,\\
	f_2 (N_k) = N_{\text{A}} - N_{\text{B}} + N_{\text{C}} - N_{\text{D}}\,,\\
	f_3 (N_k) = N_{\text{A}} - N_{\text{B}} - N_{\text{C}} + N_{\text{D}}\,.
\end{aligned}
\end{equation}
Finally, $H_3$ is an appropriate function of $z_i$ whose explicit form, which we will not need in the following, is given in~\cite{Bertolini:2001gg}.

We will use the above solution in the next subsection to uncover properties of \Ne{1} $U(N)$ SQCD. Notice that this solution has a naked singularity, a common feature of all classical solutions describing fractional branes on orbifolds. One could then proceed by examining the appearance of an enhan\c con mechanism also in this \Ne{1} case, as we did in chapter~\ref{c:8susy}. Unlike the cases of the conifold and Maldacena--N\'u\~nez solutions analyzed in sections~\ref{s:MN} and~\ref{s:conifold}, it does not seem possible to obtain a non-singular solution via a deformation of the geometry.

\subsection*{The dual gauge theory}

Let us concentrate on a configuration made up of $N$ fractional D3-branes of type A and $N_f$ fractional D3-branes of type B in $\mathbb{R}^{1,3}\times\czz$. As we already said, the theory living on the branes of type A is $U(N)$ SQCD with \Ne{1} supersymmetry, with $N_f$ ``quark'' chiral multiplets $Q^i$ and $N_f$ ``antiquark'' chiral multiplets $\tilde{Q}_{\tilde{\jmath}}$.

As we are now used to, we can immediately use the supergravity solution~\eqref{zzsol} to extract the running coupling constant of the gauge theory. Since it will be useful in the following, we will do it by computing the flux of $G_3$ along an appropriate cycle. Let us therefore identify the cycles $A_i$ and $B_i$ on this Calabi--Yau orbifold which will be relevant for the computation of the coupling constant and non-perturbative superpotential.

\begin{figure}
\begin{center}
\includegraphics[scale=.5]{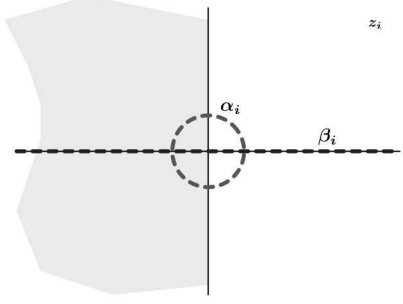}
\caption{{\small Compact 1-cycles $\alpha_i$ and noncompact 1-cycles $\beta_i$ on the $z_i$ planes.}}
\label{f:adscycles}
\end{center}
\end{figure}

As we can see from the Hodge diamond~\eqref{zzh2}, this orbifold does not have any non-trivial (1,2) or (2,1)-forms coming from the twisted sectors. All the ingredients we have at our disposal are then the (1,1)-forms $\omega_2^{(i)}$. We can therefore identify three compact cycles $A_i$ and three non-compact cycles $B_i$ by simply taking the direct product of the two-cycles $\mathcal{C}_i$ with suitable one-cycles on the $z_i$ planes. Specifically, we define:
\begin{equation}\label{ABcycles}
	A_i = \alpha_i \times \mathcal{C}_i\,,\qquad
	B_i = \beta_i \times \mathcal{C}_i\,,\qquad i=1,2,3\,,
\end{equation}
where the compact cycles $\alpha_i$ and noncompact cycles $\beta_i$ on the $z_i$ plane are the ones shown in figure~\ref{f:adscycles}.
%:Figure: A-D-S cycles
Therefore, the fluxes of $G_3$ along the cycles will be given by:
\begin{equation}
	\int_{A_i\text{ or }B_i} G_3 = \int_{A_i\text{ or }B_i} d\gamma_j \wedge \omega_2^{(j)}
		= \int_{\alpha_i\text{ or }\beta_i} d\gamma_i \int_{\mathcal{C}_i} \omega_2^{(i)}
		= \int_{\alpha_i\text{ or }\beta_i} d\gamma_i\,,
\end{equation}
where we have used the definition~\eqref{ABcycles} of the cycles and the normalizations~\eqref{zzomeganorm}.

Using~\eqref{zzgamma} and the explicit expressions~\eqref{fNk} for the functions $f_i(N_k)$, we can compute the running gauge coupling constant of the theory living on the $N$ branes of type A, which is given by:%
\footnote{Notice that there is a difference of a factor of 2 with respect to the \cz{} case considered in the previous chapter. This is due to the fact that in the present case the orbifold projector has an additional $\tfrac{1}{2}$ factor.}
\begin{equation}
	\frac{1}{\gym} = \frac{1}{8\pi\gs i} \frac{1}{(2\pi\ls)^2}
		\sum_{i=1}^{3}\int_{B_i} G_3
		= \frac{1}{8\pi^2} (3N - N_f) \ln \frac{\rho_c}{\rho_0}\,,
\end{equation}
where we have used a cut-off $\rho_c$ on the integration on $\rho_i = \abs{z_i}$ along the $\beta_i$ cycles on the $z_i$ planes. The additional cut-off $\rho_0$ at short distances is due to the fact that since we have a singular supergravity solution we cannot reach the position of the branes, namely $\rho=0$. If we proceed by implementing the usual ``stretched string'' energy/radius relation:
\begin{equation}
	\rho_c = 2\pi\ls^2\ \mu\,,\qquad \rho_0 = 2\pi\ls^2\ \Lambda\,,
\end{equation}
we get the correct one-loop running coupling constant for our $U(N)$ SQCD:
\begin{equation}
	\frac{1}{\gym} = \frac{1}{8\pi^2} (3N - N_f) \ln \frac{\mu}{\Lambda}\,.
\end{equation}

Can we extract additional information on the gauge theory? We want to turn to the vacuum structure, namely to the moduli space of the theory. It is known (see for instance~\cite{Intriligator:1995au}) that $U(N)$ SQCD has a very different behavior depending on the number of flavors. In particular, let us concentrate on the case in which the number of flavors $N_f$ is less or equal to the number of colors $N$. This is the so-called Affleck--Dine--Seiberg theory~\cite{Affleck:1983mk}, for which it is known that a non-perturbative superpotential is generated and produces a runaway behavior. Let us see how we can recover this structure from geometry.

As a starting point, consider just a single fractional D3-brane of type A together with one of type B. As single objects, they are charged under all four sectors of closed string theory (and this is the reason why they cannot move off the orbifold fixed point), one untwisted and one twisted, with charges whose sign is given by the same signs appearing in the representations~\eqref{zzirrep}. We then see that a brane of type A and one of type B have the same charge under the sector twisted by $g_1$, but opposite charge under the sectors twisted by $g_2$ and $g_3$.

Since fractional branes of different types are of course mutually BPS objects, we can then think of constructing an A+B superposition, which will be charged under the sector twisted by $g_1$ (with a charge double with respect to a single fractional brane), but will not carry any charge under the other two twisted sectors. This means that such a superposition could freely move in the $z_1$ plane, which is left fixed by the action of $g_1$. Such a movement will cause the breaking of the $U(1)\times U(1)$ gauge group of the theory living on the superposition down to $U(1)$ via the Higgs mechanism, and the full theory can be recovered by introducing images under the orbifold action in the $z_1$ plane on the covering space, which is needed for the configuration to be invariant under the orbifold action.

Therefore, starting from our $N$ branes of type A and $N_f$ of type B, we can build $N_f$ A+B superpositions and move them away from the origin at arbitrary points in the plane $z_1$. The movement of the A+B superpositions is naturally interpreted as giving arbitrary vacuum expectation values to the ``meson'' matrix $M^i_{\tilde{\jmath}} = Q^i \tilde{Q}_{\tilde{\jmath}}$, thus breaking the gauge group down to the $U(N-N_f)$ theory living on the remaining $N-N_f$ fractional branes of type A still placed at the origin.

\begin{figure}
\begin{center}
\includegraphics[scale=.6]{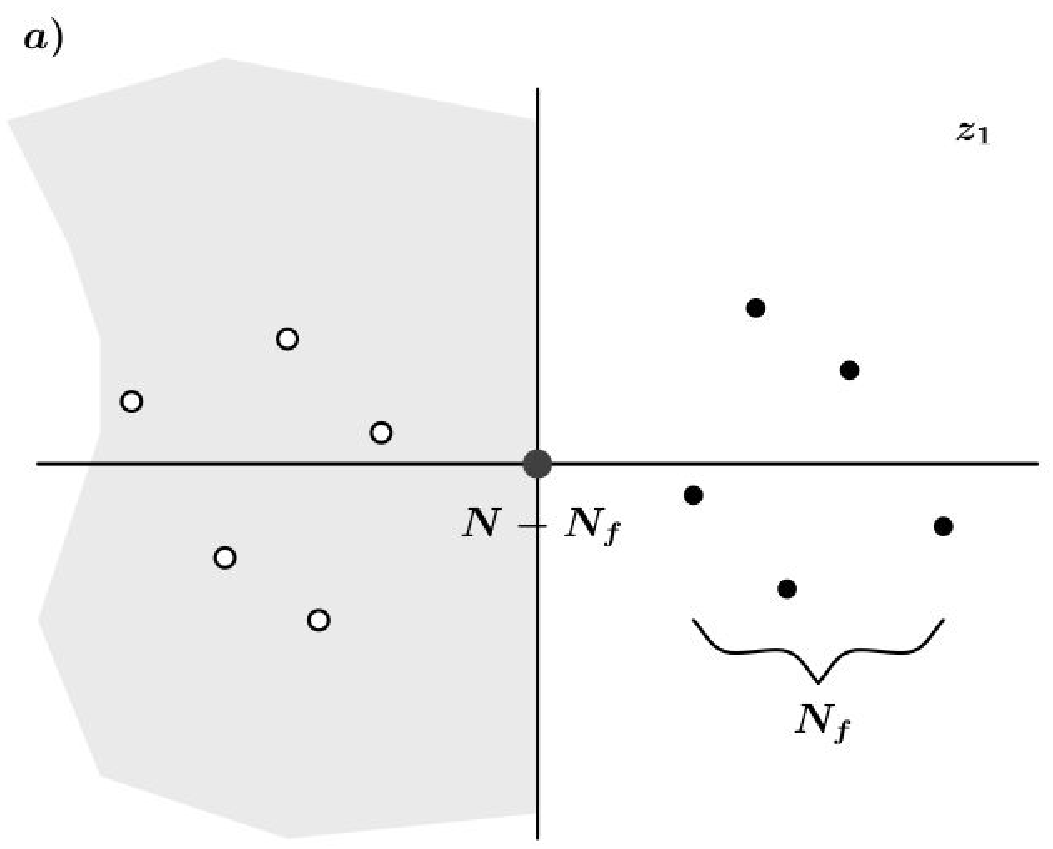}\qquad
\includegraphics[scale=.6]{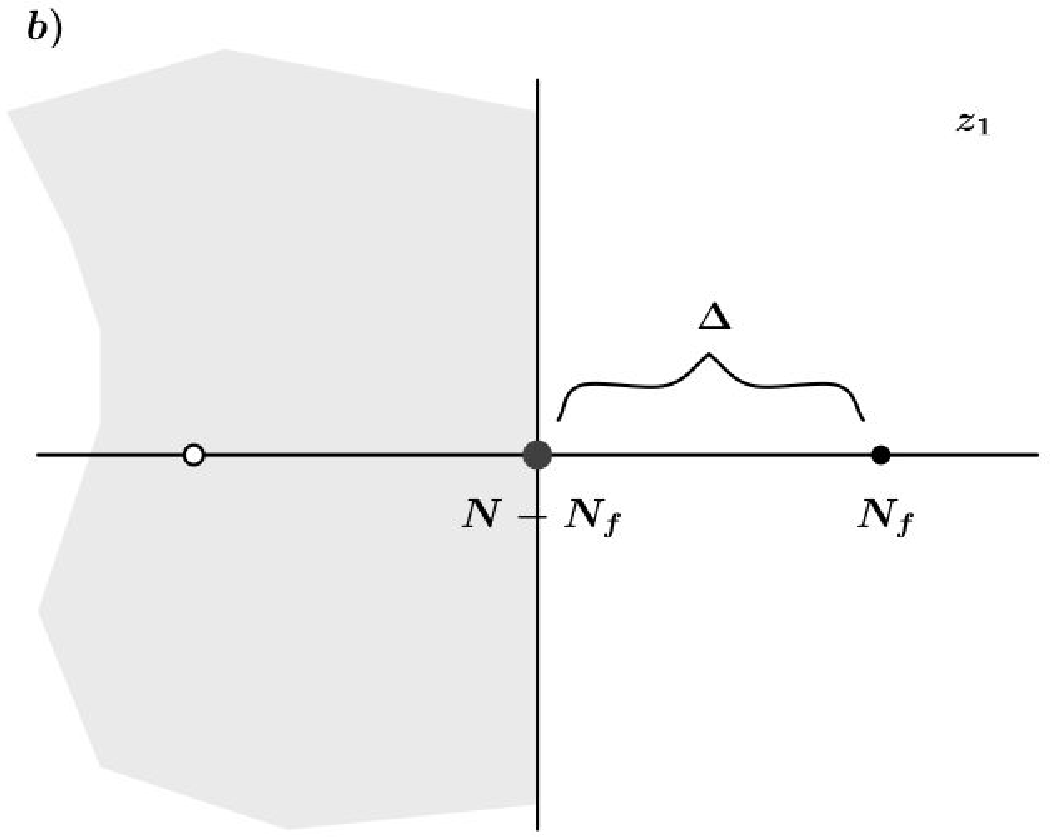}
\caption{{\small Moduli space of the Affleck--Dine--Seiberg theory via fractional branes. \emph{a)} A+B superpositions at arbitrary points of the $z_1$ plane, together with their images on the covering space. \emph{b)} The configuration which makes the ``meson'' matrix $M^i_{\tilde{\jmath}}$ proportional to the identity.}}
\label{f:adsmoduli}
\end{center}
\end{figure}

We therefore see that our D-brane construction has uncovered the correct classical moduli space of the Affleck--Dine--Seiberg theory in a very natural way~\cite{Imeroni:2003cw}, as shown in figure~\ref{f:adsmoduli}a.
%:Figure: A-D-S moduli space
Notice that the construction makes very clear that something drastic happens when passing from the case $N_f \le N$ to the one where $N_f > N$. In the latter case, all we have done is no longer valid and one is forced to look for some alternative description.

How is the supergravity solution~\eqref{zzsol} modified when we form and move the A+B superpositions? For definiteness, let us place all $N_f$ superpositions at the same point $z_1 = \Delta$ on the real axis of the $z_1$ plane, as in figure~\ref{f:adsmoduli}b (the images will then be placed at the identified point $z_1 = -\Delta$). This configuration makes the ``meson'' matrix proportional to the identity, $M^i_{\tilde{\jmath}} = m^2 \delta^i_{\tilde{\jmath}}$, where $m$ is related to $\Delta$ via $\Delta = 2\pi\ls^2\ m$ by the usual gauge/gravity identification.

We will be interested in the new expression of the twisted fields $\gamma_i$, which will be changed as follows:
\begin{equation}
	\gamma_i = i K \left[ \frac{\pi}{2\gs} + (N - N_f) \ln \frac{z_i}{\epsilon}
		+ \delta_{i,1}\ N_f\ \ln \frac{z_1-\Delta}{\epsilon}
		+ \delta_{i,1}\ N_f\ \ln \frac{z_1+\Delta}{\epsilon} \right]\,.
\end{equation}
This implies that the flux of $G_3$ along the $B_i$ cycles is given by:
\begin{equation}\label{gammab}
	\frac{1}{iK} \int_{\beta_i} d\gamma_i
		\sim ( N- N_f ) \ln \frac{\rho_c}{\rho_0} + \delta_{i,1}\ 2N_f\ \ln \frac{\Delta}{\rho_0}\,,
\end{equation}
where we have used the usual integration path from figure~\ref{f:adscycles} and we assumed $\rho_0\ll \rho_c \ll \Delta$.%
\footnote{Again there are corrections to~\eqref{gammab} that, once interpreted in the gauge theory, look like non-perturbative contributions analogous to the ones encountered in sections~\ref{s:MN} and~\ref{s:conifold}.}
The gauge coupling now reads:
\begin{equation}
	\frac{1}{\gym} = \frac{1}{8\pi\gs i} \frac{1}{(2\pi\ls)^2}
		\sum_{i=1}^{3}\int_{B_i} G_3
		= \frac{3}{8\pi^2} (N - N_f) \ln \frac{\rho_c}{\rho_0}
		+ \frac{2N_f}{8\pi^2} \ln \frac{\Delta}{\rho_0}\,,
\end{equation}
which in terms of the gauge theory scales becomes:
\begin{equation}
	\frac{1}{\gym} = \frac{3}{8\pi^2} (N - N_f) \ln \frac{\mu}{\Lambda}
		+ \frac{2N_f}{8\pi^2} \ln \frac{m}{\Lambda}\,,
\end{equation}
with $\mu \ll m$. The above coupling can be expressed in terms of the low energy effective unbroken $U(N-N_f)$ theory as:
\begin{equation}
	\frac{1}{g^2_{\text{YM}}}
		= \frac{3}{8\pi^2}\ ( N - N_f )\ \ln \frac{\mu}{\Lambda_L}\,,
\end{equation}
where the low energy scale $\Lambda_L$ is related to $\Lambda$ and $m$ via:
\begin{equation}
	\Lambda_L^{\phantom{L} 3 ( N - N_f ) }
		= \frac{\Lambda^{ 3N - N_f }}{m^{2N_f}}
		= \frac{\Lambda^{ 3N - N_f }}{\det M}\,.
\end{equation}
We then see that the supergravity computation precisely reproduces, beyond the running coupling constant, the expected matching of scales in the gauge theory~\cite{Intriligator:1995au}. This concludes our analysis of the theory at the classical and perturbative levels. In the next subsection, we will instead turn to non-perturbative phenomena, with the computation of the effective superpotential.

\subsection*{The Affleck--Dine--Seiberg superpotential}

The configuration of fractional D-branes on \czz{} that we have been studying is able to give us information also on the non-perturbatively generated effective superpotential. In section~\ref{s:conifold}, we derived the Veneziano--Yankielowicz superpotential~\cite{Veneziano:1982ah} from the warped deformed conifold solution. In the case at hand,  the supergravity solution we have at our disposal is not a smooth solution as the one in~\cite{Klebanov:2000hb}. Rather, we are in a situation, the orbifold limit, where all cycles are shrinking, in a similar way to what happens in the singular conifold solution of Klebanov and Tseytlin~\cite{Klebanov:2000nc}. However, we have seen in section~\ref{s:conifold} that the Klebanov--Tseytlin solution is sufficient for extracting the needed fluxes of $G_3$, for the simple reason that they are precisely identified with the fluxes of the Klebanov--Strassler solution. The same fact is true for our case. Notice however that this is not the case for the periods of the holomorphic three-form $\Omega$, which depend crucially on the details of the deformation of the geometry. Nevertheless, the geometric considerations which are necessary for getting the correct periods do not depend on the details of the classical solution, and we will see that one is able to obtain them also in the present case~\cite{Imeroni:2003cw}.

We already derived the flux of $G_3$ along the non-compact $B_i$ cycles. The computation of the flux along the $A_i$ cycles is also easy since one has:
\begin{equation}
	\frac{1}{iK} \int_{\alpha_i} d\gamma^i = 2\pi i\ ( N - N_f )\,.
\end{equation}

Let us now consider the periods of the holomorphic $(3,0)$-form $\Omega$. As in the case of the conifold, in order to get sensible results it is necessary to deform the singular orbifold geometry. Let us start by noting that the \czz{} orbifold can be described as the $F(x,y,z,t)=0$ hypersurface in $\mathbb{C}^4$, where:
\begin{equation}
	F = xyz + t^2\,.
\end{equation}
In fact, the invariant variables are given by:
\begin{equation}
	x = z_1^2\,,\qquad
	y = z_2^2\,,\qquad
	z = z_3^2\,,\qquad
	t = i z_1 z_2 z_3\,,
\end{equation}
whose engineering dimensions are then $[x]=[y]=[z]=L^2$, $[t]=L^3$. The simplest possibilty for a deformation of the complex structure, which also resolves the singularity completely, is given by the constant deformation:
\begin{equation}
	F(x,y,z,t) = xyz + t^2 - \xi^2
\end{equation}
(notice that $[\xi]=L^3$). In~\cite{Berenstein:2003fx}, Berenstein has shown, via holomorphy considerations strengthened by a matrix model computation, that this is indeed the correct deformation to consider. He also showed that the deformation parameter $\xi$ is related to the gaugino condensate $S$ of the dual gauge theory, as we will argue below.
We can then compute the holomorphic (3,0)-form:
\begin{equation}
	\Omega = \frac{1}{2\pi i} \oint_{F=0}
		\frac{dx \wedge dy \wedge dz \wedge dt}{F}
		= \frac{dx \wedge dy \wedge dz}
		{2\sqrt{\xi^2 - xyz}}\,.
\end{equation}

Now, in order to compute the periods of $\Omega$ along a specific $A_i$ (or equivalently $B_i$) cycle, we define (with a little abuse of notation) $x=z_i^2$, $y=u+iv$, $z=u-iv$ and $\varepsilon^2 = \xi^2 / x$, and compute:
\begin{equation}
	\int \Omega = -i \int \frac{dx \wedge du \wedge dv}
		{\sqrt{\xi^2 - x (u^2+v^2) }}
		= -i \int \frac{dx}{\sqrt{x}} \int_{\mathcal{C}_i}
		\frac{du \wedge dv}{\sqrt{\varepsilon^2 - u^2 - v^2}}\,.
\end{equation}
We already met the last integral, and computed it in insert~\ref{i:K3} on page~\pageref{i:K3}. We therefore have:
\begin{equation}
	\int \Omega = -i \int \frac{dx}{\sqrt{x}} 4\pi \frac{\xi}{\sqrt{x}}
		= - 4 \pi i \xi \int{\frac{dx}{x}}
		= - 8 \pi i \xi \int{\frac{dz_i}{z_i}}\,.
\end{equation}
The periods of $\Omega$ along the cycles $A_i$ and $B_i$ are then finally given by:
\begin{equation}
	\int_{A_i} \Omega = - 8 \pi i \xi \oint{\frac{dz_i}{z_i}}
		= 16 \pi^2 \xi\,,
\end{equation}
\begin{equation}
	\int_{B_i} \Omega
		= - 8 \pi i \xi \int_{\xi^{1/3}}^{\rho_c} {\frac{d\rho_i}{\rho_i}}
		= \frac{8 \pi i}{3}\ \xi\ \ln \frac{\xi}{\rho_c^3} \,,
\end{equation}
where in the latter we use the same cutoff $\rho_c$ that we used in the computation of the fluxes of $G_3$, while now the lower limit of integration is given by an opportune power of the deformation parameter $\xi$.

We have now all the necessary ingredients to compute the effective superpotential of the gauge theory by using Vafa's formula~\eqref{Vafa}. Let us first summarize the fluxes we have found:
\begin{equation}\label{fluxsummary}
\begin{aligned}
	\frac{1}{iK} \int_{A_i} G_3 &= 2\pi i\ ( N - N_f )\,, & 
	\frac{1}{16\pi^2} \int_{A_i} \Omega &= \xi\,,\\
	\frac{1}{iK} \int_{B_i} G_3 &= ( N - N_f )\ \ln \frac{\rho_c}{\rho_0}
		+ \delta_{i,1}\ 2 N_f\ \ln \frac{\Delta}{\rho_0}\,, &
	\frac{1}{16\pi^2} \int_{B_i} \Omega
		&= -\frac{1}{2\pi i}\ \frac{\xi}{3}\ \ln \frac{\xi}{\rho_c^3}\,.
\end{aligned}
\end{equation}
The superpotential is given by:
\begin{equation}
\begin{split}
	W &= \frac{1}{16\pi^2 iK}\ \frac{1}{(2\pi\ls^2)^3}\ \sum_{i=1}^3\
		\left[\ \int_{A_i} G_3 \int_{B_i} \Omega
		- \int_{A_i} \Omega \int_{B_i} G_3\ \right]\\
	&= -\frac{1}{(2\pi\ls^2)^3}\ 
		\left[\ 3 ( N - N_f )\ \frac{\xi}{3}\ \ln \frac{\xi}{\rho_c^3}
		+ 3( N - N_f )\ \xi\ \ln \frac{\rho_c}{\rho_0}
		+ 2 N_f\ \xi\ \ln \frac{\Delta}{\rho_0}
		\ \right]\,.
\end{split}
\end{equation}

We now re-express the geometrical quantities in terms of gauge theory quantities, by using again the ``stretched string'' gauge/gravity relation. Notice that the deformation parameter $\xi$, due to its engineering dimensions, is naturally identified by the relation with a dimension 3 operator in the gauge theory, namely the gaugino condensate $S$, in perfect analogy with the case of the conifold studied in section~\ref{s:conifold}. In summary we have: 
\begin{equation}
	\rho_c = 2\pi\ls^2\ \mu\,,\qquad
	\rho_0 = 2\pi\ls^2\ \Lambda\,,\qquad
	\Delta = 2\pi\ls^2\ m\,,\qquad
	\xi = (2\pi\ls^2)^3\ S\,,
\end{equation}
and we can write the gauge theory effective superpotential as:
\begin{equation}
	W_{\text{eff}} = - ( N - N_f )\ S\ \ln \frac{S}{\Lambda^3}
		- 2 N_f\ S\ \ln \frac{m}{\Lambda}\,.
\end{equation}
Though this result is correct, let us redefine the scales in order to write it in a more conventional way. The appropriate redefinition is $\Lambda\to e^{1/3}\Lambda$, $m\to e^{1/3}m$, and we get:
\begin{equation}\label{VYT}
	W_{\text{eff}} = ( N - N_f )\ \left[\ S - S\ \ln \frac{S}{\Lambda^3}\ \right]
		- 2 N_f\ S\ \ln \frac{m}{\Lambda}\,,
\end{equation}
which is precisely the Taylor--Veneziano--Yankielowicz superpotential~\cite{Taylor:1983bp}. At the minimum we have:
\begin{equation}
	S = \left(\frac{\Lambda^{3N-N_f}}{m^{2N_f}}\right)^{\frac{1}{N-N_f}}\,,
\end{equation}
and we get the Affleck--Dine--Seiberg superpotential~\cite{Affleck:1983mk}:
\begin{equation}\label{ADS}
	W_{\text{eff}}  = ( N - N_f )\ \left[\
		\frac{\Lambda^{3N-N_f}}{m^{2N_f}}\ \right]
		^{\frac{1}{N-N_f}}
		= ( N - N_f )\ \left[\
		\frac{\Lambda^{3N-N_f}}{\det M}\ \right]
		^{\frac{1}{N-N_f}}\,.
\end{equation}

As an aside, notice that if $N_f = 0$ the above results reproduce the Veneziano--Yankielowicz superpotential for pure \Ne{1} Super Yang--Mills theory~\cite{Veneziano:1982ah}, that we already obtained in section~\ref{s:conifold} from the conifold solution:
\begin{equation}
	W_{\text{VY}} = N \ \left[\ S - S\ \ln \frac{S}{\Lambda^3}\ \right]\,.
\end{equation}
Its value at the minimum (where $S=\Lambda^3$) is $W_{\text{VY}} = N \Lambda^3$. Let us stress that this is not just a formal limit of the result obtained for $N_f > 0$, but rather one could start from the beginning by considering only $N$ fractional branes of type I, and Vafa's formula would then yield, through the same computation we did above, the Veneziano--Yankielowicz superpotential.

Another observation concerns the case in which $N_f = N$. The above results are still valid, since the brane interpretation is clear - all branes have formed A$+$B superpositions and there are no fractional branes (and thus no effective gauge theory) left. The result for the moduli space can be read from the superpotential~\eqref{VYT} for $N_f = N$. The minimization procedure implies $\det M = \Lambda^{2N}$, which is the correct result expected from gauge theory.

We therefore see that our classical solution, together with some geometrical considerations, has been able to give us a big deal of information on the \Ne{1} Affleck--Dine--Seiberg, at the classical, perturbative and even non-perturbative level. It would be very interesting to use this system of fractional branes to analyze SQCD also in the phase where $N_f > N$, where Seiberg duality is supposed to take place~\cite{Seiberg:1994pq}. Indeed a construction of Seiberg duality for quiver theories, which is deeply linked to the situation we have studied, was presented in~\cite{Berenstein:2002fi} with the implementation of quite formal methods, and it would be interesting to look for a precise dual supergravity description.

\vspace{1.5cm}

\section*{Acknowledgments}

There is not enough space here to properly thank all people that have been crucial in my Ph.D. experience. Let me first express my deep gratitude to my supervisor Alberto Lerda, to Paolo Di Vecchia and to my external referee Costas Bachas. Among other people I wish to thank, let me at least mention Matteo Bertolini, Marco Bill\`o, Eleonora Dell'Aquila, Marialuisa Frau, Troels Harmark, Ernesto Lozano-Tellechea, Paolo Merlatti, Niels Obers, Igor Pesando, Rodolfo Russo, Stefano Sciuto and Giuseppe Vallone. My work during the last three years has been partially supported by the European Commission Marie Curie Training Site Fellowship HPMT-CT-2000-00010, by the European Commission RTN programme HPRN-CT-2000-00131, and by MIUR under contract 2001-1025492.

\cleardoublepage
\addcontentsline{toc}{chapter}{Bibliography}
\bibliographystyle{../bibtex/myutcaps}
\bibliography{../bibtex/mybib}

\end{document}